\documentclass[3p,sort&compress]{elsarticle}

\usepackage{booktabs,array}
\usepackage[table]{xcolor}

\usepackage[T1]{fontenc}
\usepackage[utf8]{inputenc}
\usepackage{lmodern}

\usepackage{xcolor}
\usepackage{amsmath}
\usepackage{amssymb}
\usepackage{amsfonts}
\usepackage{bbold}
\usepackage{microtype}
\usepackage[linktoc=page]{hyperref}
\usepackage{cleveref}
\usepackage{url}
\usepackage{epigraph}
\usepackage{booktabs}
\usepackage{multirow}
\usepackage{longtable}
\usepackage{threeparttable}

\newcommand{\Mp}{M_\mathrm{Pl}}
\newcommand{\eff}{\mathrm{eff}}
\newcommand{\dd}{\mathrm{d}}
\newcommand{\sdg}{\sqrt{-g}}

\newcommand{\mn}{{\mu\nu}}
\newcommand{\ab}{{\alpha\beta}}

\newcommand{\Ij}{{ij}}
\newcommand{\rs}{r_\mathrm{s}}
\newcommand{\rst}{{r_\star}}

\newcommand{\mL}{\mathcal{L}}

\newcommand{\ellm}{{\ell m}}
\newcommand{\lambdaEell}{{c^{(\mathrm{E})}_\ell}}
\newcommand{\lambdaBell}{{c^{(\mathrm{B})}_\ell}}
\newcommand{\lambdaEellbar}{{\bar{c}^{(\mathrm{E})}_\ell}}
\newcommand{\lambdaBellbar}{{\bar{c}^{(\mathrm{B})}_\ell}}

\newcommand{\Rstar}{R_*}
\newcommand{\X}{{X}}
\newcommand{\XX}{\mathcal{X}}
\newcommand{\ccs}{c}
\newcommand{\lambdar}{\lambda_{\mathrm{r}}}
\newcommand{\RR}{{}^{(n)} \hspace{-0.3mm} \mathcal{R}}
\DeclareMathOperator{\diag}{diag}
\DeclareMathOperator{\tho}{\text{\TH}}
\DeclareMathOperator{\affine}{\sigma}
\def\Mm{{\cal M}}
\newcommand{\Hzerf}{H_0}
\newcommand{\Honef}{H_1}
\newcommand{\Htwof}{H_2}
\newcommand{\Hf}{H}

\newcommand{\Kf}{K}
\newcommand{\Rf}{R}
\newcommand{\Vf}{V}
\newcommand{\deltaf}{\delta}

\newcommand{\teuk}{\ensuremath{\hspace{.05cm}\mathaccent\Box{\text{\tiny \textsc{T}}}}\,}

\newcommand\apj{\ref@jnl{ApJ}}%
\newcommand\icarus{\ref@jnl{Icarus}}%
\newcommand\psj{\ref@jnl{PSJ}}%

\newcommand{\Ef}{\mathbf{E}}

\def\be{\begin{equation}}
\def\ee{\end{equation}}

\biboptions{sort&compress}

\makeatletter
\DeclareRobustCommand{\rcite}[1]{%
  \rcite@aux#1,\@nil{#1}%
}
\def\rcite@aux#1,#2\@nil#3{%
  \if\relax#2\relax
    Ref.~\cite{#3}%
  \else
    Refs.~\cite{#3}%
  \fi
}
\makeatother

\numberwithin{equation}{section}

\usepackage{tikz}
\usepackage[compat=1.1.0]{tikz-feynman}

\newcommand{\gravconv}{
\begin{tikzpicture}[baseline]
\begin{feynman}
\vertex (a);
\vertex [right=1cm of a] (b);
\vertex [right=0.2cm of d, label=90:$\scriptstyle{\rho\sigma}$] (e);
\vertex [left=0.2cm of a, label=90:$\scriptstyle{\mu\nu}$] (f);
\diagram* {
(a) -- [photon, momentum={[arrow shorten=0.3pt]$\scriptstyle{k}$}
] (b)
}; 
\end{feynman}
\end{tikzpicture}
}

\newcommand{\TidalHconv}{
\begin{tikzpicture}[baseline]
\begin{feynman}
\vertex [crossed dot, minimum size=0.25cm, line width=1.5pt](a) {};
\vertex [right=1cm of a] (b);
\diagram* {
(a) -- [plain, line width=1.5pt, momentum={[arrow shorten=0.25pt]$\scriptstyle{k}$}] (b)
}; 
\end{feynman}
\end{tikzpicture}
}

\newcommand{\MassScZ}{
\begin{tikzpicture}[baseline]
\begin{feynman}
\vertex [dot, minimum size=0.25cm, label=180:$m$] (a) {};
\vertex [right=0.8cm of a] (c);
\vertex [right=0.2cm of c] (d);
\vertex [right=0.2cm of d, label=90:$\scriptstyle{\mu\nu}$] (f);
\diagram* {
(a) -- [photon, momentum={[arrow shorten=0.3pt]$\scriptstyle{k}$}
] (d),
}; 
\end{feynman}
\end{tikzpicture}
}

\newcommand{\metricsecondorder}{
\begin{tikzpicture}[baseline=1cm]
\begin{feynman}
\vertex  (a) {};
\vertex [above=0.8cm of a,  dot, minimum size=0.25cm, label=180:$m$] (aU) {};
\vertex [below=0.8cm of a,  dot, minimum size=0.25cm, label=180:$m$] (aD) {};
\vertex [above=0.5cm of a] (c);
\vertex [right=1.2cm of a] (c1);
\vertex [right=1cm of c1] (d);
\diagram* {
(aD) -- [photon] (c1),
(aU) -- [photon] (c1),
(c1) -- [photon] (d)
}; 
\end{feynman}
\end{tikzpicture}
}

\newcommand{\metricthirdorderA}{
\begin{tikzpicture}[baseline=1cm]
\begin{feynman}
\vertex [dot, minimum size=0.25cm, label=180:$m$] (a) {};
\vertex [above=1cm of a,  dot, minimum size=0.25cm, label=180:$m$] (aU) {};
\vertex [below=1cm of a,  dot, minimum size=0.25cm, label=180:$m$] (aD) {};
\vertex [right=1.2cm of a] (c1);
\vertex [right=1cm of c1] (d);
\vertex [right=0.6cm of a] (x1);
\vertex [above=0.5cm of x1] (y1);
\diagram* {
(a) -- [photon] (y1),
(aU) -- [photon] (y1),
(y1) -- [photon] (c1),
(aD) -- [photon] (c1),
(c1) -- [photon] (d)
}; 
\end{feynman}
\end{tikzpicture}
}

\newcommand{\metricthirdorderB}{
\begin{tikzpicture}[baseline=1cm]
\begin{feynman}
\vertex [dot, minimum size=0.25cm, label=180:$m$] (a) {};
\vertex [above=1cm of a,  dot, minimum size=0.25cm, label=180:$m$] (aU) {};
\vertex [below=1cm of a,  dot, minimum size=0.25cm, label=180:$m$] (aD) {};
\vertex [right=1.2cm of a] (c1);
\vertex [right=1cm of c1] (d);
\diagram* {
(a) -- [photon] (c1),
(aU) -- [photon] (c1),
(aD) -- [photon] (c1),
(c1) -- [photon] (d)
}; 
\end{feynman}
\end{tikzpicture}
}

\newcommand{\scatteringMbb}{
\begin{tikzpicture}[baseline=1cm]
\begin{feynman}
\vertex [dot, minimum size=0.25cm] (a) {};
\vertex [below=1cm of a,  dot, minimum size=0.25cm] (aD) {};
\vertex [right=0.6cm of a] (x1);
\vertex [above=0.5cm of x1] (y1);
\vertex [right=0.6cm of aD] (z1);
\vertex [below=0.5cm of z1] (w1);
\diagram* {
(a) -- [photon] (y1),
(aD) -- [photon] (w1),
(a) -- [draw=black, double, double distance=2pt, line width=0.5pt, dash pattern=on 3pt off 2pt] (aD)
}; 
\end{feynman}
\end{tikzpicture}
}

\newcommand{\MassScW}{
\begin{tikzpicture}[baseline]
\begin{feynman}
\vertex [dot, minimum size=0.25cm, label=180:$m$] (a) {};
\vertex [right=0.8cm of a] (c);
\vertex [right=0.8cm of c] (d);
\diagram* {
(a) -- [photon] (d),
}; 
\end{feynman}
\end{tikzpicture}
}

\newcommand{\LinearLoveGR}{
\begin{tikzpicture}[baseline]
\begin{feynman}
\vertex [square dot, minimum size=0.25cm, label=180:$C_{\text{E},\ell}$] (a) {};
\vertex [below=1cm of a, crossed dot, minimum size=0.25cm, line width=1.5pt] (aT) {};
\vertex [right=1cm of a] (d);
\vertex [right=0.2cm of d, label=90:$\scriptstyle{\mu\nu}$] (d2);
\diagram* {
(a) -- [photon, momentum={[arrow shorten=0.3pt]$\scriptstyle{k}$}] (d),
(aT) -- [plain, line width=1.5pt] (a)
}; 
\end{feynman}
\end{tikzpicture}
}

\begin{document}

\begin{frontmatter}

\title{\textbf{\Large Love numbers of black holes and compact objects}}

\author[USU,IFT]{María J. Rodríguez}
\ead{majo.rodriguez.b@gmail.com}

\author[Paris]{Luca Santoni}
\ead{santoni@apc.in2p3.fr}

\author[UVA]{Adam R. Solomon\corref{cor1}}
\ead{adam.solomon@gmail.com}
\cortext[cor1]{Corresponding author}

\address[USU]{Department of Physics, Utah State University, 4415 Old Main Hill Road, Logan, UT 84322, USA}

\address[IFT]{Instituto de F\'isica Te\'orica UAM-CSIC, Universidad Aut\'onoma de Madrid, Cantoblanco, 28049 Madrid, Spain}

\address[Paris]{Universit\'e Paris Cit\'e, CNRS, Astroparticule et Cosmologie, 10 Rue Alice Domon et L\'eonie Duquet, F-75013 Paris, France}

\address[UVA]{Department of Physics, University of Virginia, 382 McCormick Road, Charlottesville, VA 22904, USA}

 \vspace{.3cm}

\begin{abstract}
The Love numbers of a gravitating body are  response coefficients encoding its tidal deformability. In compact binary systems, they appear in the  gravitational waveform during the inspiral phase and will be measurable by  upcoming gravitational-wave observatories. 
This review provides a comprehensive and pedagogical account of the theoretical foundations of Love numbers and surveys the most recent advances in the study of tidal effects in compact objects, with particular emphasis on black holes and neutron stars.

We begin with a gentle introduction to tidal effects in Newtonian gravity, leading into a discussion of how to robustly define tidal responses in general relativity using the effective field theory and post-Newtonian frameworks.
After an overview of the perturbation theory of black holes and neutron stars, we review the computation of Love numbers and dissipative response coefficients in a wide range of settings, including the static, dynamical, and nonlinear tidal responses of Kerr black holes and neutron stars in four-dimensional general relativity.
We further discuss the extension of these results to charged black holes and a number of ``new physics'' scenarios, including higher-dimensional black holes and other black objects, (anti-) de Sitter black holes, supergravity black holes, and theories beyond general relativity.
Finally we provide an overview of the tantalizing zoo of hidden symmetries of general relativity that have been uncovered in the attempt to explain the famous vanishing of static black hole Love numbers.

\end{abstract}

\end{frontmatter}

\tableofcontents

\newpage

\part*{Introduction}
\addcontentsline{toc}{part}{Introduction}
\label{sec:intro}

\epigraphhead[]{
\epigraph{What is Love?
    
    Baby don't hurt me,
    
    No more.}{\textsc{Haddaway}}
}

\noindent
Astronomy was, from ancient times until 2016, practiced almost exclusively through 
the observation of electromagnetic radiation.
The only other fundamental force known to produce classical 
radiation at macroscopic distances is gravity, which generates gravitational waves 
whenever massive objects accelerate~\cite{LIGOScientific:2016aoc}. Due to the extreme weakness of 
gravity --- roughly $10^{-28}$ times weaker than electromagnetism --- gravitational 
radiation took humanity
rather
longer to detect than its electromagnetic counterpart.
Gravitational wave astronomy has since matured into a full-fledged observational 
science.
The global 
detector network LIGO--Virgo--KAGRA (LVK) has accumulated a steady stream of 
events; as of the fourth observing run (O4), the catalog stands at around 300 binary black hole mergers~\cite{LIGOScientific:2025hdt, 
LVK2025_GWTC4_Update, LIGOScientific:2025snk, LVK2025_GWTC4_Cosmology}.

The gravitational-wave sources we observe at present are binary systems of compact 
objects. By \textit{compact} we mean a body whose mass $M$ is confined within 
a radius comparable to its Schwarzschild radius,
\begin{equation}
\rs \equiv \frac{2GM}{c^2} \, ,
\nonumber
\end{equation}
where $G$ is the Newtonian gravitational constant,
so that its gravitational field is within the strong gravity r\'egime where the effects of general relativity are important.
In practice, the compact objects we observe in gravitational-wave events are black holes (BHs) and neutron stars (NSs), with characteristic radii $R_{\rm BH} = \rs$ and $R_{\rm NS} \sim 3\text{--}5\, \rs$, respectively.\footnote{White dwarfs are far less compact, $R_{\rm WD} \sim 10^3\, \rs$, and  do not contribute measurably to the tidal effects observable by current  ground-based detectors~\cite{AmaroSeoaneEtAl2017, ShapiroTeukolsky1983}.}

Compact object binaries evolve primarily through the emission of gravitational waves, which 
carry energy and angular momentum away from the orbiting bodies, causing them to 
gradually spiral inward.
The resulting signal 
comprises three characteristic phases: the adiabatic early inspiral, the highly dynamical and nonlinear
merger, and the subsequent ringdown.

The gravitational waveform during the inspiral phase is particularly sensitive to the \emph{internal structure} of the participating objects. When widely separated relative to their sizes, compact objects behave effectively as if they were point particles
\cite{Damour:1984rbx,Damour:1995kt,Goldberger:2004jt,Goldberger:2005cd,Porto:2005ac,Porto:2007qi}.
The leading-order observable finite-size effect arises from adiabatic tidal distortions: each body is tidally deformed by the gravitational field of its companion, which in turn modifies the orbital dynamics and consequently the emitted waveform.
These tidal signatures, though small in 
magnitude, carry direct information about the internal structure of the inspiraling bodies, providing a unique probe of physics in extreme environments, such as the high-density nuclear physics expected to be at play in neutron stars
\cite{Flanagan:2007ix,Hinderer:2007mb}. 

Surprising results have recently emerged for black holes, revealing that --- when probed from large distances --- they are remarkably rigid and effectively non-deformable under tidal interactions~\cite{Fang:2005qq,Damour:2009vw,Binnington:2009bb,Kol:2011vg,Chakrabarti:2013lua,Poisson:2014gka,Gurlebeck:2015xpa,Landry:2015zfa,Pani:2015hfa,Hui:2020xxx,LeTiec:2020spy,LeTiec:2020bos,Chia:2020yla,Goldberger:2020fot,Charalambous:2021mea}.\footnote{When subjected to an external tidal perturbation, a black hole's metric \textit{does} deform (see \rcite{10.1063/1.1724317,Hartle:1973zz,Hartle:1974gy,Hawking:1972hy,DEath:1975jps,DEath:1996flw} for early investigations of tidal distortions of black holes, and \rcite{Poisson:2005pi,Taylor:2008xy,Damour:2009va} for more recent work). For instance, the shape of the horizon deviates from sphericity in a tidal environment. We stress that what vanish are the induced tidal moments of the object's gravitational field, as measured by a distant observer.}
This \textit{a priori} unexpected fact turned out to be the harbinger of previously-unrecognized \emph{hidden symmetries} of general relativity itself more than a century after its initial formulation \cite{Porto:2016zng,Charalambous:2021kcz,Hui:2021vcv}, situating the physics of tidal deformations directly at the interface of cutting-edge gravitational theory and state-of-the-art observational astrophysics.

The study of tidal effects in compact object binaries has become an active area of research which is expected to undergo a direct confrontation with measurements performed by future high-precision gravitational-wave 
observatories. 
This review aims to define, in a pedagogical way, the essential concepts spanning  from effective field theory to gravitational-wave astronomy, survey the key  literature, and point toward future observational opportunities.\footnote{Although our main focus is on compact sources, the study of tidal effects and responses to external forces has broad relevance across astrophysics. Tidal deformability provides direct constraints on the internal structure and density distribution of gravitating bodies, making it a powerful tool not only in gravitational-wave astronomy but also in planetary and exoplanetary science. Observations of tidal effects on planets such as Jupiter~\cite{2017SSRv..213....5B,2017GeoRL..44.4694F,lai2021jupitersdynamicallovenumber,2020ApJ...891...42W,Idini_2022} and Saturn~\cite{2007Sci...317.1384A} inform models of their internal dynamic flows~\cite{2024SSRv..220...22F}, reduce uncertainties in the hydrogen-helium equation of state and phase separation (see, e.g., \rcite{2016ApJ...826..127K,2016A&A...596A.114M,2016ApJ...820...80H}), aid the interpretation of atmospheric properties, improve our understanding of giant-planet formation histories~\cite{2018A&A...613A..32N}, and can offer insights into the structure of exoplanets~\cite{Chernov_2017,2025MNRAS.540.1544V}.
Tidal deformations on Earth are slow and small, forming over hours as the oceans and crust respond to  the Moon and Sun. Near a black hole, by contrast, tidal forces can disrupt a star  in seconds to minutes, illustrating the vast range of tidal phenomena in the  universe.}

\vspace{0.4cm}

\noindent
\textbf{Outline:}
This review is organized into three main parts:

\medskip
\noindent\textbf{Part I}
(\Cref{sec:newton,sec:EFT})
introduces 
the essential concepts underlying gravitational tidal effects. 
\Cref{sec:newton} presents a Newtonian treatment, covering the exterior 
problem, linear response theory, tides of polytropic fluids, and spinning bodies. 
\Cref{sec:EFT} develops the corresponding framework in general relativity (GR), 
establishing the worldline effective field theory (EFT), dissipative and nonlinear tidal effects, and the 
post-Newtonian description of tidal interactions.

\medskip
\noindent\textbf{Part II}
(\Cref{sec:BHsolutions,sec:BHPT,sec:NS})
develops the relativistic framework for perturbations of compact objects. 
\Cref{sec:BHsolutions} reviews the Schwarzschild and Kerr solutions --- their 
geometry, symmetries, geodesics, and wave equations --- and introduces the Petrov 
classification and Geroch--Held--Penrose formalism. \Cref{sec:BHPT} 
covers black hole perturbation theory, including the even and odd sector 
decomposition for Schwarzschild, static modes, Chandrasekhar duality, and the 
Teukolsky formalism for Kerr. \Cref{sec:NS} turns to neutron stars, presenting the 
Tolman--Oppenheimer--Volkoff equations and stellar perturbation theory.

\medskip
\noindent\textbf{Part III}
(\Cref{sec:love-compute,sec:sym})
combines the machinery set up in the first two Parts to compute Love numbers and study their underlying symmetries.
\Cref{sec:love-compute} introduces methods for computing tidal 
response coefficients, covering linear and higher-order analysis and extensions 
beyond four-dimensional  GR, including higher-dimensional, rotating, charged, and 
AdS black holes, black branes, and modified gravity. \Cref{sec:sym} 
explores the symmetry principles that organize tidal responses, from ladder 
symmetries and near-zone expansions to nonlinear and dynamical perturbations. 

\medskip
\noindent The appendices collect the technical foundations used throughout, 
including hypergeometric and Legendre functions, spherical harmonics, and differential forms.

\medskip
\noindent
{\bf Note added:}
During the final stages of preparation, we became aware of another comprehensive review on the tidal responses of compact objects~\cite{CPreview}. While there is some overlap, both reviews have distinct emphases and should be regarded as complementary.

\vspace{0.4cm}

\noindent
\textbf{Notation and conventions:}
Throughout this review we adopt the following conventions. The spacetime metric in $D$ spacetime dimensions has signature  $(-,+,+,\cdots)$. We work in natural units $\hbar = c = 1$ unless stated otherwise, and denote the reduced Planck mass by $\Mp=  (8 \pi G)^{-\frac1{D-2}}$,
with $G$ Newton's gravitational constant. In instances where $G$ is set to unity for convenience, this will be explicitly indicated.  We use the curvature convention ${R^\rho}_{\sigma\mu\nu}=\partial_\mu\Gamma^\rho_{\nu\sigma}+\dots$ and $R_{\mu\nu}={R^\rho}_{\mu\rho\nu}$. The Levi--Civita tensor is denoted $\epsilon_{\mu_1\cdots\mu_D}$ with $\epsilon_{0\cdots D-1}=\sdg$.

We use Greek indices, $\mu, \nu, \rho, \dots$, for spacetime coordinates; Latin indices from the middle of the alphabet, $i, j, k, \dots$, for spatial coordinates;  Latin indices from the beginning of the alphabet, $a, b, c, \dots$, for local Lorentz (tangent-space) components; capital Latin indices $A, B, C, \dots$ for the angular coordinates on the two-dimensional sphere $S^2$; and capital mid-alphabet Latin indices $I,J,K,\dots$ for coordinates in the $(r,\theta)$ subspace. The Hodge star in $D$ dimensions is denoted $\star_D$, except for $\star_4$ which we write as $\star$.

We will often use multi-index notation, such as $A_\ell = (i_1 \cdots i_\ell)$, or simply $L = (i_1 \cdots i_\ell)$; for instance $\partial_L\equiv \partial_{i_1}\cdots \partial_{i_\ell}$. 
The notation $\langle{\cdots}\rangle$ indicates trace-free symmetrization over the enclosed indices. 
We (anti)symmetrize indices with weight one, so that, for example, $A_{[ij]} = \frac{1}{2} \left(A_{ij} - A_{ji}\right)$.

Our convention for the decomposition in spherical harmonics and the Fourier transform is 
\begin{equation}
    \psi(t,r, \theta,\varphi)= \sum_{\ell,m}\int\frac{\dd\omega}{2\pi} e^{-i\omega t}  \psi_{\ell m}(\omega,r)Y_{\ell m}( \theta,\varphi).
\nonumber
\end{equation}
For simplicity, we will sometimes omit the arguments and the $\ell m$ subscripts on $\psi_{\ell m}$, relying on the context to distinguish between the different meanings.

The main symbols used consistently throughout are:

\begin{center}
\renewcommand{\arraystretch}{1.45}
\begin{tabular}{p{3.2cm} p{10cm}}
\hline\hline
\textbf{Symbol} & \textbf{Meaning} \\
\hline
$\Rstar$ & Radius of the compact object \\
$M,\, M'$ & Mass of the primary and companion object \\
$\rs$ & The Schwarzschild radius $\rs=2GM$ \\
$D$ & Total number of spacetime dimensions \\
$\ell$, $m$ & Multipole order ($\ell = 2$ is quadrupolar) and azimuthal quantum number \\
$k_\ell$, $\lambda_\ell$ & Tidal Love numbers: the ratio of an induced multipole moment to the applied tidal field, defined through the falloff of the Newtonian potential or metric perturbation (see, e.g., \cref{eq:Iell,eq:intro-g-ansatz}); we use $k_\ell$ to denote the dimensionless ratio, while $\lambda_\ell$ is dimensionful and related to $k_\ell$ by a power of $\Rstar$ (see, e.g., \cref{eq:kelllambdaell}) \\
$c_\ell^{(\rm E)}$, $c_\ell^{(\rm B)}$ 
           & EFT tidal coefficients (Wilson couplings) for electric- and magnetic-parity perturbations, respectively, in the point-particle effective description of compact objects~\cite{Goldberger:2004jt,Goldberger:2005cd} (see e.g.,~\cref{eq:S-ho}) \\
$\Lambda$, $H$ &  For rotating objects, we denote the conservative and dissipative coefficients in the expansion of the response tensor~\cite{Goldberger:2020fot} by $\Lambda$ and $H$, respectively (see, e.g., \rcite{Saketh:2022xjb,Saketh:2023bul} and \cref{tab1:threshold} for a summary) \\
$\lambda_\mathrm{r}$ & Radiation wavelength of the gravitational wave 
              (distinct from the EFT couplings above) \\
$Y_{\ell m}(\theta,\varphi)$ & Scalar spherical harmonics on the two-sphere $S^2$ \\
$\varphi$ & Azimuthal angle (spacetime coordinate) \\
$\phi$ & Scalar field (unless stated otherwise) \\
$\affine$ & Affine parameter along a geodesic \\
$X^\mu(\affine)$ & Worldline of the compact object \\
$\vec{p}$ & Spatial three-vector (bold arrow notation) \\
$\mathcal{E}_{ij}$ & External quadrupole tidal tensor, 
 $\mathcal{E}_{ij} = -\partial_{\langle i}\partial_{j \rangle} U_{\rm tidal}$, with $U_{\rm tidal}$ denoting an external tidal potential  \\
$I_{ij}$ & Induced quadrupole moment tensor \\
$f(r)$ & The ``emblackening factor'' $f(r)=1-\rs/r$ appearing in the Schwarzschild metric \\
$\Delta$ & The function $\Delta=r^2f(r)=r(r-\rs)$ (Schwarzschild) and $\Delta=r(r-\rs)+a^2$ (Kerr)\\
\hline\hline
\end{tabular}
\end{center}

\noindent 
A note on potential conflicts: the symbol $\lambda$ appears both as the coefficient of the decaying branch of the tidal solution (see, e.g., \cref{eq:Uellm-2}), $\lambda_\ell$, and as the radiation wavelength, $\lambda_{\mathrm{r}}$; these quantities are always written with different subscripts to avoid ambiguity. Moreover, $\lambda$ also denotes a field in \cref{eq:Levenstatic} and below. Similarly, the symbol $\Lambda$, in addition to denoting response coefficients in \cref{sec:dissipation}, also appears in the definition of a constant in \cref{eq:LambdaellRW} and as a separation constant in the Teukolsky equation \eqref{eq:teuk-rad-boxed}. 

Note that we use three different symbols to refer to tidal response coefficients: $c_\ell$ are the (dimensionful) couplings in the worldline EFT (see, e.g., \cref{eq:S-ho}); $\lambda_\ell$ is the (dimensionful) coefficient measuring the relative amplitude of the decaying $r^{-\ell-1}$ branch with respect to the growing $r^\ell$ branch in the large-distance expansion of the field solution (see, e.g., \cref{eq:Uellm-2}); and $k_\ell$ are the dimensionless response coefficients appearing in the metric solution
(see \cref{eq:intro-g-ansatz}).
In this work, the symbol $U$ denotes different quantities: it represents the Newtonian gravitational potential in \cref{sec:newton}, and the fluid-element velocity in \cref{eq:deltau-NS}.  Each of these differences should be immediately clear from the context and should not cause confusion.
  \\

\vspace{0.4cm}

\noindent
\textbf{Terminology: }
We clarify the following terms which are used throughout this review, since 
conventions differ across the literature.

\medskip
\noindent\textit{Love numbers vs.~tidal deformability.}
Tidal deformability generically refers to the ability of a body to deform under the action of an external tidal field. The terminology \textit{Love number} is named after A.~E.~H.~Love, who originally developed  and applied the concept of tidally-induced multipole moments in the context of geodynamics~\cite{1909MNRAS..69..476L,1911spge.book.....L}. The original two numbers introduced by Love were subsequently extended by T.~Shida  \cite{Shida}. However, a relativistic formulation of Love numbers, following their Newtonian counterpart, was developed only several decades later by T.~Damour~\cite{DamourProblemeNcorps,Damour:1982wm},\footnote{We cannot resist mentioning a curious coincidence that, incidentally, relates (almost perfectly) the last names of the authors of Refs.~\cite{1909MNRAS..69..476L} and \cite{Damour:1982wm} --- pioneers of tidal effects in Newtonian and relativistic contexts, respectively ---  in their respective languages.} which subsequently led to the calculation of the relativistic tidal response --- both polar and axial --- of black holes and neutron stars~\cite{Hinderer:2007mb,Flanagan:2007ix,Damour:2009vw,Binnington:2009bb}.
In the literature, Love numbers  are often denoted by $k_\ell$ and refer to the dimensionless ratio that characterizes the tidal response at a fixed multipole order $\ell$, normalized as, for example, in \rcite{Poisson:2020vap}. In the language of modern EFT, the Love numbers correspond to EFT Wilson couplings (denoted below by $c_\ell$). These are related to $k_\ell$ --- the latter obtained from a full GR calculation --- through powers of the compactness and numerical factors, as determined via a matching procedure (see \cref{sec:EFT}). The terms \textit{Love numbers} and \textit{tidal response coefficients}, referring  to either $k_\ell$ or $c_\ell$, are often used interchangeably throughout the literature, both in the Newtonian and relativistic contexts.

\medskip
\noindent\textit{Conservative vs.\ dissipative response.}
The tidally induced response of a gravitating body is generally a complex number and can be separated into \textit{conservative} (or \textit{elastic}) and \textit{dissipative} components, corresponding respectively to the real and imaginary pieces of the response coefficient. 
The \textit{conservative} response refers to energy-conserving (i.e., time-reversal invariant) deformations of the object, while the \textit{dissipative} response 
captures effects associated with energy absorption, such as through the horizon in black holes or through viscosity and mode excitation in neutron stars. 
Although the term Love number is sometimes used in the literature to denote both the conservative and dissipative components of an object's induced response (see, e.g., \rcite{Poisson_Will_2014,2014ARA&A..52..171O} in the Newtonian context and \rcite{Chakrabarti:2013lua,Pani:2015hfa,Pani:2015nua,LeTiec:2020spy,LeTiec:2020bos} in GR), it has become increasingly common in recent years to reserve the term Love numbers for the conservative part of $k_\ell$ (or $\lambda_\ell$) only, and to refer to the dissipative contribution as \textit{tidal heating} (or simply \textit{dissipative coefficients}); see, e.g., \rcite{Chia:2020yla,Goldberger:2020fot,Charalambous:2021mea}.\footnote{The terminology \textit{tidal heating} originates from the Newtonian concept of tidal dissipation, and was first applied to black holes by Hartle~\cite{Hartle:1973zz} (see also \rcite{Thorne:1984mz}) in the context of slowly rotating holes. It was subsequently systematized and established as standard terminology in the black hole context through a series of works in the in the 2000s--2010s~\cite{Hughes:2001jr,Poisson:2009qj,Chatziioannou:2012gq}.} This is the convention that we adopt throughout this work. 
Different conventions in the literature can sometimes be a source of confusion, and we therefore warn the reader about this point. For instance, the induced response of a Kerr black hole in general relativity was computed in \rcite{Poisson:2014gka,Landry:2015zfa,Pani:2015hfa,LeTiec:2020spy,LeTiec:2020bos}. Although these references appear to use the term Love numbers generically for the induced response, the non-vanishing quantities that they obtain correspond to the dissipative response~\cite{Chia:2020yla,Goldberger:2020fot}. According to our distinction above, the conservative response --- and hence the Love numbers --- vanishes for four-dimensional black holes in general relativity.

\medskip
\noindent\textit{Static vs.\ dynamical response.}
The \textit{static response coefficients} are defined at zero tidal driving frequency, $\omega = 0$, and characterize the adiabatic, static deformation of the body. \textit{Dynamical coefficients}, in contrast, refer to the frequency-dependent response of the object, which can include mode-resonance enhancements and reduces to the static values in the $\omega \to 0$ limit.  
For both static and dynamical responses, one can distinguish between conservative and dissipative effects, per our convention above.
Note that dissipative effects can arise even at zero frequency, typically as a result of the body's rotation  \cite{LeTiec:2020spy,LeTiec:2020bos,Chia:2020yla}. 
As for the static response, some works in the literature generically refer to any induced time-dependent response as dynamical Love numbers. In what follows, we adhere to the more widely adopted convention of reserving the term \textit{dynamical Love numbers} for the 
conservative
response (see \cref{tab1:threshold} for a summary).

\medskip
\noindent\textit{Near zone vs.\ far zone.}
In the study of tidal interactions in binaries, it is useful to distinguish \emph{regions around a compact object} based on how gravitational perturbations propagate.
The \textit{near zone} is the region where $\omega r \ll 1$, in which long-distance effects are negligible and the field typically admits a power-law expansion. The \textit{far zone} (or radiation zone) is defined by $\omega r \gg 1$. The terms \textit{near-zone} and \textit{far-zone approximations} refer to truncating the perturbation equations to these respective r\'egimes. The near zone is sometimes used for matching off-shell correlation functions with the EFT description of compact sources, while the far zone is more suitable for studying scattering processes. In both cases, the tidal EFT Wilson coefficients can be obtained by matching to the full theory.

\medskip
\noindent\textit{Ladder symmetries vs.~Love symmetry.}
In the context of black hole perturbation theory, various symmetry arguments have been proposed in the literature to explain the vanishing of black hole Love numbers in general relativity. The \textit{ladder symmetries} were originally introduced in \rcite{Hui:2021vcv,Hui:2022vbh} for Schwarzschild and Kerr black holes (see~\rcite{Berens:2022ebl,Rai:2024lho} for an extension to Reissner--Nordstr\"om black holes and~\rcite{Combaluzier-Szteinsznaider:2024sgb,Kehagias:2024rtz,Gounis:2024hcm} for a generalization to nonlinear order in response theory) and correspond to a set of linearly realized symmetries of time-independent and dynamical perturbations. The \textit{Love symmetry} refers to a hidden $\mathrm{SL}(2,\mathbb{R})$ symmetry group of a near-zone approximation of the radial perturbation equation, introduced in \rcite{Charalambous:2021kcz,Charalambous:2022rre}. When present, it places the physical solution in a highest-weight representation of the algebra, which forces the static Love number to vanish.   The two sets of symmetries are inequivalent and are both distinct from the isometries of the background spacetime. An explanation of the difference between the two proposals in the case of scalar fields can be found in \rcite{Hui:2022vbh}. More recently, a symmetry-based explanation for the vanishing and non-renormalization of the nonlinear static Love numbers of spherically symmetric black holes, based on a spurion argument~\cite{Kol:2011vg}, has been proposed in \rcite{Parra-Martinez:2025bcu}. We will return to these topics in \cref{sec:sym}, where we provide a comprehensive summary of all symmetry principles discussed so far in the literature.

\begin{center}
\renewcommand{\arraystretch}{1.35}
\begin{tabular}{p{2.2cm} p{10.5cm}}
\hline\hline
\textbf{Abbreviation} & \textbf{Meaning} \\
\hline
ADM  & Arnowitt--Deser--Misner \\
AdS  & Anti-de Sitter \\
BH   & Black hole \\
BMS  & Bondi--Metzner--Sachs \\
CFT  & Conformal field theory \\
dS   & de Sitter \\
EAdS & Euclidean anti-de Sitter \\
ECO  & Exotic compact object \\
EFT  & Effective field theory \\
EoS  & Equation of state \\
IR   & Infrared \\
KK   & Kaluza--Klein \\
LIGO & Laser Interferometer Gravitational-Wave Observatory \\
LISA & Laser Interferometer Space Antenna \\
LVK  & LIGO--Virgo--KAGRA \\
NS   & Neutron star \\
GR   & General relativity \\
GW   & Gravitational wave \\
GW170817 & GW event~\cite{LIGOScientific:2017vwq} \\
PM   & Post-Minkowskian \\
PN   & Post-Newtonian \\
RG   & Renormalization group \\
RN   & Reissner--Nordstr\"{o}m \\
STU  & STU supergravity black hole family \\
TOV & Tolman--Oppenheimer--Volkoff \\
UV & Ultraviolet \\
\hline\hline
\end{tabular}
\end{center}

\newpage

\part{Tidal effects}
\label{Part:love}
\epigraphhead[]{
\epigraph{Tide goes in, tide goes out. \\You can't explain that.}{\textsc{Bill O'Reilly}}
}

Tidal effects are the first signs of gravity. According to the equivalence principle, a small laboratory in free fall behaves as if it were in flat spacetime. It is the \emph{differences} in the gravitational potential across the lab which lead to a measurable (tidal) gravitational acceleration. Around a point $P$ we can construct a normal coordinate system by using coordinate transformations to put the metric at $P$ in the form of the Minkowski metric ($g_\mn(P)=\eta_\mn$) and to eliminate its first derivatives ($\Gamma^\mu_\ab(P)=0$). This uses up the freedom to change coordinates, leading to unavoidable deviations starting at quadratic order in $x^\mu$,
\begin{equation}
g_\mn = \eta_\mn -\frac13R_{\mu\ab\nu}x^{\alpha}x^{\beta} + \mathcal{O}(x^3),
\end{equation}
reflecting the fact that the Riemann curvature cannot be gauged away, and corresponding physically to the aforementioned tidal effects.
Similarly, in a binary inspiral, tidal interactions are the first indications that the objects are extended, finite-size bodies rather than idealized point particles characterized only by their mass and spin.

In this Part we discuss how the finite size and internal structure of a gravitating object leaves an imprint on its gravitational interactions. We begin in \cref{sec:newton} with a Newtonian treatment, where Love numbers were originally defined, then describe in \cref{sec:EFT} how these concepts generalize in general relativity.
 
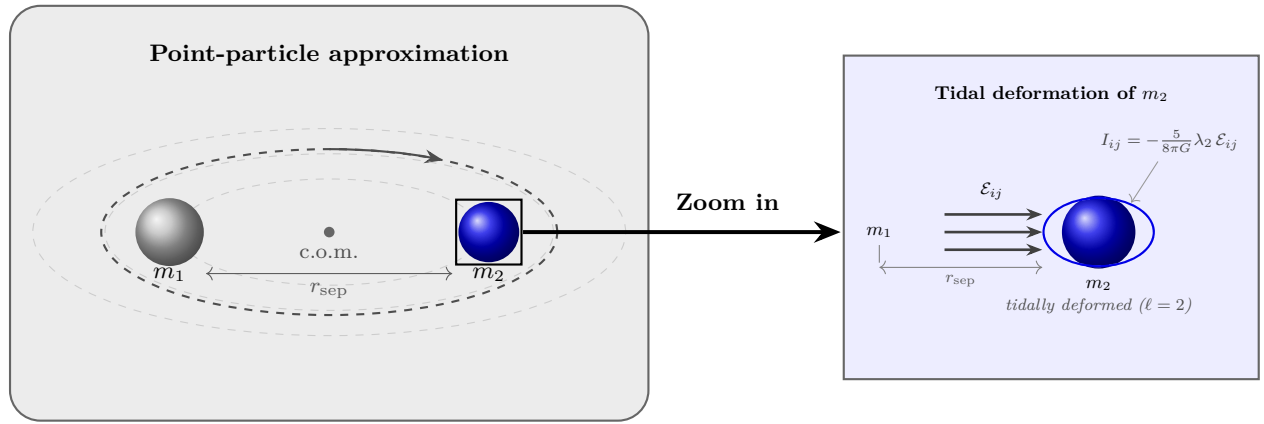
\begin{figure}[h!]
\centering
\begin{tikzpicture}[scale=1.0]

\begin{scope}
  \fill[gray!15, rounded corners=4mm] (-4.2,-2.5) rectangle (4.2,3.0);
  \draw[black!60, rounded corners=4mm, thick] (-4.2,-2.5) rectangle (4.2,3.0);
\end{scope}

\node[black, font=\small\bfseries] at (0, 2.35) {Point-particle approximation};

\draw[black!70, thick, dashed] (0,0) ellipse (3.0cm and 1.1cm);

\draw[-{Stealth[length=7pt]}, black!70, thick]
  (0,1.1) arc[start angle=90, end angle=60, x radius=3.0cm, y radius=1.1cm];

\fill[black!60] (0,0) circle (0.07cm);
\node[black!70, font=\footnotesize] at (0.0,-0.28) {c.o.m.};

\foreach \r in {0.80, 1.18, 1.56} {
  \draw[black!20, thin, dashed] (0,0) ellipse (\r*2.5cm and \r*0.88cm);
}

\shade[ball color=gray!55!white] (-2.1,0) circle (0.45cm);
\node[black, font=\small\bfseries] at (-2.1,-0.56) {$m_1$};

\shade[ball color=blue!90!black] (2.1,0) circle (0.40cm);
\node[black, font=\small\bfseries] at (2.1,-0.56) {$m_2$};

\draw[<->, black!65, thin] (-1.62,-0.55) -- (1.62,-0.55)
  node[midway, below, font=\footnotesize] {$r_{\rm sep}$};

\draw[black, thick] (1.67,-0.43) rectangle (2.53,0.43);

\draw[-{Stealth[length=10pt, width=8pt]}, black, line width=1.5pt]
  (2.55, 0.0) -- (6.73, 0.0);
\node[black, font=\small\bfseries] at (5.24, 0.40) {Zoom in};

\begin{scope}[xshift=9.5cm, scale=0.78, every node/.style={scale=0.78}]

  \fill[blue!7] (-3.5,-2.5) rectangle (3.5,3.0);
  \draw[black!60, thick] (-3.5,-2.5) rectangle (3.5,3.0);

  \node[black, font=\small\bfseries] at (0, 2.35) {Tidal deformation of $m_2$};

  \node[black, font=\small\bfseries] at (-2.9, 0.0) {$m_1$};
  \draw[black!50, thin] (-2.9,-0.22) -- (-2.9,-0.52);

  \foreach \yy in {0.30, 0.0, -0.30} {
    \draw[-{Stealth[length=6pt]}, black!75, thick, line width=1.0pt]
      (-1.80, \yy) -- (-0.15, \yy);
  }

  \node[black, font=\footnotesize\bfseries] at (-1.0, 0.70) {$\mathcal{E}_{ij}$};

  \draw[<->, black!50, thin] (-2.90,-0.60) -- (-0.15,-0.60)
    node[midway, below, font=\footnotesize, black!65] {$r_{\rm sep}$};

  \shade[ball color=blue!90!black] (0.8,0) circle (0.62cm);
  \draw[blue!40!white, thick, dashed] (0.8,0) ellipse (0.93cm and 0.58cm);
  \draw[blue!90!black, thick] (0.8,0) ellipse (0.93cm and 0.58cm);

  \node[black, font=\small\bfseries] at (0.8,-0.90) {$m_2$};
  \node[black!70, font=\footnotesize\itshape] at (0.8,-1.25) {tidally deformed ($\ell=2$)};

  \node[black!75, font=\footnotesize, align=center] at (2, 1.55)
    {$I_{ij} = -\frac{5}{8\pi G} \lambda_2\,\mathcal{E}_{ij}$};
  \draw[black!40, thin, ->] (1.8, 1.20) -- (1.35, 0.5);

\end{scope}

\end{tikzpicture}
\caption{%
  \textit{Left:} Two compact objects $m_1$ and $m_2$ in a binary orbit, treated as point particles to leading order.
  Dashed rings indicate the emitted gravitational-wave signal.
  \textit{Right:} Zooming in on $m_2$ reveals that its finite size leads to a quadrupolar ($\ell=2$) tidal bulge induced by the external tidal field $\mathcal{E}_{ij}$ sourced by the companion $m_1$.
  The induced quadrupole $I_{ij}$  defines the tidal deformability parameter $\lambda_2$, which corresponds to the $\ell=2$ Love number (see \cref{eq:Iell,eq:kelllambdaell}).
}
\label{fig:tidal_zoom}
\end{figure}

The setting we have in mind is the two-body problem in GR, namely two compact objects in a binary system during the early, adiabatic stage of the inspiral phase. To leading order, these objects interact as point particles. However, as the separation decreases, or as experimental sensitivity improves, tidally-induced deviations from point-particle behavior exert a noticeable influence on the orbital dynamics (see \cref{fig:tidal_zoom}).

Remarkably, the leading tidal effect on the orbit is universal, encoded in response coefficients called \emph{Love numbers} which are intrinsic properties of a self-gravitating object. One can therefore quantify this effect by {\it zooming in} on one body and treating its companion as an arbitrary external source. This procedure is at the heart of both the Newtonian and relativistic treatments presented in this Part.
This universality is natural in the EFT framework as a consequence of decoupling between the orbital and internal timescales,
\begin{equation}
   T_\mathrm{orb}\sim\sqrt{\frac{r_\mathrm{sep}^3}{GM}},\qquad T_\mathrm{int}\sim\sqrt{ \frac{R^3}{GM}}, 
\end{equation}
where $r_\mathrm{sep}$ is the orbital separation and $M$ and $R$ are the mass and radius of the compact body.

\section{A Newtonian introduction to tides}
\label{sec:newton}

\subsection{Exterior problem}

In Newtonian gravity the fundamental field is the gravitational potential 
$U$, the gradient of which is the gravitational acceleration,
\begin{equation}
\vec{F} = m \vec \nabla U,
\end{equation}
which in turn governs the motion of an object of mass $m$ via Newton's second law, $\vec{F} = m\vec{a}$. The potential itself is determined by the distribution of matter, namely its density $\rho(\vec x)$, via the Poisson equation,
\begin{equation}
\nabla^2 U = -4\pi G \rho \label{eq:Poisson}.
\end{equation}
Here $\nabla^2$ is the three-dimensional Laplacian, e.g., in Cartesian coordinates $\nabla^2=\partial_x^2+\partial_y^2+\partial_z^2$, and $G$ is the gravitational constant.
We will be particularly interested in the vacuum Poisson equation,
\begin{equation}
\nabla^2 U = 0. \label{eq:Poisson-vac}
\end{equation}

Typically one solves \cref{eq:Poisson-vac} outside a star and \cref{eq:Poisson} inside it, matching the two solutions at the boundary. The interior problem is often quite difficult as it requires solving, e.g., hydrodynamical equations coupled to gravity. Moreover we may have significant uncertainty about the object's interior composition, as is the case for neutron stars. We will therefore find it fruitful to think of the exterior problem as the main problem of interest, with information about the compact object itself encoded in boundary conditions imposed at the surface.

For a comprehensive treatment we refer the reader to the book by Poisson and 
Will~\cite{Poisson_Will_2014}; here we summarize the key elements.

\subsection{Nearly spherical objects}

Suppose the gravitational field is exactly spherically symmetric about the origin, as in the case of a non-rotating, undistorted star. This symmetry makes spherical coordinates $(r,\theta,\varphi)$ the natural choice for describing the system. In these coordinates, the Laplacian takes the form
\begin{equation}
r^2\nabla^2 = \partial_r(r^2 \, \partial_r) + \nabla^2_{S^2},
\end{equation}
with the Laplacian on the unit 2-sphere $S^2$ given by
\begin{equation}\label{eq:laplace-S2}
    \nabla^2_{S^2}=\frac{1}{\sin\theta}\partial_\theta(\sin\theta \partial_\theta) + \frac1{\sin^2\theta}\partial_\varphi^2.
\end{equation}
Under the assumption of spherical symmetry, $U$ depends only on the radial coordinate $r=\sqrt{x^2+y^2+z^2}$ and the vacuum Poisson equation reduces to
\begin{equation}
\partial_r(r^2\partial_rU)=0.
\end{equation}
Integrating once yields
\begin{equation}
\partial_r U = \frac{c_1}{r^2},
\end{equation}
where $c_1$ is an integration constant. Integrating a second time gives
\begin{equation}
U = -\frac{c_1}r + c_2,\label{eq:U-sph-sym}
\end{equation}
where $c_2$ is another integration constant. The value of $c_2$ is unphysical in classical gravitation, as we can only measure the gradient of the potential, not its absolute value.\footnote{In general relativity, the equivalence principle requires that all energy gravitate, and so an overall constant in the gravitational potential --- a \emph{cosmologial constant} --- is physical. In this review we will usually set the cosmological constant to zero, as the cosmological constant our Universe appears to have is negligible at the scale of a binary inspiral.} Identifying $c_1$ with the mass $M$ of the object (multiplied by $-G$) we recover the standard Newtonian gravitational potential for a point particle,\footnote{As a sanity check on our identification $c_1=-GM$, we can solve Poisson's equation \eqref{eq:Poisson} with a delta-function source $\rho(\vec x) = M\delta^{(3)}(\vec x)$, corresponding to an object of mass $M$ at $\vec x=0$, and use the identity
\[ \nabla^2r^{-1} = -4\pi\delta^{(3)}(\vec x). \]}
\begin{equation}
U = \frac {GM}r.
\end{equation}
Taking the gradient reproduces Newton's famous inverse-square law,
\begin{equation}
F=m\frac{\dd U}{\dd r} = -\frac{GMm}{r^2}.
\end{equation}
Note that the only information about the object which contributes to its gravitational field in spherical symmetry is its mass; its size and other details of its internal composition are irrelevant, and the object gravitates as if it were a point particle of mass $M$ located at $r=0$.\footnote{This is a case of the \emph{shell theorem} in Newtonian gravity. In a spherically-symmetric gravitational field (or indeed any spherically-symmetric potential scaling as $1/r$ or $r^2$, such as the electrostatic force or the force due to a cosmological constant), a particle at distance $R$ from the center only feels the force from matter at $r<R$, which gravitates as if it were concentrated at the center, while the contributions from matter at $r>R$ cancel each other perfectly. This remarkable property of Newtonian gravity persists in general relativity, in the form of \emph{Birkhoff's theorem}, which ensures that the exterior of a spherically-symmetric star is described by the same spacetime as a non-spinning black hole.}

Spherical symmetry provides a good approximation for non-rotating bodies in a binary system when the orbital separation is large compared to the sizes of the objects. Zooming in on one of the bodies, it can be viewed as a nearly spherically symmetric object immersed in an external gravitational field. The external potential, generated by the companion, slightly breaks the spherical symmetry. These small deviations from spherical symmetry are naturally described using an expansion in  
\emph{spherical harmonics}. These are harmonic functions on the unit 2-sphere $S^2$, denoted by $Y_{\ell m}(\theta,\varphi)$, which are eigenfunctions of the spherical Laplacian \eqref{eq:laplace-S2} with eigenvalues $-\ell(\ell+1)$,
\begin{equation}\label{eq:sph-harm-eq}
\frac{1}{\sin\theta}\partial_\theta(\sin\theta\, \partial_\theta Y_{\ell m}) + \frac1{\sin^2\theta}\partial_\varphi^2Y_{\ell m} = -\ell(\ell+1)Y_{\ell m}.
\end{equation}
The solutions to this equation are restricted by the periodicity of the angular coordinates $\theta$ and $\varphi$, which requires $\ell$ and $m$ to be integers satisfying
\begin{equation}
\ell \ge 0, \quad -\ell \le m \le \ell.
\end{equation}
The solutions to \cref{eq:sph-harm-eq} that satisfy these boundary conditions are the spherical harmonics, which take the form
\begin{equation}
Y_{\ell m}(\theta,\varphi) = \sqrt{\frac{2\ell+1}{4\pi}\frac{(\ell-m)!}{(\ell+m)!}} P_\ell^m(\cos\theta)e^{im\varphi},
\end{equation}
where $P_\ell^m(x)$ are the associated Legendre polynomials of degree $\ell$ and order $m$.\footnote{These functions are reviewed in \ref{app:legendre}. Strictly speaking this formula applies to spherical harmonics with $m\geq0$; the $m<0$ harmonics satisfy the relation $Y_{\ell,-m}=(-1)^m\bar{Y}_{\ell m}$.} We choose the proportionality constant so that the spherical harmonics satisfy the orthonormality property
\begin{equation}
\int\limits_{S^2} \dd \Omega \,Y_{\ell m} \bar{Y}_{\ell' m'} \equiv \int\limits_{\theta=0}^\pi\int\limits_{\varphi=0}^{2\pi} \dd \theta\, \dd\varphi\, \sin\theta \,Y_{\ell m} \bar{Y}_{\ell' m'} = \delta_{\ell\ell'}\delta_{m m'},\label{eq:orthonorm-Ylm}
\end{equation}
where an overbar denotes the complex conjugate, $\dd\Omega = \sin\theta\,\dd\theta\,\dd\varphi$ is the integration measure on $S^2$, and $\delta_{ij}$ is the Kronecker delta, which is equal to 1 when $i=j$ and 0 otherwise. For a more detailed review of the spherical harmonics, see \ref{app:sph-harm}

The spherical harmonics form a complete basis for functions on $S^2$, so a general gravitational potential $U(t,r,\theta,\varphi)$ can be decomposed into a superposition of spherical harmonic modes labelled by $\ell$ and $m$,
\begin{equation}
U(t,r,\theta,\varphi) = \displaystyle\sum_{\ell m}U_{\ell m}(t,r)Y_{\ell m}(\theta,\varphi),\qquad\displaystyle\sum_{\ell m}\equiv\sum_{\ell=0}^\infty\sum_{m=-\ell}^\ell.
\end{equation}
As a consequence of the orthonormality property \eqref{eq:orthonorm-Ylm} we can extract an individual $\ell, m$ multipole by convolving the full potential with the (complex conjugate of the) relevant spherical harmonic,
\begin{equation}
U_{\ell m}(t,r) = \int \dd\Omega \,U(t,r,\theta,\varphi) \bar Y_{\ell m}(\theta,\varphi).
\end{equation}

Because the Poisson equation is linear, every ($\ell,m$) mode decouples from the rest.
In the absence of matter,
\begin{align}
0 &= \displaystyle\sum_{\ell m}\nabla^2 (U_{\ell m}Y_{\ell m}) \nonumber\\
&= \displaystyle\sum_{\ell m}Y_{\ell m}\left(\frac{1}{r^2}\partial_r (r^2\partial_r) - \frac{\ell(\ell+1)}{r^2}\right)U_{\ell m},
\end{align}
so that each mode $U_{\ell m}(t,r)$ satisfies the radial equation\footnote{Notice that $m$ does not appear in this equation. This is due to spherical symmetry, as we review in \ref{app:sph-harm}: spherical harmonics with the same $\ell$ and different $m$ are related by rotations. When we consider spinning objects, for which the physics is not invariant under the full three-dimensional rotation group, $m$ will play a role in the dynamics.}
\begin{equation}
\boxed{\left[\partial_r(r^2\partial_r)-\ell(\ell+1)\right]U_{\ell m} = 0.}\label{eq:poisson-rad}
\end{equation}
This is a second-order ordinary differential equation for the radial profile $U_{\ell m}(r)$, so it possesses two linearly independent solutions. (The Poisson equation is independent of time and contains no time derivatives, so any time dependence in $U_\ellm(t,r)$ is ``along for the ride.'') A suitable basis can be constructed by assuming a power-law form $U\sim r^n$,
\begin{align}
0 &= \left[\partial_r(r^2\partial_r)-\ell(\ell+1)\right]r^n\nonumber\\
&= \left[n(n+1)-\ell(\ell+1)\right]r^n.
\end{align}
A natural basis of linearly independent solutions to \cref{eq:poisson-rad} corresponds to the two solutions to $n(n+1)=\ell(\ell+1)$, namely a \emph{growing mode} scaling as $r^\ell$ and a \emph{decaying mode} scaling as $r^{-\ell-1}$,
\begin{equation}
U_\ell^\mathrm{g}(r) = r^\ell,\quad U_\ell^\mathrm{d}(r) = \frac1{r^{\ell+1}}.\label{eq:poisson-basis}
\end{equation}

Generally speaking, the growing modes correspond to the external tidal field, 
which formally is located at asymptotic infinity ($r\to\infty$) and has been expanded in multipoles. 
The \emph{decaying modes}, on the other hand, are associated with the internal structure of the gravitating body; 
for $\ell\geq2$, they describe the induced response of the object to the external field. 

The general solution to \cref{eq:poisson-rad} is a linear combination of the growing and decaying solutions,
\begin{equation}
U_{\ell m}(t,r) = a_{\ell m}(t) U_\ell^\mathrm{g}(r) + b_{\ell m}(t) U_\ell^\mathrm{d}(r).
\end{equation}
The coefficients $a_{\ell m}$ and $b_{\ell m}$ are determined by boundary conditions. We have already seen how this works for the monopole $\ell=m=0$. There we obtained the spherically-symmetric potential \eqref{eq:U-sph-sym}, corresponding to $U_0^\mathrm{d}=1/r$, with the mass of the object acting as the boundary condition fixing the constant $c_1$. The other solution, $U_0^\mathrm{g} = 1$, shifts the gravitational potential by an unphysical constant. The dipole ($\ell=1$) is absent in the body's rest frame. For $\ell\geq2$ both the growing and decaying modes may be present, with boundary conditions (typically imposed at the surface of the object) fixing $a_\ellm$ and $b_\ellm$.

\subsection{Green's function solution to Poisson's equation}

We want to solve the Poisson equation in either the form \eqref{eq:Poisson},
\begin{equation}
\nabla^2 U = -4\pi G \rho,
\end{equation}
or the multipole form,
\begin{equation}\label{eq:Poisson-multipole}
\left[\partial_r(r^2\partial_r)-\ell(\ell+1)\right]U_{\ell m} = -4\pi G r^2\rho_{\ell m}.
\end{equation}
Here we have decomposed the density in spherical harmonics,
\begin{equation}
\rho(t,\vec x) = \displaystyle\sum_{\ell m}\rho_{\ell m}(t,r)Y_{\ell m}(\theta,\varphi).
\end{equation}
It is standard to solve these equations using the machinery of \emph{Green's functions}. 

The Green's function $G(\vec x,\vec x')$ for \cref{eq:Poisson} satisfies
\begin{equation}
\nabla^2G(\vec x,\vec x') = \delta(\vec x-\vec x').
\end{equation}
Then we may write the general solution formally as\footnote{We use the same symbol $G$ for the gravitational constant and for the Green's function; the meaning should be clear from context.}
\begin{equation}
U(t,\vec x) = -4\pi G\int \dd^3 x' G(\vec x,\vec x')\rho(t,\vec x').
\end{equation}
The Green's function for the 3D Laplacian $\nabla^2$, imposing as a boundary condition that the solution vanishes at infinity, is
\begin{equation}
G(\vec x,\vec x') = -\frac1{4\pi|\vec x-\vec x'|}.
\end{equation}
This is a standard result following from the identity
\begin{equation}
\nabla^2\frac1{|\vec x-\vec x'|} = -4\pi\delta(\vec x-\vec x').
\end{equation}
The solution to the Poisson equation is then written
\begin{equation}
U(t,\vec x) = G\int \dd^3 x'\frac{\rho(t,\vec x')}{|\vec x-\vec x'|}.
\end{equation}
This can be thought of simply as a superposition of $1/r$ potentials for each fluid element.

We may similarly solve the multipole Poisson equation \eqref{eq:Poisson-multipole}. Let us define the order-$\ell$ Green's function $g_\ell(r,r')$ by demanding that it satisfy
\begin{equation}\label{eq:green-multipole}
\left[\partial_r\left(r^2\partial_r\right)-\ell(\ell+1)\right]g_\ell(r,r') = \delta(r-r').
\end{equation}
and that it obey the boundary conditions $g_\ell(0,r') = g_\ell(\infty,r') = 0$.
When $r\neq r'$, the source term in \cref{eq:green-multipole} vanishes and the Green's function satisfies the vacuum Poisson equation \eqref{eq:poisson-rad}, for which we have seen the solutions are (cf.~\cref{eq:poisson-basis}) $g_\ell\sim r^\ell$ and $g_\ell\sim r^{-\ell-1}$.
\begin{equation}
g(r,r') = a_\ell r^\ell + \frac{b_\ell}{r^{\ell+1}} ,\quad r\neq r'.
\end{equation}
The boundary conditions $g_\ell(0,r') = g_\ell(\infty,r') = 0$ tell us that $a_\ell=0$ for $r>r'$ and $b_\ell=0$ for $r<r'$, so we can write $g_\ell(r,r')$ in piecewise form,
\begin{align}
g_\ell(r,r') &= \begin{cases}
f_1(r')\left(\frac r{r'}\right)^\ell, & r<r' \\
f_2(r')\left(\frac r{r'}\right)^{-\ell-1}, & r>r'
\end{cases}.
\end{align}
We also require the Green's function to be continuous at $r=r'$,\footnote{To justify this we could imagine that there is a step function discontinuity, in which case $\partial_r g(r,r')\sim \delta(r-r')$. Its derivative $\delta'(r-r')$ cannot be written in terms of delta functions alone, but the delta function is the only distributional object present in \cref{eq:green-multipole}.} so that $f_1(r')=f_2(r')\equiv f(r')$,
\begin{equation}
g_\ell(r,r') =r' f(r') \frac{r_<^\ell}{r_>^{\ell+1}},
\end{equation}
where we have defined
\begin{equation}
r_< \equiv \min(r,r'),\quad r_> \equiv \max(r,r').
\end{equation}
Finally we compute $f(r')$ by using the fact that the derivative $\partial_r g(r,r')$ is discontinuous at $r=r'$.
To this end we integrate \cref{eq:green-multipole} over an infinitesimally small region,
\begin{align}
1 &= \lim_{\epsilon\to0}\int\limits_{r'-\epsilon}^{r'+\epsilon}\dd r\left[\partial_r\left(r^2\partial_r\right)-\ell(\ell+1)\right]g_\ell(r,r')\nonumber\\
&=r'^2\left[\partial_r g_\ell(r,r')|_{r\to r'^+} - \partial_r g_\ell(r,r')|_{r\to r'^-}\right]\nonumber\\
&= -(2\ell+1)r'f(r').
\end{align}
We conclude the Green's function solving \cref{eq:green-multipole} is
\begin{equation}\label{eq:green-sol-newt-multi}
\boxed{g_\ell(r,r') = -\frac1{2\ell+1}\frac{r_<^\ell}{r_>^{\ell+1}}.}
\end{equation}
Note that this may also be obtained by expanding the real-space Green's function $G(\vec x,\vec x')$,
\begin{equation}
G(\vec x,\vec x') = \displaystyle\sum_\ellm g_\ell(r,r')Y_\ellm(\theta,\varphi)\bar Y_\ellm(\theta',\varphi'),
\end{equation}
as a consequence of the identity \cite{Poisson_Will_2014}
\begin{equation}
\frac1{|\vec x-\vec x'|} = \displaystyle\sum_\ellm\frac{4\pi}{2\ell+1}\frac{r_<^\ell}{r_>^{\ell+1}}Y_\ellm(\theta,\varphi)\bar Y_\ellm(\theta',\varphi').
\end{equation}

More generally, let us suppose that we seek to solve a differential equation of the form
\begin{equation}
\mathcal L \phi(r) = J(r),
\end{equation}
where $\mathcal{L} = \alpha(r)\partial_r^2+\cdots$ is a second-order differential operator which like the Poisson operator possesses linearly-independent solutions with growing and decaying behavior at $r\to\infty$. Performing the same procedure as above we obtain the Green's function,
\begin{equation}
g(r,r') = \frac{\phi_\mathrm{g}(r_<)\phi_\mathrm{d}(r_>)}{\alpha(r')W(r')},
\end{equation}
where the \emph{Wronskian} of the two solutions is
\begin{equation}
W(r) \equiv W[\phi_\mathrm{g},\phi_\mathrm{d}] = \phi_\mathrm{g}(r)\phi_\mathrm{d}'(r)-\phi_\mathrm{d}(r)\phi_\mathrm{g}'(r).
\end{equation}
It is straightforward to check that this reproduces \cref{eq:green-sol-newt-multi} upon setting $\alpha(r)=r^2$, $\phi_\mathrm{g}(r) = r^\ell$, and $\phi_\mathrm{d}(r) = r^{-\ell-1}$.

Finally we use the Green's function \eqref{eq:green-sol-newt-multi} to obtain the general solution to \cref{eq:Poisson-multipole} with asymptotically flat boundary conditions,
\begin{align}
U_{\ell m}(t,r) &= -4\pi G \int \limits_0^\infty \dd r' g_\ell(r,r') r'^2\rho_{\ell m}(t,r') \nonumber\\
&= \frac{4\pi G}{2\ell+1}\int \limits_0^\infty \dd r' \frac{r_<^\ell}{r_>^{\ell+1}} r'^2\rho_{\ell m}(t,r').
\end{align}
To make sense of this expression let us first consider a point outside the star (or other compact object), $r>\Rstar$. Since the integrand only has support for $r'<\Rstar$, we can write this as
\begin{align}
U_{\ell m}(t,r) &= \frac{4\pi G}{2\ell+1}\frac{1}{r^{\ell+1}}\int \limits_0^{\Rstar} \dd r' r'^{\ell+2}\rho_{\ell m}(t,r') \nonumber\\
&= \frac{4\pi G}{2\ell+1}\frac{1}{r^{\ell+1}}I_{\ell m}(t),
\end{align}
where
\begin{equation}\label{eq:star-multi-mom}
I_{\ell m}(t) \equiv \int\limits_0^{\Rstar}\dd r' r'^{\ell+2}\rho_{\ell m}(t,r')
\end{equation}
are the star's multipole moments. Inside the star we may similarly write the solution as
\begin{equation}\label{eq:Ulm-interior}
U_{\ell m}(t,r) = \frac{4\pi G}{2\ell+1}\left(p_{\ell m}(t,r)r^\ell +\frac{q_{\ell m}(t,r)}{r^{\ell+1}}\right),
\end{equation}
where
\begin{equation}
p_{\ell m}(t,r) \equiv \int\limits_r^{\Rstar}\dd r' \frac{\rho_{\ell m}(t,r')}{r'^{\ell-1}},\qquad q_{\ell m}(t,r) \equiv \int\limits_0^r\dd r' r'^{\ell+2}\rho_{\ell m}(t,r').
\end{equation}
Note that $I_{\ell m}(t)=q_{\ell m}(t,\Rstar)$. We have thus obtained the internal and external potential of a star deformed away from spherical symmetry.

If the problem at hand has different boundary conditions, as the tidal response problem does, then its effect is (as it must be) accounted for by adding a homogeneous term to the Green's function, i.e.,
\begin{equation}
G(\vec x,\vec x') = -\frac1{4\pi|\vec x-\vec x'|} + G^\mathrm{hom}(\vec x,\vec x'),\qquad\nabla^2G^\mathrm{hom}(\vec x,\vec x') = 0,
\end{equation}
for the real-space Green's function, and
\begin{equation}
g_\ell(r,r') = -\frac1{2\ell+1}\frac{r_<^\ell}{r_>^{\ell+1}} + g^\mathrm{hom}_\ell(r,r'),\qquad \left[\partial_r\left(r^2\partial_r\right)-\ell(\ell+1)\right]g^\mathrm{hom}_\ell(r,r')=0.
\end{equation}
for the spherical harmonic Green's function. We have already seen how to solve $\left[\partial_r\left(r^2\partial_r\right)-\ell(\ell+1)\right]g^\mathrm{hom}_\ell(r,r')=0$: it has power-law growing and decaying solutions, one of which is already present in $g_\ell(r,r')$ (depending on whether $r>r'$ or $r<r'$).

\subsection{Tides and linear response theory}

Consider a spherically symmetric gravitating body immersed in an external gravitational field. 
In isolation, the body’s potential is simply the Newtonian monopole, $GM/r$, but the presence of a companion or distant mass introduces a tidal field, $U_\mathrm{tidal}$, which deforms the body. 
These deformations generate an additional gravitational field that back-reacts on the environment.  

The objective of this section is to systematically construct the full gravitational potential outside the body, capturing both the externally imposed tidal field and the body's induced response. 
Along the way, we will introduce the relevant multipole moments and define the \emph{tidal Love numbers}, which quantify the strength of the induced response. 
For now, we focus on the static, adiabatic r\'egime, where the tidal field varies slowly compared to the body’s internal dynamical timescale. 
Later sections will relax these assumptions to include dynamical and relativistic effects. 

For now, we focus on \emph{static} tides, {i.e.}, any time dependence in $U_\mathrm{tidal}$ is sufficiently slow that an adiabatic approximation is valid: the body remains in hydrostatic equilibrium.

\subsubsection{Tidal moments and multi-index notation}

The tidal moments are defined by differentiating the external potential $\ell$ times, removing traces, and evaluating at the origin~\cite{Poisson_Will_2014}:
\begin{equation}
\mathcal E_{i_1\cdots i_\ell} \equiv -\partial_{\langle i_1}\cdots \partial_{i_\ell\rangle} U_\mathrm{tidal}\Big|_{r=0}.
\end{equation}
Here, the use of partial derivatives implies that $x^i$ are Cartesian coordinates. 
It is convenient to introduce multi-index notation $L = (i_1, \dots, i_\ell)$, so that we can write
\begin{equation}
\mathcal E_L \equiv -\partial_{\langle L \rangle} U_\mathrm{tidal}\Big|_{r=0}.
\end{equation}
We also adopt the shorthand $\partial_L \equiv \partial_{i_1}\cdots \partial_{i_\ell}$. The tidal potential can then be expanded as
\begin{equation}
U_\mathrm{tidal}(\vec{x}) = - \sum_{\ell m} \frac{1}{\ell!} \mathcal E_L x^L,
\end{equation}
where $x^L \equiv x^{i_1}\cdots x^{i_\ell}$. Details of the Cartesian-to-spherical dictionary are given in \ref{app:sph-harm}.

Introducing the unit radial vector $n^i = x^i/r$, with $n^L = n^{i_1}\cdots n^{i_\ell}$, the tidal multipoles in spherical coordinates are
\begin{equation}
n^L \mathcal E_L = \sum_{m=-\ell}^{\ell} \mathcal E_{\ell m} Y_{\ell m}(\theta,\varphi), \qquad
\mathcal E_{\ell m} = \mathcal E_L \int \dd\Omega\, n^L \bar Y_{\ell m}.
\end{equation}
In this way, we may write the tidal potential in the familiar spherical-harmonic decomposition:
\begin{equation}
U_{\ell m,\mathrm{tidal}}(r) = - \frac{1}{\ell!} \mathcal E_{\ell m} r^\ell.
\end{equation}

\subsubsection{Induced response and Love numbers}

When the external tidal field breaks the spherical symmetry of the body, the body develops multipole moments in response. 
For an initially unperturbed spherical star, only the monopole $I_{00}$ is non-zero. The tidal field $\mathcal E_{\ell m}$ induces a non-vanishing $I_{\ell m}$:
\begin{equation}\label{eq:Uellm}
U_{\ell m}(r) = - \frac{1}{\ell!} \mathcal E_{\ell m} r^\ell + \frac{4 \pi G}{2\ell+1} \frac{I_{\ell m}}{r^{\ell+1}}.
\end{equation}

In the linear response and adiabatic approximation (weak field, slow evolution), the induced multipoles are proportional to the applied tidal field~\cite{Poisson_Will_2014,Flanagan:2007ix, Binnington:2009bb, Damour:2009vw}: 
\begin{equation}\label{eq:Iell}
\frac{4 \pi G}{2\ell+1}I_{\ell m} = -\frac{2 k_\ell}{\ell!} \Rstar^{2\ell+1} \mathcal E_{\ell m},
\end{equation}
where $k_\ell$ are the (Newtonian) \emph{tidal Love numbers}.

According to the definition \eqref{eq:Iell}, the $k_\ell$ are dimensionless. In the remainder of this review, it will sometimes be convenient to absorb $\Rstar^{2\ell+1}$ into the response coefficient, introducing a dimensionful parameter $\lambda_\ell$ related to $k_\ell$ via\footnote{We will interchangeably refer to both $\lambda_\ell$ and $k_\ell$ as Love numbers.}
\begin{equation}
\lambda_\ell = 2 k_\ell \Rstar^{2\ell+1},
\label{eq:kelllambdaell}
\end{equation}
so that the full external potential of the body can be expressed as
\begin{equation}\label{eq:Uellm-2}
U_{\ell m}(r) = - \frac{1}{\ell!} \mathcal E_{\ell m} \left(r^\ell + \frac{\lambda_\ell}{r^{\ell+1}}\right).
\end{equation}
The form \eqref{eq:Uellm-2} --- or, equivalently, the one expressed in terms of $k_\ell$ --- is convenient, as it separates the imposed tidal field from the body's induced response.

The Love numbers characterize an object's susceptibility to tidal deformation. Importantly, they are intrinsic to a given object: once $\lambda_\ell$ is measured for a particular $\ell$, that value will govern the body's linear response to any small tidal deformation.

To compute a body's Love number, one may explicitly solve the gravitational field equation in the exterior, impose boundary conditions at the surface, and then simply read off the relative size of the decaying term to the growing term at infinity. For the Poisson equation we know the full solution for all $r$,
\begin{equation}
U_\ellm(r) = \alpha_\ellm r^\ell + \frac{\beta_\ellm}{r^{\ell+1}},
\end{equation}
so it is straightforward to see that the Love number is simply 
\begin{equation}\label{eq:Newton-Love}
\lambda_{\ell} = \frac{\beta_{\ell m}}{\alpha_{\ell m}}.
\end{equation}
If we have a model for the star's interior, we may compute an interior profile for $U_\ellm$ and read off $\alpha_\ellm$ and $\beta_\ellm$. Alternatively, we may choose to ``parametrize our ignorance'' and take $\alpha_\ellm$ and $\beta_\ellm$ as inputs characterizing a given object.

Away from the cozy confines of Newtonian gravity, the gravitational potential in vacuum will not follow simple power laws of the radius. In the applications we consider in this work, the solutions will be more involved functions of $r$ which possess the same asymptotic behavior at infinity but behave differently as $r\to \Rstar$.
In this case, to which \cref{sec:love-compute} is devoted, the Love number is defined in the asymptotic r\'egime, 
\begin{equation}
U_{\ell m}(r) \overset{r\to\infty}\longrightarrow -\frac1{\ell!}\mathcal{E}_{\ell m}\left(r^\ell + \frac{\lambda_\ell}{r^{\ell+1}}\right).
\end{equation}

\subsection{Stellar tides}

Astrophysical stars are usually well modeled as \emph{perfect fluids}, which are completely characterized by their density $\rho$, pressure $p$, and velocity field $\vec{v}$.  
By definition, perfect fluids are idealized as viscosity-free, non-conductive, and isotropic, with no internal shear stresses or heat conduction.  
This idealization captures the dominant hydrodynamical behavior of stars while simplifying the treatment of their tidal and gravitational responses. The  variables $(U,\rho,p,v^i)$ satisfy a system of equations consisting of the Poisson equation \eqref{eq:Poisson}, the \emph{Euler equation},\footnote{Note that we are using the Eulerian picture of fluid mechanics.}
\begin{equation}\label{eq:Euler-Newt}
\partial_t\vec v +(\vec v \cdot \vec\nabla)\vec v = \vec\nabla U - \frac1\rho\vec\nabla p,
\end{equation}
and the \emph{continuity equation},
\begin{equation}\label{eq:cont-Newt}
\partial_t\rho + \nabla\cdot(\rho \vec v) = 0.
\end{equation}
To complete the system we must specify a relation between $\rho$ and $p$, known as the \emph{equation of state} (EoS). Typically this is of the form
\begin{equation}
p = p(\rho).
\end{equation}
A widely used equation of state, derived under the assumptions that neighboring fluid elements are in thermal equilibrium and that the energy density is proportional to the pressure, is the polytrope:
\begin{equation}
p(\rho) = K \rho^\Gamma = K \rho^{1 + 1/n},
\end{equation}
where $K$, $\Gamma$, and $n$ are constants. Here, $\Gamma$ is the adiabatic index and $n$ is the polytropic index. The case $n = 0$ corresponds to an incompressible fluid, for which $\rho$ is constant.
Realistic neutron‑star EoS are often well approximated by polytropic models with effective indices in the range $n\in[0.5,1]$, consistent with current constraints from nuclear physics and multimessenger observations (see, e.g., the review in \rcite{Chatziioannou:2020pqz} for a recent summary of EoS parameter ranges).  
In the literature, EoS are commonly implemented as piecewise polytropes \cite{Yagi:2016bkt,Chatziioannou:2020pqz} to capture the stiffening of matter at supranuclear densities.\\

\paragraph{Unperturbed configuration}
Let us consider a non-rotating star (or a non-rotating black hole) in equilibrium, isolated from any external matter. 
The background solution is spherically symmetric and contained within a sphere of radius $\Rstar$, which we identify as the object's surface. 
Within this radius, the density is $\rho(t,r,\theta,\varphi)=\rho(t,r)$, i.e., it depends only on time and the radial coordinate. 
For an equilibrium configuration, we will typically take the strictly static limit, $\rho(t,r) \to \rho(r)$, though a slow time dependence may be allowed in the context of adiabatic tidal variations, as discussed in \cref{sec:love-compute}.

The fluid velocity vanishes, $\vec v=0$, so the continuity equation tells us $\partial_t\rho=0$, while the Euler and Poisson equations are
\begin{align}
\partial_rp &= \rho\partial_rU, \\
\partial_r(r^2\partial_rU) &= -4\pi Gr^2\rho.
\end{align}
Let us introduce the mass function $m(r)$, which satisfies
\begin{equation}
m'(r) = 4\pi r^2\rho.
\end{equation}
Inserting this into the Poisson equation and integrating (demanding that $m(0)=0$) we see that
\begin{equation}
\partial_rU = -\frac{Gm(r)}{r^2},
\end{equation}
and plugging this into the Euler equation we find
\begin{equation}
\partial_rp = -\frac{Gm(r)\rho}{r^2}.
\end{equation}
This is the condition for hydrostatic equilibrium in spherical symmetry.

\paragraph{Perturbed configuration}

Now let us suppose that the body is not quite spherically symmetric. We will, however, assume it remains within hydrostatic equilibrium. Assuming that the deviations from spherical symmetry are small, we may employ perturbation theory, and assume that $\rho_\ellm$ and $U_\ellm$ for $\ell\geq2$ are small compared to their unperturbed values $\rho_{00}$ and $U_{00}$. Note also that $\vec v$ (or $v^i$) is a perturbative quantity of the same size as $U_\ellm$ and $\rho_\ellm$ in principle, though the assumption of hydrostatic equilibrium for the perturbed system implies $\vec v=0$.

We thus expand the density and gravitational potential as  
\begin{equation}
\rho = \bar{\rho} + \delta \rho, 
\qquad 
U = \bar{U} + \delta U,
\end{equation}
where the background quantities $\bar{\rho} \equiv \rho_{00}$ and $\bar{U} \equiv U_{00}$ are spherically symmetric, while the perturbations $\delta\rho$ and $\delta U$ contain  multipoles with $\ell \geq 2$. 

The Euler equation, when linearized around the background, gives
\begin{equation}
\vec\nabla\delta U - \frac1{\bar\rho}\left(\vec\nabla\delta p - \frac{\delta\rho}{\bar\rho}\vec\nabla\bar p\right)=0.
\label{eq:Euler-pert}
\end{equation}
Using the background relation, this can be rewritten as, for the radial component,
\begin{equation}
\partial_r\delta U - \frac1{\bar\rho}\partial_r\delta p - \frac{Gm(r)}{r^2}\frac{\delta\rho}{\bar\rho}=0 .
\end{equation}
Decomposing all perturbations in spherical harmonics,
\begin{equation}
\partial_r U_\ellm - \frac1{\bar\rho}\partial_r p_\ellm - \frac{Gm(r)}{r^2}\frac{\rho_\ellm}{\bar\rho}=0,
\end{equation}
the angular components of \cref{eq:Euler-pert} imply
\begin{equation}
U_\ellm - \frac1{\bar\rho}p_\ellm = 0.
\end{equation}
i.e.,
\begin{equation}
p_\ellm = \bar\rho U_\ellm.
\end{equation}
Taking a radial derivative yields
\begin{equation}
\partial_r p_\ellm = \bar\rho\partial_r U_\ellm + \partial_r\bar\rho U_\ellm
\end{equation}
and substituting this into the radial component of \cref{eq:Euler-pert}, one finds
\begin{equation}
\frac{Gm(r)}{r^2}\delta\rho =- \partial_r\bar\rho U_\ellm.
\end{equation}

To relate these quantities to the fluid displacement, we introduce the  vector $\vec{\xi}$ via $\vec{v} = \partial_t \vec{\xi}$. Integrating the continuity equation in time and linearizing, one obtains
\begin{equation}
\delta\rho=-\partial_i(\bar\rho\xi^i).
\end{equation}
In addition, from $\partial_i\xi^i=0$~\cite{Poisson_Will_2014}, this simplifies to
\begin{equation}
\delta\rho=-\xi^i\partial_i\bar\rho.
\end{equation}
Assuming an adiabatic perturbation, the pressure perturbation satisfies
\begin{equation}
\frac{\delta p}{\bar p}\propto\frac{\delta\rho}{\bar\rho},
\end{equation}
which implies~\cite{Poisson_Will_2014}
\begin{equation}
\delta p = - \xi^i \partial_i \bar{p}.
\end{equation}
We will only need the radial component of $\xi^i$ which we shall expand in spherical harmonics as
\begin{equation}
\xi^r = \displaystyle\sum_\ellm r f_\ellm(r)Y_\ellm(\theta,\varphi).
\end{equation}

Then the equations above become
\begin{align}
f_\ellm &= -\frac{\rho_\ellm}{r\bar\rho'},\\
p_\ellm &= \frac{Gm(r)\bar\rho}{r}f_\ellm.
\end{align}
primes denote $\partial_r$. Note that $m(r)$ here is given by its background value to this order in perturbation theory. We can now write
\begin{equation}\label{eq:Euler-cont-PW}
U_\ellm = \frac{p_\ellm}{\bar\rho} = \frac{Gm(r)}{r}f_\ellm.
\end{equation}
Notice that the perturbed density, pressure, and potential are all determined by $f_\ellm(r)$, so we will focus on solving for this quantity.
Substituting \cref{eq:Euler-cont-PW} into \cref{eq:Poisson-multipole} we obtain the \emph{Clairaut equation},
\begin{equation}
r^2f_\ellm''+6\mathcal{D}\left(rf_\ellm'+f_\ellm\right)-\ell(\ell+1)f_\ellm = 0,\label{eq:clairaut}
\end{equation}
where we have defined
\begin{equation}
\mathcal{D}(r) \equiv \frac{4\pi r^3\rho}{3m(r)}.
\end{equation}
If we perform a further change of variables,\footnote{We omit an $m$ subscript on $\eta_\ell$ because this quantity turns out not to depend on the azimuthal quantum number. This is a consequence of the fact that Clairaut's equation is independent of $m$, so the only dependence on $m$ that $f_\ellm$ can have is via an $m$-dependent multiplicative factor, which drops out of the logarithmic derivative.}
\begin{equation}\label{eq:etaell}
\eta_\ell \equiv \frac{\dd \ln f_{\ell m}}{\dd \ln r},
\end{equation}
then we may rewrite \cref{eq:clairaut} as the first-order \emph{Radau equation},
\begin{equation}\label{eq:radau}
r \eta_\ell'+\eta_\ell(\eta_\ell-1)+6\mathcal{D}\left(\eta_\ell+1\right)-\ell(\ell+1)= 0.
\end{equation}
This form is more suitable for numerical integration.

To relate $f_\ellm$ and $\eta_\ell$ to Love numbers, we may compute \cref{eq:Euler-cont-PW} and its first derivative and evaluate each at $r=\Rstar$. Using the fact that the density vanishes at the surface, $\rho(\Rstar)=0$, and the requirement that $U_\ellm$ and its first derivative both be continuous at $r=\Rstar$, it is straightforward to show that
\begin{align}
\frac{GM}{\Rstar}f_\ellm(\Rstar) &= U_\ellm(\Rstar) ,\\
\frac{GM}{\Rstar}f_\ellm'(\Rstar) &= U'_\ellm(\Rstar)+\frac{U_\ellm(\Rstar)}\Rstar.
\end{align}
Using the functional form of $U_\ellm(r)$, cf.~\cref{eq:Uellm}, and the definition \eqref{eq:etaell}, we can combine these into an expression for $\eta_\ell$ evaluated at the surface,
\begin{align}
\eta_\ell(\Rstar) &= \frac{\Rstar U_\ellm'(\Rstar)}{U_\ellm(\Rstar)}+1 \nonumber\\
&= \frac{-\frac{\ell+1}{\ell!}\mathcal{E}_{\ell m} \Rstar^\ell -  \frac{4\pi G\ell}{2\ell+1}\frac{I_\ellm}{\Rstar^{\ell+1}}}{-\frac1{\ell!}\mathcal{E}_{\ell m} \Rstar^\ell + \frac{4\pi G}{2\ell+1}\frac{I_\ellm}{\Rstar^{\ell+1}}}.
\end{align}
We can solve this for the ratio of $I_\ellm$ to $\mathcal{E}_\ellm$, 
\begin{equation}
\frac{I_\ellm}{\mathcal{E}_\ellm}  = -\frac{2\ell+1}{\ell!}\frac{\Rstar^{2\ell+1}}{4\pi G}\frac{\ell+1-\eta(\Rstar)}{\ell+\eta(\Rstar)},
\end{equation}
which is related to $\lambda_\ell$ via \cref{eq:Iell,eq:kelllambdaell}.

It is important to note that this is a constant and depends only on the composition of the body; in particular, the applied tidal field $\mathcal E_\ellm$ drops out of the right-hand side. This is in agreement with our earlier assumption, based on linear response theory, that the induced response $I_\ellm$ is proportional to the tidal field $\mathcal E_\ellm$, cf.~\cref{eq:Newton-Love}.

To compute the Love numbers for a given equation of state, one integrates \cref{eq:radau} outward from $r=0$ to $r=\Rstar$, using the boundary condition $\eta_\ell(r=0)=\ell-2$ \cite{Poisson_Will_2014}.\footnote{This boundary condition follows from a local analysis of \cref{eq:clairaut} near $r=0$, where $\mathcal D$ goes to unity and the solutions go as $r^{\ell-2}$ and $r^{-\ell-3}$. The latter is singular as $r\to0$, so focusing on the growing solution we see that $\eta_\ell\to\ell-2$.}

\subsection{Spinning bodies}

Realistic stars in isolation --- before accounting for tidal interactions --- are nearly, but not perfectly, spherical. Gravity naturally tends to pull matter into a spherical configuration, but rotation introduces deviations from perfect symmetry that must be considered. The primary effect of rotation on tidal interactions is through \textit{dissipation}. However, if we neglect rotationally induced deformations and approximate the star as a rigidly rotating sphere, the tidal response in the rotating frame takes the same form as that of a non-rotating object.

For a discussion of rotational deformations and dissipative tides in the Newtonian context, we refer the reader to \rcite{Poisson_Will_2014}. 
We will return to a fully relativistic treatment of rotating bodies in \cref{sec:EFT}, where we discuss, in particular, rotation and dissipative effects within the effective description of spinning compact objects.

\subsection{Beyond the static and linear response}

In physical setups, \cref{eq:Iell}  generally captures only the leading order tidal response of the object. This relation assumes a time-independent and weak external tidal field. In the context of binary systems, however, these assumptions become increasingly inaccurate as the inspiral progresses, and both frequency-dependent corrections and nonlinear effects in the tidal field's amplitude can become significant. In the Newtonian r\'egime, one would therefore like to replace \cref{eq:Iell} with an expression that schematically looks like
\begin{equation}
\begin{split}
I_{\ell m} & \propto   k_\ell\mathcal{E}_\ellm 
+ {\nu}_\ell  \dot{\mathcal{E}}_\ellm
+ \ddot{k}_\ell  \ddot{\mathcal{E}}_\ellm
 + {\cal O}(\dddot{\mathcal{E}})
\\
& \quad  + \sum_{\ell_1 m_1 \ell_2 m_2} k_{\ell\ell_1\ell_2}^{(2)} {\cal I}_{\ellm}^{\ell_1 m_1 \ell_2 m_2} \mathcal{E}_{\ell_1 m_1}\mathcal{E}_{\ell_2 m_2}
+ {\cal O}(\mathcal{E}^3)
\end{split}
\label{eq:Iell2}
\end{equation}

The second and third terms on the first line of \cref{eq:Iell2} capture time-dependent effects that are linear and quadratic in the tidal field frequency, respectively (with the overdots on ${\mathcal{E}}_{\ell m}$ corresponding to time derivatives).
For non-rotating bodies, corrections involving an odd number of time derivatives break time-reversal invariance and are associated with dissipative effects. The leading dissipative correction is captured by the coefficient $\nu_\ell$, which can be interpreted also as a viscosity-induced delay between the action of the tidal field
and the body's response~\cite{Poisson_Will_2014}. 
At one higher order in derivatives, the term $\ddot{k}_\ell  \ddot{\mathcal{E}}_\ellm$ captures the first (conservative) dynamical correction  to the object's tidal response, and its coefficient $\ddot{k}_\ell$ is related to what is often referred to as the \emph{dynamical Love number}.\footnote{Note the convention that the overdots on $\ddot k_\ell$ do not denote time derivatives but just label it as the coefficient of the $\mathcal{O}(\partial_t^2)$ term.}

On the second line, the terms account  instead for nonlinear response  corrections in the weak field expansion. The leading term is quadratic in the external field's amplitude, and its coefficient ${k}_{\ell\ell_1\ell_2}^{(2)}$ captures the second-order induced response.  The symbol ${\cal I}_{\ell m}^{\ell_1 m_1 \ell_2 m_2}$ is included to  enforce standard angular momentum selection rules.  We will come back to the expansion \eqref{eq:Iell2} in the relativistic context in \cref{sec:EFT} below, where we will provide a fully general and systematic description of  post-adiabatic and nonlinear corrections to the tidal response in terms of an effective expansion.

In the Newtonian context, the expansion \eqref{eq:Iell2} has been applied to systems such as neutron stars --- where the dynamical~\cite{Andersson:2019dwg,Passamonti:2020fur,Steinhoff:2021dsn,Pratten:2021pro,Passamonti:2022yqp,Flanagan:2006sb,Poisson:2020eki,Gupta:2020lnv,Gupta:2023oyy} and nonlinear~\cite{Schenk:2001zm,2012ApJ...751..136W,Weinberg:2013pbi,Weinberg:2015pxa,Yu:2022fzw} coefficients have been linked to the star's normal modes (see also~\rcite{Poisson_Will_2014,Poisson:2020vap,Pitre:2023xsr}) --- as well as to planets, see, e.g.,~\rcite{2020ApJ...891...42W,lai2021jupitersdynamicallovenumber,Idini_2022,2022EGUGA..24.5440D}.

\newpage
\section{Tides in general relativity}
\label{sec:EFT}

\subsection{Love, covariantly}
\label{sec:Lovecovariantly}

The above discussion defines Love numbers unambiguously in Newtonian gravity. But gravity is not really Newtonian --- if it were, we would not be concerned about gravitational radiation from compact-object binaries. What defines the Love numbers in general relativity?

As soon as one attempts to apply the same logic as in \cref{sec:newton} to general relativity, several difficulties arise. First, the Poisson equation \eqref{eq:Poisson} is replaced by the Einstein field equations and, instead of the Newtonian potential $U$, one must work with the metric tensor, which carries two independent degrees of freedom.
One approach is to work with the $g_{tt}$ metric component, which in the Newtonian limit is related to the potential $U$ by $g_{tt} = -e^{2U}$, and the $g_{t\varphi}$ component, which captures the other independent mode in the static limit (see \cref{sec:BHPT} for details). The drawback of this approach is its manifest coordinate dependence, leading to ambiguities: in what coordinate system does the falloff at infinity accurately capture the body's intrinsic static response and not other physical effects? The ideal definition would be a \emph{generally covariant} one, with a clear physical
interpretation.
A further complication arises from the nonlinearity of the Einstein equations: unlike in the Poisson equation, the tidal field and the response now couple through nonlinearities, making an unambiguous extraction of the response coefficients from the full solution less straightforward.

Let us start with a simple ansatz that provides a natural parametrization for the relativistic generalization of \cref{eq:Uellm-2} to the $g_{tt}$ and $g_{t\varphi}$ metric components of a perturbed Schwarzschild black hole~\cite{Thorne:1980ru},
\begin{subequations}\label{eq:intro-g-ansatz}
\begin{align}
g_{tt} &= -1 + \frac{\rs}{r} - \sum_{\ell,m} \frac{2\mathcal{E}_{\ellm}^{(\mathrm{E})}}{\ell(\ell-1)}  \left[r^\ell\left( 1 + a_{\ell,1}^{\mathrm{E}} \frac{\rs}{r} +\dots \right) + 2k_\ell^{(\mathrm{E})}
\frac{\Rstar^{2\ell+1}}{r^{\ell+1}}   
\left(1 + b_{\ell,1}^{\mathrm{E}} \frac{\rs}{r} +\dots\right) \right] Y_{\ellm} + {\cal O}(\dot{{\cal E}},{\cal E}^2) ,
\label{eq:introgtt0}\\
g_{t\varphi} &=  \sum_{\ell ,m}  \frac{2\mathcal{E}_{\ellm}^{(\mathrm{B})}}{3\ell(\ell-1)}  \left[ r^{\ell+1} \left( 1 + a_{\ell,1}^{\mathrm{B}} \frac{\rs}{r} +\dots \right) - 2 k_\ell^{(\mathrm{B})}
 \frac{\ell+1}{\ell} \frac{\Rstar^{2\ell+1}}{r^{\ell}} \left(1 + b_{\ell,1}^{\mathrm{B}} \frac{\rs}{r} +\dots\right)  
\right] \sin\theta\partial_\theta Y_{\ellm} + {\cal O}(\dot{{\cal E}},{\cal E}^2),
\label{eq:introgtphi0}
\end{align}
\end{subequations}
up to subleading time-dependent and nonlinear corrections in the tidal source, which we omit here for simplicity. The (dimensionless) coefficients $k_\ell^{(\mathrm{E})}$ and $k_\ell^{(\mathrm{B})}$ are intended to generalize the response coefficients $k_\ell$ in \cref{eq:Iell} (see also \cref{eq:Uellm-2})~\cite{Hinderer:2007mb,Damour:2009vw,Binnington:2009bb}. The growing and decaying profiles $r^\ell$ and $r^{-\ell-1}$, which in the Newtonian description were unambiguously associated respectively with tidal and response fields in \cref{eq:Uellm-2}, are now multiplied by  prefactors in the form of power laws in $1/r$. The coefficients $a_i$ and $b_i$ in the expansions are determined by the Einstein equations and correspond to subleading relativistic corrections to the solution for $g_{\mu\nu}$.

From \cref{eq:intro-g-ansatz} it is immediately clear that a subtlety arises in general relativity which is not present in Newtonian gravity: at sufficient subleading order in the expansion, namely for $i\geq2\ell+1$, corrections to the tidal field begin to overlap with the induced response terms, leading in particular to an ambiguity between the relativistic correction $a_{\ell,2\ell+1}^{(\mathrm{E/B})}$ and the putative tidal response $k_\ell^{(\mathrm{E/B})}$.
This raises the question of how to unambiguously distinguish the external tidal field contribution from the induced response.
An  approach in this direction was proposed in \rcite{Pani:2015hfa,LeTiec:2020spy,LeTiec:2020bos}, where the body's response --- obtained from the perturbed metric with the tidal field subtracted out --- is calculated using the Geroch--Hansen method~\cite{Geroch:1970cd,Hansen:1974zz}. 
In particular, in \rcite{LeTiec:2020spy,LeTiec:2020bos} (see also \rcite{Charalambous:2021mea}) the tidal field is identified by analytically continuing $\ell\in\mathbb Z$ to real values $\ell\in\mathbb R$.
For a generic non-integer $\ell$, there is no ``$(2\ell+1)$th term'' in the corrections to the external field,
and the tidal response coefficient can be unambiguously extracted. The restriction to $\ell\in\mathbb Z$ is taken at the end of the computation.
Similarly in spirit, an alternative way of disentangling subtle coordinate-dependent effects in the gravitational potential was proposed in \rcite{Kol:2011vg} (see also \rcite{Hui:2020xxx,Rodriguez:2023xjd,Hadad:2024lsf,Charalambous:2023jgq,Glazer:2024eyi,Charalambous:2025ekl}), based on an analytic continuation in the number of spacetime dimensions $D$. Analogously to the analytic continuation in $\ell$, the presence of a tunable parameter $D$ provides a prescription for defining the tidal field, which can then be subtracted from the perturbed metric. A different perspective was taken in \rcite{Gralla:2017djj}, where it is argued that individual Love numbers may not be well defined, but that one can nevertheless extract a body’s response by computing the difference with respect to a reference body of the same mass.

The most common approaches proposed in the literature to unambiguously define and compute relativistic Love numbers in the strong-gravity r\'egime, obtained by solving the Einstein equations for the perturbations of gravitating objects~\cite{Fang:2005qq,Hinderer:2007mb,Damour:2009vw,Binnington:2009bb}, are based on the post-Newtonian (PN) approach~\cite{Poisson:2009qj,Taylor:2008xy,Poisson:2018qqd,Poisson:2020vap,Pitre:2023xsr,Pitre:2025qdf} and effective field theory (EFT)~\cite{Goldberger:2004jt,Goldberger:2005cd}. 
In the following we describe these two methods. First, we focus on the EFT description of gravitating objects. The EFT method is useful because gauge invariance and physical observables are built in, providing a systematic framework to resolve gauge ambiguities that may affect relativistic calculations of Love numbers, which we review in~\cref{sec:love-compute}. The post-Newtonian approach, in which the gravitational moments are obtained from a post-Newtonian matching --- appearing as a property of the compact body when viewed as a skeletonized object moving in a post-Newtonian spacetime --- is summarized in~\cref{sec:post-newtonian}.

\subsection{Effective field theory approach to binary dynamics}
\label{sec:EFTapproach}

Effective field theories are ubiquitous in physics, arising whenever there is a separation of scales. The idea is to parametrize our ignorance about the (possibly unknown, or highly complicated) small-scale physics by identifying a set of low-energy degrees of freedom and symmetries. The strategy is  to write down the most general Langrangian including all possible local operators  constructed from the low-energy degrees of freedom and allowed by the symmetries. These operators are organized as a series expansion  in powers  of derivatives and fields. The symmetries allow one in turn to identify a set of rules to consistently power count  the unknown coefficients in the  Lagrangian. These  coefficients --- or \textit{Wilson couplings} --- can either be constrained experimentally, or determined by matching the EFT to an explicit   ultraviolet completion, when such a theory is available. EFTs have been widely applied across many areas of physics, and it is not too much of a stretch to say that every experimentally successful physical theory to date is an effective field theory.
In the following, we will not review EFTs in full generality, but rather focus on the worldline effective action for gravitating objects.
For comprehensive reviews of effective field theory in a variety of physical contexts relevant to the current problem, see the recent textbook by Burgess \cite{Burgess:2020tbq}, as well as the review articles by Pich~\cite{Pich:1998xt}, Donoghue~\cite{Donoghue:1995cz,Donoghue:2012zc}, Kaplan~\cite{Kaplan:2005es}, Rothstein~\cite{Rothstein:2003mp,Rothstein:2014sra,Rothstein:2015}, Burgess~\cite{Burgess:2007pt}, Goldberger~\cite{Goldberger:2007hy,Goldberger:2022rqf,Goldberger:2022ebt}, Porto~\cite{Porto:2016pyg},  Manohar~\cite{Manohar:2018aog}, Levi~\cite{Levi:2018nxp}, and Penco~\cite{Penco:2020kvy}.

A note on language: The EFT approach was pioneered in particle physics. In particle physics the expansion parameter is typically a ratio of energy scales, namely the energy $E$ probed by some process and the ``cutoff'' $\Lambda$. The theory is built as a systematic expansion in $E/\Lambda$; this may be implemented at the level of the action by, e.g., a derivative expansion $\partial/\Lambda$. Often $\Lambda$ is the mass scale of heavy particles unaccounted for in the EFT, and the breakdown of the effective theory at $E\sim\Lambda$ is interpreted as the fact that these new states can spontaneously pair produce at such energies.

In the worldline effective theory we discuss below, the expansion parameter is, strictly speaking, a ratio of length scales rather than energy scales, although the two are of course closely related. 
Nevertheless we will use the particle physics jargon frequently: we will often refer to the unknown small-scale physics as ``high-energy'' or ``ultraviolet'' (UV) and the large-scale physics described by the EFT as ``low-energy'' or ``infrared'' (IR).

The fundamental idea behind the EFT approach to compact object binaries is essentially the ``zoom-in'' picture discussed at the start of this section, made concrete using the modern language of (quantum) field theory. 
This ``zoom-in'' is made feasible in practice due to the widely separated physical scales involved in the system. To see why, consider the inspiral of two self-gravitating objects, initially evolving adiabatically along non-relativistic bound orbits. Let $r$ denote the orbital separation, and $\Rstar$  the typical size of the bodies in the binary, which we take to be $\Rstar\ll r$.  For  compact objects, we have $\Rstar\gtrsim  \rs\equiv 2GM$. 
The hierarchy $\Rstar\ll r$ is valid typically in the early inspiral, when the evolution of the system is slow. 
Using the virial theorem from Newtonian gravity, one can translate this into a 
condition on the  typical relative velocity:
\begin{equation}
     v^2 \sim \frac{GM}{r} \lesssim \frac{\Rstar}{r} \ll 1 .
\label{eq:virial}
\end{equation}
This  is the key condition underlying the  post-Newtonian  approach (see e.g.~\rcite{Blanchet:2006zz,Blanchet:2013haa} for some reviews, and~\cref{sec:post-newtonian} below), which consists in seeking a solution to the Einstein equations for a compact binary system as an expansion in powers of $v\ll1$.

Note that, in addition to the object's typical size $\Rstar$  and the relative separation $r$, there is another scale in the problem: the wavelength $\lambdar$ of the emitted radiation. This can be roughly estimated as the inverse of the orbital frequency, $\lambdar\sim r/v$. 
The EFT approach to gravitationally bound systems relies on the crucial observation that these three length scales satisfy a strong parametric separation: $\Rstar\ll r\ll \lambdar$, with the relative ratios controlled by the same parameter $v$.
This hierarchy allows one to effectively  decouple physical effects that enter at different orders in $v$, leading to significant conceptual and technical simplifications.
As a result, the two-body problem can be reformulated in terms of a hierarchy (or ``tower'') of effective field theories~\cite{Goldberger:2006bd}, each associated with a different scale and defined by a finite set of degrees of freedom and symmetries.
For instance, at distances $d$ such that $\Rstar\ll d\ll r$, 
orbital dynamics can be neglected, and one can construct an EFT for a single, isolated compact object. Assuming the object's internal dynamics is gapped, its microscopic structure can be integrated out. The resulting effective Lagrangian describes low-energy degrees of freedom, which correspond to the Goldstone bosons of spontaneously broken Poincaré symmetries.
Zooming out to orbital scales, one can construct an EFT that includes radiation and potential gravitons coupled to non-relativistic point particles. Further away, integrating out potential gravitons yields an effective description of the binary as a composite particle. Finally, integrating out radiation modes at the scale $\lambdar$ leads to an EFT  of dynamical multipole moments.
This multi-stage EFT framework for perturbative gravitational dynamics --- which was pioneered in \rcite{Goldberger:2004jt,Goldberger:2005cd}   (see  \rcite{Goldberger:2006bd,Foffa:2013qca,Rothstein:2014sra,Porto:2016pyg,Levi:2018nxp,Goldberger:2022ebt,Goldberger:2022rqf} for some reviews) --- provides a modern reformulation of the PN expansion, in a style more closely aligned with techniques used in high-energy physics. In what follows, we will focus on the r\'egime suited to the early inspiral, namely the first stage of this framework: the EFT of an isolated compact object.

\subsubsection{Worldline effective field theory for compact objects}
\label{sec:WEFT}

Viewed from a sufficient distance relative to a given experimental accuracy, any self-gravitating object looks like a point particle: it sources the gravitational field of a point particle, and follows a geodesic through spacetime~\cite{Damour:1984rbx,Damour:1995kt,Goldberger:2004jt,Goldberger:2005cd}. These of course are not necessarily true of extended objects, and as one moves closer to the object (or increases their experimental sensitivity) there will be deviations from the point-particle picture due to the finite size and structure of the object. These corrections are controlled by a small (dimensionless) parameter, in this case the ratio of the size of the object to the length scale probed experimentally, e.g., the radius of curvature of the background spacetime or the orbital separation between binary companions. 

Effective field theory consists of a set of tools for systematically computing these finite-size effects by taking maximal advantage of the separation of scales. In essence this is done by Taylor expanding in the small parameter as early as possible. To a given level of experimental accuracy, the point-particle picture needs to be supplemented with only a finite number of additional interactions. These are encoded by local operators in the Lagrangian, whose Wilson coefficients are universal, meaning that once they are measured or calculated once, the same values can be used in any other setup. Moreover these interactions can be determined using the standard rules of ``bottom-up'' EFT: one simply writes down every local term allowed by the symmetries of the system, organized in powers of the ratio of scales.

Let us demonstrate this concretely.\footnote{We will be laser focused on the application of worldline effective field theory to tidal effects. For a comprehensive and pedagogical overview of the EFT approach to the binary inspiral across a range of scales, we refer to \rcite{Porto:2016pyg,Goldberger:2022ebt,Levi:2018nxp}.} 
First of all, following the standard lore of EFTs, we will need to identify the relevant low-energy degrees of freedom. Assuming non-rotating objects and gapped internal dynamics, the low-energy degrees of freedom are: (i) the gravitational metric tensor $g_{\mu\nu}$;  and (ii) the compact object’s worldline ${\X}^\mu(\affine)$, parametrized by an affine parameter $\affine$. 
Should the point particle possess more quantum numbers (such as angular momentum, electric/magnetic charge, etc.), additional degrees of freedom (such as a spin vector, electromagnetic field, etc.) must also be included. For the initial presentation, we assume that the point particle has no conserved charges other than its mass. Since astrophysical compact objects typically have non-trivial angular momentum, we discuss the generalization to spinning objects in \cref{sec:includingspin}.

The dynamics  then follows from a worldline action that couples $\X^\mu(\affine)$  to the gravitational field $g_{\mu\nu}$. 
To construct the operators in the EFT using the ingredients above,  we need to identify the symmetries of the problem.
These will  dictate the  form of the EFT operators, and provide the rules for power counting. 
The compact object spontaneously breaks Poincar\'e symmetries, which are nonlinearly realized by the worldline degrees of freedom. The symmetries  required in the action are thus those associated with gauge redundancy in the  variables  $\{g_{\mu\nu},\X^\mu(\affine)\}$, i.e.,~: (i) spacetime diffeomorphism invariance, $\delta \X^\mu = \xi^\mu(x)$, acting on the metric, as usual, as $\delta g_{\mu\nu}(x)=\nabla_\mu \xi_\nu+ \nabla_\nu \xi_\mu$;
and (ii) reparametrization invariance  of the worldline, $\affine\rightarrow \tilde \affine(\affine)$ \cite{Goldberger:2004jt,Goldberger:2020fot,Porto:2016pyg,Levi:2018nxp,Goldberger:2022ebt,Goldberger:2022rqf}.
To enforce reparametrization invariance, it is convenient to define an \textit{einbein} $e(\affine)$, such that $\tilde  e(\tilde \affine)\dd \tilde  \affine = e(\affine) \dd{\affine}$.\footnote{The name einbein comes from the fact that it is a one-dimensional vielbein along the worldline. A standard motivation for the einbein formulation is that the action \eqref{eq:PP} does not apply to massless particles, although this is of course not relevant in the present context, where the point particle describes a massive compact object.}

Equipped with the fields and the symmetries of the problem, we can now construct an effective theory around the point-particle approximation.
At leading order, the dynamics of a point particle in a background spacetime with metric $g_\mn$ is governed by the usual Nambu--Goto action on the worldline,
\begin{align}
S_\mathrm{pp} &= -m\int\dd \tau \nonumber\\
&= -m\int \dd {\affine}\sqrt{-g_\mn \frac{\dd \X^\mu}{\dd{\affine}}\frac{\dd \X^\nu}{\dd{\affine}}},\label{eq:PP}
\end{align}
where $\tau$ is the proper time along the particle's worldline, and $m$ is a free parameter with dimensions of  energy, which we later relate  to the mass of the object through matching (cf. \cref{sec:matching}).\footnote{Note that extremizing $S_{\rm pp}$ gives the usual geodesic equation for a test particle in a gravitational field:
\[
0= \delta \left[m \int \dd \tau \right] 
= - m \int \dd \tau \left[  
 g_{\mu\nu} \frac{\dd{\X}^\mu}{\dd \tau} \frac{\dd}{\dd \tau}
 + \frac{1}{2}\partial_\nu g_{\rho\sigma} \frac{\dd{\X}^\rho}{\dd \tau}\frac{\dd{\X}^\sigma}{\dd \tau}
\right]\delta \X^\nu,
\]
implying
\begin{equation}
a^\mu \equiv \frac{\dd^2 \X^\mu}{\dd \tau^2} + %
\Gamma^\mu_{\rho\sigma }\frac{\dd{\X}^\rho}{\dd \tau}\frac{\dd{\X}^\sigma}{\dd \tau} = \frac{\dd \X^\rho}{\dd \tau} \nabla_\rho \frac{\dd \X^\mu}{\dd \tau} =0 .
\label{eq:acc}
\end{equation}
}

The point-particle action can also be written without square roots by integrating in the einbein, 
\begin{equation}
\dd \tau^2 = m^2 e^2\dd{\affine}^2.
\label{eq:dtau0}
\end{equation}
The action for the point particle of mass $m$ in the einbein formalism takes the Polyakov form
\begin{equation}
S_\mathrm{pp} = \frac12\int\dd{\affine} \left(e^{-1}g_\mn \frac{\dd \X^\mu}{\dd{\affine}}\frac{\dd \X^\nu}{\dd{\affine}}-em^2\right),
\label{eq:PP-ein}
\end{equation}
which is equivalent to the original action \eqref{eq:PP}, as one can easily check by simply integrating out the einbein using its equation of motion:
\begin{equation}
\frac1{e^2}g_\mn\frac{\dd {\X}^\mu}{\dd{\affine}}\frac{\dd {\X}^\nu}{\dd{\affine}}+m^2=0.
\end{equation}

Describing the coupling of the point particle to the perturbations of the metric $g_{\mn}$ requires specifying the gravitational dynamics in the ``bulk.'' In general relativity, this is encoded by the Einstein--Hilbert   action. The point-particle term \eqref{eq:PP-ein} is thus supplemented  by\footnote{To add these two actions, which have different integration measures, we can use a delta function,
\[
\int\dd\sigma = \int\dd^4x\;\delta^{(4)}(x-\X(\sigma)).\]}
\begin{equation}\label{eq:S-EH}
S_\mathrm{EH} = \frac{\Mp^2}2\int\dd^4x\sdg R,
\end{equation}
with $R$ the Ricci scalar associated to $g_\mn$.
In addition, because the leading-order term  $S_{\rm pp}$ is universal,  to capture the object's finite size and its model-dependent  internal structure  we need to augment $S_{\rm pp}$ with additional higher-order terms. These are organized as an expansion in derivatives. As such, they are ``irrelevant'' at large distances, where they provide only small corrections to the point-particle approximation. In contrast, they all become  comparable in size at the cutoff of the EFT --- set by the inverse radius of the object --- where the effective expansion breaks down. We will denote the higher-order operators collectively as $S_\mathrm{ho}$.

All in all, the  effective action can be written as
\begin{equation}\label{eq:S-EFT}
S_\mathrm{EFT} = S_\mathrm{EH} + S_\mathrm{pp} + S_\mathrm{h.o.},
\end{equation}
where $S_\mathrm{ho}$ contains the EFT corrections to $S_\mathrm{pp}$.\footnote{There can also be higher-order corrections to the Einstein--Hilbert action due to gravitational physics beyond general relativity, which we will ignore  (see \cref{sec:love-beyond-gr} for a discussion on Love numbers in theories beyond general relativity).} 

All possible worldline scalars appearing in $S_\mathrm{ho}$ can be classified using the relativistic theory of tidal expansions~\cite{Zhang:1986cpa,Damour:1990pi,Damour:1991yw}.
The only constraint on the terms appearing in $S_\mathrm{ho}$ is that they be local operators built out of the low-energy degrees of freedom --- the worldline $\X^\mu$ and the metric $g_\mn$, as well as a local Lorentz frame along the worldline if we wish to describe spinning particles (cf. \cref{sec:includingspin}) --- and invariant under the symmetries of the system, namely spacetime diffeomorphism invariance for $g_\mn$ and reparametrization invariance of the worldline. We can ensure the latter by using the proper time as our worldline parameter, ${\affine}=\tau$. In this parametrization the particle's 4-velocity $u^\mu$ is properly normalized,
\begin{equation}
u^\mu = \frac{\dd \X^\mu}{\dd\tau},\quad g_\mn u^\mu u^\nu = -1.
\label{eq:umu1}
\end{equation}

To implement diffeomorphism invariance we simply work with tensorial quantities built out of the fields available to us, namely the metric $g_\mn$, the covariant derivative $\nabla_\mu$, the Riemann tensor $R_{\mn\ab}$, and the 4-velocity $u^\mu$. Because $u^\mu$ picks out a preferred time direction (corresponding to the particle's rest frame), we can decompose tensors into space and time components using $u^\mu$ and the projector onto spatial slices orthogonal to the worldline,
\begin{equation} \label{eq:spat-prog}
P^\mu_\nu=\delta^\mu_\nu+u^\mu u_\nu,
\end{equation}
without breaking diffeomorphism invariance. This is crucial because temporal and spatial derivatives are not on equal footing in the EFT we wish to construct; indeed, the leading-order physics is time-independent. We will therefore write the EFT action in terms of spatial indices in the particle's rest frame $(i,j,\cdots)$ rather than spacetime indices $(\mu,\nu,\cdots)$. This does not break spacetime diffeomorphism invariance: the spatial indices should be thought of as shorthand for projection with $P^\mu_\nu$, so that 4D covariant expressions can be recovered using, e.g.,
\begin{equation}
\nabla_i \to P^\mu_\nu\nabla_\mu .
\end{equation}

Rather than the Riemann tensor it is more convenient to work with its tracefree counterpart, the Weyl tensor $C_{\mn\ab}$, which differs from $R_{\mn\ab}$ by terms involving the Ricci tensor $R_\mn$ and Ricci scalar $R$,
\begin{equation}\label{eq:Weyl-def}
    C_{\mn\ab} = R_{\mn\ab} -2g_{[\mu|[\alpha}R_{\beta]|\nu]} + \frac13 R g_{\mu[\alpha}g_{\beta]\nu}.
\end{equation}
These vanish when the vacuum Einstein equation $R_\mn=0$ is satisfied, and it is well known that EFT operators which vanish on shell (i.e., when the lowest-order equations of motion are satisfied) can be removed by a field redefinition \cite{Arzt:1993gz}. The reason is straightforward. An EFT action consists of the lowest-order piece $S_0$ (here $S_\mathrm{EH}+S_\mathrm{pp}$) plus corrections $\epsilon S_1+\epsilon^2S_2+\cdots$ (here $S_\mathrm{ho}$) which are suppressed relative to $S_0$ by some small parameter $\epsilon$ (here for instance the ratio of the size of the object to the orbital scale). Now suppose we perform an $\mathcal{O}(\epsilon)$ field redefinition, $Y\to Y+\epsilon\,\delta Y$, where $Y$ stands for some field(s) in the EFT and $\delta Y$ is some functional of the field(s). To $\mathcal{O}(\epsilon)$ the action transforms as
\begin{align}
S_\mathrm{EFT}[Y] &= S_0[Y] + \epsilon S_1[Y] +\mathcal{O}(\epsilon^2)\nonumber\\
&\to S_0[Y+\epsilon\delta Y] + \epsilon S_1[Y]+\mathcal{O}(\epsilon^2) \nonumber\\
&= S_0[Y] + \epsilon\Bigg( S_1[Y] + \underbrace{\frac{\delta S_0}{\delta Y}}_\text{eom}\delta Y\Bigg) +\mathcal{O}(\epsilon^2).
\end{align}
The effect of the field redefinition is to add to $S_1$ a term proportional to the equation of motion obtained by varying $S_0$ with respect to $Y$. Therefore if such a term is already present in $S_1$, we can remove it by performing a field redefinition at $\mathcal{O}(\epsilon)$ and suitably choosing the functional $\delta Y$. The reader may object that there are additional terms generated at $\mathcal{O}(\epsilon^2)$. However these will by construction only shift the coefficients of terms already present in $S_2$. Any terms in $S_2$ which vanish on shell can be removed by a field redefinition $Y\to Y+\epsilon^2\delta Y$, and so on to whatever order in the EFT expansion is necessary for a given practical purpose.

Since we have a preferred time direction picked out by $u^\mu$, we can decompose the Weyl tensor
into electric and magnetic components,
\begin{equation}
E_\mn \equiv C_{\mu\alpha\nu\beta}u^\alpha u^\beta,\qquad B_\mn \equiv \frac12\epsilon_{\mu\alpha\beta\rho}C^{\alpha\beta}{}_{\nu\sigma}u^\rho u^\sigma.
\end{equation}
These are traceless, $g^\mn E_\mn = g^\mn B_\mn = 0$, and are purely spatial in the sense that $E_\mn u^\mu = B_\mn u^\mu = 0$, or in the rest frame of the particle, $E_{0\mu}=B_{0\mu}=0$. We will accordingly write them as $E_\Ij$ and $B_\Ij$. That the Weyl tensor can be fully characterized by its electric and magnetic components follows from the symmetries of $C_{\mn\ab}$.\footnote{A useful relation  expressing the Weyl tensor in terms of its electric and magnetic components is:
\begin{equation}
C_{\mu\nu\alpha\beta} = \left(g_{\mu\nu\rho\sigma}g_{\alpha\beta\lambda\tau}-\epsilon_{\mu\nu\rho\sigma}\epsilon_{\alpha\beta\lambda\tau}\right)u^\rho u^\lambda E^{\sigma\tau} + \left(\epsilon_{\mu\nu\rho\sigma}g_{\alpha\beta\lambda\tau}+g_{\mu\nu\rho\sigma}\epsilon_{\alpha\beta\lambda\tau}\right)u^\rho u^\lambda B^{\sigma\tau},
\label{eq:weyltoEB}
\end{equation}
where $\epsilon_{\mu\nu\rho\sigma}$ is the Levi--Civita tensor  %
and $g_{\mu\nu\rho\sigma}\equiv g_{\mu\rho}g_{\nu\sigma}-g_{\mu\sigma}g_{\nu\rho}$. We stress that \cref{eq:weyltoEB} is valid only in $D=4$ spacetime dimensions. In $D>4$, the Weyl tensor cannot be expressed  in terms of just $E_{\mu\nu}$ and $B_{\mu\nu}$.
In fact, recall that the Weyl tensor has the same symmetries as the Riemann tensor, while being  in addition trace-free, i.e.,~${C^\rho}_{\mu\rho\nu}=0$. This implies that the number of independent components of the Weyl tensor in $D$ dimensions is
\begin{equation}
\frac{D^2(D^2-1)}{12}- \frac{D(D+1)}{2} = \frac{D(D+1)}{2} \left[ \frac{D(D-1)}{6} -1 \right] .
\label{eq:countD}
\end{equation}
For instance, in $D=4$  the Riemann tensor has $20$ independent components, $10$ of which are contained  in the Weyl tensor. Moreover, the counting \eqref{eq:countD} vanishes for $D=3$, implying that $C_{ik jl}$ is not independent from $C_{0i0j}$ and  $C_{0ijk}$; that is, in $D=4$ the EFT is a functional of just $E_{\mu\nu}$ and $B_{\mu\nu}$. In $D>4$, one would instead need to include in the EFT also operators involving   $C_{ik jl}$ (see, e.g., \rcite{Hui:2020xxx}).
}
This can also be seen geometrically: holding one pair of indices fixed, Weyl is an antisymmetric $(0,2$) tensor, or 2-form (cf. \ref{app:diff-forms}). In four dimensions, any 2-form can be written as a sum of self-dual and anti-self-dual pieces, corresponding to the electric/magnetic decomposition.

The task of constructing $S_\mathrm{ho}$ then boils down to constructing local operators built out of the ingredients
\begin{equation}
g_\mn,\quad\nabla_\mu,\quad u^\mu,\quad E_\Ij,\quad B_\Ij,
\end{equation}
and organizing them in an expansion in the number of (spatial and temporal) derivatives per field. For instance, the leading order correction is built out of the squares of the electric and magnetic parts of the Weyl tensor,\footnote{A coupling between the electric and magnetic tensors $c^{(\mathrm{EB})}_2E_\Ij B^\Ij$ is also allowed by the symmetries; we will exclude such terms by additionally imposing parity invariance, i.e., invariance under time reversal $t\to-t$, which is a good approximation for the dynamics of compact objects like neutron stars which are composed of Standard Model matter. This assumption may be violated by parity-violating physics beyond the Standard Model, in which case $c^{(\mathrm{EB})}_2$ can be comparable in size to $c^{(\mathrm{E})}_2$ and $c^{(\mathrm{B})}_2$ \cite{Modrekiladze:2022ioh}.}
\begin{equation}
S_\mathrm{h.o.} = \int\dd\tau\left(c_2^{(\mathrm{E})} E_\Ij E^\Ij + c_2^{(\mathrm{B})} B_\Ij B^\Ij + \cdots\right),
\label{eq:E2B2-0}
\end{equation}
where $c_2^{(\mathrm{E})}$ and $c_2^{(\mathrm{B})}$ are Wilson coefficients encoding information about the size and shape of the object. These particular operators turn out to describe the $\ell=2$ multipole moments. The higher multipoles are encoded in  operators involving spatial derivatives of the $E_{ij}$ and $B_{ij}$ fields. It is common to choose the following basis for these higher-derivative operators:
\begin{equation}
\left(\nabla_{\langle i_1}\cdots \nabla_{i_{\ell-2}}E_{i_{\ell-1}i_\ell\rangle}\right)^2,\quad \left(\nabla_{\langle i_1}\cdots \nabla_{i_{\ell-2}}B_{i_{\ell-1}i_\ell\rangle}\right)^2,
\label{eq:E2B2}
\end{equation}
where angular brackets denote traceless symmetrization. In the large-distance limit described by the EFT, the metric is a perturbation of flat space, $g_\mn = \eta_\mn + 2h_\mn/\Mp$.
To leading order in the metric fluctuations $h_\mn$, the operators \eqref{eq:E2B2} are the only ones we are allowed to write down, so that the leading-order correction to the EFT action is
\begin{equation}
S_\mathrm{ho} = \int\dd\tau\displaystyle\sum_{\ell=2}^\infty\left[\lambdaEell E_{i_{1}\cdots i_\ell}E^{i_{1}\cdots i_\ell}
+ \lambdaBell 
B_{i_{1}\cdots i_\ell} B^{i_{1}\cdots i_\ell}
\right] 
+ \mathcal{O}(h^2),
\label{eq:S-ho}
\end{equation}
where we have introduced the notation 
\begin{equation}
E_{i_{1}\cdots i_\ell}\equiv\partial_{\langle i_1}\cdots \partial_{i_{\ell-2}}E_{i_{\ell-1}i_\ell\rangle},
\qquad
B_{i_{1}\cdots i_\ell}\equiv\partial_{\langle i_1}\cdots \partial_{i_{\ell-2}}B_{i_{\ell-1}i_\ell\rangle} .
\label{eq:EEEBBB}
\end{equation}

The Wilson coefficients $\lambdaEell$ and $\lambdaBell$ are nothing other than the static Love numbers, up to a normalization constant to be determined \cite{Goldberger:2004jt,Damour:2009vw,Kol:2011vg}. These provide the sought-after generalization of the Love numbers to general relativity; they are manifestly covariant because the effective theory is, and they encode precisely the information we want Love numbers to encode, namely the internal structure of the object beyond the point-particle approximation. 

To understand the scaling of the Love number coupling, it is convenient to momentarily restore the factors of the speed of light in \cref{eq:S-ho} such that the limit $c \to \infty$ reproduces the Newtonian result~\cite{Henry:2019xhg}. This will allow us to make the power counting more explicit. The action of the body becomes
\begin{equation}
	\int \dd \tau \left\{ - m c^2+ c^4\sum_{\ell=2}^\infty\left[\lambdaEell E_{i_{1}\cdots i_\ell}E^{i_{1}\cdots i_\ell}
+ \lambdaBell 
B_{i_{1}\cdots i_\ell} B^{i_{1}\cdots i_\ell}
\right]  
	\right\} \; .
    \label{scalingac}
\end{equation}
The combinations $c^2 E_{\mu\nu}$ and $c^2 B_{\mu\nu}$ have dimensions $T^{-2}$, where $T$ is a characteristic timescale. Taking $T$ to be the  dynamical timescale of the object, $T \sim (\Rstar^3 / G m)^{1/2}$, and requiring \cref{scalingac} to have the dimension of an action, one  obtains the scaling~\cite{Henry:2019xhg}
\begin{equation}
\lambdaEell = \frac{1}{\ell! (2\ell-1)!!} \frac{\Rstar^{2\ell+1}}{G}\lambdaEellbar,
\qquad
\lambdaBell = \frac{\ell(\ell-1)}{2(\ell+2)!(2\ell-1)!!} \frac{\Rstar^{2\ell+1}}{G}\lambdaBellbar ,
\label{eq:lambdakell}
\end{equation}
where $\lambdaEellbar$ and $\lambdaBellbar$ are dimensionless numbers, with numerical prefactors chosen to match standard conventions in the literature~\cite{Henry:2019xhg,Poisson:2020vap}.

At subleading order in the EFT expansion, one can add time-derivative operators, e.g., $\dot{E}_\mn^2$,
where $\dot E_\mn \equiv u^\alpha\nabla_\alpha E_\mn$, which describe subdominant time-dependent corrections to the adiabatic evolution of the system. We will return to this in \cref{sec:dissipation} below.
One can also consider operators nonlinear in the metric perturbation $h_\mn$, such as $E_\mn E^{\nu\alpha} E^\mu_\alpha$.
These types of operators capture nonlinear deformations and responses of the point particle, which we will briefly discuss in \cref{sec:nonlinearities}.

\subsubsection{Including spin}
\label{sec:includingspin}

We have reviewed the construction  of  the worldline EFT for non-rotating objects.
Since astrophysical sources typically possess spin, it is important to generalize the EFT \eqref{eq:S-EFT} to account for rotation.

The treatment of spin in general relativity has a long and well-developed history. In the context of compact binary inspirals, considerable effort over several decades has gone into constructing a systematic description of gravitating spinning bodies. An effective action approach --- building on the foundational works of Refs.~\cite{Hanson:1974qy} and \cite{Bailey:1975fe} in flat and curved spacetimes, respectively, and extending the EFT framework for non-spinning, spherically symmetric objects introduced in \rcite{Goldberger:2004jt} --- was first formulated in \rcite{Porto:2005ac}. This framework was subsequently refined in \rcite{Porto:2008tb,Porto:2008jj}, where a Routhian formulation following \rcite{Yee:1993ya} was adopted.
In this approach, spin effects are incorporated perturbatively, with power counting based on a small $v/c$ expansion. A further extension within the conservative sector, achieving fourth post-Newtonian order (4PN) accuracy for rapidly rotating compact objects (valid until dissipative effects, which we discuss in \cref{sec:dissipation}, become relevant~\cite{Poisson:2004cw}), was developed in \rcite{Levi:2015msa} (see also~\rcite{Levi:2008nh,Levi:2010zu,Hergt:2011ik,Levi:2014sba,Levi:2014gsa,Levi:2015uxa,Levi:2015ixa} for related work).

In what follows, we review how to incorporate spin into the point-particle EFT, focusing for the moment on conservative effects only. For more comprehensive reviews, see, e.g., \rcite{Porto:2016pyg,Levi:2018nxp,Goldberger:2022rqf}.

{
In order to describe spinning objects, the point-particle description must be supplemented with the additional structure of a local frame $e_a^\mu(\affine)$, which encodes the orientation of the object~\cite{Goldberger:2020fot,Hanson:1974qy,Bailey:1975fe,Porto:2005ac,Steinhoff:2011sya,Delacretaz:2014oxa} (see also \rcite{Porto:2016pyg,Levi:2018nxp,Charalambous:2021mea,Goldberger:2022ebt,Goldberger:2022rqf}):\footnote{Although we will mainly focus on four-dimensional spacetimes, the construction has an analogue in higher-spacetime dimensions, see, e.g., \rcite{Glazer:2024eyi}.}
\begin{equation}
e_a^\mu(\affine),
\qquad
\qquad  a = 0,1,2,3  .
\end{equation}
The vielbein $e^\mu_a(\affine)$
carries both a spacetime index, $\mu$, and internal SO$(3,1)$ particle frame index $a = 0,1,2,3$.
It can be thought of as a mapping between the fixed background spacetime and the instantaneous orientation of the object.
The vielbein is related to the spacetime metric $g_{\mu \nu}$ and the internal metric $\eta_{ab}$ through
\begin{equation}
\label{eq:restframevielbein}
g_{\mu\nu}e_a^\mu e_b^\nu = \eta_{ab},\qquad
\eta_{ab}e^a_\mu e^b_\nu = g_{\mu\nu},
\end{equation}
i.e., $e^\mu_a e^a_\nu = \delta^\mu_\nu$  and $e^\mu_a e^b_\mu = \delta^b_a$.
In the particular case of Minkowski spacetime, $g_{\mu\nu} = \eta_{\mu\nu}$, the vielbein at any given instant is simply the Lorentz transformation into the particle's rest frame.\footnote{In the absence of gravity, the formalism was first introduced by Regge and Hanson \cite{Hanson:1974qy}, to treat the classical motion of relativistic spinning particles coupled to electromagnetic fields.  The extension of the Regge--Hanson formalism to curved spacetime, and its applications to perturbative binary dynamics first appeared in \rcite{Porto:2005ac}. A modern treatment of spinning particles from the point of view of nonlinearly realized symmetries  can be found in \rcite{Delacretaz:2014oxa}.}
}

{
The rotation of the particle relative to a fixed inertial frame is encoded in
\begin{equation}
\Omega^{ab} =g^{\mu \nu}e^a_\mu \frac{D}{D{\affine}}e^b_\nu\,,
\label{Omegadef}
\end{equation}
where we have defined
\begin{align}
    \frac{D}{D{\affine}}e^a_\mu &= \frac{\dd{\X}^{\rho}}{\dd{\affine}}\nabla_\rho e^a_\mu \nonumber\\
    &= \frac{\dd{e}^a_\mu}{\dd{\affine}} -\Gamma^\rho_{\sigma \mu}\frac{\dd{\X}^\sigma}{\dd {\affine}} e^a_\rho,
\end{align}
corresponding to the rotation of the vielbein along the worldline's trajectory. $\Omega^{ab}$ represents the angular velocity of the particle, and is antisymmetric, $\Omega^{ab} = -\Omega^{ba}$.
}

{
Compared to the construction for non-rotating objects, the worldline EFT now includes an additional spin degree of freedom. Operators in the EFT encompass all possible interactions among the low-energy fields (i.e., the metric, the worldline, and the spin) and, in addition to general covariance and reparametrization invariance, must also respect the internal Lorentz invariance of the local frame field:
\begin{equation}
e^a_\mu (\affine) \rightarrow \tilde{e}^a_\mu(\affine) = {\Lambda^a}_b e^b_\mu(\affine)\, ,
\end{equation}
with ${\Lambda^a}_b$  a constant Lorentz matrix.\footnote{Note that because of Lorentz invariance, the definition of the tetrad is unique up to a local Lorentz transformation. One can transform the local `$a$' index by a local Lorentz transformation, i.e.,~$\tilde{e}^a_\mu = {\Lambda^a}_b e^b_\mu$, without changing  $g_{\mu\nu}$:
\[
\tilde{g}_{\mu\nu} = \eta_{ab}\tilde e^a_\mu \tilde e^b_\nu
= \eta_{ab} {\Lambda^a}_c {\Lambda^b}_d  e^c_\mu  e^d_\nu
= \eta_{cd}   e^c_\mu  e^d_\nu = g_{\mu\nu} .
\]
}
}

{
A common approach to describing the dynamics of a spinning point particle  is  to use a
first-order form,\footnote{There is a different way of proceeding by employing the Routhian formalism, e.g., \cite{Porto:2006bt,Porto:2016pyg}.} where the point-particle action is given by \cite{Porto:2005ac,Porto:2016pyg,Goldberger:2020fot}
\begin{equation}
    S_{\rm pp}= \int\dd{\affine}\left[\frac{\dd \X^\mu}{\dd {\affine}} p_a e_\mu^a+\frac{1}{2}S^{ab}\Omega_{ab}-\frac{1}{2}e\big(p_ap^a+m^2(p,S)\big)+e\lambda_aS^{ab}p_b
    \right],
    \label{eq:pointparticleaction}
\end{equation}
where the momentum $p^a$ and the spin $S_{ab}$, 
\begin{equation}
p_\mu \equiv \frac{\delta S_{\rm pp}}{\delta \dot \X^\mu}\,,
\qquad
S^{ab} \equiv 2\frac{\delta S_{\rm pp}}{\delta \Omega_{ab}}\,,
\end{equation}
are conjugate variables to the particle's velocity $\dot{\X}^\mu\equiv \dd\X^\mu/\dd {\affine}$  and angular momentum $\Omega_{ab}$.\footnote{Note that $\dot \X^\mu$ and $u^\mu$ defined in \cref{eq:umu1} coincide in flat spacetime. However, they can differ once gravitational corrections are included. We therefore prefer to use different notation to distinguish them.} As usual we use the vielbein and its inverse to convert between spacetime and Lorentz indices,
\begin{equation}\label{eq:spacetime-p-s}
p_\mu = e^a_\mu p_a ,
\qquad
S_{\mu\nu} = e^a_\mu e^b_\nu S_{ab} .
\end{equation}

In the formulation \eqref{eq:pointparticleaction}, the quantity $m^2$ is in principle an arbitrary function of all possible scalars constructed out of   $p_a$, $S_{ab}$ and $g_{\mu\nu}$. The form of this function is not predicted by the point-particle EFT, but rather is determined by matching to the ultraviolet theory of the extended object. Note that $m^2(p,S)$  determines the Regge trajectory \cite{Hanson:1974qy} of the spinning particle, i.e., the relation between the invariant mass $p^2$ and the spin, which follows from the variation of $S_{\rm pp}$ with respect to $e(\affine)$.
Finally, $\lambda_a$ is a Lagrange multiplier, which is necessary to enforce a supplementary constraint on $S^{ab}$: 
\begin{equation}
S^{ab}p_b =0 ,
\label{eq:ssc}
\end{equation}
which is usually referred to as the \textit{spin-supplementary condition}.\footnote{Recall that information about the particle's angular momentum is encoded in the angular velocity matrix $\Omega_{ab}$, which   is antisymmetric. As such, it has    $D(D-1)/2$ independent components in $D$ spacetime dimensions.  However,  the angular momentum of classical objects belongs to  the adjoint representation of the spatial rotation group SO$(D-1)$, whose dimension is  $(D-1)(D-2)/2$ independent components. 
The spin-supplementary condition is thus necessary to reduce the number of independent components of $\Omega_{ab}$ from $D(D-1)/2$ down to $(D-1)(D-2)/2$,
\[
\frac{D(D-1)}{2}- (D-1) = \frac{(D-1)(D-2)}{2},
\]
as required by Poincar\'e invariance for a physical spin degree of freedom.}

The variation of \cref{eq:pointparticleaction} with respect to the kinematic variables $(\X^\mu, p_a,e^a_\mu,S_{ab},\lambda_a,e)$ yields the Papapetrou--Mathison--Dixon equations of motion \cite{Mathisson:1937zz,Dixon:1970zza,Papapetrou:1951pa}.
For instance, varying with respect to $e^a_{\mu}$, we have
\begin{equation}
\frac{\dd \X^\mu}{\dd {\affine}}p_a +\frac{1}{2}S_{ab} g^{\mu\alpha} \frac{D}{D {\affine}} e^b_\alpha +\frac{1}{2}  \frac{D}{D {\affine}}(S_{ab} e^b_\alpha g^{\alpha\mu}) =0.
\end{equation}
Contracting with $e^a_\beta g^{\beta \nu}$, and rewriting the last term as 
\begin{equation}
    \frac{1}{2} e^a_\beta g^{\beta\nu} \frac{D}{D {\affine}}(S_{ab} e^b_\alpha g^{\alpha\mu}) =  \frac{1}{2}\frac{D}{D {\affine}} S^{\nu\mu} + \frac{1}{2} S_{ab} e^a_\alpha g^{\alpha\mu} \frac{D}{D {\affine}} (e^b_\beta g^{\beta\nu}) 
\end{equation}
yields, after taking the antisymmetric combination,
\begin{equation}
\frac{D}{D{\affine} } S^{\mu\nu} = \frac{\dd \X^\mu}{\dd {\affine}} p^\nu - \frac{\dd \X^\nu}{\dd {\affine}}p^\mu .
\label{eq:DSmunu}
\end{equation}
Analogously, the variation with respect to $\X^\mu({\affine})$ yields \cite{Blanchet:2013haa}
\begin{equation}
\frac{D}{D{\affine}} p^\mu = -\frac{1}{2} {R^\mu}_{\nu\rho\sigma} \frac{\dd \X^\nu}{\dd {\affine}} S^{\rho\sigma}, 
\label{eq:accspin}
\end{equation}
which generalizes \cref{eq:acc} for non-zero spin. The right-hand side of \cref{eq:accspin}  corresponds to the usual Papapetrou--Mathison--Dixon force of a spinning point particle, in the absence of other interactions.

In the absence of dissipation, the variation with respect to $p^\mu$ and $S^{\mu\nu}$ gives a relation between  $(p^\mu, S^{\mu\nu})$ and $(\dot{\X}^\mu, \Omega^{\mu\nu})$. %
On the other hand, the relation between $\dot{\X}^\mu$ and $p^\mu$ follows in general from \cref{eq:DSmunu,eq:accspin} and the constraint $S^{ab}p_b=0$. In fact, from $\frac{D}{D{\affine} }(S^{\mu\nu}p_\nu)=p_\nu\frac{D}{D{\affine} }S^{\mu\nu}+ {S^{\mu}}_\nu\frac{D}{D{\affine} }p^\nu$, as well as \cref{eq:DSmunu,eq:accspin}, it follows that~\cite{Goldberger:2020fot} 
\begin{equation}
 p^2\frac{\dd \X^\mu}{\dd {\affine}}  - p^\mu p_\nu\frac{\dd \X^\nu}{\dd {\affine}}
 -\frac{1}{2} R_{\nu\lambda\rho\sigma} \frac{\dd \X^\lambda}{\dd {\affine}} S^{\rho\sigma}  S^{\mu\nu} = 0 .
\end{equation}
For an object at rest (and in the absence of additional interactions), one recovers that $p^\mu$ is proportional to $\dd \X^\mu/\dd {\affine}$. Then the equations of motion imply that
\begin{equation}
p^2 = M^2,
\qquad
S^2\equiv \frac{1}{2}S^{\mu\nu}S_{\mu\nu}
\end{equation}
are conserved along the worldline,
\begin{equation}
\frac{D}{D {\affine}} p^2 = \frac{D}{D {\affine}} S^2 = 0.
\end{equation}

As in the case of non-rotating objects, finite-size effects are captured by supplementing the point-particle action with derivative operators. Remaining within the conservative sector (we will include dissipation in \cref{sec:dissipation}), and working at leading order in the derivative expansion, the dominant quadratic operators that one can write down are\footnote{Although we do not mention them explicitly, as our focus is on induced responses, the EFT also contains operators that are linear in the Weyl tensor, contracted with some multi-index tensor (see, e.g., \rcite{Levi:2018nxp}). The coefficients of these operators describe the intrinsic multipole moments of the object, as opposed to induced moments.  An intriguing feature of $D=4$ Kerr black holes is that they have the gravitational multipole moments of an elementary spinning particle~\cite{Vines:2017hyw,Guevara:2018wpp,Chung:2018kqs,Arkani-Hamed:2019ymq}.}
\begin{equation}
S_\mathrm{h.o.} = \int\dd\tau\left(c^{(\mathrm{E})}_{i_1i_2 \vert j_1 j_2}E^{i_1i_2} E^{j_1 j_2} + c^{(\mathrm{B})}_{i_1i_2 \vert j_1 j_2}B^{i_1i_2} B^{j_1 j_2} + \cdots\right),
\label{eq:Shospin2}
\end{equation}
where  we have replaced the scalar coefficients $c_2^{(\mathrm{E})}$ and $c_2^{(\mathrm{B})}$ of \cref{eq:E2B2-0} by the tensors $c^{(\mathrm{E})}_{i_1i_2 \vert j_1 j_2}$ and $c^{(\mathrm{B})}_{i_1i_2 \vert j_1 j_2}$, respectively.
The tensorial structures arise from the fact that, for rotating objects, different combinations involving the spin can be constructed~\cite{Goldberger:2020fot,Charalambous:2021mea,Saketh:2022xjb,Saketh:2023bul}.
The most general parametrization of the tensors $c^{(\mathrm{E/B})}_{i_1i_2 \vert j_1 j_2}$  can be obtained by combining, in all symmetry-compatible ways, the available building blocks: the metric $g_{\mu\nu}$, the totally antisymmetric tensor $\epsilon^{ijk}$, and the spin degree of freedom. 
We return to this in \cref{sec:dissipation} below, where we discuss dissipative effects and  provide a general expression for the spin-dependent response tensors $c^{(\mathrm{E/B})}_{i_1i_2 \vert j_1 j_2}$.}

\subsubsection{Dissipative tidal effects}
\label{sec:dissipation}

An action like \cref{eq:Shospin2} can only model conservative finite-size effects. However,  compact objects are known to dissipate energy. Examples of dissipative dynamics in astrophysical compact sources are   absorption of gravitational energy by the horizons of black holes, or  dissipative processes of neutron stars due to internal effective fluid viscosity. Since rotation, as well as finite-frequency effects, are typically associated with dissipation, to obtain the most general parametrization of the response tensors $c^{(\mathrm{E/B})}_{i_1i_2 \vert j_1 j_2}$ in \cref{eq:Shospin2}, it is thus convenient  to first understand how the previous analysis extends to incorporate dissipation.

A modern EFT description of dissipative effects was first developed in \rcite{Goldberger:2005cd} for non-rotating compact objects. In this framework, dissipation through the horizon is attributed  to a set of  modes  localized on the worldline. The dynamics of these modes is encoded in appropriate correlation functions, which can be determined via a matching procedure to graviton absorption and emission processes. 
Subsequent work~\cite{Goldberger:2019sya,Goldberger:2020wbx} extended this approach to analyze horizon dissipation and quantum effects in scattering processes involving black hole asymptotic states, with direct applications to relativistic binary black hole systems.

In the context of spinning black holes, the EFT description of dissipation was initially developed for slowly rotating systems. Early work~\cite{Porto:2007qi} analyzed tidal couplings to gravitons in the small-spin limit, while~\rcite{Endlich:2015mke} extended the framework --- still within the slow-rotation r\'egime --- to include couplings to more general external fields. In particular, the effective approach employed in~\rcite{Endlich:2015mke} builds on an action derived in~\rcite{Delacretaz:2014oxa} for a relativistic spinning object coupled to gravity, based on a coset construction for spontaneously broken spacetime symmetries (see also~\rcite{Endlich:2016jgc} for related work). 
The formalism was later generalized to Kerr black holes with arbitrary spin~\cite{Goldberger:2020fot}, including the near-extremal r\'egime.

To model dissipation we introduce some degrees of freedom localized on the worldline of the object, following \rcite{Goldberger:2004jt,Goldberger:2005cd}.
These additional degrees of freedom --- which we will collectively denote  by $\XX$ --- can absorb energy and provide an effective description of dissipative processes.
A useful analogy is the hydrogen atom: when photons interact with it, the atom can absorb the incoming radiation by exciting the electron to a higher energy level. In this picture, where the point particle is analogous to a hydrogen atom, the various microscopic $\XX$ states correspond to the atom’s energy levels.

The point-particle action, including energy exchange between the worldline and external fields in  the bulk, can be obtained by promoting the ingredients in \cref{eq:pointparticleaction} to  functions of the $\XX$ degrees of freedom: 
\begin{equation}
    S_{\rm pp}= \int\dd{\affine}\left[\frac{\dd \X^\mu}{\dd{\affine}} p_a(\XX) e_\mu^a+\frac{1}{2}S^{ab}(\XX)\Omega_{ab}-\frac{1}{2}e\Big(p_a(\XX)p^a(\XX)-L_\XX(\XX,e^{-1}\dot \XX)\Big)+e\lambda_aS^{ab}(\XX)p_b(\XX)
    \right],
    \label{eq:disspointparticleaction}
\end{equation}
where the  Lagrangian $L_\XX$ captures  the internal dynamics. The momentum $p^a(\XX)$ and spin $S^{ab}(\XX)$   account for the excitation of the internal degrees of freedom $\XX$, and  are  now interpreted as composite operators~\cite{Goldberger:2020fot}. 
Note that, in the presence of dissipation, the momentum variables depend on the internal degrees of freedom $\XX$ and cannot be obtained in a model-independent way. In other words, one needs in principle to know the dynamics of $\XX$, i.e.,~$L_\XX(\XX,e^{-1}\dot \XX)$. In some cases, this requirement can be sidestepped if  the microscopic  theory is known, and one can obtain the relation between $S^{\mu\nu}$ and $\Omega^{\mu\nu}$ by performing an explicit matching \cite{Goldberger:2020fot}.

We now wish to couple the worldline to external fields. Beyond the conservative charge/multipole terms, interactions between the point particle and bulk fields can be written as \cite{Goldberger:2004jt,Goldberger:2005cd,Goldberger:2009qd}
\begin{equation}
S_{\rm h.o.} = \int\dd\tau \left[ Q_{\text{E}}^{i_1\cdots i_\ell}(\XX,e) E_{i_{1}\cdots i_\ell}
+ Q_{\text{B}}^{i_1\cdots i_\ell}(\XX,e) B_{i_{1}\cdots i_\ell}
\right],
\label{eq:sint}
\end{equation}
where $Q_{\text{E},\text{B}}^{i_1\cdots i_\ell}(\XX,e)$ are composite operators built from  $\XX$ and the vielbein $e$.\footnote{Note that \cref{eq:sint} can be straghtforwardly extended to the case of a point particle coupled to electromagnetic and scalar fields. We have, respectively,
\begin{equation}
S_{\rm h.o.}^{\mathrm{EM}} = \int\dd\tau \left[ Q_{\mathcal{E}}^{i_1\cdots i_\ell}(\XX,e) \partial_{\langle i_1}\cdots \partial_{i_{\ell-1}} \mathcal{E}_{i_\ell\rangle}
+ Q_{\mathcal{B}}^{i_1\cdots i_\ell}(\XX,e)  \partial_{\langle i_1}\cdots \partial_{i_{\ell-1}}  \mathcal{B}_{ i_\ell\rangle}
\right],
\label{eq:sintEM}
\end{equation}
\begin{equation}
S_{\rm h.o.}^\phi = \int\dd\tau \, Q_{\phi}^{i_1\cdots i_\ell}(\XX,e)  \partial_{\langle i_1}\cdots \partial_{i_{\ell}\rangle}\phi
,
\label{eq:sintscalar}
\end{equation}
where $\phi$ is the scalar field, and $\mathcal{E}_i$ and $\mathcal{B}_i$ are the electric and magnetic fields, respectively.
}
From a bottom-up perspective, we generally do not have  access to the explicit form of $Q_{\text{E},\text{B}}^{i_1\cdots i_\ell}(\XX,e)$.
Nevertheless, we can  analyze the structure of its correlation functions. In practice, we will parametrize the correlators of $Q_{\text{E},\text{B}}^{i_1\cdots i_\ell}(\XX,e)$ with some unknown coefficients, which encode all the information about the unknown microscopic physics, and determine them by  matching to explicit models.

Operationally, since we typically cannot track the internal detailed microstate of the object from the perspective of a long-distance observer,  we  need to average over all possible microscopic configurations that the $\XX$ degrees of freedom could be in. In field theory, this is tantamount  to integrating out the $\XX$ degrees of freedom using the path intergral. 
Since these can be gapless, the fields that we actually track --- e.g., the point particle itself and the external probes, namely those associated with low-energy observables ---  form an effective open quantum system.  
They can  exchange energy and dissipate, whereas  the $\XX$ sector can be excited by arbitrarily small amounts of energy.  As is typical for dissipative systems, because we cannot specify the final state, the appropriate framework for integrating out the $\XX$ degrees of freedom is the Schwinger--Keldysh formalism --- see, e.g., \rcite{kamenev2011field,Akyuz:2023lsm,Haehl:2015foa,Crossley:2015evo,Liu:2018kfw,Haehl:2024pqu,Glorioso:2016gsa,Weinberg:2005vy,Glazer:2024eyi} for comprehensive discussions. 

In short, the  Schwinger--Keldysh approach consists in  doubling the fields and following the evolution of the system along a closed-time contour that runs from $t=-\infty$ to the time of interest and then back to $t = -\infty$. Concretely, the Schwinger--Keldysh effective action with dissipative couplings can be obtained by performing  the following in-in path integral:
\begin{equation}
    \exp\Big(i \Gamma^{\text{in-in}}(F_1,F_2)\Big)=\int \mathcal{D}\XX_1 \mathcal{D}\XX_2 \,e^{i S[\XX_1,F_1]-i S[\XX_2,F_2]}\,,
    \label{eq:ininint}
\end{equation}
where $ F_{1,2}= \big\{\X_{1,2}^\mu, (e_{1,2})_\mu^A, e_{1,2},h_{1,2}\big\}$ are the long-distance degrees of freedom along the two branches of the in-in contour. 
Integrating out the  $\XX$ degrees of freedom introduces some effective interactions between the fields defined on the two-branch contour,  leading to dissipative effects in $\Gamma^\text{in-in}$. 
It is usually convenient to define  the combinations
\begin{equation}
h_{+}\equiv \frac{1}{2}\left(h_1+h_2\right) ,
\qquad
h_{+}\equiv h_1-h_2 ,
\end{equation}
and similarly for all other fields, which form   the Keldysh basis.
Evaluating the path integral \eqref{eq:ininint} yields an effective interaction for the low-energy degrees of freedom.

As an illustration, let us focus for the moment on the purely electric sector.
In linear response theory, performing the path integral above effectively corresponds to  replacing the $Q$ operators by  
\be
\langle Q_{\text{E},I}^{i_1\cdots i_\ell }(\tau) \rangle = \int\dd \tau' K^{(\mathrm{E})}_{IJ}(\tau - \tau^\prime)^{i_1\cdots i_\ell | j_1\cdots j_{\ell'}} E^J_{j_1\cdots j_{\ell'}}(\tau^\prime)\,,
\label{eq:QElinearR}
\ee
where $K^{(\mathrm{E})}_{IJ}(\tau - \tau^\prime)$ is a response kernel,  which leads  to the in-in effective action
\be
\label{eq:gammaintgeneric}
\Gamma_{\rm int}^\text{in-in}\supset \int\dd{\tau}_1\dd{\tau}_2\, K^{(\mathrm{E})}_{IJ} ({\tau}_1-{\tau}_2)
^{i_1\cdots i_\ell\vert j_1\cdots j_{\ell'}} \left(
\partial_{\langle i_1}\cdots\partial_{i_{\ell-2}}E^{I}_{i_{\ell-1}i_\ell\rangle} ({\tau}_1)\right)\left(
\partial_{\langle j_1}\cdots\partial_{j_{\ell-2}}E^{J}_{j_{\ell-1}j_{\ell'}\rangle} ({\tau}_2) \right),
\ee
for the electric sector, and similarly for the magnetic one. Note that the presence of spin allows for a more general structure, including mixed couplings between the gravito-electric and gravito-magnetic sectors; we will return to this below.
The indices $I,J$ run over the Keldysh basis, i.e.,~$I,J=\{+,-\}$. %
The effective couplings $K_{IJ}({\tau}_1,{\tau}_2)$, which are in general aribitrary functions of $\tau_1$ and $\tau_2$, depend on time only through the difference $\tau_1-\tau_2$ for stationary solutions.  In particular, they  generalize the interaction term \eqref{eq:Shospin2} in the presence of dissipative dynamics. Note also that they can be interpreted as  the  Green's function of the $Q_I$ operators in the appropriate basis. 
In particular, %
suppressing indices for simplicity, $K^{(\mathrm{E})}_{+-}$ is related to the retarded Green's function of the $Q$ operators via (see, e.g.,~\rcite{Goldberger:2020fot,Saketh:2022xjb,Saketh:2023bul,Combaluzier--Szteinsznaider:2025eoc})\footnote{Note a slight difference in convention with respect to~\rcite{Goldberger:2020fot,Saketh:2022xjb,Saketh:2023bul}.}
\begin{equation}
K^{(\mathrm{E})}_{+-}({\tau}_1-{\tau}_2)= - G_R^{({\text{E}})}(\tau_1 - \tau_2) \equiv i \langle [Q_{\text{E},+}(\tau_1), Q_{\text{E},-}(\tau_2)] \rangle\theta(\tau_1-\tau_2) .
\label{eq:KpmGR}
\end{equation}

As in any EFT, this approach is mostly useful when there is a hierarchical separation of scales in the system. Assuming that the characteristic timescale of the $\XX$ degrees of freedom is parametrically faster than the timescale on which we probe the system, we can effectively model the internal dynamics and response  as instantaneous. This means that we can expand the couplings $K_{IJ}({\tau}_1-{\tau}_2)$ as a series of Dirac delta functions,\footnote{See e.g.,~\rcite{Saketh:2022xjb,Saketh:2023bul} for a covariantized parametrization of the electric and magnetic tidally induced multipole moments, and their respective couplings to the tidal fields in the action of a spinning point particle. In particular, we refer to appendix A of \rcite{Saketh:2023bul} for a discussion of different bases for the induced multipole moments~\cite{Goldberger:2020fot,Saketh:2022xjb}.}
\begin{equation}
K({\tau}_1-{\tau}_2)
^{i_1\cdots i_\ell\vert j_1\cdots j_{\ell'}}
= \ccs_0^{i_1\cdots i_\ell\vert j_1\cdots j_{\ell'}} \delta(t-t')+
\ccs_1^{i_1\cdots i_\ell\vert j_1\cdots j_{\ell'}}\delta'(t-t')+\cdots\,,
\label{eq:expCcc}
\end{equation}
where we suppressed $I,J$ indices for simplicity. Equivalently, in Fourier space this is
\begin{equation}
K(\omega)
^{i_1\cdots i_\ell\vert j_1\cdots j_{\ell'}}
= \ccs_0^{i_1\cdots i_\ell\vert j_1\cdots j_{\ell'}} +
\ccs_1^{i_1\cdots i_\ell\vert j_1\cdots j_{\ell'}}i\omega+\cdots\,.
\label{eq:Comega}
\end{equation}
Under this assumption, the interactions \eqref{eq:gammaintgeneric} become  \textit{local} operators, leading to an ordinary EFT describing the response to external fields. 
Notice that \cref{eq:Comega} contains terms with odd powers of the frequency. As promised, the effective expansion \eqref{eq:Comega} can describe dissipative effects (which are odd under time reversal), in addition to conservative dynamics. We stress that not all terms that are odd (even) in $\omega$ correspond necessarily to dissipative (conservative) effects. The characteristics of the object may introduce other time-reversal odd quantities. One such example is  the spin: in this case the separation into even and odd under time reversal --- which continues to correspond to conservative and dissipative effects, respectively --- is dictated by the total number of powers of $\omega$ and the spin, rather than the frequency alone \cite{Saketh:2023bul}.\footnote{Note that terms that are  odd under time reversal can not in general be mimicked by any conservative local couplings between the worldline and external fields without doubling the number of fields. The local description is a byproduct of the Schwinger--Keldysh approach (in addition to the assuption of separation of scales).}

From a symmetry viewpoint, the presence of  dissipative terms in the Schwinger--Keldysh effective action can be seen   as a consequence of  integrating out the $\XX$ variables. This  breaks the two independent time translation symmetries  --- which act separately on the two copies of the fields on the double contour ---  to the diagonal combination \cite{Akyuz:2023lsm,Haehl:2015foa,Crossley:2015evo,Liu:2018kfw}. Note that one can bypass the introduction of the auxiliary worldline degrees of freedom $\XX$ and the path integral \eqref{eq:ininint} by directly writing down the  couplings \eqref{eq:gammaintgeneric} between the fields on the two branches of the in-in contour, following the standard general rules of the Schwinger--Keldysh approach~\cite{Feynman:1963fq,Caldeira:1982iu,Calzetta:1986cq,Kamenev:2009jj}. Both approaches are clearly equivalent and lead to the same description of the open system.\footnote{See, e.g., \rcite{Salcedo:2025ezu} for an application in the cosmological context. In the same context, an open EFT formulation in terms of composite operators --- reminiscent of  the $Q$'s in  \cref{eq:sint} --- was developed in \rcite{LopezNacir:2011kk} (see also \rcite{Creminelli:2023aly} for an explicit example).}

The coefficients $\ccs^{i_1\cdots i_\ell\vert j_1\cdots j_{\ell'}}$  in \cref{eq:Comega} are so far generic constant tensors. However, not all components of these tensors are  independent. Their structure is dictated by the symmetries of the system (e.g., invariance under rotations). To avoid redundancies in the EFT description, it is convenient to choose a basis in  which the tensors $\ccs^{i_1\cdots i_\ell\vert j_1\cdots j_{\ell'}}$ can be decomposed. There are of course various possibilities. A convenient one is a decomposition  in sums of \textit{Thorne tensors} of the form
\be
\ccs^{i_1\cdots i_\ell\vert j_1\cdots j_{\ell'}}= \sum_{m,m'} \ccs^{ \ell\ell'mm'} \,\mathcal{Y}^{i_1\cdots i_\ell}_{\ell m} \mathcal{Y}_{\ell'm'}^*{}^{j_1\cdots j_{\ell'}}\,,
 \label{eq:thornetensorprod}
\ee
where $\ccs^{\ell\ell'mm'} $ parameterize the linear combination --- these are the quantities that one would typically try to determine by  matching between the EFT and the ultraviolet --- and where $\mathcal{Y}^{i_1\cdots i_\ell}_{\ell m}$ are the Thorne tensors.
The Thorne tensors are traceless symmetric tensors, usually defined in terms of a reference vector, and have the property that their contraction with the unit vector $\hat x^i \equiv x^i/\lvert \vec x\rvert$ is a spherical harmonic with quantum numbers $\ell,m$~\cite{trautman1965lectures,Thorne:1980ru,Charalambous:2021mea}: 
\be
{\cal Y}_{\ell m}^{i_1\cdots i_\ell} \hat x_{ i_1}\cdots \hat x_{i_\ell} = Y_{\ell m}(\theta,\varphi).
\label{eq:thornetensorsY}
\ee
For rotating systems, the reference vector can be taken to just be the spin $s^i$.\footnote{It is useful to mention that, in three-dimensional space, the spin vector $s^i$ is dual to a 2-form $S^{ij} = \epsilon^{ijk}s_{k}$, which are also the spatial components of the tensor $S_\mn$ \eqref{eq:spacetime-p-s}.} In the following, we show how one can derive the relation to the spherical harmonics on the 2-sphere $S^2$, and obtain a parametrization for the $\ccs^{i_1\cdots i_\ell\vert j_1\cdots j_{\ell'}}$ coefficients (see also appendix B of \rcite{Glazer:2024eyi}).

Given a spin vector $s^i$, let $m$ (ranging from $-\ell$ to $\ell$) be the angular momentum quantum number  with respect to the $s^i$ direction.
One can  identify the plane orthogonal to $s^i$.  Denoting it  by $P^{ij}$, one has, for instance, $P_{xy}^{ij} = x^{[i}y^{j]}$ for  $s^i = z^i$ (as the $xy$ plane is orthogonal to $s^i$). This orthogonality can be expressed as
\be
\epsilon_{ijk}P^{ij}s^k = s\,,
\ee
which defines the normalization of the object's spin.
A basis of orthogonal vectors can then be chosen in the plane defined by $P_{ij}$.  We will call such vectors $e_a^i$, satisfying $e_a^i s_i = 0$, and $e_a^i e_b^i = 0$ for $a\neq b$.\footnote{One can always orthonormalize the basis in such a way that $e_a^i e_b^i = \delta_{ab}$.} 
The Thorne tensors can be constructed by taking all possible traceless combinations of the vectors $s^i$ and $e^i_a$ with the right number of indices. Let us start by considering some explicit examples, before presenting the general expression.

We begin with the dipole, $\ell=1$. In this case,  the three possible tensors we can construct are \cite{Glazer:2024eyi}
\begin{align}
m &= 0: \hspace{.5cm}s^i\,,\\
m &=\pm1:  \hspace{.18cm}e_a^i\,.
\end{align}
As a concrete basis one can take for instance, in spherical coordinates,
\be
\label{eq:ThornetensorbasisS2}
\begin{aligned}
\hat x^i = 
\left(
\begin{array}{c}
\sin\theta\cos\varphi\\
\sin\theta\sin\varphi\\
\cos\theta
\end{array}
\right)
\hspace{.75cm}
s^i = 
\left(
\begin{array}{c}
0\\
0\\
1
\end{array}
\right)
\hspace{.75cm}
e_\pm^i= 
\left(
\begin{array}{c}
1\\
\pm i\\
0
\end{array}
\right)\,,
\end{aligned}
\ee
where  $s^i$ has been normalized unity, while $e_\pm^i$ satisfy $e_{\pm}\cdot e_{\pm} = 0$ and $e_{\pm}\cdot e_{\mp} =2$. 
Then one can construct the usual $\ell=1$ spherical harmonics by contracting the vectors above,
\begin{align}
Y_{10} &= \hat C_{10}\,s^i \hat x_i =  \frac{1}{2}\sqrt\frac{3}{\pi}\cos\theta\,,\\
Y_{11} &= \hat C_{11}\,e_+^i \hat x_i = \frac{1}{2}\sqrt\frac{3}{2\pi}e^{i\varphi}\sin\theta\,,\\
Y_{1-1} &= \hat C_{1-1}\,e_-^i \hat x_i = -\frac{1}{2}\sqrt\frac{3}{2\pi}e^{-i\varphi}\sin\theta\,,
\end{align}
where we have introduced the  constants
\be
\hat C_{lm} =
\begin{cases}
 (-1)^m\left(\frac{2l+1}{4\pi}\frac{(l-\lvert m\rvert)!}{(l+\lvert m\rvert)!}\right)^\frac{1}{2} \frac{(2l)!}{2^l l!(l-\lvert m\rvert)!}& m\geq 0,\\
\left(\frac{2l+1}{4\pi}\frac{(l-\lvert m\rvert)!}{(l+\lvert m\rvert)!}\right)^\frac{1}{2} \frac{(2l)!}{2^l l!(l-\lvert m\rvert)!}& m<0 .
 \end{cases}
\ee
The $\ell=1$ Thorne tensors can then be defined as
\begin{align}
{\cal Y}_{10}^i &= \hat C_{10}\,s^i\,,\\
{\cal Y}_{11}^i &= \hat C_{11}\,e_+^i\,,\\
{\cal Y}_{1-1}^i &= \hat C_{1-1}\,e_-^i\,.
\end{align}

The story proceeds similarly for $\ell=2$, although the construction  is slightly more involved. As before we can define
\begin{align}
m &= 0: \hspace{.5cm}{\cal Y}_{20}^{ij} = \hat C_{20}\,s^{\langle i}s^{j \rangle}\,,\\
m &=\pm1:  \hspace{.18cm}{\cal Y}_{2\pm 1}^{ij} = \hat C_{2\pm 1} e_\pm^{\langle i}s^{j\rangle}\,,\\
m &=\pm2:  \hspace{.18cm}{\cal Y}_{2\pm 2}^{ij} = \hat C_{2\pm 2} e_\pm^{\langle i}e_\pm^{j\rangle}\,,
\end{align}
in such a way that we get the usual spherical harmonics when we contract  with $\hat x_i\hat x_j$.
Note that we chose above  the combination $s^{\langle i}s^{j \rangle}$ to define the ${\cal Y}_{20}^{ij}$ Thorne tensor. However, thanks to the identity 
\be
e_+^i e_-^j = \delta^{ij}- s^i  s^j - i\epsilon^{ijk} s_k\,,
\label{eq:epemidentity}
\ee
we could have equivalently used $e_+^{\langle i} e_-^{j\rangle}$.

By induction, one can easily generalize the construction of the Thorne tensors at arbitrary $\ell$:
\be
{\cal Y}_{\ell\,\pm m}^{i_1\cdots i_\ell} = \hat C_{\ell \pm m}\,e_{\pm}^{\langle i_1}\cdots e_{\pm}^{i_m} s^{i_{m+1}}\cdots s^{i_\ell\rangle}\,,
\ee
which yield the usual  spherical harmonics \eqref{eq:thornetensorsY} when contracted with $\hat x^i$, and   provide a basis of symmetric traceless tensors for the quantities $\ccs^{i_1\cdots i_\ell\vert j_1\cdots j_{\ell'}}$ \eqref{eq:thornetensorprod}.\footnote{Note that the reality property of the spherical harmonics,
\be
Y^*_{\ell m} = (-1)^m Y_{\ell\,-m}\,,
\ee
gets translated into the following  property for the Thorne tensors:
\be
{\cal Y}^*{}_{\ell\,\pm m}^{i_1\cdots i_\ell} = (-1)^m {\cal Y}{}_{\ell\,\pm m}^{i_1\cdots i_\ell} \,.
\ee
}
Note that, if the physics of system preserves both angular momentum and azimuthal angular momentum, then  $\ell=\ell'$ and $m=m'$, and the decomposition in terms of Thorne tensors reads
\be
\ccs^{i_1\cdots i_\ell\vert j_1\cdots j_{\ell'}}=  \delta^{\ell\ell'}\sum_{m} \ccs^{ \ell m} \,\mathcal{Y}^{i_1\cdots i_\ell}_{\ell m} \mathcal{Y}_{\ell m}^*{}^{j_1\cdots j_{\ell}}\,.
 \label{eq:thornetensorprod2}
\ee

Clearly, the Thorne tensors are not the only basis that can be used to decompose the effective couplings $\ccs^{i_1\cdots i_\ell\vert j_1\cdots j_{\ell'}}$. In the literature, different choices have been made for the decomposition. For instance, in \rcite{Goldberger:2020fot,Saketh:2023bul,Chia:2024bwc} responses are parameterized in terms of the basis of tensors built from $\delta_{ij},  s_i$ and $\epsilon_{ijk} s^k$. This choice is completely equivalent, but it is instructive  to see explicitly how it is related to  the decomposition \eqref{eq:thornetensorprod2}.

To understand the relation between the two bases, let us focus on each term  in the sum~\eqref{eq:thornetensorprod2}. Restricting to $m\geq0$ (the terms with $m<0$  can be obtained by complex conjugation), we have
\be
\mathcal{Y}^{i_1\cdots i_\ell}_{\ell m} \mathcal{Y}^*{}^{j_1\cdots j_\ell}_{\ell m}= \lvert\hat C_{\ell m}\rvert^2\,e_{+}^{\langle i_1}\cdots e_{+}^{i_m} s^{i_{m+1}}\cdots s^{i_\ell\rangle}\, e_{-}^{\langle j_1}\cdots e_{-}^{j_{m'}} s^{j_{m'+1}}\cdots s^{j_\ell\rangle}\,.
\ee
Because both Thorne tensors on the left-hand side have the same value of  $m$, this expression involves the same number of $e_+^i$ and $e_-^j$ vectors. We can thus use the identity \eqref{eq:epemidentity} to get rid of pairs of $e_\pm^i$ vectors  on the right-hand side, and  trade them  for the objects $\delta^{ij}$,  $s^i s^j$ and $S_{ij}\equiv \epsilon_{ijk}s^k$, which are the building blocks used in \rcite{Goldberger:2020fot,Saketh:2023bul,Chia:2024bwc} for the decomposition \eqref{eq:thornetensorprod2}.

As an illustration, let us see explicitly what the relation between the two bases is for $\ell =2$. We start by writing explicitly the five terms appearing in the sum~\eqref{eq:thornetensorprod2}:
\begin{align}
\label{eq:YY1}
\mathcal{Y}^{ij}_{22} \mathcal{Y}^*_{22}\,{}_{kl} &= \lvert\hat C_{22}\rvert^2~e_{+}^{\langle i}e_{+}^{j\rangle }\, e^{-}_{\langle k}e^{-}_{l\rangle }\\
\mathcal{Y}^{ij}_{21} \mathcal{Y}^*_{21}\,{}_{kl} &= \lvert\hat C_{21}\rvert^2~e_{+}^{\langle i}s^{j\rangle}\, e^{-}_{\langle k}s_{l\rangle}\\
\mathcal{Y}^{ij}_{20} \mathcal{Y}^*_{20}\,{}_{kl} &= \lvert\hat C_{20}\rvert^2~s^{\langle i}s^{j\rangle}\, s_{\langle k}s_{l\rangle}\,,
\label{eq:YY2}
\end{align}
where $\mathcal{Y}^{ij}_{2-2} \mathcal{Y}^*_{2-2}\,{}_{kl}$ and $\mathcal{Y}^{ij}_{2-1} \mathcal{Y}^*_{2-1}\,{}_{kl} $ are  related to $\mathcal{Y}^{ij}_{22} \mathcal{Y}^*_{22}\,{}_{kl}$ and $\mathcal{Y}^{ij}_{21} \mathcal{Y}^*_{21}\,{}_{kl} $, respectively, by  complex conjugation. 
In terms of the building blocks  $\delta_{ij}, s_i$, and $S_{ij}$, one can construct the following  five-dimensional basis~\cite{Saketh:2023bul,Chia:2024bwc}:
\begin{align}
B^{(1)}{}^{ij}_{kl} &= \delta^{\langle i}_{\langle k}\delta^{j\rangle}_{l\rangle} , & B^{(4)}{}^{ij}_{kl} &=  S^{\langle i}_{\langle k}s^{j\rangle}s_{l\rangle} , \\
B^{(2)}{}^{ij}_{kl} &=  S^{\langle i}_{\langle k}\delta^{j\rangle}_{l\rangle} , & B^{(5)}{}^{ij}_{kl} &=  s^{\langle i}s_{\langle k}s^{j\rangle}s_{l\rangle} . \\
B^{(3)}{}^{ij}_{kl} &=  s^{\langle i}s_{\langle k}\delta^{j\rangle}_{l\rangle}\,,
\end{align}
Using  the identity~\eqref{eq:epemidentity}, we can rewrite \cref{eq:YY1,eq:YY2}  in the new basis as \cite{Glazer:2024eyi}
\begin{align}
\mathcal{Y}^{ij}_{22} \mathcal{Y}^*_{22}\,{}_{kl} &= \lvert\hat C_{22}\rvert^2 \left(
2B^{(1)}{}^{ij}_{kl} -2iB^{(2)}{}^{ij}_{kl} -4B^{(3)}{}^{ij}_{kl} +2iB^{(4)}{}^{ij}_{kl} +B^{(5)}{}^{ij}_{kl} 
\right),\\
\mathcal{Y}^{ij}_{21} \mathcal{Y}^*_{21}\,{}_{kl} &= \lvert\hat C_{21}\rvert^2\left(
B^{(3)}{}^{ij}_{kl} -iB^{(4)}{}^{ij}_{kl} -B^{(5)}{}^{ij}_{kl} 
\right),\\
\mathcal{Y}^{ij}_{20} \mathcal{Y}^*_{20}\,{}_{kl} &= \lvert\hat C_{20}\rvert^2\,B^{(5)}{}^{ij}_{kl} ,
\end{align}
where once again the terms  with $m=-2,-1$ can be obtained by complex conjugation.

These relations can be inverted, and one can easily express  the $B^{(n)}$ in terms of ${\cal Y}{\cal Y}^*$ if desired. 
Note that there are three real and two imaginary linear combinations of ${\cal Y}{\cal Y}^*$ products.
Plugging the decomposition of each $\ccs_k^{i_1\cdots i_\ell\vert j_1\cdots j_{\ell'}}$ into \cref{eq:Comega}, we see that one gets a real (imaginary) combination whenever the total sum of $\omega$, $s^i$ and $S^{ij}$ is even (odd).

Following \rcite{Saketh:2022xjb,Saketh:2023bul}, one can write
\begin{align}
\ccs_0^{ij | kl} & = \Lambda_{\omega^0,S^0} \delta^{\langle i}_{\langle k}\delta^{j \rangle}_{l \rangle} + 
H_{\omega^0,S^1} {S^{\langle i}}_{\langle k}\delta^{j \rangle}_{l \rangle} + 
\Lambda_{\omega^0,S^2} {s^{\langle i}} s_{\langle k}\delta^{j\rangle}_{l \rangle} + 
H_{\omega^0,S^3} {s^{\langle i}} s_{\langle k}{S^{j\rangle}}_{l\rangle} + 
\Lambda_{\omega^0,S^4} {s^{\langle i}} s_{\langle k} s^{j\rangle} s_{l\rangle},
\label{eqlambdas}
\\
\ccs_1^{ij | kl} & = H_{\omega^1,S^0} \delta^{\langle i}_{\langle k}\delta^{j\rangle}_{l\rangle} + 
\Lambda_{\omega^1,S^1} {S^{\langle i}}_{\langle k}\delta^{j\rangle}_{l\rangle} + 
H_{\omega^1,S^2} {s^{\langle i}} s_{\langle k}\delta^{j\rangle}_{l\rangle} + 
\Lambda_{\omega^1,S^3} {s^{\langle i}} s_{\langle k}{S^{j\rangle}}_{l\rangle} + 
H_{\omega^1,S^4} {s^{\langle i}} s_{\langle k} s^{j\rangle} s_{l\rangle},
\label{eqlambdas1} 
\\
\ccs_2^{ij | kl} & = \Lambda_{\omega^2,S^0} \delta^{\langle i}_{(k}\delta^{j\rangle}_{l\rangle} + 
H_{\omega^2,S^1} {S^{\langle i}}_{(k}\delta^{j\rangle}_{l\rangle} + 
\Lambda_{\omega^2,S^2} {s^{\langle i}} s_{\langle k}\delta^{j\rangle}_{l\rangle} + 
H_{\omega^2,S^3} {s^{\langle i}} s_{\langle k}{S^{j\rangle}}_{l\rangle} + 
\Lambda_{\omega^2,S^4} {s^{\langle i}} s_{\langle k} s^{j\rangle} s_{l\rangle},
\label{eqlambdas2}
\end{align}
and similarly at higher orders in $\omega$. In the previous expressions --- which apply to both electric and magnetic sectors (with $E$ and $B$ subscripts suppressed for simplicity, see, e.g., \cref{eq:gammaintgeneric}) --- $\Lambda$ and $H$ are constant coefficients. Specifically, $\Lambda$ multiplies terms where the total  sum of $\omega$, $s^i$ and $S^{ij}$ is even, whereas $H$ is associated with terms where the   sum is  odd. Accordingly, $\Lambda$ captures the conservative response, while $H$ characterizes dissipative finite-size effects. 

Note that, for non-spinning objects, the tensor structure in the expressions above simply boils down to the symmetric-trace-free product of Kronecker symbols. In this case, $\Lambda_{\omega^0,S^0}$  corresponds to the static Love numbers, while $\Lambda_{\omega^{2n},S^0}$ for $n\geq1$ are commonly referred to as the dynamical Love numbers.

A summary of the main conservative and dissipative coefficients in the worldline EFT is given in \cref{table: summary tidal number 1} (see also \rcite{Saketh:2023bul,Chia:2024bwc} for more details). We also indicate the PN order at which each coefficient (in the electric sector) first contributes \cite{Blanchet:2013haa}.
The first two rows correspond to the conservative sector, while the last three show the leading dissipation coefficients. Note that, within each (conservative or dissipative) sector, increasing the power of $\omega$ by one corresponds to a shift of  $+1.5$PN. This follows roughly from the scaling $\omega\sim \frac{v}{r} \sim v^3$ (see \cref{eq:virial}).
This explains the increasing PN order from top to bottom within each sector, as shown in the rightmost column of \cref{table: summary tidal number 1}.\footnote{For rotating bodies, the PN counting  is implicitly done assuming large spin \cite{Blanchet:2013haa}.}

\begin{table}[h]
  \centering
  \begin{threeparttable}
  \small
    \begin{tabular*}{0.75\textwidth}{l@{\extracolsep{\fill}}cc}
    \toprule
      \textit{Response type} & \textit{Notation} & \textit{Leading PN order} \\
    \midrule\midrule
      \multicolumn{3}{c}{\textbf{Tidal Love numbers ($\Lambda$ coefficients)}} \\
    \midrule
      Static tidal Love numbers
        & $\Lambda^E_{\omega^0, S^{2n}}$
        & 5 \\
      \midrule
      \multirow{2}{*}{Dynamical tidal Love numbers}
        & $\Lambda^E_{\omega^1,S^{2n+1}}$
        & 6.5 \\
        & $\Lambda^E_{\omega^2,S^{2n}}$
        & 8 \\
    \midrule\midrule
      \multicolumn{3}{c}{\textbf{Tidal dissipation numbers ($H$ coefficients)}} \\
    \midrule
      LO (static) tidal dissipation numbers
        & $H^E_{\omega^0, S^{2n+1}}$
        & 2.5 \\
      \midrule
      NLO tidal dissipation numbers
        & $H^E_{\omega^1,S^{2n}}$
        & 4 \\
      \midrule
      NNLO tidal dissipation numbers
        & $H^E_{\omega^2,S^{2n+1}}$
        & 5.5 \\
    \bottomrule
    \end{tabular*}
  \end{threeparttable}
\caption{Leading-order (LO) and (next-to-)next-to-leading-order ((N)NLO) conservative and dissipative tidal response coefficients in the worldline EFT, together with the PN order at which they first contribute. We introduced an extra superscript $E$ on $\Lambda$ and $H$ to emphasize that the PN order shown in the  rightmost column refers to the electric sector $E$. Magnetic tidal dissipation effects  appear at 1PN order higher than their electric counterparts, due to the relative velocity suppression, $B\sim v E$) \cite{Chia:2024bwc}. }
\label{tab1:threshold}
  \label{table: summary tidal number 1}
\end{table}

The generalization to higher orders in spatial gradients is almost straightforward: one simply adds more indices.\footnote{Note that adding spatial derivatives is equivalent to introducing additional powers of $1/r$, thereby raising the PN order at which the effect first contributes. In particular, each additional spatial derivative in an EFT operator increases its contribution by $+1$PN relative to operators at the previous order within the same conservative or dissipative sector.} 
However, the presence of an extra vector --- namely, the spin --- allows for a richer set of structures. In particular, beyond expressions such as \cref{eq:QElinearR}, one can construct more general tidally induced moments, including mixing terms between the electric and magnetic sectors~\cite{Goldberger:2020fot,Saketh:2022xjb,Saketh:2023bul}. For instance, at leading order in derivatives, the electric quadrupole moment $Q_{\mathrm{E}}^{ij}$, in addition to terms proportional to $E_{ij}$ (cf. \cref{eq:QElinearR,eq:Comega}), also admits contributions proportional to the (octupolar) magnetic field $B^{ijk}$. In frequency space,
\be
\langle Q_{\mathrm{E}}^{ij} \rangle \supset \nu_{\mathrm{E}}^{ij \langle kl} s^{m\rangle} B_{klm}
\,,
\label{eq:QElinearRmixed}
\ee
and similarly for the tidally induced magnetic moments. The coupling in \cref{eq:QElinearRmixed} is parity-even and correctly captures, for instance, the dissipative dynamics of spinning black holes in general relativity.\footnote{Recall that, under parity, the $E$ and $B$ fields transform as
\begin{equation}
    E_{i_1\cdots i_\ell} \rightarrow (-1)^\ell E_{i_1\cdots i_\ell} ,
    \qquad
    B_{i_1\cdots i_\ell} \rightarrow (-1)^{\ell+1} B_{i_1\cdots i_\ell} .
\end{equation}
From a general EFT perspective, there is no fundamental principle requiring parity invariance. Allowing for more general systems --- including UV completions that violate parity  --- would admit additional couplings beyond those discussed above (see, e.g., \rcite{Modrekiladze:2022ioh}).
}
Similarly to \cref{eqlambdas}, the response tensor $\nu_{\mathrm{E}}^{ij \langle kl}$ admits an expansion in the building blocks as
\begin{equation}
    (\nu_{\mathrm{E}})^{ij }{}_{kl}  = \tilde \Lambda_{S^0} \delta^{\langle i}_{\langle k}\delta^{j \rangle}_{l \rangle} + 
\tilde H_{S^1} {S^{\langle i}}_{\langle k}\delta^{j \rangle}_{l \rangle} + 
\tilde \Lambda_{S^2} {s^{\langle i}} s_{\langle k}\delta^{j\rangle}_{l \rangle} + 
\tilde H_{S^3} {s^{\langle i}} s_{\langle k}{S^{j\rangle}}_{l\rangle} + 
\tilde \Lambda_{S^4} {s^{\langle i}} s_{\langle k} s^{j\rangle} s_{l\rangle},
\label{eqnuEBs}
\end{equation}
where again the $\tilde{\Lambda}$ coefficients encode the conservative tidal response and the $\tilde{H}$ coefficients describe the dissipative response  (see~\rcite{Saketh:2022xjb,Saketh:2023bul} for a more detailed classification, including more mixing terms).

\subsubsection{Nonlinear tidal response}
\label{sec:nonlinearities}

The expression \eqref{eq:QElinearR} captures the leading-order, induced linear response of the point object. However, general relativity is inherently a nonlinear theory. Field nonlinearities generally affect the tidal deformability of the object at subleading order and, much like the linear response, they admit a natural interpretation within the framework of EFT.

In the literature, the nonlinear extension of the point-particle EFT has  been employed within the post-Minkowskian (PM) framework to compute leading-PM contributions to the two-body  Hamiltonian \cite{Bern:2020uwk}. It was subsequently applied, for the first time within the  EFT and using off-shell matching, to compute the nonlinear (conservative)  Love numbers of Schwarzschild black holes and neutron stars in  \rcite{Riva:2023rcm,Iteanu:2024dvx,Combaluzier-Szteinsznaider:2024sgb}  and \rcite{Pani:2025qxs}, respectively.

The nonlinear response of the object can be studied by solving for $Q_I$ at nonlinear order. In practice, one generalizes \cref{eq:QElinearR} as follows~\cite{Bern:2020uwk,Riva:2023rcm,Iteanu:2024dvx,Combaluzier-Szteinsznaider:2024sgb,Hadad:2024lsf}: 
\begin{multline}
\langle Q_{E,B}^{i_L}(\tau) \rangle 
= \sum_{n=1}^\infty \sum_{k=0}^n \int \dd \tau_1 \cdots \dd\tau_n   \, \RR_{E,B}^{i_L\vert j_{L_1} \cdots   \,  j_{L_k} \vert j_{L_{k+1}} \cdots   \,  j_{L_n}  }(\tau-\tau_1,\dots,\tau-\tau_n) 
\\
 \times E_{j_{L_1}}(\tau_1) \cdots E_{j_{L_k}}(\tau_k)   
B_{j_{L_{k+1}}}(\tau_{k+1}) \cdots B_{j_{L_n}}(\tau_n)   ,
\label{QREn}
\end{multline}
where we have introduced the multi-index notation $i_L\equiv i_1\cdots i_\ell$, and   $\RR_{E,B}$ represents the $n^{\text{th}}$-order response function. 
Note that the logic we are following here is analogous  to nonlinear polarization theory in electromagnetism, where the nonlinear polarization of an optical medium is parametrized, similarly to \cref{QREn}, as an expansion in powers of the external field (see, e.g., \rcite{10.1093/acprof:oso/9780198702764.001.0001}). In that context, the $E$ and $B$ components of the Weyl tensor in \cref{QREn} are replaced by the electric and magnetic fields, respectively, and $Q$ corresponds to the nonlinear induced polarization of the medium.

The response function $\RR_{E,B}$ in \cref{QREn} is completely general, although it must satisfy a set of consistency requirements.  Beyond the obvious symmetry properties under the exchange of its multi-indices, an important non-trivial constraint arises from causality:  $\RR$ must vanish  whenever any of its arguments $\tau-\tau_j$, for $j=1,\dots,n$, turns negative.
Furthermore, the tensorial structure of $\RR$ is  dictated by the  symmetries of the system, in analogy with \cref{eq:expCcc}. In particular, it can be decomposed into sums of Thorne tensors, which provide a convenient basis.

Substituting \cref{QREn} back into the effective action \eqref{eq:sint} results in a series of nonlinear local interactions on the worldline. As an example, let us consider  non-rotating objects in the static r\'egime. To leading order in the gradient expansion in the purely even sector, the interaction terms read
\begin{equation}
S_\mathrm{h.o.} = \sum_{n\geq 1}\int\dd\tau \, c^{(\mathrm{E})}_{22\cdots2}{E_{i_1}}^{i_2}{E_{i_2}}^{i_3} \cdots {E_{i_{n+1}}}^{i_{1}} ,
\label{eq:E2B2-0nonlinear}
\end{equation}
which generalizes the electric sector of \cref{eq:E2B2-0} beyond linear order. Similar expressions hold in the odd case and for operators that mix $E$ and $B$ fields. 
For instance, at  cubic order, the only possible operators compatible with the symmetries  are\footnote{Parity symmetry implies that the couplings of the operators ${B_{i_1}}^{i_2}{B_{i_2}}^{i_3} {B_{i_{3}}}^{i_{1}}$ and ${E_{i_1}}^{i_2}{E_{i_2}}^{i_3} {B_{i_{3}}}^{i_{1}}$ vanish identically.}
\begin{equation}
S_\mathrm{h.o.} = \int\dd\tau \left( c^{(\mathrm{E})}_{222}{E_{i_1}}^{i_2}{E_{i_2}}^{i_3} {E_{i_{3}}}^{i_{1}} 
+ c^{(\mathrm{EB^2})}_{222}{E_{i_1}}^{i_2}{B_{i_2}}^{i_3} {B_{i_{3}}}^{i_{1}} 
\right).
\label{eq:E2B2-0cubic}
\end{equation}

The coefficients in \cref{eq:E2B2-0nonlinear,eq:E2B2-0cubic} couple quadrupolar ($\ell=2$) fields only. The generalization of \cref{eq:E2B2-0nonlinear} to higher multipoles is straightforward and is obtained by simply adding gradients on each field, analogously to \cref{eq:S-ho} for the linear theory. In these cases, the quadrupolar coefficients $c^{(\mathrm{E})}_{22\cdots2}$ get replaced by $c^{(\mathrm{E})}_{\ell_1 \cdots \ell_{n+1}}$, where the set of numbers $(\ell_1\cdots\ell_{n+1})$ defines a multiplet, with  each $\ell_j$ denoting  the number of gradients acting on the $j$th $E$ field in the multiplet. 

The couplings $c_{\ell_1 \cdots \ell_{n+1}}$, with $n>1$, correspond to the \textit{nonlinear Love numbers}~\cite{Bern:2020uwk,Poisson:2020vap,Riva:2023rcm,Iteanu:2024dvx,Combaluzier-Szteinsznaider:2024sgb,Hadad:2024lsf}.
As discussed above, at a given order in the perturbative expansion, the number of independent coefficients $c_{\ell_1 \cdots \ell_{n+1}}$  is determined by the symmetries of the system --- in the present case, by  spherical symmetry. For instance, let us consider a generic cubic  operator with arbitrary number of spatial derivatives, involving the $E$ field only. Such an operator is obtained by contracting the indices of three different $E$ fields \eqref{eq:EEEBBB}, say 
$E_{i_1 i_2 \cdots i_\ell}$, $E_{j_1 j_2 \cdots j_{\ell_1}}$, and $E_{k_1 k_2\cdots k_{\ell_2}}$.
The contraction of indices between two of them 
gives rise to several possibilities, depending on the number of contracted indices. At one extreme, when the number of contractions is maximal, we are left with  as few as $|\ell_2 - \ell_1|$ free indices; at the other, with no contractions at all,  there are $\ell_1 + \ell_2$ free indices in total. The resulting object is then contracted with the remaining $E$ field to obtain a scalar under  rotations.  One recognizes that this corresponds precisely to the standard  selection rule associated with the addition of angular momenta \cite{Bern:2020uwk,Riva:2023rcm,DeLuca:2023mio}.
At this perturbative order, it is straightforward to see (e.g., \cite{Combaluzier-Szteinsznaider:2024sgb}) that  --- similarly to the linear response --- there can be at most a single non-trivial independent contraction of indices, and therefore only one coupling $c^{(\mathrm{E})}_{\ell_1 \ell_2 \ell_{3}}$  for given multiplet $(\ell_1\ell_2\ell_3)$. The same pattern repeats at higher orders.
For instance, the cubic response operators involve four $E$ fields with indices $\ell_1, \ell_2, \ell_3, \ell_4$, subject to specific constraints among them. Importantly, the sum of all angular momenta, $\ell_1 + \ell_2 + \ell_3 +\cdots$, must be even. Starting from $\mathcal{O}(E^4)$, however, for a fixed set of angular momenta $(\ell_1\ell_2\cdots
\ell_{n+1})$, multiple independent coefficients may arise.
The precise counting of inequivalent contractions and independent Wilson coefficients follows from standard group-theoretic arguments, see e.g., \rcite{Bern:2020uwk,Haddad:2020que,Ruhdorfer:2019qmk}.

\subsection{Examples of EFT calculations and matching}
\label{sec:matching}

The effective field theory \eqref{eq:S-EFT} is completely general and, as long as one is concerned with low-energy processes, it fully captures the relevant physics of the system. The consistency and predictivity of the theory rely on the fact that, up to a certain energy scale and desired level of precision, only a finite number of terms in the expansion \eqref{eq:S-EFT} are needed to provide an accurate description of the dynamics. In particular, the EFT can be employed to compute any type of low-energy observable --- such as scattering amplitudes or correlation functions --- and, as long as one restricts attention to observables involving low-energy fields, the EFT is as predictive as the full theory, whether or not the latter is known.
As discussed earlier, the effective couplings encode all information about the underlying ultraviolet  physics. If the microscopic description is unknown, the best one can often do is to constrain these couplings experimentally. Conversely, if a high-energy  completion is available, the couplings can be computed explicitly by comparing suitable observables in the EFT with their counterparts in the full theory. This procedure, commonly referred to as \textit{matching}, will be the focus of the present section. We will present some explicit EFT calculations, to be compared later with analogous results in general relativity, and illustrate a simple example of matching.

In contrast to many situations in physics, in our case the ``UV  theory'' is known: general relativity.\footnote{One may also envision scenarios in which general relativity is modified at energy scales accessible to gravitational-wave observations of merging compact objects~\cite{Endlich:2017tqa,Yunes:2016jcc,Franciolini:2018uyq,Hui:2021cpm,Cano:2020cao} (see \rcite{Donoghue:1994dn,Donoghue:1995cz,Burgess:2003jk,Donoghue:2012zc,Donoghue:2017pgk,Donoghue:2022eay} for an introduction to general relativistic EFT). In such cases, the ``full theory'' is different from general relativity, and the matching procedure will determine the Wilson coefficients in terms of the parameters of the alternative description (e.g.,~\cite{Barbosa:2025uau,Endlich:2017tqa,Cano:2025zyk}). See \Cref{part:part3Love} below for more details.} Exact solutions in general relativity are notoriously hard to come by, and typically only possible in the presence of simplifying assumptions \cite{Stephani:2003tm}. Fortunately we will not need to solve the nonlinear Einstein equations exactly: during the inspiral phase it is generally the case that the spacetime metric in the vicinity of the object is the Schwarzschild or Kerr spacetime plus a small correction, so it is appropriate to describe the UV by \emph{black hole perturbation theory}, the study of small fluctuations around a black hole spacetime. This theory will be the subject of \Cref{Part:BHPT}; here we concentrate on the low-energy EFT.

In the following, we study several examples of worldline EFT calculations that can be used to compute low-energy gravitational observables. We begin with the gravitational potential sourced by a point particle, before turning to tidal effects. In the literature, EFT calculations and matching of tidal couplings are commonly classified as either \textit{off-shell} or \textit{on-shell}. We briefly go through both approaches in turn, providing examples of off-shell potential calculations and on-shell scattering amplitudes.

\subsubsection{ADM mass matching from background metric reconstruction}
\label{sec:ADMmassex}

The point particle couples to gravity and sources a gravitational potential. To provide a concrete illustration of an EFT calculation and the associated matching procedure, it is instructive to show how this potential can be computed in the EFT. The calculation can be performed in several equivalent ways. To be as  pedagogical as possible, we will present two complementary approaches: the first one evaluates the graviton one-point function using a diagrammatic method; the second one, somewhat more direct, involves solving the Einstein field equations in the presence of a source term.

\paragraph{Mass matching via a diagrammatic approach}
In \cref{eq:PP,eq:pointparticleaction} we introduced the parameter $m$ to define the point-particle action. We now show that this parameter matches the asymptotic Arnowitt--Deser--Misner (ADM) mass of the object. In particular, computing the potential sourced by the point particle amounts to reconstructing the full background metric order by order in $Gm$, which we will see explicitly up to next-to-leading order.

Our starting point is   the EFT  \eqref{eq:S-EFT} for a non-rotating body, to which we will add a gauge-fixing term following the standard Faddeev--Popov procedure in order to define a propagator for the graviton. We expand the metric tensor as 
\begin{equation}
	g_{\mu\nu}  = \bar{g}_{\mu\nu} + \frac{2}{\Mp} h_{\mu \nu} \, ,
	\label{eq:exp_BG}
\end{equation}
where $h_{\mu\nu}$ denotes the canonically normalized graviton field fluctuation  around the background metric $\bar{g}_{\mu\nu}$, which satisfies  the vacuum Einstein equations, 
\begin{equation}
\bar{R}_{\mu\nu}-\frac{1}{2}\bar{g}_{\mu\nu} \bar{R} = 0,
\end{equation}
with $\bar{R}_{\mu\nu}$ and $\bar{R}$ representing the Ricci tensor and scalar of $\bar{g}_{\mu\nu}$, respectively.

Following the standard background-field method \cite{DeWitt:1967ub,tHooft:1974toh,Abbott:1980hw}, we can compute the effective action of the system, $W[\mathcal{J}]$, by formally performing the  path integral
\begin{equation}
	e^{i W[\mathcal{J}]} = e^{i S_{0}[\bar{g}, x({\affine})]} \int \mathcal{D}h \exp\left\{ i S_{h}[\bar{g}, h, x({\affine})] + i S_{\textrm{GF}}[\bar{g}, h] + i \int \!\!\dd^4 y \sqrt{-\bar{g}} \mathcal{J}^{\mu\nu}(y) h_{\mu\nu}(y)\right\} \, .
	\label{eq:PIexp}
\end{equation}
Here $\mathcal{J}$ is an external source, and  we have split the EFT action \eqref{eq:S-EFT} as  $S_{\text{EFT}}[g, x(\tau)] = S_{0}[\bar{g}, x({\affine})] + S_{h}[\bar{g}, h, x({\affine})]$, with  $S_{0}$ defined as $S_{\text{EFT}}$ at $h_{\mu \nu}=0$, and $S_{h}[\bar{g}, h, x(\tau)]$ containing all corrections involving the fluctuation $h_{\mu \nu}$. 
In particular, the quadratic action in $h_{\mu\nu}$ from $S_{h}$ is 
\begin{align}
S_\mathrm{EH}^{(2)}  &= \frac12 \int \dd^4x \sqrt{-\bar g}  \,  h \mathcal E h \nonumber \\
&= \int\dd^4x \sqrt{-\bar g} \left(-\frac12 \bar \nabla_\alpha h_\mn \bar\nabla^\alpha h^\mn +\frac12 \bar g^{\mu\nu}\partial_\mu h \partial_\nu h -\bar\nabla_\mu h\bar\nabla_\nu h^\mn + \bar\nabla_\alpha h_\mn \bar\nabla^\nu h^{\mu\alpha}\right),
\label{eq:EHaction2}
\end{align}
where the \emph{Lichnerowicz operator} is
\begin{align}\label{eq:lich}
\mathcal{E}^{\mn\ab}h_\ab &= \epsilon^{\mu\alpha\rho\gamma}\epsilon^{\nu\beta\sigma}{}_\gamma \bar\nabla_\alpha \bar \nabla_\beta h_{\rho\sigma} \nonumber \\
&= \bar \Box h^\mn  + \bar \nabla^\mu \bar \nabla^\nu h - 2 \bar \nabla^\alpha \bar \nabla^{(\mu}h^{\nu)}_\alpha +\left(\bar \nabla_\alpha \bar \nabla_\beta h^\ab - \bar \Box h\right) \bar g^\mn,
\end{align}
with $h\equiv \bar g ^{\mu\nu}{h}_{\mu\nu}$, and $\bar{\nabla}_\mu$  the covariant derivative compatible with $\bar{g}_{\mu\nu}$. The term $S_{\textrm{GF}} $  is the gauge-fixing action 
\begin{equation}
	S_{\textrm{GF}} 
	\equiv -\int \dd^4 x \sqrt{-\bar{g}} \,  \bar{g}^{\mu\nu}
		\left( \bar{g}^{\alpha\beta}\bar{\nabla}_{\alpha} h_{\beta\mu} - 
		\frac{\bar{g}^{\alpha\beta}}{2}\bar{\nabla}_\mu h_{\alpha\beta} \right) \left( \bar{g}^{\rho\sigma}\bar{\nabla}_{\rho} h_{\sigma\nu} - 
		\frac{\bar{g}^{\rho\sigma}}{2}\bar{\nabla}_\nu h_{\rho\sigma} \right) \, ,
		\label{eq:BGGF}
\end{equation}
which enforces the harmonic (de Donder) gauge choice  in the background-field formulation \cite{DeWitt:1967ub,tHooft:1974toh,Abbott:1980hw} (see also \rcite{Riva:2023rcm,Iteanu:2024dvx}; in addition, a useful reference for a list of Feynman rules is~\rcite{Riva:2022uxb}).

In general, it is the physical situation that determines the choice of background metric $\bar{g}_{\mu\nu}$.
For instance, $\bar{g}_{\mu\nu}$ can correspond to some external potential or tidal field. In the present case, since we are interested in computing the vacuum-induced potential, we will take $\bar{g}_{\mu\nu}$ to be just the Minkowski metric, i.e.,~$\bar g_{\mu\nu}\equiv \eta_{\mu\nu}$. In this case, the vacuum equations of motion for $h_{\mu\nu}$ in de Donder (dD) gauge, obtained from $S_\mathrm{EH}^{(2)}+S_\mathrm{GF}$, boil down simply to 
\begin{equation}
\square h_{\mu\nu} - \frac{1}{2} \eta_{\mu\nu} \square h =0 ,
\end{equation}
while the gauge condition reads
\begin{equation}
\partial_\mu\left(h^\mn-\frac12h\eta^\mn\right) = 0.
\label{deDonderG}
\end{equation}

We are interested in computing the one-point function  of the field $h_{\mu\nu}$. 
In terms of  the effective action $W[\mathcal{J}]$ this corresponds to\footnote{We shall drop the exponential of   $S_0$, which is field-independent, from \cref{eq:PIexp}.}
\begin{equation}
	\langle h_{\mu\nu}(x)\rangle = \frac{1}{\sqrt{-\bar{g}}}  \frac{\delta  W[\mathcal{J}]}{i\delta \mathcal{J}^{\mu\nu}(x)}  \bigg|_{\mathcal{J}=0} \, ,
\end{equation}
which amounts to evaluating all Feynman diagrams with one external $h_{\mu\nu}(x)$ field. 
We stress that, although the framework is formulated in terms of a path integral, we are ultimately interested  in the classical limit. This limit can be enforced from the outset by applying a saddle-point approximation \cite{Goldberger:2004jt}. In diagrammatic terms, this corresponds to neglecting all diagrams that contain closed loops of the graviton field $h_{\mu\nu}$.  This is also the reason we do not need to add any ghost fields in \cref{eq:PIexp}. 

The diagrams with one external graviton leg and no bulk loops are~\cite{Goldberger:2004jt,Goldberger:2022rqf}
\begin{equation}
    \raisebox{3pt}{\MassScW} 
    + \raisebox{31pt}{\metricsecondorder} + \raisebox{31pt}{\metricthirdorderA} +\raisebox{31pt}{\metricthirdorderB}
    +\cdots,
\label{fig:metricreconstruction}
\end{equation}
where the ellipsis indicates diagrams with an increasing number of mass-$m$ insertions on the worldline (denoted by  black circle). Curly lines represent  graviton propagators. 
Bulk graviton vertices are obtained by expanding the Einstein--Hilbert action $S_{\text{EH}}$ \eqref{eq:S-EH} to cubic, quartic, and higher orders in perturbation theory around Minkowski space. In the following, we explicitly compute  the first diagram in the series \eqref{fig:metricreconstruction} and show that it reproduces the leading Schwarzschild  correction to the background Minkowski metric. Subleading corrections in $Gm$ can be obtained by evaluating the remaining diagrams in the expansion \eqref{fig:metricreconstruction}. Resumming the series ultimately reconstructs the full Schwarzschild metric~\cite{Goldberger:2004jt,Duff:1973zz,Jakobsen:2020ksu,Mougiakakos:2020laz,Damgaard:2024fqj,Mougiakakos:2024nku}.

The first diagram of \cref{fig:metricreconstruction} is obtained by combining the vertex in $S_{\rm pp}$   that couples $h_{\mu\nu}$ to the point particle, with a graviton bulk propagator.
The  Feynman rules for the point-particle coupling and the graviton propagator in de Donder (dD) gauge are, respectively,
\begin{align}
    \raisebox{3pt}{\MassScZ} & \quad  = \quad  i  \frac{m}{\Mp} \int\dd{\affine} \, e^{-i k\cdot \X({\affine})} \frac{\dd \X^\mu}{\dd {\affine}} \frac{\dd \X^\nu}{\dd {\affine}}  \; , \\
	\raisebox{3pt}{\gravconv} & \quad \equiv \quad  G^{\rm dD}_{\mu\nu\rho\sigma }(k^2)=\frac{i}{k^2} P_{\mu\nu\rho\sigma } =  -\frac{i}{2k^2}\left(\eta_{\mu\rho}\eta_{\nu\sigma} + \eta_{\mu\sigma}\eta_{\nu\rho} - \eta_{\mu\nu}\eta_{\rho\sigma}  \right) \, ,
    \label{gravitonprop}
\end{align}
from which we can compute the graviton one-point function.  In the rest frame of the point particle, we have
\begin{align}
	\langle h_{\mu\nu}(x) \rangle & = -\frac{m}{\Mp} \int \dd {\affine} \, \frac{\dd \X^\rho}{\dd {\affine}} \frac{\dd{\X}^\sigma}{\dd{{\affine}}} \int \frac{\dd^4 k}{(2\pi)^4}
			\frac{e^{-i k \cdot ( \X(\affine) - x )}}{k^2}P_{\rho\sigma\mu\nu} \\
			& = \frac{m}{\Mp} 
			\left(\frac{\dd{\X}_\mu}{\dd {\affine}}\frac{\dd{\X}_\nu}{\dd {\affine}} +\frac{1}{2}\eta_{\mu\nu}\right)
			\int \frac{\dd^3 \vec{k}}{(2\pi)^3}
			\frac{e^{i \vec{k} \cdot \vec{x} }}{\vec{k}^2} \, .
			\label{eq:SCHLO}
\end{align}
Using the formula
\begin{equation}
\int \frac{\dd^d \vec{k}}{(2\pi)^d}
			\frac{e^{i \vec{k} \cdot \vec{x} }}{\vec{k}^2} 
= \frac{\Gamma(\frac{d}{2}-1)}{(4\pi)^{d/2}}\left( \frac{\vec{x}^2}{4}\right)^{1-\frac{d}{2}} ,
\end{equation}
we obtain
\begin{equation}
	\langle h_{\mu\nu}(x) \rangle  =  \frac{\Mp}{2} \frac{\rs}{r}
		\left(\eta_{\mu\nu} + 2 \frac{\dd{\X}_\mu}{\dd{\affine}} \frac{\dd {\X}_\nu}{\dd{\affine}} \right) ,
	\label{eq:Schwarz1-D}
\end{equation}
where $r = \sqrt{\vec{x}^2}$.
Choosing  ${\affine} = t$ as the affine parameter, so that $\X^\mu(t) = (t, \vec{0})$ and $\frac{\dd \X^\mu(t)}{\dd {\affine}} = (1, \vec{0})$, \cref{eq:Schwarz1-D} reproduces the Schwarzschild metric in isotropic coordinates \cite{Ivanov:2024sds,Dennison:2010wd} up to the first non-trivial  order  in $G$:\footnote{A note of caution: here, $r$ denotes the radial distance in isotropic coordinates and should not be confused with the Schwarzschild coordinate $r$ that we use in \cref{sec:BHPT} (see, e.g., \cref{eq:sch}).} 
\begin{equation}
	\dd s^2 = \big( \eta_{\mu\nu} + 2\Mp^{-1} \langle h_{\mu\nu}(x) \rangle \big) 
	\dd x^\mu \dd x^\nu = 
	- \left(1 -  \frac{\rs}{r}\right) \dd t^2 
	+ \left(1+  \frac{\rs}{r}\right) \dd \vec{x}^2 + \mathcal{O}(\rs^2) \, ,
\label{Schmetriclinearrs}
\end{equation}
provided one identifies the parameter $m$ with the ADM mass $M$, i.e., 
\begin{equation}
2Gm=2GM\equiv \rs .
\label{ADMmassmatch}
\end{equation}
We stress that, for the identification $m=M$, it is crucial that the comparison between the EFT solution   and the full metric linearized in $\rs$ \eqref{Schmetriclinearrs} be performed in the same gauge. Alternatively,  the same result can be obtained by comparing gauge-invariant quantities in the EFT  and in the full theory.

\paragraph{Solving the equations of motion with a point source}
For completeness, we note that a more mundane way of obtaining the matching \eqref{ADMmassmatch} is to solve    the Einstein equations directly in the static limit. Varying $S_{\rm EFT}=  S_{\rm EH}+  S_{\rm pp}$ yields,  in the rest frame of the point particle,
\begin{equation}
\Mp^2 G_{\mu\nu}^{(1)}[h] = - m \, \delta_\mu^0\delta_\nu^0 \, \delta^{(3)}(\vec{x}),
\label{eq:Gm}
\end{equation}
where $G_{\mu\nu}^{(1)}$ is the linearized Einstein tensor, 
\begin{equation}
G_{\mu\nu}^{(1)}[h] = - \frac{1}{\Mp} \left[  \partial^\rho \partial_\mu h_{\nu\rho} + \partial^\rho \partial_\nu h_{\mu\rho}  -\square h_{\mu\nu}  -\partial_{\mu}\partial_\nu h -\left( \partial^\rho\partial^\sigma h_{\rho\sigma} -\square h \right)\eta_{\mu\nu}   \right] .
\end{equation}
We now look for static, spherically-symmetric  solutions  of  \cref{eq:Gm}.
In the de Donder gauge \eqref{deDonderG}, which implies $\partial^\rho\partial^\sigma h_{\rho\sigma} =\frac{1}{2}\square h$, the linearized Einstein tensor reduces to
\begin{equation}
G_{\mu\nu}^{(1)}[h] =  \frac{1}{\Mp} \left(  \square h_{\mu\nu}  - \frac{1}{2}\eta_{\mu\nu}   \square h  \right) .
\label{GmunudD}
\end{equation}
Taking the trace of \cref{eq:Gm} with $\eta^{\mu\nu}$ and using  \cref{GmunudD}, one finds  $\Mp \square h = - m \,  \delta^{(3)}(\vec x)$. Plugging this back into \cref{eq:Gm} gives the following equation for $h_{\mu\nu}$:
\begin{equation}
\nabla^2 h_{\mu\nu}  = -\frac{m}{\Mp} \left(\delta_\mu^0\delta_\nu^0 + \frac{1}{2}\eta_{\mu\nu} \right) \delta^{(3)}(\vec{x}),
\end{equation}
where the d'Alembert operator has been replaced by  the spatial  Laplacian for time-independent perturbations.
Integrating over a sphere of radius $r$, and applying the divergence theorem and  spherical symmetry, one finds the following equation in $r$: $4\pi  r^2\partial_r h_{\mu\nu} = -\frac{m}{\Mp} \left( \delta_\mu^0\delta_\nu^0 + \frac{1}{2}\eta_{\mu\nu} \right)$. Dividing by $r^2$ and integrating over $r$ yields
\begin{equation}
2\Mp^{-1}  h_{\mu\nu} =  \frac{2G m}{r} \left( 2 \delta_\mu^0\delta_\nu^0 + \eta_{\mu\nu} \right) ,
\end{equation}
in agreement with \cref{Schmetriclinearrs}, upon identifying $m=M$.

\subsubsection{Tidal response from one-point function}
In the previous section, we illustrated the computation of a one-point function within the EFT framework, using the action  $S_\mathrm{EH} + S_\mathrm{pp}$. We now slightly modify the physical setup and assume that, in addition to the flat metric $\eta_{\mu\nu}$, the background $\bar{g}_{\mu\nu}$ includes an external tidal field, which we will denote by $H_{\mu\nu}$. On this background, we again compute the one-point function of $h_{\mu\nu}$. In this case, rather than describing the gravitational potential sourced by the point particle, the one-point function represents the response induced by the external  field $H_{\mu\nu}$. As before, for pedagogical clarity, we will present different equivalent methods of performing this calculation.

\paragraph{Tidal response from a diagrammatic approach}
Let us begin with a diagrammatic approach. Starting from the EFT action \eqref{eq:S-EFT}, we are now interested in couplings to the point particle involving two graviton lines. Focusing on the static tidal response at tree level, the relevant interactions are those arising from $S_\mathrm{h.o.}$ without time derivatives. The couplings generated by the expansion of $S_{\rm pp}$ in $h_{\mu\nu}$ are important for reconstructing the tidal source at subleading order in $\rs$ and for performing matching calculations at loop level (see, e.g., \rcite{Kol:2011vg,Ivanov:2022hlo} for some examples), but they do not contribute to a tree-level calculation.
We may therefore neglect contributions from the point-particle action $S_{\rm pp}$ at this stage. We will comment on how to extend the calculation to loop order at the end of this section.

As mentioned, we consider the following background metric:
\begin{equation}
	\bar{g}_{\mu\nu}(x) = \eta_{\mu\nu} +  H_{\mu\nu}(x) \, , 
	\label{eq:exp_flat}
\end{equation}
where $H_{\mu\nu}(x)$ is assumed to be static. In particular, it solves the flat-space Einstein equations with tidal field boundary conditions at infinity, and is  regular at the origin. In  Fourier space, we  write it at zero frequency as
\begin{equation}
H_{\mu\nu}(k) = 2\pi \delta(k^0) H_{\mu\nu}(\vec{k}) \, ,
\end{equation}
and introduce the following  diagrammatic convention:
\begin{equation}
	\raisebox{3pt}{\TidalHconv} \quad \equiv\quad H_{\mu\nu} \notag \, .
\end{equation}
The advantage of working with the background field method is that the final result for $h_{\mu\nu}$ is
covariant under diffeomorphisms of the external metric $\bar g_{\mu\nu}$.
This allows us to choose the tidal field in any convenient gauge of our choice. In what follows, we will choose $H_{\mu\nu}$ to satisfy the vacuum Einstein equation on a flat background consistent with the Regge--Wheeler gauge~\cite{Regge:1957td}, which we discuss in \cref{sec:BHPT}. In particular, focusing for simplicity  on the gravito-electric sector,
\begin{equation}
	H_{\mu\nu}(\vec{k}) = \sum_{\ell, m}\mathcal{E}^{({\rm E})}_{\ell m}(\eta_{\mu\nu} +2u_\mu u_\nu)i^\ell {\cal Y}_{\ell m}^{i_1\cdots i_\ell}\frac{\partial}{\partial k^{i_1}}\cdots \frac{\partial}{\partial k^{i_\ell}} (2\pi)^3 \delta{}^{(3)}(\vec{k}) \, ,
\label{eq:tidalHmunu}
\end{equation}
where $\mathcal{E}^{({\rm E})}_{\ell m}$ is the amplitude of the  tidal $E$ field, and similarly for the gravito-magnetic sector~\cite{Iteanu:2024dvx}. 

We start from the tidal action \eqref{eq:S-ho}, where we express the $E$ component of the Weyl tensor in terms of the canonically-normalized graviton field $h_{\mu\nu}$ in the static limit as $E_{ij}=-\partial_i\partial_jh_{00}/\Mp$.  We can then extract the  Feynman rule for the Love number vertex (at fixed $\ell$):
\begin{equation}
	\LinearLoveGR = 2 i \lambdaEell  2\pi\delta(k^0)\int \frac{\dd^3 \vec{q}}{(2\pi)^3} \, q_{\langle i_1} \cdots q_{i_{\ell-2}} \bar{E}(\vec{q})_{i_{\ell-1} i_\ell\rangle} k^{i_{1}}\cdots k^{i_{\ell-2}}\mathfrak{E}^{(0)i_{\ell-1} i_\ell}_{\mu\nu}(k) \, ,
\label{eq:linearresponseFD}
\end{equation}
where we have introduced the functional derivative $\mathfrak{E}^{(0)ij}_{\mu\nu}\equiv \delta E^{ij}/\delta h^{\mu\nu}$,   denoted  the electric components of the  Weyl tensor  computed on the tidal solution \eqref{eq:tidalHmunu} by  $\bar{E}_{ij}$, and represented the coupling $\lambdaEell$ with a black square. We are following here the conventions of \rcite{Riva:2023rcm,Iteanu:2024dvx,Riva:2022uxb} for the Feynman diagrams.
Note that the overall factor of 2 in \cref{eq:linearresponseFD} arises from the mixed  term in $H$ and $h$ in the expansion of the action at quadratic order in the metric fluctuation, while the $2\pi\delta(k^0)$ factor results from the worldline time integral. Note also that, while the momentum $\vec k$ is taken to be outgoing, the momentum $\vec q$ is ingoing in the vertex; in particular, this implies an extra $(-1)^\ell$ factor, which canceled against the  $i^{2\ell-4}$  from the momentum-space Fourier transform. 
From \cref{eq:linearresponseFD}, the linear response is readily computed as 
\begin{equation}
	2\Mp^{-1}\langle h_{\mu\nu}(k)\rangle = -8\pi \frac{\lambdaEell}{\Mp} \delta(k^0)\frac{k^{i_1}\cdots k^{i_{\ell-2}}}{\vec{k}^2} \mathfrak{E}^{(0)i_{\ell-1} i_\ell}_{\rho\sigma}(k)P^{\rho\sigma}{}_{\mu\nu}\int \frac{\dd^3 \vec q}{(2\pi)^3} q_{\langle i_1}\cdots q_{i_{\ell-2}} \bar{E}(\vec{q})_{i_{\ell-1} i_\ell\rangle} \, ,
\label{eq:kappahm2}
\end{equation}
where we used the expression \eqref{gravitonprop} for the de Donder propagator.
Using \cref{eq:tidalHmunu} and the definition of the Weyl tensor \eqref{eq:Weyl-def}, we find 
\begin{equation}
	\bar{E}(\vec{q})_{ij} =  \frac{1}{2}\sum_{\ell, m}\mathcal{E}^{({\rm E})}_{\ell m} q_i q_j i^\ell {\cal Y}_{\ell m}^{i_1\cdots i_\ell}\frac{\partial}{\partial q^{i_1}}\cdots \frac{\partial}{\partial q^{i_\ell}} (2\pi)^3 \delta{}^{(3)}(\vec{q}) \, .
\end{equation}
Transforming to real space, \cref{eq:kappahm2} becomes
\begin{equation}
	2\Mp^{-1} \langle h_{\mu\nu}(x)\rangle = - 2   \frac{\lambdaEell}{\Mp^2} \sum_m\mathcal{E}^{({\rm E})}_{\ell m} \ell!  {\cal Y}_{\ell m}^{i_1\cdots i_\ell} 
	 P^{00}{}_{\mu\nu} (-i)^\ell \int \frac{\dd^3 \vec k}{(2\pi)^3}   \frac{k_{\langle i_1}\cdots k_{i_\ell \rangle}}{\vec{k}^2} e^{i \vec{k}\cdot \vec{x}}
	 \, .
\end{equation}
Finally, using  the identity
\begin{equation}
	(-i)^\ell\int \frac{\dd^d \vec k}{(2\pi)^d} \frac{k^{\langle i_1} \cdots k^{i_\ell \rangle}}{\vec{k}^2}e^{i \vec{k}\cdot\vec{x}} =  (-1)^\ell \frac{\Gamma(\frac{d}{2}-1)\Gamma(2-\frac{d}{2})}{2^{2-d-\ell}(4\pi)^{d/2}\Gamma(2-\ell-\frac{d}{2})}
    \frac{x^{\langle i_1} \cdots x^{i_\ell \rangle}}{(r^2)^{\ell+d/2-1}} \, ,
	\label{eq:FTgenl}
\end{equation}
we can derive the linear response field $\langle h_{\mu\nu}(x)\rangle$ for generic $\ell$. For instance, for the $t$--$t$ component we find
\begin{align}
	2\Mp^{-1} \langle h_{00}(x)\rangle  &=  \frac{\lambdaEell}{\Mp^2}\sum_m\mathcal{E}^{({\rm E})}_{\ell m} \frac{\ell \,\Gamma(2\ell)}{2^{\ell+1}\pi}  {\cal Y}_{\ell m}^{i_1\cdots i_\ell} 
    \frac{x^{\langle i_1} \cdots x^{i_\ell \rangle}}{r^{2\ell+1}} \nonumber\\
    &    =    \frac{\lambdaEell}{\Mp^2} \sum_m \mathcal{E}^{({\rm E})}_{\ell m} \frac{\ell\, \Gamma(2\ell)}{2^{\ell+1}\pi}  
    \frac{Y_{\ell m} }{r^{\ell+1}} ,
	\label{eq:hLR00}
\end{align}
where in the second equality we used \cref{eq:thornetensorsY}. 
Note that $h_{00}$ does not transform under a time-independent gauge transformation (in Minkowski spacetime). It is therefore a genuinely gauge-invariant quantity, which we can use to match directly to a given UV solution. For later convenience, we use \cref{eq:hLR00} to evaluate the Weyl $E$ field, which remains gauge-invariant beyond the static limit at linear order in perturbation theory. Using the definition $H_{00}(x)=r^\ell \mathcal{E}^{({\rm E})}_{\ell m} Y_{\ell m}$ and summing  over all multipoles, we obtain
\begin{equation}
E_{rr}(x) = \bar E_{rr}(x) +	 \langle E_{rr}(x)\rangle  = -\frac{1}{2} \sum_{\ell, m}\ell(\ell-1) \mathcal{E}^{({\rm E})}_{\ell m} \left(
r^{\ell-2} + \frac{\lambdaEell}{\Mp^2} \frac{(\ell+2)(\ell+1) \Gamma(2\ell)}{2^{\ell+1}(\ell-1)\pi}  
    r^{-\ell-3}
\right)Y_{\ell m} 
\, .
	\label{eq:ERij}
\end{equation}

The expression \eqref{eq:ERij} can be compared with the $r$--$r$ component of the $E$-field obtained from the solution \eqref{eq:introgtt0}. The result of this  matching is 
\begin{equation}
\bar c^{(\mathrm{E})}_\ell=k^{(\mathrm{E})}_\ell ,
\label{eq:matchingcEl}
\end{equation}
with $\bar c^{(\mathrm{E})}_\ell$ defined in \cref{eq:lambdakell}, thereby identifying the coefficients $k^{(\mathrm{E})}_\ell$ --- computed from a full-theory calculation (see \cref{part:part3Love}) --- with the EFT Love number couplings.\footnote{We stress  that this is only a ``tree-level'' comparison. A complete matching requires accounting for gravitational nonlinearities, as we discuss in \cref{sec:matchingsummary}.}

\paragraph{Solving the equations in the presence of a tidal source}
The same result can be obtained from directly solving the static equations perturbatively in the metric in the presence of  a tidal field source. For simplicity, we will again focus on the gravito-electric sector (see, e.g., \rcite{Kol:2011vg,Hui:2020xxx,Hadad:2024lsf} for more details).

Once again, we start from the EFT action~\eqref{eq:S-EFT}.
In what follows, we will perturb around flat space and denote the metric fluctuation by $h_{\mu\nu}$, i.e.,~$g_\mn = \eta_\mn + 2h_\mn/\Mp$. Unlike in the previous section, it will be convenient to include  the tidal field in $h$; we will  separate tidal and response fields later. 
The quadratic Einstein--Hilbert action can be read off from \cref{eq:EHaction2}, with covariant derivatives replaced by  derivatives on a flat background in Cartesian coordinates, so that $(\bar g,\bar \nabla)\to(\eta,\partial)$.
The field $h_{\mu\nu}$ is taken to satisfy the de Donder gauge \eqref{deDonderG}. (Unlike the diagrammatic approach, we assume  here that the tidal and response fields satisfy the same gauge condition.)
We can also decompose the metric into its space and time components using the particle's four-velocity and the associated projector, so the de Donder gauge is\footnote{Because we are working in Cartesian coordinates we do not need to distinguish between upper and lower spatial indices.}
\begin{equation}
\partial_i h_{0i} = 0,\quad \partial_j h_\Ij = \frac12\left(\partial_i h_{jj}-\partial_ih_{00}\right),
\end{equation}
and the Einstein--Hilbert action is 
\begin{align}
S_\mathrm{EH} &= \int\dd^4x \left(-\frac12\partial_\alpha h_\mn \partial^\alpha h^\mn +\frac14(\partial h)^2\right) \nonumber \\
&= \int\dd^4x\left(-\frac14(\partial_i h_{00})^2-(\partial_ih_{0j})^2-\frac12(\partial_ih_{jk})^2 + \frac14(\partial_i h_{jj})^2 -\frac12\partial_i h_{00}\partial_i h_{jj} +\dots\right) ,
\end{align}
where the ellipsis denotes terms containing time derivatives acting on the fields, which will not contribute in what follows since we restrict the analysis to the static response limit.
It useful to further decompose $h_\mn$ as 
\begin{equation}
h_{00} = \phi,\quad h_{0i} = A_i,\quad h_{ij} = \bar h_{ij} + \phi \delta_\Ij,
\end{equation}
where  the fields $(\phi,A,\bar h_\Ij)$ do not depend on time. The de Donder gauge condition then translates to
\begin{equation}
\partial_i A_i = 0,\qquad \partial_j \bar h_\Ij = \frac12\partial_i \bar h.
\end{equation}
The Einstein--Hilbert Lagrangian density, defined from the action by $S_\mathrm{EH}=\int \dd^4x\mL_\mathrm{EH}$, is conveniently computed using the double-$\epsilon$ expression,
\begin{align}
\mL_\mathrm{EH}  &= -\frac12   \epsilon^{\mu\alpha\rho\gamma}\epsilon^{\nu\beta\sigma}{}_\gamma \partial_\rho h_\mn \partial_\sigma h_\ab \nonumber \\
&= -\frac12 \epsilon^{\mu\alpha i \gamma}\epsilon^{\nu\beta j}{}_\gamma \partial_i h_\mn \partial_j h_\ab \nonumber \\
&= \frac12\epsilon^{ikm}\epsilon^{jln} \partial_i h_{kl} \partial_j h_{mn} + \epsilon^{ikm}\epsilon^{jl}{}_m\left(\partial_iA_k\partial_jA_l-\partial_i\phi\partial_jh_{kl}\right)  \nonumber \\
&= \frac12\epsilon^{ikm}\epsilon^{jln}\partial_i \bar h_{kl} \partial_j \bar h_{mn}-(\partial\phi)^2 + \frac12F_\Ij F^\Ij,
\end{align}
where $F_\Ij = \partial_iA_j-\partial_jA_i$, and where we have integrated by parts in going to the third line.
Our convention is $\epsilon_{0\Ij k} = \epsilon_{\Ij k}$.  We can now see why we work with $\bar h_\Ij$ rather than $h_\Ij$: we have simply performed a field redefinition to demix $\phi$ and $h_\Ij$.

The Riemann tensor to linear order in $h_\mn$ is (recalling that $h_\mn$ is related to the metric perturbation $\delta g_\mn$ by $\Mp\delta g_\mn = 2h_\mn$)
\begin{align}
\Mp R_{\mn\ab} &= -4\partial_{[\mu|}\partial_{[\alpha}h_{\beta]|\nu]} \nonumber \\
&= \partial_\mu\partial_\beta h_{\nu\alpha} + \partial_\nu\partial_\alpha h_{\mu\beta} - \partial_\nu\partial_\beta h_{\mu\alpha} - \partial_\mu\partial_\alpha h_{\nu\beta}.
\end{align}
On shell (i.e., when the linearized Einstein equations hold) the Riemann and Weyl tensors coincide, so it is straightforward to compute the electric and magnetic fields,
\begin{subequations}
\begin{align}
\Mp E_\Ij &= \Mp R_{0i0j} \nonumber\\
&= -\partial_i\partial_j \phi,\\
\Mp B_\Ij &= \Mp \tilde R_{0i0j} = \frac\Mp2\epsilon_{0ikl}R^{kl}{}_{0j} \nonumber\\
&= \frac12\epsilon_{ikl}\partial_jF^{kl}.
\end{align}
\end{subequations}
Thus, the EFT action for $\phi$, including the higher-order terms \eqref{eq:S-ho} ($E$ sector only), is
\begin{equation}
S_{\mathrm{EH}}^\phi+ S_\mathrm{ho}^\phi = - \int\dd^4x \,  (\partial\phi)^2 + \int \dd \tau \sum_{\ell=2}^\infty \frac{\lambdaEell}{\Mp^2}  \partial_{\langle i_1} \cdots \partial_{i_\ell\rangle}\phi \partial^{\langle i_1} \cdots \partial^{i_\ell\rangle}\phi
\label{eq:phiactionE}
\end{equation}

Let us now compute the response induced by an external gravito-electric tidal field. For that, it is enough to focus on the $\phi$ action~\eqref{eq:phiactionE}.
We can split
\begin{equation}
\phi = \phi^{(0)} + \varepsilon \phi^{(1)} + \mathcal{O}(\varepsilon^2)
\end{equation}
where $\varepsilon$ is a formal expansion parameter (which we will set to one in final results).   $\phi^{(0)}$ represents the tidal field while $\phi^{(1)}$ corresponds to the induced response solution, i.e.,~we can identify:
\begin{equation}
\phi^{(0)} = \phi^\mathrm{tidal},\quad \phi^{(1)} = \phi^\mathrm{resp}.
\end{equation}
By definition, the tidal field satisfies the bulk  equations of motion (i.e., those in the absence of the point-particle interaction terms $S_\mathrm{ho}$), which in de Donder gauge and in the static limit reduce simply to $\nabla^2\phi=0$. 
The  corresponding  tidal solution, satisfying regular boundary conditions at the location of the point particle,   is 
\begin{align}\label{eq:phi-tidal}
    \phi^\mathrm{tidal} &= \frac{\Mp}{2}\sum_{\ell=2}^\infty\sum_{m=-\ell}^\ell \mathcal{E}^{({\rm E})}_{\ell m} r^\ell Y_{\ell m}(\theta,\varphi) \nonumber\\
    &= \frac{\Mp}{2}\sum_{\ell=2}^\infty\sum_{m=-\ell}^\ell \mathcal{E}^{({\rm E})}_{\ell m}{\mathcal{Y}}^{\ell m}_{i_1\cdots i_\ell}x^{i_1}\cdots x^{i_\ell},
\end{align}
where in the second line we have used \cref{eq:thornetensorsY}. Using the identity 
$\partial_{i_1\cdots i_\ell}(x^{j_1}\cdots x^{j_\ell}) = \ell!\delta^{(j_1}_{i_1}\cdots\delta^{j_\ell)}_{i_\ell}$, we further have
\begin{equation}
\partial_{\langle i_1} \cdots \partial_{i_\ell\rangle}\phi^\mathrm{tidal} = \frac{\Mp}{2}\sum_{\ell m} \mathcal{E}^{({\rm E})}_{\ell m} \ell! {\mathcal{Y}}^{\ell m}_{i_1\cdots i_\ell}.
\end{equation}
Focusing on a single multipole, the equation for the response $\phi^\mathrm{resp}$ obtained from the action~\eqref{eq:phiactionE} at linear  order in $\varepsilon$ is
\begin{align}
\nabla^2\phi^\mathrm{resp} &= (-1)^{\ell+1} \frac{\lambdaEell}{\Mp^2} \int\dd \tau \, \partial^{i_1}\cdots \partial^{i_\ell}\left(\delta^{(4)}(x-\X(\tau))\partial_{\langle i_1}\cdots \partial_{i_\ell\rangle}\phi^\mathrm{tidal}\right) \nonumber\\
&= (-1)^{\ell+1}\frac{\lambdaEell}{2\Mp}  \mathcal{E}^{({\rm E})}_{\ell m} \ell! {\mathcal{Y}}^{\ell m}_{i_1\cdots i_\ell}\int\dd \tau \, \partial^{ i_1}\cdots \partial^{i_\ell}\delta^{(4)}(x-\X(\tau)) \nonumber\\
&\equiv J_\eff(x).
\end{align}
Choosing the rest frame of the object as our coordinate system, with the object located at the origin, we can write $\X^\mu(\tau)=(\tau,0,0,0)$.
Then we can integrate over the worldine in  $J_\eff$, 
\begin{equation}
J_\eff = (-1)^{\ell+1} \frac{\lambdaEell}{2\Mp}  \mathcal{E}^{({\rm E})}_{\ell m} \ell! {\mathcal{Y}}_{\ell m}^{i_1\cdots i_\ell}\partial_{ i_1}\cdots \partial_{i_\ell}\delta^{(3)}(\vec x).
\end{equation}
The solution for $\phi^\mathrm{resp}$ can be obtained by convolving the effective source $J_\eff$ with the Green's function,
\begin{equation}
\phi^\mathrm{resp}(\vec x) = [G\ast J_\eff](\vec x) \equiv \int\dd^3\vec x' G(\vec x-\vec x')J_\eff(\vec x'),
\end{equation}
where the Green's function for the three-dimensional Euclidean Laplace operator is
\begin{equation}
G(\vec x-\vec x') = -\frac{1}{4\pi |\vec x-\vec x'|}.
\end{equation}
To perform this integral it is especially convenient to go to Fourier space, where convolution becomes multiplication,
\begin{equation}
\phi^\mathrm{resp}(\vec p) = G(\vec p)J_\eff(\vec p).
\end{equation}
The Fourier transform of the Green's function is simply
\begin{equation}
G(\vec p) = -\frac1{\vec p^2},\label{eq:Green-mom-space}
\end{equation}
and it is similarly straightforward to calculate the momentum-space source function,
\begin{align}
J_\eff(\vec p) &= \int \dd^3 \vec{x} \,  e^{-i\vec p\cdot\vec x} J_\eff(\vec x) \nonumber\\
&= (-1)^{\ell+1}\frac{\lambdaEell}{2\Mp}  \mathcal{E}^{({\rm E})}_{\ell m} \ell!  {\mathcal{Y}}_{\ell m}^{i_1\cdots i_\ell} \int \dd^3 \vec{x} \,   e^{-i\vec p\cdot\vec x}\partial_{ i_1}\cdots \partial_{i_\ell}\delta^{(3)}(\vec{x}) \nonumber\\
&= -\frac{\lambdaEell}{2\Mp}  \mathcal{E}^{({\rm E})}_{\ell m} \ell!  {\mathcal{Y}}_{\ell m}^{i_1\cdots i_\ell} \int \dd^3\vec{x} \,   \delta^{(3)}(\vec{x})\partial_{ i_1}\cdots \partial_{i_\ell}e^{-i\vec p\cdot\vec x} \nonumber\\
&= -(-i)^{\ell}\frac{\lambdaEell}{2\Mp}  \mathcal{E}^{({\rm E})}_{\ell m} \ell!  {\mathcal{Y}}_{\ell m}^{i_1\cdots i_\ell} p_{ i_1}\cdots p_{i_\ell}
\end{align}
and therefore the Fourier transform of $\phi^\mathrm{resp}$,
\begin{align}
\phi^\mathrm{resp}(\vec p) &=  (-i)^{\ell}\frac{\lambdaEell}{2\Mp}  \mathcal{E}^{({\rm E})}_{\ell m} \ell!  {\mathcal{Y}}_{\ell m}^{i_1\cdots i_\ell} \frac{p_{ i_1}\cdots p_{i_\ell}}{\vec p^2}.
\end{align}
Finally we go back to real space,
\begin{align}
\phi^\mathrm{resp}(\vec x) &=  \int \frac{\dd^3\vec{p}}{(2\pi)^3}  e^{i\vec p\cdot\vec x} \phi^\mathrm{resp}(\vec p) \nonumber \\
&= (-i)^{\ell}\frac{\lambdaEell}{2\Mp}  \mathcal{E}^{({\rm E})}_{\ell m} \ell!  {\mathcal{Y}}_{\ell m}^{i_1\cdots i_\ell} \int \frac{\dd^3\vec{p}}{(2\pi)^3} e^{i\vec p\cdot\vec x}\frac{p_{ i_1}\cdots p_{i_\ell}}{\vec p^2}.
\end{align}
To evaluate the momentum integral we use \cref{eq:FTgenl}:
\begin{equation}
\phi^\mathrm{resp}(\vec x) = \frac{\lambdaEell}{\Mp}  \mathcal{E}^{({\rm E})}_{\ell m}   
\frac{\Gamma(2\ell+1)}{2^{\ell+3}\pi}
{\mathcal{Y}}_{\ell m}^{i_1\cdots i_\ell} 
    \frac{x^{ i_1} \cdots x^{i_\ell}}{r^{2\ell+1}} \, .
\end{equation}
Putting everything together, we have
\begin{equation}
\phi = \frac{\Mp}{2}\sum_{\ell=2}^\infty\sum_{m=-\ell}^\ell \mathcal{E}^{({\rm E})}_{\ell m}   \left(  r^\ell + \frac{\lambdaEell}{\Mp^2}    
\frac{\Gamma(2\ell+1)}{2^{\ell+2}\pi}
r^{-\ell-1} \right) Y_{\ell m}(\theta,\varphi).
\end{equation}
Plugging into the definition of the Weyl tensor we find
\begin{equation}
\begin{split}
E_{rr} & = -\frac{1}{\Mp}\partial_r^2 \phi
\\
& =
-\frac{1}{2} \sum_{\ell=2}^\infty\sum_{m=-\ell}^\ell  \ell(\ell-1) \mathcal{E}^{({\rm E})}_{\ell m} \left(  r^{\ell-2} + \frac{\lambdaEell}{\Mp^2}   
\frac{(\ell+2)(\ell+1)\Gamma(2\ell)}{2^{\ell+1} (\ell-1)\pi}
r^{-\ell-3} \right) Y_{\ell m}(\theta,\varphi),
\end{split}
\label{eq:Errm2}
\end{equation}
which recovers \cref{eq:ERij}.

\paragraph{Tidal response from in-in path integral}
The calculation of the tidal response, as shown above using either a diagrammatic approach or a direct solution of the equations of motion, must be modified in the presence of dissipative phenomena. As discussed in \cref{sec:dissipation}, the appropriate framework for accounting for dissipation is the in-in (Schwinger--Keldysh) formalism. In what follows, we streamline this calculation by showing how to compute the tidal response of a point particle to linear order in the frequency. For simplicity, we again focus on the parity-even sector of the perturbations.

From the in-in effective action \eqref{eq:gammaintgeneric}, we aim to compute the one-point function of the  electric component $E_{+ ij}$ of the Weyl tensor, in the presence of a background field $\bar{E}_{ij}$:
\begin{equation}
    \langle E_{+ij}(t,\vec{x})\rangle_\text{in-in}=\int \mathcal{D}h_+\mathcal{D}h_- E_{+\,ij}(t,\vec{x})e^{i\Gamma^\text{in-in}_\text{int}[h_\pm,X_\pm]} .
\end{equation}
We compute the expectation value
of $E_{+ ij}$ because this is the field  that has a classical interpretation. For the external
classical source we fix $h^+_1 = h^+_2 \equiv h^+$  (equivalently, $E^+_1 = E^+_2 \equiv E^+$  for the linearized Weyl tensor).
Using the explicit form \eqref{eq:gammaintgeneric}, we obtain the following expression for
the one-point function \cite{Goldberger:2020fot,Saketh:2023bul,Glazer:2024eyi,Combaluzier--Szteinsznaider:2025eoc,Kobayashi:2025vgl} (see \rcite{Combaluzier--Szteinsznaider:2025eoc} for a detailed derivation):
\begin{equation}
    \langle E_{+ij}(t,\vec{x})\rangle_\text{in-in}=i\sum_{\ell=2}^\infty\int \dd\tau_1 \dd\tau_2\, K^{(\mathrm{E})}_{\ell}(\tau_2-\tau_1)\langle E_{+ij}(t,\vec x) E_{-\,A_\ell}\left(\tau_2\right) \rangle\bar{E}_+^{A_\ell }\left(\tau_1\right)\,,
    \label{eq:1-pt_fct_electric}
\end{equation}
where we used the multi-index notation $A_\ell\equiv i_1\cdots i_\ell$ (see \cref{eq:EEEBBB}), and where $K^{(\mathrm{E})}_{\ell}$ is defined by $K^{(\mathrm{E})}_{\ell}(\tau)\delta^\ell_{\ell'} \delta^{\langle i_1}_{\langle j_1}\cdots \delta^{ i_\ell\rangle}_{ j_\ell\rangle} \equiv  {K^{(\mathrm{E})}_{+-}(\tau)^{i_1\cdots i_\ell}}_{ j_1\cdots j_{\ell'}}$
and is related to the retarded Green's function of the $Q$ operators via \cref{eq:KpmGR}.

The response kernel can then be expanded in Fourier space similarly to \cref{eq:Comega}, 
\begin{equation}
    K^{(\mathrm{E})}_{\ell}(\tau_2-\tau_1) = \int \frac{\dd {\omega}}{2\pi}e^{-i{\omega}(\tau_2-\tau_1)} K^{(\mathrm{E})}_{\ell}(\omega) =
    \int \frac{\dd{\omega}}{2\pi}e^{-i{\omega}(\tau_2-\tau_1)}\left( \ccs_{0,\ell}+i{\omega}\ccs_{1,\ell}+\cdots\right),
\label{eq:EFTLNseven}
\end{equation}
where $\ccs_{0,\ell}$ are related to the static Love numbers, $\ccs_{1,\ell}$ capture the linear-in-frequency dissipative response of the object, and so on. In the following, we will systematically neglect terms of order $\omega^2$ and higher. See \rcite{Combaluzier--Szteinsznaider:2025eoc, Kobayashi:2025vgl} for a calculation at order $\omega^2$. 
From the definition of the Weyl tensor,  we have on shell
\begin{equation}
\begin{aligned}
    E_{ij} &= C_{0i0j}  = \frac{2}{{\Mp}} \left[ \partial_i\partial_{[0}h_{j]0}-\partial_0\partial_{[0}h_{j]i} \right] = {\frac{1}{\Mp}}\left[ -\partial_i\partial_j h_{00}+2\partial_0\partial_{(i}h_{j)0}+\mathcal{O}(\omega^2) \right], \\
     E_{A_\ell} &= {\frac{1}{\Mp}} \left[-\partial_{A_\ell}h_{00}+2\partial_0\partial_{\langle  A_{\ell-1}}h_{i_\ell\rangle 0}+\mathcal{O}(\omega^2)\right].
    \end{aligned}
\end{equation}
We next fix the de Donder gauge and expand the correlator $\langle E^+_{ij}(t,\vec x) E^-_{A_\ell}\left(\tau_2\right) \rangle\bar{E}^{+\,A_\ell }\left(\tau_1\right)$ in \cref{eq:1-pt_fct_electric} up to order $\mathcal{O}(\omega)$. Using the gauge condition $\partial^\mu(h_{\mu\nu}-\frac{1}{2}\eta_{\mu\nu}h)=0$ together with the equations of motion  $\square h_{\mu\nu}=0$,  we note that subtracting traces in the definitions of $E^-_{A_\ell}$ and $\bar{E}^{+\,A_\ell }$  yields contributions of order $\mathcal{O}(\omega^2)$ to the product~\eqref{eq:1-pt_fct_electric}. Consequently, in the following we may replace $\langle {\cdots} \rangle$ by $({\cdots})$.
Performing the contraction of the Weyl  $E$ operators then gives
\begin{equation}
     E_{A_\ell} \bar{E}^{A_\ell} = {\frac{1}{\Mp^2}} \left[\partial_{A_\ell}h_{00}\partial^{A_\ell}\bar h^{00}-2\partial_{A_\ell}h_{00}\partial^0\partial^{A_{\ell-1}}\bar{h}^{i_\ell0}-2\partial_0\partial_{A_{\ell-1}}h_{i_\ell0}\partial^{A_\ell}\bar{h}^{00} +\mathcal{O}(\omega^2) \right] .
     \label{Electric_square}
\end{equation}
Putting this together, one obtains
\be
   \langle E_{+ij}(t,\vec x) E_{-A_\ell}\left(\tau_2\right) \rangle\bar{E}_+^{A_\ell }\left(\tau_1\right) 
   = -{\frac{1}{\Mp^3}}\langle  \partial_i\partial_j h_{+00} \partial_{A_\ell}h_{-00}\rangle\left(\partial^{A_\ell}\bar{h}_+^{00}-2\partial^0\partial^{A_{\ell-1}}\bar{h}_+^{i_\ell0} \right)+\mathcal{O}(\omega^2),
\label{eq:Simp_corr_electric}
\ee
where we dropped terms that are either $\mathcal \mathcal{O}(\omega^2)$ or proportional to $P_{0i00}=0$, with $P_{\mu\nu\rho\sigma}$ defined as in \cref{gravitonprop}. We emphasize that $h_{\mu\nu}$ and $\bar{h}_{\mu\nu}$ correspond, respectively, to the response and tidal fields in the de Donder gauge. It is also useful to note that the combination in parentheses is
\begin{equation}
\partial_{A_\ell}\bar{h}_{00}-2\partial_0\partial_{A_{\ell-1}}\bar{h}_{i_\ell0} = \partial_{A_\ell}\bar{h}_{00}^{\text{RW}} + \mathcal{O}(\omega^2),
\label{eq:dDtoRWeven}
\end{equation}
where $\bar{h}_{00}^{\text{RW}}$ is the $t$--$t$ component of the tidal field metric perturbation in Regge--Wheeler gauge (see \rcite{Combaluzier--Szteinsznaider:2025eoc} for a derivation), which we can express  as~\cite{Poisson_Will_2014}
\begin{align}
\bar{h}_{00}^{\text{RW}} (t,\vec{x})&= \frac{\Mp}{2} e^{-i\omega t}\sum_{\ell,m} \mathcal{E}^{(\mathrm{E})}_{\ell m} Y_\ell^m r^\ell+ \mathcal{O}(\omega^2) \nonumber\\
&= \frac{\Mp}{2}  e^{-i\omega t}\sum_{\ell} \mathcal{E}^{(\mathrm{E})}_{j_1\cdots j_\ell}x^{j_1}\cdots x^{j_\ell} + \mathcal{O}(\omega^2) ,
\end{align}
where $\mathcal{E}^{(\mathrm{E})}_{i_1\cdots i_\ell}$ is a traceless symmetric tensor.
Substituting this into~\cref{eq:1-pt_fct_electric} for the one-point function and employing the expression for the instantaneous graviton propagator in the de Donder gauge,
\begin{equation}
    \langle h_{+\mu\nu}(t,\vec{x})h_{-\rho\sigma}(\tau_1, \vec{0})\rangle=i\delta(t-\tau_1)P_{\mu\nu\rho\sigma}\int \frac{\dd^{3}\vec{p}}{(2\pi)^{3}}\frac{e^{i\vec{p}\cdot \vec{x}}}{\vec{p}^2},
    \label{graviton_propagator}
\end{equation}
we obtain  
\be
    \langle E_{+ij}(t,\vec x)\rangle_\text{in-in} 
    = -\frac{e^{-i\omega t}}{4\Mp^2}\sum_{\ell=2}^\infty \ell! (-i)^\ell K^{(\mathrm{E})}_{\ell}(\omega) \mathcal{E}^{(\mathrm{E})}_{i_1\cdots i_\ell}  \partial_i\partial_j\int \frac{\dd^{3}\vec{p}}{(2\pi)^{3}}e^{i\vec{p}\cdot\vec{x}}\frac{p^{i_1}\cdots p^{i_\ell}}{\vec{p}^2},
\ee
which, in position space, reads
\be
    \langle E_{+ij}(t,\vec x)\rangle_\text{in-in} 
    = -\frac{e^{-i\omega t}}{2{\Mp^2}}\sum_{\ell=2}^\infty K^{(\mathrm{E})}_{\ell}(\omega)\frac{ \Gamma\left(2\ell+1\right)}{2^{\ell+3}\pi}\mathcal{E}^{(\mathrm{E})}_{i_1\cdots i_\ell}\partial_i\partial_j  \frac{x^{i_1}\cdots x^{i_\ell}}{\vert \vec{x}\vert^{2\ell+1}} ,
\label{eq:evenEplus}
\ee
where we used \cref{eq:FTgenl}.

Similarly, the background Weyl tensor can be expressed as follows. Focusing on the $r$--$r$ component, we have
\begin{equation}
\begin{aligned}
    \bar{E}_{rr} & =\frac{1}{{\Mp^2}}\left(-\partial_r^2\bar{h}_{00}+2\partial_0 \partial_r   \bar{h}_{0r}\right)+\mathcal{O}(\omega^2)
    \\
& = -\frac{1}{{\Mp^2}}\partial_r^2 \bar{h}_{00}^{\text{RW}}  +\mathcal{O}(\omega^2)
    = -\frac{1}{2}e^{-i\omega \tau}\sum_{\ell=2}^\infty \mathcal{E}^{(\mathrm{E})}_{\ell m} \,  \ell(\ell-1) r^{\ell-2} Y_\ell^m  +\mathcal{O}(\omega^2).
\end{aligned}
    \label{tidal_Weyl_even}
\end{equation}
Using $\mathcal{E}^{(\mathrm{E})}_{j_1\cdots j_\ell}x^{j_1}\cdots x^{j_\ell} = \sum_m\mathcal{E}^{(\mathrm{E})}_{\ell m} r^\ell Y_\ell^m $ with $r\equiv \vert \vec x\vert$, and noting that   $\nabla_r\nabla_r=\partial_r\partial_r$ in \cref{eq:evenEplus}, then adding together \cref{eq:evenEplus,tidal_Weyl_even} we obtain 
\begin{equation}
    \bar{E}_{rr}+\langle E_{+\,rr}\rangle_\text{in-in}=-\frac{1}{2}e^{-i\omega t}\sum_{\ell,m}  \ell(\ell-1)\mathcal{E}^{(\mathrm{E})}_{\ell m}Y_{\ell}^{m} 
    \left( r^{\ell-2} + \frac{K^{(\mathrm{E})}_{\ell}(\omega)}{\Mp^2}\frac{(\ell+2)(\ell+1)\Gamma(2\ell)}{2^{\ell+2}(\ell-1)\pi} r^{-\ell-3} \right) +\mathcal{O}(\omega^2),
    \label{EFT_BC_solution}
\end{equation}
which agrees with \cref{eq:ERij,eq:Errm2} upon identifying $\lambdaEell$ with $\ccs_{0,\ell}/2$ in \cref{eq:EFTLNseven} (the relative factor of $2$ arises from comparing the in-in action~\eqref{eq:gammaintgeneric} with the action~\eqref{eq:S-ho} when the latter is expressed in the Keldysh basis).

\subsubsection{Tidal response from scattering amplitudes}

In the previous subsection, we showed how to extract the static and dynamical responses of a compact object in the worldline EFT framework from an off-shell calculation of a graviton one-point function. One potential disadvantage of off-shell methods is that the metric computed from the graviton one-point function is a coordinate-dependent quantity. As a result, as we discussed, one must either extract a gauge-invariant quantity (such as the linearized Weyl tensor) to perform the matching, or apply an additional gauge transformation to match whatever gauge is used in the full-theory calculation.

Here, we focus on a different, on-shell observable, namely a scattering amplitude. The system we have in mind consists of a probe field scattering off a compact source:
$$\text{worldline} + \text{particle }X \longrightarrow \text{worldline} + \text{particle }X .$$
The amplitude derived in the EFT can then be matched against an analogous calculation in a fully relativistic setup, providing an independent --- yet equivalent --- determination of the Love numbers and dissipative couplings.

We begin by introducing the definition of the scattering matrix $S$ (adopting the conventions of \rcite{Saketh:2023bul}). The transition amplitude between asymptotic states is given by
\begin{equation}
{}_{\rm out}\!\langle \vec k',h' \vert \vec k, h \rangle_{\rm in}
\equiv
\langle \vec k',h' \vert S \vert \vec k,h \rangle ,
\label{eq:Sdef}
\end{equation}
where $h$ denotes the helicity of the particle. The single-particle states are normalized according to
\begin{equation}
\langle \vec k,h \vert \vec k',h' \rangle
=
2|\vec k|\,(2\pi)^3 \delta^{(3)}(\vec k-\vec k')\,\delta_{hh'}
\end{equation}
in four spacetime dimensions. For massless particles, the on-shell condition reads $\omega^2 = \vec k^2$.
Following the standard procedure,  we further decompose the scattering matrix as
\begin{equation}
S = 1 + iT \, .
\end{equation}
It is then convenient to make time-translation invariance manifest by isolating the energy-conserving delta function and introducing the dimensionful scattering amplitude $\mathcal{M}$ through
\begin{equation}
\langle \vec k',h'|iT| \vec k,h\rangle
= (2\pi)\delta(\omega'-\omega)\, i\mathcal{M}(\omega, \vec k \to \vec k',h \to h') .
\label{eq:Mdef}
\end{equation}
Following \rcite{Saketh:2023bul}, to facilitate comparison with fully relativistic calculations, it is convenient to work in a spherical basis. Introducing the spherical-wave states $|\omega,\ell,m,h\rangle$, characterized by frequency $\omega$, angular momentum numbers $\ell,m$, and helicity $h$, with normalization 
\begin{equation}
\langle \omega,\ell,m,h | \omega',\ell',m',h' \rangle
= (2\pi)\delta(\omega-\omega')\delta_{\ell\ell'}\delta_{mm'}\delta_{hh'} ,
\end{equation}
the scattering amplitude takes the form~\cite{Saketh:2023bul}
\begin{equation}
\langle \omega',\ell',m',h'| iT |\omega,\ell,m,h\rangle
= (2\pi)\delta(\omega'-\omega)\, i\mathcal{A}(\omega,\ell,m,h ,\ell',m',h') ,
\label{eq:Adef}
\end{equation}
where the (dimensionless) scattering matrix $\mathcal{A}$ is related to  $\mathcal{M}$ via
\begin{equation}
\mathcal{A}(\omega,\ell,m,h,\ell',m',h')
=
\int \frac{\dd^3 \vec k_1\dd^3 \vec k_2}{4(2\pi)^6|\vec k_1||\vec k_2|}
\sum_{h_1,h_2}
\langle \omega,\ell',m',h'|\vec k_2,h_2\rangle
\,\mathcal{M}(\omega,\vec k_1\to\vec  k_2,h_1\to h_2)\,
\langle \vec k_1,h_1|\omega,\ell,m,h\rangle
\label{eq:A_M_relation}
\end{equation}
with the transformation matrix  given by \cite{Goldberger:2020fot}
\begin{equation}
\langle \omega,\ell,m,h|\vec k,h'\rangle
=
(2\pi)^2\sqrt{\frac{2\ell+1}{2\pi\omega}}
\,\delta(\omega-|\vec k|)\,\delta_{hh'}\,
D^{\ell}_{mh}(\varphi,\theta,0) ,
\label{eq:spherical_transform}
\end{equation}
where $D^{\ell}_{mh}(\varphi,\theta,0)$ is the Wigner-D matrix and $(\theta,\varphi)$ denotes the orientation of $\vec k$.

In a diagonal basis,  the amplitude can be related to the scattering phase shift as
\begin{equation}
i\mathcal{A} = 1 - \eta_{\ell m} e^{2i\delta_{\ell m}} ,
\label{eq:phase_shift}
\end{equation}
where $\delta_{\ell m}$ is the elastic scattering phase shift and
$1-\eta_{\ell m}^2$ is the absorption probability.
For gravitational perturbations, the expression for the  graviton polarization  in the spherical basis is (see \rcite{Saketh:2023bul} for details)
\begin{equation}
\varepsilon^{h=\pm2}_{ij}(\vec k)
=
\sum_{m}
\langle i,j|\ell=2,m\rangle
D^{\ell=2}_{m,h=\pm2}(\varphi,\theta,0) .
\end{equation}

In the worldline theory or EFT (see \cref{eq:sint}), the leading-order (zero-frequency) gravitational Compton amplitude due to induced quadrupolar tides is given by
\vspace{-0.35cm}
\begin{equation}
i\mathcal{M}(\vec k_{\rm in} \to \vec k_{\rm out}, h \to h)
\equiv
\raisebox{45pt}{\scatteringMbb}
= 
i\frac{\omega^{4}}{4 \Mp^{2}}
c^{(\mathrm{E})}_{0\, ij,kl}\,
\varepsilon^{kl}_{h}(\vec k_{\rm in})\,
\varepsilon^{*\,ij}_{h}(\vec k_{\rm out})
+ \text{magnetic} + \text{h.o.~terms} ,
\end{equation}
where $c^{(\mathrm{E})}_{0\, ij,kl}$ is obtained from the expansion of the Green's function \eqref{eq:Comega}. The black dot denotes the coupling to the worldline in \cref{eq:sint}, the internal dashed line represents the (retarded) propagator of the composite operator $Q$ discussed in \cref{sec:dissipation} (see \cref{eq:KpmGR}), and the curly lines correspond to the external gravitons.

In the case of a rotating point particle, we can take advantage of the decomposition \eqref{eqlambdas}~\cite{Saketh:2023bul}. Assuming the spin to be oriented along the $\hat{z}$-axis, we have $ \hat{s}_i = \delta_{i3}$, and
\begin{equation}
{J_z}^i_j = -i{\epsilon^i}_{jk} \hat{s}^k = -i \hat{S}^i{}_j,
\end{equation}
(Recall that $S_{ij} \equiv \epsilon_{ijk}\,\hat{s}_k$, where $\hat{s}_k$ is the unit vector pointing along the spin of the particle in its rest frame.)
The amplitude can be thus  transformed from  the plane-wave basis into the spherical-wave basis as shown in \rcite{Saketh:2023bul},
\begin{multline}
i\mathcal{A}(\omega,\ell=2,m, \ell=2,m,h)
= i \frac{\omega^5}{40\Mp^2\pi}\bigg[
\Lambda^{(\mathrm{E})}_{\omega^0,S^0} + \frac{1}{2} i m H^{(\mathrm{E})}_{\omega^0,S^1} 
\\
+ \frac{1}{6} (4-m^2 ) \left(  \Lambda^{(\mathrm{E})}_{\omega^0,S^2}
+ i m H^{(\mathrm{E})}_{\omega^0,S^3}
- \Lambda^{(\mathrm{E})}_{\omega^0,S^4}(m^2-1) \right) \bigg]
+ \text{magnetic} + \text{h.o.~terms}
,
\end{multline}
which holds for spinning objects at leading order in the small-frequency expansion \cite{Goldberger:2020fot,Ivanov:2022qqt,Saketh:2023bul}. When extended to the next-to-leading  order in $\omega$, a similar calculation yields~\cite{Saketh:2023bul,Saketh:2024juq,Ivanov:2026icp}
\begin{multline}
i\mathcal{A}(\omega,\ell=2,m, \ell=2,m,h)
= - \frac{\omega^6}{40\Mp^2\pi}\bigg[
H^{(\mathrm{E})}_{\omega^1,S^0} + \frac{1}{2} i m \Lambda^{(\mathrm{E})}_{\omega^1,S^1} 
\\
+ \frac{1}{6} (4-m^2 ) \left(  H^{(\mathrm{E})}_{\omega^1,S^2}
+ i m \Lambda^{(\mathrm{E})}_{\omega^1,S^3}
- H^{(\mathrm{E})}_{\omega^1,S^4}(m^2-1) \right) \bigg]
+ \text{magnetic} + \text{h.o.~terms}
.
\end{multline}
This result has been matched to the phase shift computed in full black hole perturbation theory~\cite{Saketh:2022wap,Saketh:2023bul} (see also \rcite{Chia:2020yla,Charalambous:2021mea,Saketh:2022xjb}). We will return to this discussion in \cref{subsection:DynLN}.

\subsubsection{Summary: matching Love numbers in the worldline EFT}
\label{sec:matchingsummary}

Off-shell and on-shell approaches provide complementary ways of computing low-energy observables and performing a matching that allows one to determine the Wilson coefficients, once a complete model or theory is specified. In practice, as illustrated in the example in \cref{sec:ADMmassex}, performing the matching amounts to taking some EFT quantity --- such as an in-in correlation function or a phase shift --- and comparing it with an analogous calculation in the full theory. Schematically:
\begin{equation}
    E_{ij}^\mathrm{GR} = \langle E_{ij} \rangle_{\text{in-in}}^\mathrm{EFT}[c_i] \,,
    \label{eq:offshellmatchingM}
\end{equation}
\begin{equation}
    \mathcal{M}^{\mathrm{GR}} = \mathcal{M}^{\mathrm{EFT}}[c_i] \,,
    \label{eq:amplitudematchingM}
\end{equation}
where on the left-hand side we have the GR quantities (the $E$ field and the amplitude $\mathcal{M}$), equated on the right-hand side with the corresponding EFT quantities, which depend on the effective
couplings $c_i$. This allows one to determine $c_i$ as functions of the parameters in the microscopic model.

In recent years, the combination of EFT, perturbation theory, in-in formalism, and  scattering amplitudes has proven
a powerful tool to perform the matching in \cref{eq:offshellmatchingM,eq:amplitudematchingM} for the
tidal response of compact objects.
As we review in \Cref{part:part3Love}, examples of scattering-amplitude applications include: an on-shell proof of the vanishing of static black hole Love numbers~\cite{Ivanov:2022qqt}; the extraction of the leading dissipation coefficients and dynamical Love numbers~\cite{Saketh:2022wap,Saketh:2023bul} through a near-far factorization of GR scattering amplitudes; the calculation of both static and dynamical Love numbers for a massless scalar field through gravitational Raman scattering~\cite{Ivanov:2024sds,Caron-Huot:2025tlq} --- i.e., the inelastic scattering of massless fields off compact gravitating objects --- including tidal heating effects beyond leading order and their renormalization through a non-trivial anomalous dimension \cite{Ivanov:2024sds,Goldberger:2012kf,Ivanov:2025ozg}; and the calculation of spin-1 and spin-2 Raman amplitudes~\cite{Ivanov:2026icp}. Similarly, off-shell matching has enabled the determination of the tidal response of black holes and compact objects in a variety of situations, ranging from static to dynamical tides, from linear to nonlinear response, and across different types of fields, e.g.,~\cite{Kol:2011vg,Chakrabarti:2013lua,Hui:2020xxx,Goldberger:2020fot,Charalambous:2021mea,Rai:2024lho,Hadad:2024lsf,Glazer:2024eyi,Capuano:2024qhv,Riva:2023rcm,Iteanu:2024dvx,Combaluzier-Szteinsznaider:2024sgb,Pani:2025qxs} (see \Cref{part:part3Love} for a more complete list of references).

One advantage of the amplitude approach is that results are immediately gauge-invariant and can be readily used in a UV matching~\cite{Ivanov:2022qqt,Saketh:2022wap,Saketh:2023bul,Ivanov:2024sds,Caron-Huot:2025tlq,Ivanov:2026icp}. By contrast, a potential technical disadvantage is that it requires computing far-zone quantities and accounting for the recoil of the object and tail effects, in order to ensure consistency of the in-out calculation of the dynamical gravitational response~\cite{Cheung:2023lnj,Cheung:2024byb,Ivanov:2026icp}. On the other hand, off-shell correlators can be evaluated within a near-zone expansion (e.g., \cite{Charalambous:2021mea,Hui:2020xxx,Kol:2011vg,Combaluzier--Szteinsznaider:2025eoc}), which is often technically simpler. A possible  drawback, however, is that they are not necessarily gauge-invariant, which requires either extracting a gauge-invariant quantity (such as the linearized Weyl tensor, as in \cref{eq:offshellmatchingM}) or performing the matching in a fixed set of coordinates.
We return to these aspects in \Cref{part:part3Love}. %

We stress that the calculations illustrated above apply to tree-level matching only. Since gravity is a nonlinear gauge theory, a precise determination of the Wilson coefficients requires accounting for gravitational nonlinearities, which in the diagrammatic language above corresponds to evaluating higher-loop worldline diagrams. This in turn requires computing UV-divergent integrals that need to be regulated.\footnote{The presence of infinities arising at 5PN order was argued long ago in~\rcite{Damour:1984rbx} and later fully established within the point-particle EFT~\cite{Goldberger:2004jt}.} When divergences are logarithmic, their regularization necessarily introduces a renormalization scale~\cite{Rothstein:2014sra}. Dimensional regularization is the natural choice here, as it respects the symmetries of the underlying action. Upon renormalization, the divergences are absorbed into the matching coefficients, which consequently acquire a scale dependence governed by the renormalization group.

As we review in \Cref{part:part3Love}, the static Love numbers of black holes in four-dimensional general relativity not only vanish, but also exhibit no renormalization group running --- a property that reinforces the well-known naturalness puzzle of black hole tides, which recently motivated a series of symmetry proposals discussed in \cref{sec:sym}. At finite frequency, by contrast, the Love numbers fulfill the standard Wilsonian expectation and display a non-trivial running. Determining the dynamical response in this case requires extending the previous calculations to loop order; see \rcite{Saketh:2023bul,Ivanov:2024sds,Caron-Huot:2025tlq,Combaluzier--Szteinsznaider:2025eoc,Ivanov:2026icp} for examples of EFT calculations in this direction. We provide a detailed account of the relevant literature and of what remains to be done in \Cref{part:part3Love}.

\subsection{Post-Newtonian description of tidal effects}
\label{sec:post-newtonian}

\begin{figure}[h]
\centering
\includegraphics[width=0.9\textwidth]{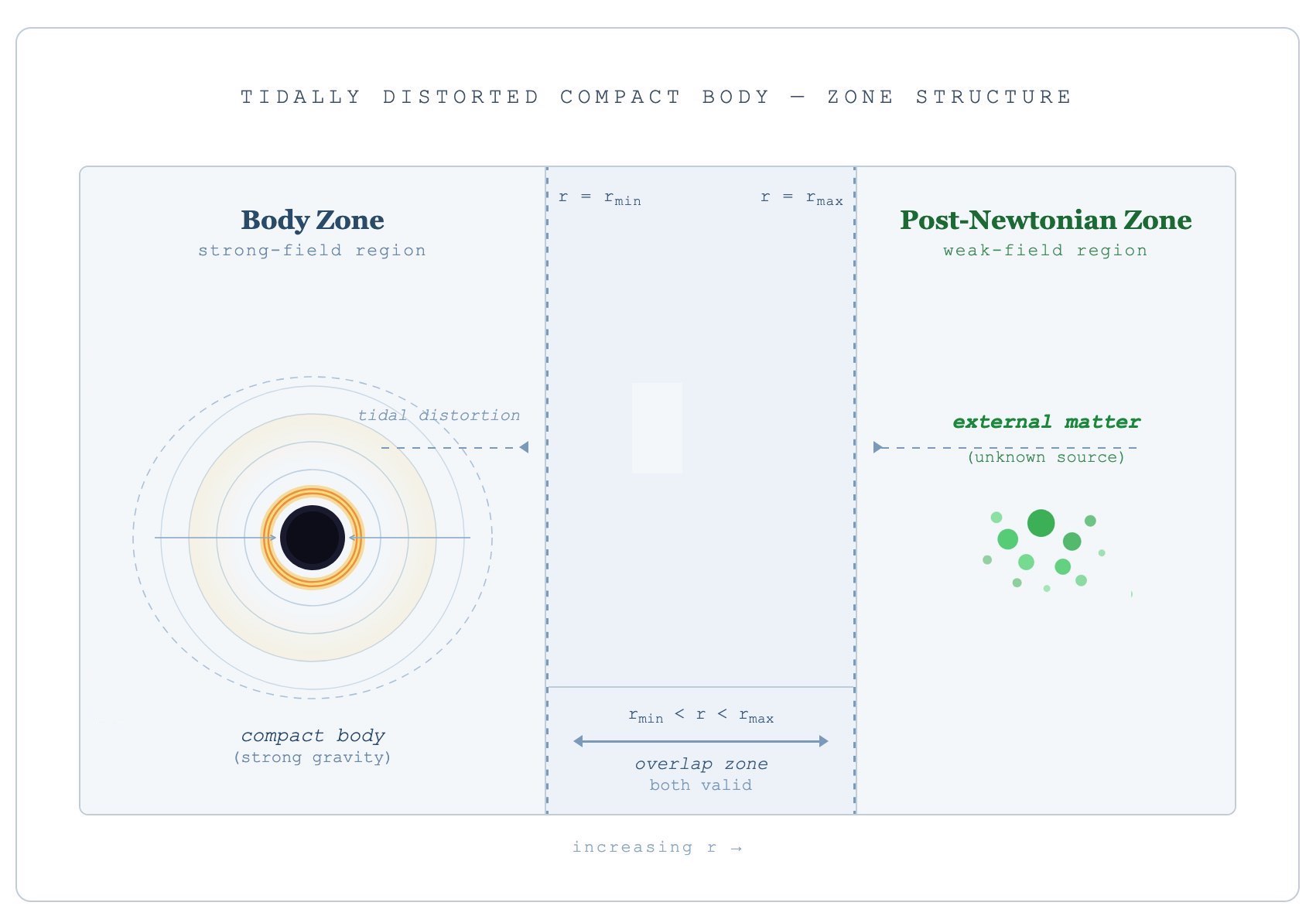}
\caption{A schematic illustration the physical setup described in the main text: {\it Black hole (left)} --- tidally distorted with an oblate event horizon, photon ring glow, 
and equipotential field lines showing the tidal deformation. The figure schematically illustrates 
the physical setup described in the main text: a compact body on the left interacting weakly with 
a distribution of external matter on the right.
{\it Body zone} --- the strong-gravity region extending from the compact body out to 
    $r = r_{\max}$  (blue dashed boundary), where gravity is strong near the body and gradually 
    weakens toward the outer boundary. {\it Post-Newtonian zone} --- the weak-field region from the external source inward 
    to $r = r_{\min}$ (purple dashed boundary), with gravity remaining weak throughout. The unknown 
    matter distribution is shown as a green diffuse cloud, though we remind the reader that in inspirals it is typically another compact object. The Love numbers and dissipative response coefficients are independent of the nature of the external matter.
{\it Overlap zone} --- the shaded box covering $r_{\min} < r < r_{\max}$, where both 
    descriptions are simultaneously valid and can be asymptotically matched.
}
\label{fig:PoissonPN}
\end{figure}

An alternative operational perspective on the tidal Love numbers  of compact objects,  traceable back to \rcite{Damour:1984rbx}, was proposed in \rcite{Poisson:2020vap}. In this approach, the induced tidal deformation is defined in terms of the moments of the body when viewed as a skeletonized object moving in a \emph{post-Newtonian} spacetime (see \rcite{Blanchet:2013haa} and references therein). The idea is based on identifying three distinct regions, defined using key scales: the mass $M$ and radius $\Rstar$ of the compact body whose induced moments we wish to compute; the mass scale $M'$ of the external matter responsible for the tidal environment; and the separation distance $a$ between the compact body and the external matter, with the assumption that $a \gg GM, \Rstar$. In this notation, the three zones are~\cite{Poisson:2020vap} (see \cref{fig:PoissonPN} for a sketch):
\begin{enumerate}
    \item \textit{Body zone:} This is the immediate neighborhood of the compact object. 
    It includes distances $ r <  r_{\max}$, where $ r_{\max}$ is much larger 
    than $GM$ but much smaller than $a$. Gravity is strong deep inside this zone and is treated using full general relativity, becoming weaker toward the outer boundary.
    
    \item \textit{Post-Newtonian zone:} This region surrounds the compact body and includes 
    the external matter but excludes the innermost strong-gravity region. It covers 
    $ r_{\min} <  r < \lambda_\mathrm{r}$, where $ r_{\min}$ is also much larger than $M$ 
    and much smaller than $a$, and $\lambda_\mathrm{r} \gg a$ represents the characteristic wavelength 
    of the emitted gravitational radiation. Gravity here is weak and is described 
    using a post-Newtonian approximation.
    
    \item \textit{Overlap zone:} The region $ r_{\min} <  r <  r_{\max}$ lies 
    between the two zones. Both the fully relativistic 
    and post-Newtonian descriptions are valid here, allowing the two approaches to be 
    consistently matched.
\end{enumerate}

In the body zone, the metric is required to satisfy the vacuum Einstein field equations. The tidal environment of the body is described by a set of tidal multipole moments $\mathcal{E}_{i_1\cdots i_\ell}$, which, in this zone, are treated as freely specifiable functions of time and are not fixed by the field equations.
A representative sample of the  $t$-$t$ component of the body-zone metric is given by \cref{eq:introgtt0}, which we rewrite here as (suppressing some indices for simplicity)\footnote{We mainly follow here the notation of \rcite{Poisson:2020vap}. Overdots on time-dependent functions, such as $\mathcal{E}_{ab}(t)$, genuinely denote time derivatives, while overdots on coefficients, such as $\dot{k}_2$, are purely notational, serving as a reminder that they are associated with time-derivative terms in the metric; in the latter case, the overdot does not imply that these coefficients are themselves differentiated with respect to $t$.}
\begin{align}
g_{tt} =&\, -1 + \frac{2GM}{r}
- \left[
    r^2(1+\cdots) + 2k_2\Rstar^5\frac{1}{r^3}\,(1+\cdots) \right]\mathcal{E}_{ab}\Omega^a\Omega^b
\notag\\
& \, -\left[\frac{2}{7} r^4(1+\cdots)   +2p_2\frac{\Rstar^8}{GM}\frac{1}{r^3}(1+\cdots)
    +\frac{8}{7}k_2\Rstar^5\frac{1}{r}(1+\cdots)
    +\frac{8}{7}k_2^2\Rstar^{10}\frac{1}{r^6}(1+\cdots)
  \right] \mathcal{E}_{c\langle a}{\mathcal{E}^c}_{b\rangle}\Omega^a\Omega^b
\notag\\
&\,
- 2\dot{k}_2GM\Rstar^5\frac{1}{r^3}\,(1+\cdots)\,\dot{\mathcal{E}}_{ab}\Omega^a\Omega^b
- \left[
    \frac{11}{42}r^4(1+\cdots)
    + 2\ddot{k}_2\frac{\Rstar^8}{GM}\frac{1}{r^3}(1+\cdots)
    + k_2\Rstar^5\frac{1}{r}(1+\cdots)
  \right] \ddot{\mathcal{E}}_{ab}\Omega^a\Omega^b
\notag\\
&\, + \cdots 
\label{eq:gttPN}
\end{align}
where $\Omega^a \equiv ( \sin\theta \cos\varphi, \sin\theta \sin\varphi, \cos\theta )$, and  $\mathcal{E}_{ab}$ are time-dependent, symmetric,
and tracefree Cartesian tensors that define the multipole moments of the external tidal field. The ellipses in the parentheses indicate higher-order terms in $GM/r$, as well as possible non-analytic contributions in $r$, while the ellipses in the last line of \cref{eq:gttPN} refer to higher-order terms in either the (spatial or temporal) derivative expansion or the field expansion. The (constant) coefficients $k_2$, $p_2$, $\dot{k}_2$, $\ddot{k}_2$, and similarly those appearing at higher orders in the multipolar expansion, are determined by the Einstein equations in the body zone for given external moments $\mathcal{E}_{ab}$, $\dot{\mathcal{E}}_{ab}$, $\ddot{\mathcal{E}}_{ab}$, and so on.
At this level, the tidal multipole moments appear in the metric as freely specifiable functions, which can be viewed as components of the external Weyl tensor, i.e., the curvature induced by the external matter, and its derivatives.

For practical use, the metric \eqref{eq:gttPN} should be embedded within a larger spacetime that includes the external matter, allowing the tidal moments $\mathcal{E}_{ab}(t)$ to be determined over this extended domain. 
The external matter could be a companion star, multiple bodies, unbound orbits, or a diffuse distribution. In a binary system, the companion may be weakly self-gravitating, allowing a post-Newtonian description, or strongly self-gravitating, requiring full general relativity. 
For instance, if the system in \cref{fig:PoissonPN} is a two-body system, the body metric could be matched to a global metric obtained from a numerical simulation of the field equations.
The advantage of the approach in \rcite{Poisson:2020vap} is that no specific scenario needs to be specified; the essential point is that the gravitational interaction with the compact body is weak and can be accurately described using post-Newtonian theory. While the post-Newtonian data fix the tidal moments in the body zone, the relativistic information specifies the multipole structure of the post-Newtonian zone.

In practice, while the solution \eqref{eq:gttPN} is commonly obtained in standard Schwarzschild coordinates with a suitable gauge choice --- e.g., the Regge--Wheeler gauge (see \cref{sec:BHPT} for more details on black hole perturbation theory) --- the body’s equations of motion in the post-Newtonian zone are usually solved in harmonic coordinates attached to the system’s barycenter. A detailed, general implementation of this program can be found in \rcite{Poisson:2020vap}; see \rcite{Poisson_Will_2014,Blanchet:2013haa} for an introduction to post-Newtonian gravity, and \rcite{Damour:1990pi,Damour:1991yw,Damour:1992qi,Racine:2004hs,Poisson:2009qj,Taylor:2008xy} for previous relevant work in  this context.

In the post-Newtonian zone, the metric tensor $g_{\mu\nu}$ is written in standard post-Newtonian form~\cite{Taylor:2008xy,Poisson:2009qj,Poisson:2020vap},
\begin{align}
g_{tt} & = -1 +2 U +2(\Psi -U^2) + \mathcal{O}(2{\text{PN}}) , \\
g_{ti} & = -4U_i + \mathcal{O}(2{\text{PN}}) , \\
g_{ij} & = (1 +2 U)\delta_{ij}  + \mathcal{O}(2{\text{PN}}) , 
\end{align}
where $U$ is the Newtonian potential, $\Psi$ is a post-Newtonian potential, and $U_i$ is a vector potential, each determined consistently up to 1PN order. The metric is written in harmonic coordinates $x^\mu = (t, x^i)$,\footnote{We are slightly abusing notation here. Note that $\vert\vec{x}\vert$ is not the $r$ introduced in \cref{eq:gttPN}, which corresponds to the standard Schwarzschild radial coordinate in the body frame of the object. See \rcite{Poisson:2020vap} for a relation between them.} centered on the post-Newtonian barycenter, which serves as the origin of an inertial reference frame.
The post-Newtonian potential $\Psi$ is often conveniently expressed as
\begin{equation}
\Psi \equiv \psi + \frac{1}{2}\frac{\partial^2 X}{\partial t^2},
\end{equation}
in terms of two auxiliary potentials $\psi$ and $X$. This decomposition is useful because, in a region of spacetime devoid of matter (i.e., away from the external sources), the Einstein field equations reduce to \cite{Racine:2004hs,Taylor:2008xy,Poisson:2020vap}
\begin{equation}
\nabla^2 U = 0, \qquad
\nabla^2 U^i = 0, \qquad
\nabla^2 \psi = 0, \qquad
\nabla^2 X = 2U,
\end{equation}
where $\nabla^2$ denotes the usual Laplace operator in three-dimensional flat space. These equations are supplemented by the harmonic coordinate condition
\begin{equation}
\partial_t U + \partial_i U^i = 0,
\end{equation}
which acts as a gauge condition on the gravitational potentials.
Each of the field equations is linear, so a solution describing both a compact body and an external matter distribution can be constructed by linear superposition. The object is modeled as a post-Newtonian monopole of mass $m$, with mass dipole and quadrupole moments~\cite{Poisson:2020vap}
\begin{align}
Q_i(t)&=Q_i[{\footnotesize \text{PN}}], \\
Q_{ij}(t)&=Q_{ij}[{\footnotesize \text{N}}]+Q_{ij}[{\footnotesize \text{PN}}] ,
\end{align}
respectively, with the Newtonian and post-Newtonian contributions highlighted. The point mass moves along a trajectory $\vec{x} = \vec{z}(t)$ in the harmonic coordinates. Its velocity is defined by $\vec{v} = \dd\vec{z}/\dd t$, and its acceleration by $\vec{a} = \dd\vec{v}/\dd t$. 
Introducing the vectors $\vec{s}\equiv \vec x-\vec{z}(t)$ and $\vec{n}\equiv \vec{s}/s$, with $s=\vert \vec{s}\vert$, the potentials are expressed as
\begin{align}
U & = \frac{GM}{s} + \frac{1}{2}Q^{ij}[{\footnotesize \text{N}}]\partial_{ij}\frac{1}{s} + U_{\mathrm{ext}} ,
\label{eq:PNU}
\\
U^i & = \left(\frac{GM}{s} + \frac{1}{2}Q^{jk}[{\footnotesize \text{N}}]\partial_{jk}\frac{1}{s} \right)v^i - \frac{1}{2}\dot{Q}^{ij}[{\footnotesize \text{N}}]\partial_{j}\frac{1}{s} + U^i_{\mathrm{ext}} , \\
\psi & =  \frac{GM\mu}{s} - Q^{i}[{\footnotesize \text{PN}}]\partial_{i}\frac{1}{s}   + \frac{1}{2}{Q}^{ij}[{\footnotesize \text{PN}}]\partial_{ij}\frac{1}{s} + \psi_{\mathrm{ext}} , 
\label{eq:PNpsi}
\\
X & =  GMs   + \frac{1}{2}{Q}^{ij}[{\footnotesize \text{N}}]\partial_{ij}s + X_{\mathrm{ext}} , 
\label{eq:PNX}
\end{align}
consistently with the skeletonized description of the compact body. In \cref{eq:PNpsi}, the quantity $\mu(t)$ appearing in $\psi$ is a post-Newtonian correction to the mass parameter $M$, to be determined by matching the post-Newtonian metric with the body-zone solution. Derivatives of $s$ can, in turn, be re-expressed in terms of $\vec{n}$ using the identities
\begin{equation}
\partial_i s= n_i,
\qquad 
\partial_i\frac{1}{s} = -\frac{n_i}{s^2} ,
\qquad
\partial_{ij}\frac{1}{s} = \frac{3}{s^3}n_{\langle ij\rangle}  .
\end{equation}
(See \rcite{Poisson:2020vap} for more identities.)
The potentials \eqref{eq:PNU}--\eqref{eq:PNX} depend on several unknowns, namely the multipole moments and the trajectory $\vec{z}(t)$ of the skeletonized object. These are determined by matching, which also yields the equations of motion for the position vector. Note that the potentials become singular in the limit $s \to 0$; this singularity is however only apparent, since $s = 0$ lies outside the post-Newtonian zone, where $s \gg GM$.

In order to determine the unknowns, one has to compare the expressions above with corresponding quantities computed in the body zone. To perform this matching, it is first necessary to ensure that the quantities being matched are expressed in the same coordinate system. It is worth recalling that, while the post-Newtonian zone solutions are obtained in harmonic coordinates in a barycentric frame, the body-zone metric is more easily determined in a frame in which the body is positioned at the origin of the coordinate system.
The next step is therefore to perform a change of coordinates that brings the compact body to the spatial origin of the new coordinate system, referred to as the ``body frame''.
The details  of this transformation are worked out in \rcite{Poisson_Will_2014,Racine:2004hs,Poisson:2009qj,Poisson:2018qqd,Poisson:2020vap}.

After transforming the barycentric potentials \eqref{eq:PNU}--\eqref{eq:PNX} from the barycentric frame --- where the compact body moves in the gravitational field of the external matter --- to the body frame --- where the body is at rest at the spatial origin --- and performing the matching in the buffer zone,\footnote{Depending on how the general-relativistic solution for the metric was obtained in the body zone, an additional gauge transformation may be required to express everything in the same set of (e.g., harmonic) coordinates~\cite{Poisson:2020vap}.} one obtains a relation between the mass quadrupole moment $Q_{ij}$ of the skeletonized object and the tidal moments $\mathcal{E}_{ij}$. All in all~\cite{Poisson:2020vap,Racine:2004hs},
\begin{equation}
Q_{ij} =  - \frac{2}{3}k_2\Rstar^5 \mathcal{E}_{ij} -\frac{2}{3}p_2\frac{\Rstar^8}{GM} \mathcal{E}_{k\langle i}{\mathcal{E}^k}_{j\rangle}- \frac{2}{3}\ddot{k}_2\frac{\Rstar^8}{GM}\frac{1}{r^3} \ddot{\mathcal{E}}_{ij}+ \text{PN corrections} .
\label{eq:QabPoisson}
\end{equation}
This procedure also fixes all remaining freedom. In particular, it determines the mass dipole moment $Q_i$ of the barycentric post-Newtonian potential, the post-Newtonian correction $\mu$ to the body’s mass parameter, the relations between the tidal moments and the barycentric potentials $U_\mathrm{ext}$, $U_\mathrm{ext}^i$, and $\psi_\mathrm{ext}$ generated by the external matter, the equations of motion for the vector $z(t)$, and all other details of the coordinate transformation. For a complete account, including the expression for  the post-Newtonian corrections in~\cref{eq:QabPoisson}, we refer the reader to \rcite{Poisson:2020vap}.

In \cref{eq:QabPoisson}, the Love numbers $k_2$, $p_2$, and $\ddot{k}_2$ are computed in full general relativity using the tools that we will  introduce in \cref{sec:BHPT}, which provide an alternative operational definition of the induced response rooted in post-Newtonian theory. As summarized above and discussed in \rcite{Poisson:2020vap}, the multipole moments are defined in the specific (harmonic) coordinate system used to describe the post-Newtonian metric. Once the boundary conditions on the metric are fully specified --- in particular, regularity at the spatial origin, vanishing ADM momentum at spatial infinity, and outgoing waves at future null infinity --- all remaining freedom is fixed, up to  possibly a global rotation \cite{Poisson:2020vap}. In this sense, the post-Newtonian framework provides a practical and coordinate-invariant method for  extracting the induced moments~\eqref{eq:QabPoisson}.

\newpage

\part{Perturbations of black holes and neutron stars}
\label{Part:BHPT}

\epigraphhead[]{
\epigraph{The black holes of nature are the most perfect macroscopic objects there are in the universe:\\the only elements in their construction are our concepts of space and time.}{\textsc{Chandrasekhar} \cite{Chandrasekhar:1985kt}}
}

Black holes are key objects of gravitational-wave astronomy and among the most fascinating subjects in theoretical physics. Their Love numbers provide a new window into their physical and mathematical properties, as well as the fundamental structure of general relativity itself. Black holes are not, of course, the only compact objects of astrophysical relevance. Gravitational-wave observations now give us direct access to neutron stars as well, providing a unique opportunity to probe their interior structure through their tidal deformation and response in inspiraling binary systems.

In the previous section, we introduced tools that enable a robust definition of tidal responses and Love numbers in a relativistic setting, including the point-particle EFT and the PN framework. To actually compute the response coefficients, one must then match the EFT or the PN metric to a full strong-field calculation performed in the UV theory.
In the case of black hole Love numbers, this is simply vacuum general relativity in four spacetime dimensions with vanishing cosmological constant --- the simplest setting and the one in which the remarkable result of vanishing Love numbers was  established. For more complex objects such as neutron stars, the Einstein equations are sourced by a matter stress-energy tensor, which couples fluid perturbations inside the star to the exterior metric perturbations.

To describe tidal deformations of black holes, we consider perturbations of the Schwarzschild and Kerr metrics. In this Part, after briefly reviewing black hole solutions in general relativity (\cref{sec:BHsolutions}), we work within linear perturbation theory, which will suffice to calculate the leading-order Love numbers in \Cref{part:part3Love}. The results derived in \cref{sec:BHsolutions,sec:BHPT} describe the exterior geometry of any compact object in vacuum general relativity --- including Schwarzschild and Kerr black holes as well as the exterior of neutron stars or other compact bodies with a physical surface.\footnote{This is strictly true for non-spinning objects, where Birkhoff's theorem guarantees that the exterior of a non-rotating star is described by the Schwarzschild metric. There is no analog of Birkhoff's theorem for rotating objects, but in practice the exterior spacetime of neutron stars and other compact objects is extremely close to Kerr.} The distinction between object types enters entirely through the boundary conditions: for black holes one imposes ingoing-wave conditions at the event horizon, whereas for objects with a surface one matches to an interior solution at the stellar radius. The Teukolsky formalism~\cite{Teukolsky:2014vca} provides a unified framework for both cases, as we discuss in \cref{sec:BHPT}.

We review the basics of stellar perturbation theory in \cref{sec:NS}, where we derive the equations governing perturbations of a perfect fluid coupled to gravity, which are needed to describe the interior physics and to determine the boundary conditions required to compute the tidal response (cf.~\cref{sec:NS_Love}). 

General relativity is an intrinsically nonlinear theory, and a complete treatment of tidal response beyond leading order requires higher-order perturbation theory. Known results for nonlinear Love numbers, obtained within both perturbation theory and effective field theory, are discussed in \cref{sec:love-nonlin} of \Cref{part:part3Love}. Extensions beyond the four-dimensional vacuum baseline --- including charged black holes, higher dimensions, non-zero cosmological constant, and modifications of GR --- will be covered in \cref{sec:highD}.

\section{Black hole solutions in general relativity}
\label{sec:BHsolutions}

We begin by reviewing the ``background'' solutions, namely the famous Schwarzschild and Kerr spacetimes of four-dimensional general relativity. This review will necessarily be brief and focused on aspects of these geometries that are relevant for the problem at hand, such as their symmetries and the behavior of massless test particles and waves. For a more thorough introduction to black holes, see, e.g., \rcite{Townsend:1997ku,Chandrasekhar:1985kt}.

\subsection{Schwarzschild}

\subsubsection{Geometry}

The Schwarzschild spacetime, which describes a non-rotating black hole, was the first exact, non-trivial solution to Einstein's field equations of general relativity to be discovered. As a consequence of Birkhoff's theorem, it describes the exterior of all spherically-symmetric bodies. In Schwarzschild coordinates $(t,r,\theta,\varphi)$,\footnote{These coordinates only cover the exterior $r>\rs$, and this is all we will need for computing tidal responses.} the Schwarzschild metric is
\begin{equation}\label{eq:sch}
\dd s^2 = -f(r)\dd t^2 + \frac1{f(r)}\dd r^2 + r^2\dd \Omega^2_2.
\end{equation}
Here we have defined the function
\begin{equation}
f(r) = 1-\frac\rs r,
\end{equation}
with the \emph{Schwarzschild radius} related to the mass $M$ by
\begin{equation}
\rs = 2GM,
\end{equation}
as well as the metric on the unit 2-sphere,
\begin{equation}
\dd \Omega^2_2 = \dd \theta^2 + \sin^2\theta\dd\varphi^2.
\end{equation}
We will also find it convenient to work with the function
\begin{equation}\label{eq:Delta}
\Delta(r)\equiv r(r-\rs) = r^2f(r).
\end{equation}
At the origin, $r=0$, and the Schwarzschild radius, $r=\rs$, this metric is singular,
\begin{subequations}
\begin{align}
\dd s^2 \overset{r\to\rs}{\longrightarrow} 0\,\dd t^2 + \frac10\dd r^2+\rs^2\dd\Omega_2^2, \\
\dd s^2 \overset{r\to0}{\longrightarrow} \frac10\,\dd t^2 + 0\dd r^2+\rs^2\dd\Omega_2^2.
\end{align}    
\end{subequations}
The singularity at $\rs$ is not physical, but rather a coordinate singularity, which can be removed by a suitable coordinate transformation. By contrast the singularity at $r=0$ is a genuine curvature singularity, signalling a breakdown of the spacetime description.

The 2-sphere at $r=\rs$ is known as the \emph{event horizon}. It is the point of no return: an observer starting at $r<\rs$ cannot escape the gravitational field, and reaches the central singularity within finite proper time. For this reason objects contained within $r=\rs$ are called \emph{black holes}: past the event horizon, the ``escape velocity'' exceeds the speed of light. Consider for simplicity a null geodesic at fixed angle,
\begin{equation}
\dd s^2 = 0 = -f\dd t^2 + \frac1f\dd r^2 \quad\Longrightarrow\quad \frac{\dd r}{\dd t} = \left|1-\frac\rs r\right|,
\end{equation}
which exceeds unity when $r<\rs$.
Objects with radius $R>\rs$ are not described by the Schwarzschild metric for $r<R$, since the Einstein equation in the interior is sourced by the stress-energy tensor of the matter.

\subsubsection{Symmetries}
\label{sec:Sch-syms}

The Schwarzschild solution is highly symmetric. The metric is manifestly invariant under rotations and time translations. The isometries of a spacetime are encoded in its Killing vectors, that is, directions $\xi^\mu$ along which the Lie derivative of the metric vanishes,
\begin{equation}\label{eq:Killing}
\mathcal L_\xi g_\mn = \nabla_{(\mu}\xi_{\nu)} = 0.
\end{equation}
The lack of explicit $t$ and $\varphi$ dependence in the metric tells us that the unit vectors in those directions are Killing,
\begin{equation}
\xi_1 = \partial_t,\quad \xi_2 = \partial_\varphi,
\end{equation}
where $v\equiv v^\mu\partial_\mu$, while spherical symmetry gives us two more Killing directions by rotating $\xi_2$, explicitly,
\begin{equation}
\xi_3 = \cos\varphi\partial_\theta - \cot\theta\sin\varphi\partial_\varphi,\quad \xi_4 = \sin\varphi\partial_\theta + \cot\theta\cos\varphi\partial_\varphi.
\end{equation}
We say that the Schwarzschild metric is \emph{static} and \emph{spherically symmetric}. Here a static metric means one which possesses a timelike Killing vector ($\partial_tg_{\mu\nu}=0$) and is invariant under
time reversal
($g_{ti}=0$).
We note for future reference that if the latter condition is not satisfied, the spacetime is instead called \emph{stationary}.

\subsubsection{Geodesics}

Linear perturbations of the metric propagate as massless waves. Before dealing with linearized general relativity, it is instructive to anticipate some of its qualitative features by investigating the behavior of massless particles and waves on the background.

Test particles follow geodesics, whose parametrized path $x^\mu(\affine)$ satisfies the geodesic equation,
\begin{equation}\label{eq:geodesic}
\frac{\dd^2 x^\mu}{\dd{\affine}^2} + \Gamma^\mu_\ab \frac{\dd x^\alpha}{\dd {\affine}} \frac{\dd x^\beta}{\dd{\affine}} = 0,
\end{equation}
with $\affine$ an affine parameter along the path and $\Gamma^\mu_\ab$ the Christoffel symbols,
\begin{equation}
\Gamma^\mu_\ab = g^\mn\left(\partial_{(\alpha}g_{\beta)\nu} - \frac12\partial_\nu g_\ab\right).
\end{equation}
A useful shortcut to avoid calculating the Christoffel symbols is to consider the point-particle Lagrangian,\footnote{This is not precisely the point-particle action \eqref{eq:PP} but rather (up to terms not involving $\affine$) its square. The equations of motion following from variation are the same.}
\begin{align}\label{eq:pp-lag-geo-sch}
\mathcal L &= g_\mn \dot x^\mu \dot x^\nu \nonumber\\
&= -f \dot t^2 + \frac{\dot r^2}f + r^2\left(\dot\theta^2+\sin^2\theta\dot\varphi^2\right),
\end{align}
where overdots denote derivatives with respect to $\affine$. The components of the geodesic equation \eqref{eq:geodesic} are the Euler-Lagrange equations with respect to each of the spacetime coordinates,
\begin{equation}
\frac{\dd}{\dd{\affine}}\frac{\partial\mathcal L}{\partial \dot x^\mu} - \frac{\partial\mathcal L}{\partial x^\mu} = 0.
\end{equation}

Geodesic motion in the Schwarzschild spacetime has a special property: it is completely integrable, and therefore can be characterized fully by first-order constraints.
First we notice that geodesic motion in Schwarzschild is planar. Consider a particle placed at an arbitrary point in the spacetime, and use the spherical symmetry to rotate to a coordinate system in which the particle is located on the equator, $\theta=\pi/2$, and its instantaneous velocity in the $\theta$ direction is also along the equator, $\dot\theta=0$. From the Euler-Lagrange equation for $\theta$,
\begin{align}
0 &= \frac{\dd}{\dd{\affine}}\left(r^2\dot\theta\right) - r^2\sin\theta\cos\theta\dot\varphi^2,
\end{align}
it is clear that we may take $\theta=\pi/2$ without loss of generality. For the $t$ and $\varphi$ directions, the geodesic equation can be integrated, since
\begin{equation}
\frac{\partial \mathcal{L}}{\partial t} = \frac{\partial \mathcal{L}}{\partial \varphi} = 0.
\end{equation}
The $t$ and $\varphi$ components of the geodesic equation are therefore each the first derivative of a conservation law,
\begin{equation}
\frac{\partial \mathcal{L}}{\partial \dot t} = -f\dot t \equiv -E,\qquad\frac{\partial \mathcal{L}}{\partial \dot \varphi} = r^2 \dot\varphi \equiv h.
\end{equation}
Physically, $E$ is the energy per unit mass of the particle at $r=\infty$, while $h$ is the angular momentum per unit mass. These conserved quantities follow directly from the existence of the Killing vectors $\partial_t$ and $\partial_\varphi$, due to an avatar of Noether's theorem: along a geodesic $x^\mu(\affine)$ the product $\xi_\mu\dot x^\mu$ is conserved,
\begin{align}
\frac{\dd}{\dd{\affine}}\left(\xi_\mu \dot{x}^\mu\right) &= \dot{x}^\nu\nabla_\nu\left(\xi_\mu \dot{x}^\mu\right) \nonumber\\
&= \dot x^\mu \dot x^\nu \nabla_\nu \xi_\mu + \xi_\mu \dot x^\nu\nabla_\nu\dot x^\mu = 0.
\end{align}
The first term vanishes due to symmetry and the Killing equation \eqref{eq:Killing}, and the second by virtue of the geodesic equation \eqref{eq:geodesic}, which we can write as $\dot x^\nu \nabla_\nu \dot x^\mu=0$.

To completely characterize the geodesics, all that remains is to solve for $r(\affine)$. Normally this would require us to solve the $r$ component of the geodesic equation. However on Schwarzschild we have enough conserved quantities that geodesic motion is integrable, with $r(\affine)$ being determined by the normalization of the 4-velocity,\footnote{This can also be seen as a conservation law, namely conservation of mass.}
\begin{align}
g_\mn \dot x^\mu \dot x^\nu &= -f \dot t^2 + \frac{\dot r^2}f + r^2\dot\varphi^2 \nonumber\\
&= \frac{\dot r^2-E^2}f+\frac{h^2}{r^2} = \begin{cases}
1&\mathrm{spacelike}\\
0&\mathrm{lightlike}\\
-1&\mathrm{timelike}
\end{cases}.
\end{align}
In particular we will be interested in perturbations of massless fields such as the gravitational or electromagnetic fields, which in the geometric optics limit propagate along lightlike ($g_\mn \dot x^\mu \dot x^\nu=0$) geodesics. Solving for $\dot r$ we obtain, in effect, a first integral of the $r$ geodesic equation,
\begin{equation}\label{eq:rdot}
\dot r^2 = E^2 - \frac f{r^2}h^2.
\end{equation}
This is precisely analogous to the conservation of energy for a particle in an effective potential,
\begin{equation}\label{eq:rdot-Veff}
\frac12\dot r^2 + V_\eff(r) = \frac12E^2,\qquad V_\eff(r) = \frac12\frac{h^2 f}{r^2}.
\end{equation}
We plot $V_\eff(r)$ in \cref{fig:scalar-geo-pot}.
The potential has no minima and a single maximum, located at the \emph{photon sphere},
\begin{equation}\label{eq:photon-sphere}
r_\mathrm{ps}=\frac32\rs.
\end{equation}
Physically this means that massless particles can only orbit a Schwarzschild black hole at the radius $r=r_\mathrm{ps}$, and that moreover such orbits are unstable; a photon orbiting just inside or outside the photon sphere will fall into the black hole or be scattered out to infinity, respectively.

\subsubsection{Wave equation}
\label{sec:Sch-KG}

Closely related to the problem of massless test particle motion is the
behavior of massless waves. In the geometric optics approximation
these precisely follow null geodesics, and one might expect the
full wave equation to behave similarly.
For
simplicity\footnote{The spin-2 wave equation, which results from linearizing Einstein's equations around  Schwarzschild, is studied in \cref{sec:BHPT-Sch}. The utility of the massless scalar as a simplified model displaying the qualitative features of gravitational perturbations will be a theme throughout \Cref{part:part3Love}.} we consider the spin-0 wave equation, i.e., the
Klein--Gordon equation for a massless scalar $\phi$,
\begin{equation}
\Box\phi = 0.
\end{equation}
Again we do not need to compute Christoffel symbols, as the d'Alembertian is simply
\begin{align}
0 = \Box\phi &= \frac1\sdg\partial_\mu\left(\sdg\partial^\mu\phi\right) \nonumber\\
&= \frac1{r^2\sin\theta}\partial_\mu\left(r^2\sin\theta g^\mn \partial_\nu \phi\right) \nonumber\\
&= -\frac{r^4}{\Delta}\partial_t^2\phi + \partial_r\left(\Delta\partial_r\phi\right) +
\nabla^2_{S^2}\phi,\label{eq:KG-sch}
\end{align}
where in the last line we multiplied by $r^2$ for convenience, and the Laplacian on the 2-sphere is
\begin{equation}
\nabla^2_{S^2}=\frac1{\sin\theta}\partial_\theta\left(\sin\theta\partial_\theta\right) + \frac1{\sin^2\theta}\partial_\varphi^2.
\end{equation}
We solve \cref{eq:KG-sch} by assuming a separable form for the field:
\begin{equation}
\phi(x^\mu) = e^{-i(\omega t - m \varphi)} \, \phi(r) \, S(\theta),
\end{equation}
where $\phi(r)$ and $S(\theta)$ are functions of the radial and polar coordinates, respectively. The simple exponential dependence on $t$ and $\varphi$ is a direct
result of their being Killing directions, in analogy to the
conservation laws discussed above: in the case of the wave equation, the statement is that $\phi$ is an eigenfunction of the Killing vectors $\partial_t$ and $\partial_\varphi$. Stationarity yields a continuous
Fourier decomposition in $t$ with frequency $\omega$, while
axisymmetry yields a discrete decomposition in $\varphi$ with integer
azimuthal number $m$.

Plugging this ansatz into the Klein--Gordon equation and dividing out an overall factor of $\phi(x^\mu)$ we find
\begin{equation}
0 = \frac1{\phi(r)}\left[\partial_r\left(\Delta\partial_r\phi(r)\right)+\frac{\omega^2r^4}{\Delta}\phi(r)\right] + \frac1{\sin\theta S(\theta)}\left[\partial_\theta\left(\sin\theta\partial_\theta S(\theta)\right)-m^2 \csc\theta S(\theta)\right].
\end{equation}
The first term depends only on $r$ while the second term depends only on $\theta$, so they must both be constants,
\begin{align}
\partial_r\left(\Delta\partial_r\phi\right)+\frac{\omega^2r^4}{\Delta}\phi &= \ell(\ell+1)\phi,\\
\partial_\theta\left(\sin\theta\partial_\theta S\right)-m^2 \csc\theta S &= -\ell(\ell+1)\sin\theta S.
\end{align}
Here $\ell$ is a constant which, as we will soon see, is required to be a non-negative integer. The $\theta$ equation is the general Legendre differential equation \eqref{eq:legendre-assoc} in slight disguise, as can be seen by defining $x = \cos\theta$,
\begin{equation}
\frac{\dd}{\dd x}\left[(1-x^2)\frac{\dd S}{\dd x}\right] - \left[\ell(\ell+1)-\frac{m^2}{1-x^2}\right]S = 0.
\end{equation}
The solutions are the associated Legendre functions \cite{DLMF,Abramowitz:1965},
\begin{equation}
S(\theta) = c_1 P_\ell^m(\cos\theta) + c_2 Q_\ell^m(\cos\theta).\label{eq:S-sch}
\end{equation}
The associated Legendre function of the second kind, $Q_\ell^m(x)$, diverges at the poles, $x=\pm1$, so on physical grounds we set $c_2=0$. The Legendre functions of the first kind also diverge at the pole $\theta=\pi$ ($x=-1$) when $\ell$ is not an integer. We see that the angular eigenfunctions are nothing other than the standard spherical harmonics, as the reader might have anticipated from the spherical symmetry of the system or from the fact that $e^{im\varphi}S(\theta)$ is clearly an eigenfunction of $\nabla^2_{S^2}$,
\begin{equation}\label{eq:sph-harm}
e^{im\varphi}S(\theta) \propto Y_\ell^m(\theta,\varphi) = Ne^{im\varphi}P_\ell^m(\cos\theta),
\end{equation}
with $\ell\in\mathbb{Z}$, $-\ell\leq m\leq \ell$, and $N$ a normalization constant whose form is given in \cref{eq:N-sph-harm}. For more on the spherical harmonics see \ref{app:sph-harm}.

The remaining equation of motion is for the radial field,
\begin{equation}\label{eq:sch-KG}
\boxed{\partial_r\left(\Delta\partial_r\phi\right)+\frac{\omega^2r^4}{\Delta}\phi - \ell(\ell+1)\phi = 0.}
\end{equation}
The reader is invited to compare this to the Poisson equation \eqref{eq:poisson-rad} in Newtonian gravity, recalling that in the flat-space limit $\Delta=r^2$. To canonically normalize this equation we rescale $\phi$,
\begin{equation}
\psi \equiv r \phi,
\end{equation}
in terms of which
\begin{equation}\label{eq:sch-KG-tort}
\boxed{\partial_{\rst}^2\psi +\left(\omega^2-\frac{ff'}r- \ell(\ell+1)\frac f{r^2}\right)\psi = 0.}
\end{equation}
Here $\rst$ is the \emph{tortoise coordinate}, defined by
\begin{equation}\label{eq:tortoise}
\dd \rst = \frac{\dd r}f \quad\Longrightarrow\quad f\partial_r = \partial_{\rst}.
\end{equation}
This integrates to
\begin{equation}
\rst = r+\rs\ln\left(\frac r\rs-1\right),
\end{equation}
with the horizon located at $\rst=-\infty$ and spatial infinity at $\rst=+\infty$.

\begin{figure}[h]
\centering
\includegraphics[width=.8\textwidth]{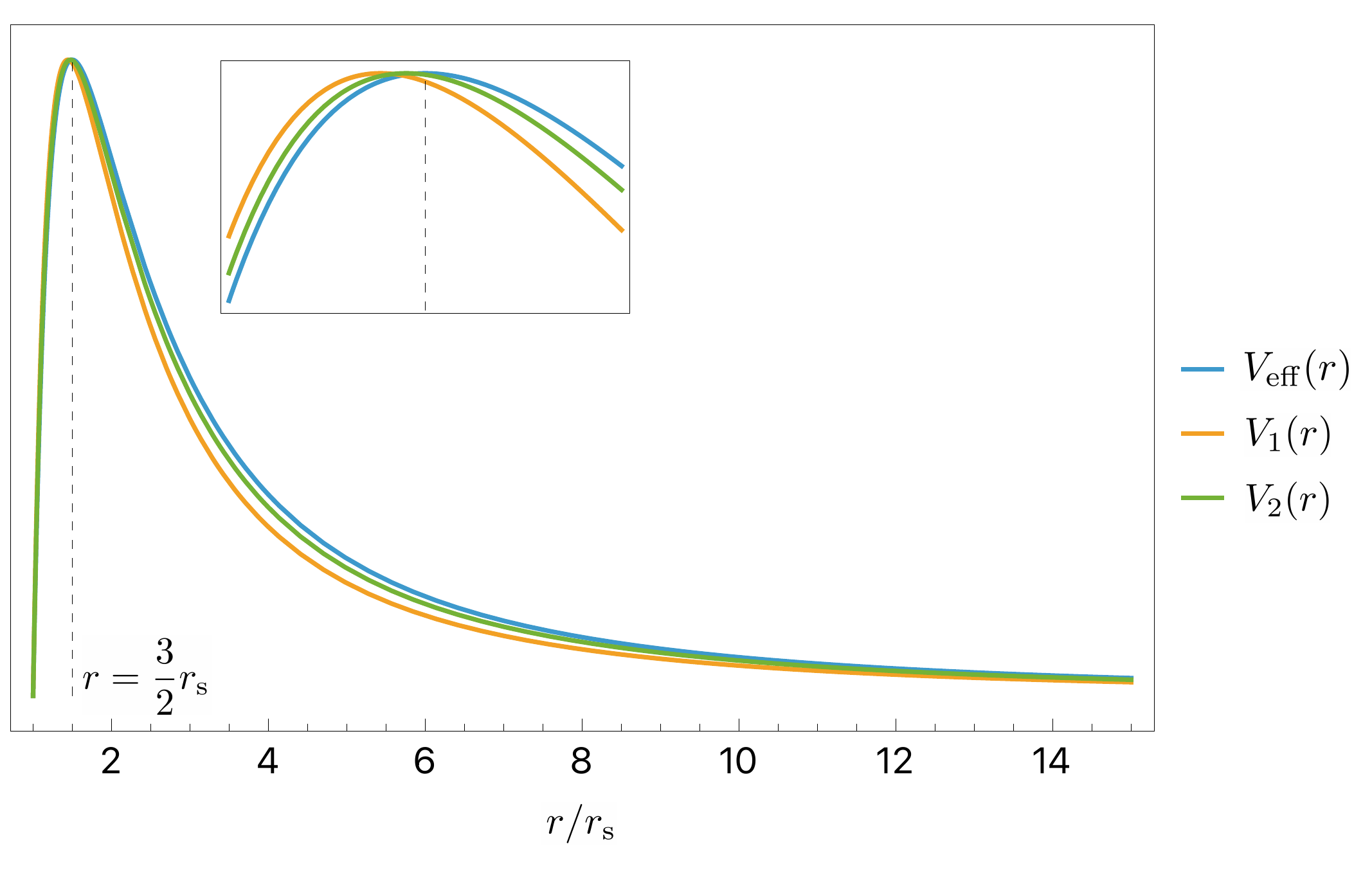}
\caption{The effective potential \eqref{eq:rdot-Veff} for null geodesics and the spin-0 effective potential \eqref{eq:V-s0} for $\ell=1,2$, normalized to unity at their maxima. The geodesic potential peaks at the photon sphere \eqref{eq:photon-sphere}, $r=3\rs/2$, while the scalar potentials peak at $r=3\rs/2-\epsilon$ with $\epsilon$ going quickly to zero with increasing $\ell$, as we see from only the first two modes.}
\label{fig:scalar-geo-pot}
\end{figure}

The Klein--Gordon equation $\Box\phi=0$ on Schwarzschild therefore reduces to the 1D ODE \eqref{eq:sch-KG} or \eqref{eq:sch-KG-tort}.
The field $\psi(r)$ obeys a Schr\"odinger-type equation in an effective potential,
\begin{equation}\label{eq:spin-0-rad}
\frac{\dd^2}{\dd r_\star^2}\psi = (V_\ell(r)-\omega^2)\psi
\end{equation}
with
\begin{align}\label{eq:V-s0}
V_\ell(r) &= \frac f{r^2}\left(\ell(\ell+1)+rf'\right) \nonumber\\ 
& = \frac f{r^2}\left(\ell(\ell+1)+\frac\rs r\right).
\end{align}
The potentials $V_\ell(r)$ are plotted for $\ell=1,2$ in \cref{fig:scalar-geo-pot}.
The general solution to \cref{eq:sch-KG} is built by integrating over all frequencies and summing over all values of $\ell$ and $m$,
\begin{equation}
    \phi(x^\mu) = \int_{-\infty}^{\infty} \sum_{\ell=0}^\infty \sum_{m=-\ell}^\ell
    \frac{\dd\omega}{2\pi}\,
    e^{-i(\omega t - m\varphi)}\,
    \phi_{\omega\ell m}(r)\,P_\ell^m(\cos\theta),
    \label{eq:KG_decomp}
\end{equation}
with $\psi_{\omega\ell m}(r) \equiv r \phi_{\omega\ell m}(r)$ solving \cref{eq:spin-0-rad}.

The spin-0 effective potential \eqref{eq:V-s0} is qualitatively very similar to the geodesic effective potential \eqref{eq:rdot-Veff}, and indeed the two potentials, suitably normalized, coincide in the limit $\ell\to\infty$, as shown in \cref{fig:scalar-geo-pot}. This reflects the fact that massless waves follow null geodesics in the geometric optics limit. Note in particular that $V_\ell(r)$ peaks very near the photon sphere \eqref{eq:photon-sphere}, with the peak approaching $r_\mathrm{ps}=3\rs/2$ as $\ell\to\infty$. The link between the null geodesic equation \eqref{eq:rdot} and the radial spin-0 equation \eqref{eq:sch-KG} can be made precise as follows \cite{Andersson:2016hmv}. Let us write \cref{eq:rdot} as
\begin{equation}
r^4\dot r^2 + \mathcal{R}(r,E,h) = 0,
\end{equation}
with
\begin{equation}\label{eq:cal-R}
    \mathcal{R}(r,E,h) = -r^4E^2+\Delta h^2.
\end{equation}
If we send $E\to i\partial_t$ and $h^2\to-\nabla^2_{S^2}$ we can write $\mathcal{R}$ as a differential operator,
\begin{equation}
    \mathcal{R}(r,\partial_t,\nabla^2_{S^2}) = r^4\partial_t^2 - \Delta\nabla^2_{S^2},
\end{equation}
such that the Klein--Gordon equation \eqref{eq:KG-sch} is
\begin{equation}
    \left[\Delta\partial_r(\Delta\partial_r) - \mathcal{R}(r,\partial_t,\nabla^2_{S^2})\right]\phi = 0.
\end{equation}
The radial equation \eqref{eq:sch-KG} follows from replacing $\partial_t$ and $\nabla^2_{S^2}$ by their eigenvalues $-i\omega$ and $-\ell(\ell+1)$.

\subsection{Kerr}

\subsubsection{Geometry}

Black holes in the real world rotate, which is not accounted for in the spherically-symmetric Schwarzschild solution \eqref{eq:sch}. Schwarzschild had derived his metric within months of the publication of Einstein's field equations, while it took nearly a half century for Kerr to find its spinning generalization \cite{Kerr:1963ud}. Here we will summarize salient features of the Kerr solution; see, e.g., \rcite{Teukolsky:2014vca,Visser:2007fj} for pedagogical reviews.

In Boyer--Lindquist coordinates,\footnote{Both Karl Schwarzschild and Robert Boyer died young within months of completing the work for which their names are immortalized. Schwarzschild discovered the first exact solution to Einstein's equations at the end of 1915 --- barely a month after Einstein published them --- while a soldier on the Russian front in World War I at the age of 42. He succumbed to a disease acquired at the front in May of 1916. Boyer was killed at the age of 33 in a mass shooting at the University of Texas in August 1966, along with fourteen other people, days before he was set to take up a position in Liverpool. His paper with Lindquist was submitted just 11 days earlier, and published posthumously in 1967.} which are a generalization of the Schwarzschild coordinates $(t,r,\theta,\varphi)$, the metric for a black hole of mass $M$ and angular momentum $J$ is written
\begin{align}\label{eq:Kerr}
\dd s^2 &= -\left(1-\frac{\rs r}\Sigma\right)\dd t^2 + \frac\Sigma\Delta\dd r^2+\Sigma\dd\theta^2 + \left(r^2+a^2+\frac{\rs r a^2}\Sigma\sin^2\theta\right)\sin^2\theta\dd\varphi^2 - \frac{2\rs r a\sin^2\theta}\Sigma\dd t\dd\varphi \nonumber\\
&= -\frac\Delta\Sigma\left(\dd t-a\sin^2\theta\dd\varphi\right)^2 +\frac{\sin^2\theta}\Sigma\left[(r^2+a^2)\dd\varphi-a\dd t\right]^2 + \frac\Sigma\Delta \dd r^2 + \Sigma\dd\theta^2.
\end{align}
Here, as for Schwarzschild, we define $\rs = 2GM$. In the following, we set $G=1$ for simplicity.  
The spin is parametrized by
\begin{equation}
a = \frac J M,
\end{equation}
which is expected (though not proven) to always be in the range $0\leq a<M$ in physical scenarios \cite{Penrose:1969pc,Hawking:1970zqf,Wald:1974hkz,Giacomazzo:2011cv,Espino:2019xcl}. The functions $\Delta$ and $\Sigma$ are given by
\begin{align}
\Delta &= r(r-\rs)+a^2 \nonumber\\&= (r-r_+)(r-r_-),\\
\Sigma &= r^2 + a^2\cos^2\theta.
\end{align}
These generalize, from the Schwarzschild case ($a=0$), the quantities $\Delta$ \eqref{eq:Delta} and $r^2$, respectively.

The \emph{inner} and \emph{outer horizons} $r_\pm$ are located at the zeros of $\Delta$, where $g_{rr}$ diverges; they are related to the parameters $\rs$ and $a$ by
\begin{align}
\rs&=r_++r_-,\\
a^2&=r_+r_-,
\end{align}
or explicitly
\begin{equation}
r_\pm = M\left(1\pm\sqrt{1-\frac{a^2}{M^2}}\right).
\end{equation}
When $a=0$ these reduce to $r_+=\rs$ and $r_-=0$. Meanwhile, the $g_{tt}$ component vanishes at $\Sigma=\rs r$; this defines the location of the outer and inner \emph{ergospheres},
\begin{equation}
r_{\mathrm{E},\pm} = M\left(1\pm\sqrt{1-\frac{a^2}{M^2}\cos^2\theta}\right).
\end{equation}
At the poles, $r_{\mathrm{E},\pm} = r_\pm$, and at the equator $(r_{\mathrm{E},+}, r_{\mathrm{E},-})=(\rs,0)$. The five different length scales introduced have the hierarchy
\begin{equation}
r_+ \geq r_{\mathrm{E},+} \geq \rs \geq r_- \geq r_{\mathrm{E},-}.
\end{equation}
The singularity at $r=r_\pm$ is a coordinate singularity, just as in Schwarzschild. The zero of $g_{tt}$ is not even a coordinate singularity due to the non-vanishing of $g_{t\varphi}$. Kerr has a genuine curvature singularity at $(r,\theta)=(0,\pi/2)$, where $\Sigma$ vanishes. In contrast to Schwarzschild, this singularity is not a point but rather a ring in the equatorial plane.

\begin{figure}[h]
\centering
\includegraphics[width=\textwidth]{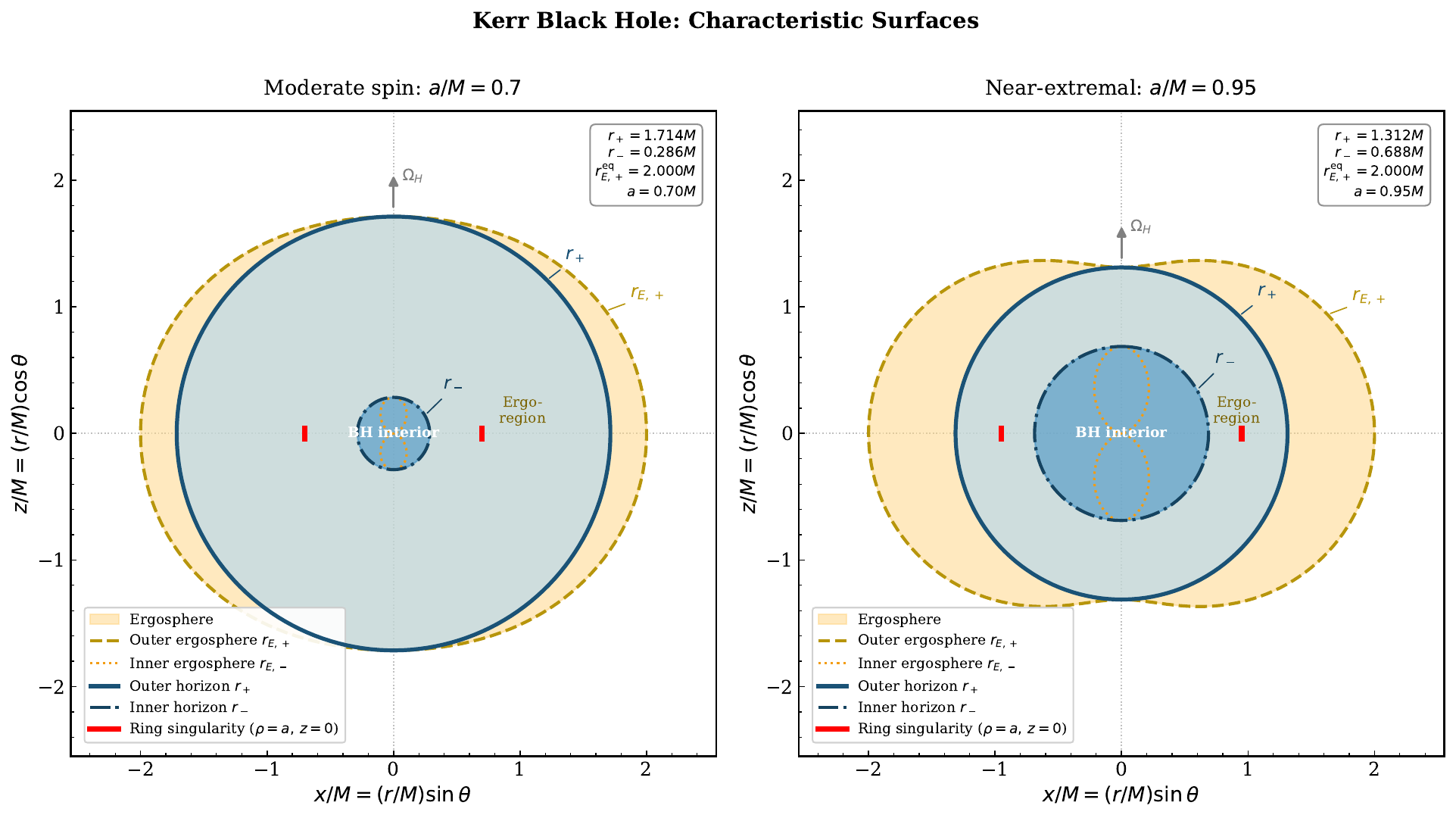}
\caption{Cross-sectional view of the characteristic surfaces of the
Kerr spacetime in Boyer--Lindquist coordinates $(x,z) =
(r\sin\theta,\, r\cos\theta)$, for moderate spin $a/M=0.7$ (left)
and near-extremal spin $a/M=0.95$ (right). The outer and inner event
horizons $r_\pm = M \pm \sqrt{M^2 - a^2}$ are shown as solid blue
curves; the outer and inner ergospheres $r_{E,\pm} =
M\pm\sqrt{M^2-a^2\cos^2\theta}$ as dashed and dotted gold curves.
The ergoregion (yellow shading) lies between the outer ergosphere
and the outer horizon; it is the region in which no static observer
can exist, since $g_{tt}>0$ there. The ring singularity (red marks)
sits at $\rho = a$, $z=0$ in the equatorial plane. At the poles
the ergosphere and horizon coincide ($r_{E,+}=r_+$), while at the
equator $r_{E,+} = \rs = 2M$ for all spins. In the extremal limit $a/M\to 1$, the two
horizons approach each other ($r_+\to r_-\to M$) and the ergoregion
grows, consistent with the hierarchy
$r_+\geq r_{E,+}\geq \rs \geq r_-\geq r_{E,-}$.}
\label{fig:kerr_surfaces}
\end{figure}

The
characteristic surfaces of the Kerr spacetime discussed above ---
the outer and inner horizons $r_\pm$, the ergospheres $r_{E,\pm}$,
and the ring singularity --- are depicted in \cref{fig:kerr_surfaces}
for two representative values of the spin parameter.

We will at times find it convenient to work with thermodynamics parameters of the Kerr black hole. These include the angular velocity at the outer horizon,
\begin{equation}\label{eq:ang-vel-hor}
    \Omega_\mathrm H=\frac a{\rs r_+},
\end{equation}
and the Hawking temperature $T_\mathrm H$, which we will equivalently express in terms of the inverse surface gravity $\beta$,
\begin{equation}\label{eq:inv-hawking-temp}
    \beta = \frac1{2\pi T_\mathrm H} = \frac{2\rs r_+}{r_+-r_-}.
\end{equation}

\subsubsection{Symmetries}
\label{sec:Kerr-syms}

Kerr is clearly less symmetric than Schwarzschild --- for one thing, it breaks spherical symmetry. Nevertheless it retains a surprisingly high degree of symmetry which ensures that the geodesic and wave equations remain integrable, and will turn out to have remarkable consequences for gravitational perturbations in the next section.

The timelike and azimuthal Killing vectors $\partial_t$ and $\partial_\varphi$ are isometries of Kerr, as is evident by the lack of explicit $t$ or $\varphi$ dependence in $g_\mn$. In contrast to the static and spherically symmetric Schwarzschild metric, Kerr is \emph{stationary} and \emph{axisymmetric}.\footnote{Recall that a stationary spacetime is one with a timelike Killing vector, while a static spacetime additionally is invariant under time reversal. The presence of the $g_{t\varphi}$ cross term in the Kerr metric means it is not static. Axisymmetry, the remnant of full spherical symmetry, is symmetry under rotations about a given axis, corresponding to the $\partial_\varphi$ Killing vector.}

In addition to these two Killing vectors, Kerr possesses a remarkable \emph{hidden symmetry}. This plays the same role for Kerr that spherical symmetry does for Schwarzschild. It is encoded in the existence of a non-degenerate rank-2 Killing \emph{tensor} $K_\mn$, which is a symmetric tensor satisfying
\begin{equation}\label{eq:Killingtensor}
\nabla_{(\mu}K_{\nu\alpha)}=0.
\end{equation}
This hidden symmetry can be seen more readily in Carter canonical coordinates $(\tau,r,y,\psi)$ \cite{Carter:1968rr,Frolov:2017kze},
\begin{equation}\label{eq:canonicalcoords}
y=a\cos\theta,\qquad \psi = \frac\varphi a,\qquad \tau = t-a\varphi.
\end{equation}
The Kerr metric \eqref{eq:Kerr} takes the compact form
\begin{equation}
    \dd s^2 = -\frac{\Delta}\Sigma\left(\dd\tau+y^2\dd\psi\right)^2 + \frac{\Delta_y}\Sigma\left(\dd\tau-r^2\dd\psi\right)^2 + \frac\Sigma{\Delta}\dd r^2+\frac\Sigma{\Delta_y}\dd y^2,
\end{equation}
where
\begin{equation}
    \Delta_y(y)=a^2-y^2.
\end{equation}
Notice the remarkable ``duality'' between the $(t,r)$ and $(\theta,\varphi)$ submanifolds in these coordinates.
A further remarkable fact is that the hidden symmetry structure we will discuss for Kerr holds for any functions $\Delta(r)$ and $\Delta_y(y)$, that is, these functions can be taken to be ``off shell.''

Following \rcite{Frolov:2017kze} we construct a potential 1-form $b$,\footnote{Here we use the language of differential forms, reviewed in \ref{app:diff-forms}.}

\begin{equation}
    b = -\frac12\left[\left(r^2-y^2\right)\dd\tau+r^2y^2\dd\psi\right],
\end{equation}
or in coordinates,
\begin{equation}
    b_\mu = -\frac12\left[\left(r^2-y^2\right)\delta_{\mu\tau}+r^2y^2\delta_{\mu\psi}\right].
\end{equation}
The associated ``field strength'' $h=\dd b$,
\begin{equation}\label{eq:principal}
    h = y\dd y\wedge\left(\dd \tau-r^2\dd\psi\right)-r\dd r\wedge\left(\dd\tau+y^2\dd\psi\right),
\end{equation}
is called the \emph{principal tensor} as it generates both the Killing vector isometries and the Killing tensor hidden symmetry. The principal tensor is a non-degenerate, closed conformal Killing--Yano form, i.e., it obeys
\begin{equation}
    \nabla_\alpha h_\mn = 2g_{\alpha[\mu}\xi_{\nu]},
\end{equation}
where the \emph{principal vector} $\xi^\mu$ is
\begin{equation}
     \xi_\nu=\frac1{D-1}\nabla^\mu h_\mn.
\end{equation}
The principal vector is always a Killing vector (the \emph{primary Killing vector}); for Kerr it is the natural timelike Killing vector in canonical coordinates,
\begin{equation}
    \xi=\dd\tau.
\end{equation}
The Hodge dual $f=\star h$ of the principal tensor is a Killing--Yano tensor, i.e., its covariant derivative is purely antisymmetric,
\begin{equation}
    \nabla_\alpha f_{\mn} = \nabla_{[\alpha} f_{\mn]}.
\end{equation}
A Killing--Yano tensor $f_\mn$ can be viewed as a ``square root'' of a Killing tensor $K_\mn$,
\begin{equation}
    K_\mn = f_{\mu\alpha}f_\nu{}^\alpha,
\end{equation}
which satisfies \cref{eq:Killingtensor} by construction. For Kerr, the Killing tensor descending from the principal tensor \eqref{eq:principal} takes the form
\begin{equation}
    K^\mu{}_\nu = \begin{pmatrix}
    0&0&0&-r^2y^2\\
    0&-y^2&0&0\\
    0&0&r^2&0\\
    -1&0&0&r^2-y^2
    \end{pmatrix}
\end{equation}
in canonical coordinates and
\begin{equation}\label{eq:KT-BL}
    K^\mu{}_\nu = \begin{pmatrix}
    -a^2&0&0&a(r^2+a^2)\sin^2\theta\\
    0&-a^2\cos^2\theta&0&0\\
    0&0&r^2&0\\
    -a&0&0&r^2+a^2\sin^2\theta
    \end{pmatrix}
\end{equation}
in Boyer--Lindquist coordinates.
From this we can construct a second Killing vector, $\eta^\mu=-K^\mn \xi_\nu$. In Kerr this is the azimuthal Killing vector in canonical coordinates,
\begin{equation}
    \eta = \dd\psi.
\end{equation}
We see that both the Killing isometries of Kerr as well as its Killing tensors ($K_\mn$ and $g_\mn$ itself) descend from the principal tensor $h_\mn$. The symmetries associated to Killing tensors are referred to as \emph{hidden symmetries}.

\subsubsection{Geodesics}

As in Schwarzschild we may study geodesics by varying the point-particle Lagrangian, cf.~\cref{eq:pp-lag-geo-sch},
\begin{align}
    \mathcal{L} &= g_\mn \dot x^\mu \dot x^\nu\nonumber\\
    &= -\left(1-\frac{\rs r}\Sigma\right)\dot t^2 + \frac\Sigma\Delta\dot r^2+\Sigma\dot\theta^2 + \left(r^2+a^2+\frac{\rs r a^2}\Sigma\sin^2\theta\right)\sin^2\theta\dot\varphi^2 - \frac{2\rs r a\sin^2\theta}\Sigma\dot t\dot\varphi
\end{align}
This is rather more complicated than \cref{eq:pp-lag-geo-sch}. Nevertheless the geodesic motion is still analytically tractable --- indeed it is \emph{integrable} --- as a result of Kerr's hidden symmetry structure.

As with Schwarzschild we have a timelike and a spacelike Killing vector,
\begin{equation}    \xi^\mu_{(t)}=\delta^\mu_t,\qquad\xi^\mu_{(\varphi)}=\delta^\mu_\varphi,
\end{equation}
a consequence of which is that $\partial\mathcal{L}/\partial t = \partial\mathcal L/\partial\varphi=0$. Noether's theorem associates to these isometries conserved quantities,
\begin{align}
    -E &= \frac{\partial\mathcal L}{\partial\dot t} = \xi_{(t)\mu}\dot x^\mu,\\
    h &= \frac{\partial\mathcal L}{\partial\dot \varphi} = \xi_{(\varphi)\mu}\dot x^\mu.
\end{align}
The conservation of $E$ and $h$ is a consequence of the geodesic equation $\dot x^\nu\nabla_\nu \dot x^\mu=0$ and the Killing equation $\nabla_{(\mu}\xi_{\nu)}=0$.

The Killing tensor $K_\mn$
also leads to a conserved quantity, called the \emph{Carter constant},
\begin{equation}
    Q = K_\mn \dot x^\mu \dot x^\nu.
\end{equation}
It is conserved by virtue of the geodesic equation and the Killing tensor equation $\nabla_{(\alpha}K_{\mn)}=0$,
\begin{align}
    \frac{\dd Q}{\dd{\affine}} &= \dot x^\alpha\nabla_\alpha(K_\mn \dot x^\mu\dot x^\nu) \nonumber\\
    &= \dot x^\alpha \dot x^\mu \dot x^\nu \nabla_{(\alpha}K_{\mn)} +2 K_\mn \dot x^\alpha\dot x^{(\mu}\nabla_\alpha\dot x^{\nu)} = 0.
\end{align}
The Carter constant depends quadratically on the momenta, in contrast to the energy per unit mass $E$ and angular momentum per unit mass $h$, which are linear in the momenta. It is not obvious from the equations of motion that this quantity is conserved, which partially explains why the symmetry associated to $K_\mn$ is called hidden. In the Schwarzschild limit $a\to0$ and fixing to the equatorial plane $\theta=\pi/2$, then $Q=h^2$.

Finally, we note that $g_\mn$ itself is a Killing tensor, and its associated conserved quantity is the quantity $\sigma=g_\mn \dot x^\mu \dot x^\nu$ which is normalized to $\pm1$ or $0$.
The existence of four conserved quantities, corresponding to the four coordinates of the geodesic, implies that the geodesic motion is completely integrable.

\subsubsection{Wave equation}
\label{sec:Kerr-KG}

The form of the wave operator $\Box$ in Boyer--Lindquist coordinates is complicated and not particularly enlightening. It is easier to work with in canonical coordinates (note the useful relation $\sdg=\Sigma$) \cite{Frolov:2017kze},
\begin{align}\label{eq:KG-kerr}
    \sdg \,\Box &= \partial_\mu\left(\sdg g^\mn\partial_\nu\right) \nonumber\\
    &= \partial_r\left(\Delta\partial_r\right) + \partial_y\left(\Delta_y\partial_y\right) - \frac1\Delta\left(r^2\partial_\tau+\partial_\psi\right)^2 + \frac1{\Delta_y}\left(y^2\partial_\tau-\partial_\psi\right)^2.
\end{align}
As in Schwarzschild, this operator admits a separable ansatz,
\begin{align}
    \phi &= e^{i(-\omega \tau+L_\psi\psi)}R(r)Y(y)\\
    &= e^{i(-\omega t+m\varphi)}R(r)S(\theta).
\end{align}
Note that $S(\theta)=Y(a\cos\theta)$ and $L_\psi=am-a^2\omega$. The massless Klein--Gordon equation $\Box\phi=0$ separates into
\begin{subequations}\label{eq:KG-kerr-sep}
\begin{align}
    \left[\partial_r\left(\Delta \partial_r \right) + \frac{(L_\psi-\omega r^2)^2}\Delta \right]R(r) &= KR(r),\label{eq:KG-kerr-R}\\
    \left[\partial_y\left(\Delta_y \partial_y\right) - \frac{(L_\psi+\omega y^2)^2}{\Delta_y}\right]Y(y) &= -KY(y).\label{eq:KG-kerr-S}
\end{align}
\end{subequations}

The separability of \cref{eq:KG-kerr} is due to the existence of four mutually-commuting differential operators, each constructed out of one of the Killing vectors or Killing tensors,
\begin{equation}
    \Box = \nabla_\mu(g^\mn\nabla_\nu),\qquad
    \mathcal K = \nabla_\mu(K^\mn\nabla_\nu),\qquad
    \mathcal{L}_t = i \xi^\mu_{(t)}\nabla_\mu,\qquad
    \mathcal{L}_\varphi = i \xi^\mu_{(\varphi)}\nabla_\mu.
\end{equation}
This is precisely analogous to the four conserved quantities of geodesics leading to integrability.
The separable solution $\phi$ is the common eigenfunction of these operators; indeed we have
\begin{equation}
    \Box\phi = 0,\qquad \mathcal K \phi = -K\phi,\qquad\mathcal{L}_t \phi = \omega\phi,\qquad \mathcal{L}_\varphi\phi=-m\phi.
\end{equation}
We see that in much the way that $\omega$ and $m$ are analogues of $E$ and $h$ for lightlike geodesics, the separation constant $K$ is the analogue of the Carter constant $Q$. (The eigenvalue of $\Box$ is the mass, corresponding to $\sigma=g_\mn \dot x^\mu \dot x^\nu$ for geodesics, which vanishes for massless particles and waves.)

The equations of motion \eqref{eq:KG-kerr-sep} for the radial and angular modes are nearly identical, but there is a crucial difference in how we solve them, as we have already seen in the $a=0$ case (cf. the discussion around \cref{eq:S-sch}). Namely, because we require regularity at the endpoints $\theta=(0,\pi)$ or $y=\pm a$, there is a basis in which one of the two linearly-independent solutions to \cref{eq:KG-kerr-S} can be dropped. The resulting solutions $S(\theta)$ are called \emph{spheroidal
harmonics}~\cite{Berti:2005gp,Press:1973zz,DLMF}. In the limit $a\to0$ they reduce to the spherical harmonics \eqref{eq:sph-harm}. Accordingly the separation constant $K$ is of the form
\begin{equation}
    K = \ell(\ell+1) + \mathcal{O}(a).
\end{equation}

\subsection{Algebraic specialness and the GHP formalism}
\label{sec:alg-spec}

We have seen that the black hole solutions of general relativity are very special, with properties that can appear, particularly in the case of Kerr, almost magical. Underlying these is a rich symmetry structure, including a non-trivial and highly useful ``hidden symmetry,'' as discussed in \cref{sec:Kerr-syms}.

At a deeper level, these symmetries and their associated magical properties are the result of the fact that the Schwarzschild and Kerr spacetimes are \emph{algebraically special}, of type D in the Petrov classification. Indeed, the Kerr spacetime is the \emph{unique} asymptotically-flat, positive-mass vacuum solution of Petrov type D in four dimensions \cite{Kinnersley:1969zza,Walker:1970un,Penrose:1985bww,Penrose:1986ca,Mars:2000gb,Stephani:2003tm,Andersson:2016hmv}. 

In this subsection we will review the Petrov classification, before moving on to discuss the \emph{Geroch--Held--Penrose (GHP) formalism}, a powerful calculus that takes maximal advantage of the symmetries of Kerr. This will enable us in the following section to deal with Kerr perturbations.

\subsubsection{Complex null tetrads and the Petrov classification}
\label{sec:null-tetrad}

The metric tensor of general relativity can be written in terms of its ``square root'' or \emph{tetrad}\footnote{Equivalently, vierbein and, in general dimension, vielbein.} $e^a_\mu$, related to the metric $g_\mn$ by
\begin{equation}\label{eq:vierbein}
    g_\mn = \eta_{ab}e^a_\mu e^b_\nu.
\end{equation}
Here lower-case Latin indices $(a,b,...)$ are \emph{internal} or \emph{Lorentz indices}, and $\eta_{ab}$ is the Minkowski metric. It can be thought of as a basis of vectors labeled by $a$, which specify a local coordinate frame at each point. Under diffeomorphisms $e^a_\mu$ transforms as a vector, and the metric is invariant under Lorentz transformations acting on the internal index. For instance, in an orthonormal basis, $\eta_{ab} = \diag(-1,1,1,1)$ and the metric is written as
\begin{equation}
    g_\mn = -e^0_\mu e^0_\nu + e^1_\mu e^1_\nu + e^2_\mu e^2_\nu + e^3_\mu e^3_\nu.
\end{equation}
For our purposes it is more convenient to consider a \emph{complex null basis},
\begin{equation}
    e^a_\mu = \left(l_\mu,n_\mu,m_\mu,\bar m_\mu\right),
\end{equation}
where $(l,n,m,\bar m)_\mu$ are null vectors, and $m_\mu$ and its conjugate $\bar m_\mu$ are complex. All inner products $e^a_\mu e^{b\mu}$ vanish with the exception of
\begin{equation}
    l\cdot n = -1,\qquad m\cdot\bar m = 1.
\end{equation}
These properties fix the metric as
\begin{equation}
    g_\mn = -2 l_{(\mu} n_{\nu)}+2 m_{(\mu}\bar{ m}_{\nu)},
\end{equation}
while the internal Minkowski metric in this basis is
\begin{equation}\label{eq:int-mink-null}
    \eta_{ab} = \begin{pmatrix}
    0&-1&0&0\\
    -1&0&0&0\\
    0&0&0&1\\
    0&0&1&0
    \end{pmatrix}.
\end{equation}

For the Schwarzschild and Kerr spacetimes, $l_\mu$ and $n_\mu$ correspond to outgoing and ingoing null vectors, respectively, while $m_\mu$ and $\bar m_\mu$ live on the 2-sphere. This effectively factorizes the full spacetime into a ``$t$--$r$ subspace'' spanned by $(l,n)$ and a ``$\theta$--$\varphi$ subspace'' spanned by $(m,\bar m)$.
In the Kinnersley frame\footnote{This frame is not unique. The Kinnersley tetrad is particularly convenient, and by far the most common, for deriving the Teukolsky equation for Kerr perturbations, which is our purpose for introducing this formalism.} \cite{Kinnersley:1969zza} the tetrad vectors take the form, in Boyer--Lindquist coordinates,
\begin{subequations}\label{eq:kinnersley}
\begin{align}
    l^\mu &= \frac1\Delta\left(r^2+a^2,\Delta,0,a\right),\\
    n^\mu &= \frac1{2\Sigma}\left(r^2+a^2,-\Delta,0,a\right),\\
    m^\mu &= \frac1{\sqrt2\bar\zeta}\left(ia\sin\theta,0,1,\frac i{\sin\theta}\right),
\end{align}
\end{subequations}
where we have introduced the Killing spinor coefficient
\begin{equation}
    \zeta \equiv r-ia\cos\theta.
\end{equation}
Notice that $\Sigma=\zeta\bar\zeta$.

In general we can package the tetrad vectors into \emph{bivectors} or \emph{2-forms}, i.e., antisymmetric rank-2 tensors, by taking exterior products,\footnote{This particular choice is motivated by the fact that the 2-forms $U$, $V$, and $W$ are \emph{self-dual}, i.e.,
$(\star X)_\mn \equiv \frac 12\epsilon_{\mn\ab}X^\ab = i X_\mn$
for $X\in(U,V,W)$. See \ref{app:diff-forms} for a review of differential forms.}
\begin{subequations}
\begin{align}
    U_\mn &= -2n_{[\mu}\bar m_{\nu]},\\
    V_\mn &= 2l_{[\mu}m_{\nu]},\\
    W_\mn &= -2l_{[\mu}n_{\nu]} + 2m_{[\mu}\bar m_{\nu]}.
\end{align}
\end{subequations}
These bivectors and their complex conjugates form a six-dimensional basis for the space of 2-forms \cite{Stephani:2003tm}.
In particular, the Weyl tensor $C_{\mn\ab}$ \eqref{eq:Weyl-def}, having two pairs of antisymmetric indices, can be written entirely in terms of $(U,V,W)$ and their complex conjugates,\footnote{This expression follows from the symmetries of the Weyl tensor and the fact that the only non-vanishing inner products of the six bivectors are $U^\mn V_\mn=2$ and $W^\mn W_\mn = -4$ and their conjugates.}
\begin{align}
    C_{\mn\ab} &= \Psi_0U_\mn U_\ab + \Psi_1(U_\mn W_\ab +W_\mn U_\ab) \nonumber\\&\hphantom{{}=}+ \Psi_2(V_\mn U_\ab+U_\mn V_\ab +W_\mn W_\ab) \nonumber\\&\hphantom{{}=}+ \Psi_3 (V_\mn W_\ab+W_\mn V_\ab) + \Psi_4 V_\mn V_\ab + \mathrm{c.c.},
\end{align}
where the \emph{Weyl scalars} $\Psi_i$ (and their complex conjugates) are components of the Weyl tensor in the complex null basis,
\begin{subequations}\label{eq:Weyl-scalars}
\begin{align}
    \Psi_0 &\equiv C_{\mn\ab}l^\mu m^\nu l^\alpha m^\beta,\\
    \Psi_1 &\equiv C_{\mn\ab} l^\mu n^\nu l^\alpha m^\beta,\\
    \Psi_2 &\equiv C_{\mn\ab} l^\mu m^\nu \bar m^\alpha n^\beta,\\
    \Psi_3 &\equiv C_{\mn\ab} l^\mu n^\nu \bar m^\alpha n^\beta,\\
    \Psi_4 &\equiv C_{\mn\ab} \bar m^\mu n^\nu \bar m^\alpha n^\beta.
\end{align}
\end{subequations}
For the Kerr spacetime only $\Psi_2$ is non-vanishing,
\begin{equation}\label{eq:Psi2}
    \Psi_2 = -\frac{M}{\zeta^3}.
\end{equation}

It is natural to consider the eigenvalue problem
\begin{equation}\label{eq:Petrov-eigenvalue}
    \frac12 C^\mn{}_\ab X^\ab = \lambda X_\mn,
\end{equation}
with $X_\mn$ an ``eigenbivector.'' The symmetries of the Weyl tensor reduce the basis of eigenbivectors from six components to four, so there are at most four linearly independent eigenbivectors, or equivalently, four \emph{principal null directions}. In vector notation, \cref{eq:Petrov-eigenvalue} is equivalent to
\begin{equation}
k_{[\rho}C_{\mu]\nu\alpha[\beta}k_{\sigma]}k^\nu k^\alpha = 0
\end{equation} with $k^\mu$ a principal null direction.

Petrov's algebraic classification of spacetimes is based on the number of independent principal null directions and the characters of any degeneracies. A spacetime is said to be \emph{algebraically special} if any of the four are degenerate, and \emph{algebraically general} if all four are independent. Each case is classified into \emph{Petrov types}.

We will be interested in spacetimes of Petrov \emph{type D}, in which there are two pairs of coinciding principal null directions, corresponding to $l^\mu$ and $n^\mu$. The type D vacuum solutions in four-dimensional general relativity have been fully catalogued \cite{Kinnersley:1969zza,Plebanski:1976gy}. These contain the Schwarzschild and Kerr metrics, as well as generalizations thereof including a cosmological constant, NUT charge, and acceleration. The non-Kerr members of the family of type D solutions violate asymptotic flatness and/or the positive mass theorem \cite{Andersson:2016hmv}. In all spacetimes of type D, the only non-vanishing Weyl scalar is $\Psi_2$.

\subsubsection{Geroch--Held--Penrose formalism}
\label{sec:GHP}

The closely-related Newman--Penrose (NP) and Geroch--Held--Penrose (GHP) formalisms decompose tensorial quantities into their components in the $(l,n,m,\bar m)$ basis, just as we encoded the Weyl tensor $C_{\mn\ab}$ into the five complex spacetime scalars $\Psi_i$ \eqref{eq:Weyl-scalars}. Similarly the connection is replaced by twelve complex ``spin coefficients'' and covariant derivatives by the four directional derivatives $\{l^\mu \nabla_\mu,n^\mu \nabla_\mu,m^\mu \nabla_\mu,\bar m^\mu \nabla_\mu\}$. %

The difference between the NP and GHP formalisms is that the latter goes a step further, classifying fields by their behavior under the residual Lorentz transformations which preserve the tetrad normalization and the directions of the principal null vectors $(l,n)$. These conditions reduce the six-parameter Lorentz group to two parameters, corresponding to the ability to rescale the tetrad vectors so that $n$ scales inversely to $l$, and similarly for $ m$ and $\bar m$,
\begin{align}
    &&&&l^\mu &\to \alpha  l^\mu, & m^\mu &\to e^{i\beta}  m^\mu &&&&\nonumber\\
    &&&&n^\mu &\to \frac1\alpha  n^\mu, & \bar m^\mu &\to e^{-i\beta} {\bar m}^\mu,&&&&
\end{align}
for some real scalar functions $\alpha$ and $\beta$, or equivalently a single complex function,
\begin{equation}
    \lambda^2 \equiv \alpha e^{i\beta}.
\end{equation}
A tensorial object $\Phi$ that transforms under residual Lorentz transformations as
\begin{equation}
    \Phi \to \lambda^p{\bar\lambda}^q\Phi
\end{equation}
is said to be \emph{weighted} of type $(p,q)$; the weight of a field is often denoted $\Phi:(p,q)$.

The GHP formalism works exclusively with tensors of definite weight. It is common to define the \emph{boost weight} $b$ and \emph{spin weight} $s$ by
\begin{equation}
    p = b+s,\quad q=b-s.
\end{equation}
Tensor fields like $g_\mn$ and $R_{\mn\ab}$ are \textit{a priori} unweighted, while the null vectors each have definite weights,
\begin{equation}
    l^\mu: (1,1),\quad n^\mu:(-1,-1),\quad m^\mu:(1,-1),\quad \bar m^\mu:(-1,1).
\end{equation}
Using these it is straightforward to compute the weights of objects like the Weyl scalars, simply by counting the factors of $l^\mu$, $n^\mu$, $m^\mu$, and $\bar m^\mu$ and adding their weights together. In the GHP calculus, all terms in an equation must have the same weight, in much the way that more familiar equations require all terms to have the same units and tensorial structure.

It is common to define a \emph{spin-$s$ field} as a quantity of weight $(2s,0)$, so that the boost and spin weights are both equal to the spin. The extreme-spin Weyl scalars $\Psi_0$ and $\Psi_4$, and multiples thereof by unweighted quantities, meet this definition for $s=2$ and $s=-2$, respectively.

There are three additional discrete transformations which leave the GHP setup invariant. One is the familiar \emph{complex conjugation}, which sends $m\leftrightarrow\bar m$. There is the \emph{prime} operation (${}^\prime$), which interchanges $l\leftrightarrow n$ and $m\leftrightarrow\bar m$. Finally the \emph{star} operation, which is less frequently used than the other two, sends $(l,n,m,\bar m)\to(m,-\bar m,-l,n)$.\footnote{For the purposes of this review, the star operation will not be needed.} Acting on an object of weight $(p,q)$, these three operations yield objects of types $(q,p)$, $(-p,-q)$, and $(p,-q)$, respectively.
The GHP formalism is especially compact because any equation implies its conjugated, primed, and starred versions (as well as combinations thereof), significantly reducing the number of independent equations for a given system.

The ordinary covariant derivative does not send a weighted object to one with well-defined weight, similar to how the partial derivative of a tensor is not itself a tensor. The same remedy is available, namely, using non-GHP-covariant components of the connection to form derivative operators with definite GHP type.

The Levi--Civita connection has twenty-four independent components, which in the GHP formalism are encoded into twelve complex scalars, known as spin coefficients.\footnote{This name comes from the spinorial approach to the GHP formalism \cite{Penrose:1985bww,Penrose:1986ca}.} These come in six pairs of quantities related by the GHP prime operation. Of these six, four have a definite GHP weight,
\begin{equation}
    \rho \equiv \bar m^\mu m^\nu \nabla_\mu l_\nu, \quad \tau \equiv n^\mu m^\nu \nabla_\mu l_\nu,\quad\kappa \equiv l^\mu m^\nu \nabla_\mu l_\nu, \quad \sigma \equiv m^\mu m^\nu \nabla_\mu l_\nu,
\end{equation}
and two are not properly GHP weighted,
\begin{equation}
    \beta \equiv \frac12\left(m^\mu n^\nu \nabla_\mu l_\nu - m^\mu \bar m^\nu \nabla_\mu m_\nu\right),\quad
    \epsilon \equiv \frac12\left(l^\mu n^\nu \nabla_\mu l_\nu - l^\mu \bar m^\nu \nabla_\mu m_\nu\right).
\end{equation}
The Goldberg--Sachs theorem implies that $\kappa = \sigma = 0$ \cite{Goldberg-Sachs,Stephani:2003tm}. This leaves us with eight independent, \textit{a priori} non-vanishing spin coefficients: $(\rho,\tau,\beta,\epsilon)$ and their primed variants.
For the Kerr spacetime in the Kinnersley frame they are given by
\begin{align}\label{eq:Kerr-spin-coeffs-kinnersley}
    &&&&\rho &= -\frac{2\Sigma}{\Delta}\rho' = -\frac{1}{\zeta},& 
    \tau &= \frac{\zeta^2}{\Sigma}\tau' = -\frac{ia\sin\theta}{\sqrt{2\Sigma}},&&&&\nonumber\\
    &&&&\beta &= \bar\beta'-\bar\tau' = \frac{\cot\theta}{2\sqrt{2}\,\zeta},& 
    \epsilon &= 0,\quad \epsilon' = \rho' - \frac{r-M}{2\Sigma}.&&&&
\end{align}
Note that $\tau$ and $\tau'$ are non-vanishing only for spinning objects.

We are now in a position to construct GHP-covariant derivative operators using the improperly-weighted scalars $(\beta,\epsilon,\beta',\epsilon')$. To begin we define the connection 1-form \cite{Ehlers:1974,Aksteiner:2010rh}
\begin{align}\label{eq:omega-GHP}
    \omega_\mu &\equiv \frac12\left(n^\nu\nabla_\mu l_\nu + m^\nu\nabla_\mu \bar m_\nu\right)\nonumber\\
    &= -\epsilon'l_\mu +\epsilon n_\mu +\beta' m_\mu -\beta \bar m_\mu.
\end{align}
A properly-weighted covariant derivative acting on a quantity of type $(p,q)$ is then
\begin{equation}\label{eq:Theta-GHP}
    \Theta_\mu \equiv \nabla_\mu - p\omega_\mu -q \bar\omega_\mu.
\end{equation}
These derivatives are best known in terms of their individual components, which for historical reasons
we are compelled to write
using Icelandic runes,
\begin{equation}
    \tho = l\cdot \Theta,\quad\tho' = n\cdot\Theta,\quad\eth = m\cdot\Theta,\quad\eth'=\bar m\cdot\Theta.
\end{equation}
Roughly speaking, derivatives with respect to $(t,r)$ are encoded in $(\tho,\tho')$, while $(\eth,\eth')$ are operators on the 2-sphere. The former raise and lower the boost weight, while the latter raise and lower the spin weight.

\newpage
\section{Black hole perturbation theory}
\label{sec:BHPT}

Love numbers characterize small deformations away from a perfectly spherically- or axisymmetric object. Having established these idealized solutions in general relativity, i.e., the Schwarzschild and Kerr geometries, we now discuss their linear perturbations. Black hole perturbation theory will serve as the UV theory when calculating Love numbers. See \rcite{Pound:2021qin} for a comprehensive review of black hole perturbation theory.

We are interested in the dynamics of small fluctuations about a black hole spacetime,\footnote{We include the factor of $2/\Mp$ to canonically normalize the metric fluctuation.}
\begin{equation}
g_\mn = \bar g_\mn+\frac2\Mp h_\mn,
\end{equation}
with $\bar g_\mn$ the background Schwarzschild \eqref{eq:sch} or Kerr \eqref{eq:Kerr} metric. To linear order in $h_\mn$ there is a simple, closed-form expression for the Einstein equations,
\begin{equation}
G_\mn[\bar g + 2\Mp^{-1}h] = \bar G_\mn + \frac2\Mp G[h]_\mn + \mathcal{O}\left(\frac{h^2}{\Mp^2}\right),
\end{equation}
where
\begin{equation}
G[h]_\mn = \nabla_\alpha \nabla_{(\mu}h_{\nu)}^\alpha - \frac12\Box h_\mn - \frac12\nabla_\mu\nabla_\nu h - \frac12\left(\nabla_\mu\nabla_\nu h^\mn - \Box h\right)g_\mn
\end{equation}
is the linearized Einstein tensor.\footnote{We encountered this expression earlier in the form of the Lichnerowicz operator \eqref{eq:lich}. They are the same up to an overall factor, $(\mathcal E h)_\mn = -2G[h]_\mn$.} The equations of motion governing the dynamics of $h_\mn$ can be obtained by computing the individual components of $G[h]_\mn$, or equivalently by expanding the Einstein--Hilbert action to quadratic order,\footnote{We need to go to quadratic order at the level of the action in order to obtain linear equations of motion. The linear part of the action, $\delta_1S$, vanishes when the background Einstein equations $\bar G_\mn=0$ are imposed.}
\begin{align}
S &= \frac{\Mp^2}{2}\int\dd^4x\sqrt{-g}R[g] \nonumber\\
&= \bar S+\delta_1S+\delta_2S+\mathcal{O}(h^3). \label{eq:EH-NL}
\end{align}
To compute $\delta_2S$ without going through the considerable algebra required to expand $\sdg R$ to quadratic order in $h$, we may take a generic metric variation $g\to \bar g+\delta g$ and Taylor expand,
\begin{equation}
S[\bar g+\delta g] = S[\bar g] + \delta S + \frac12\delta^2S+\cdots,
\end{equation}
so that (dropping overbars from background quantities)
\begin{align}
\delta_2 S = \frac12\delta^2S &= \frac{\Mp^2}{4}\delta^2\int\dd^4x\sdg R \nonumber\\
&= \frac{\Mp^2}{4}\delta\int\dd^4x \sdg G_\mn \delta g^\mn \nonumber\\
&= \frac{\Mp^2}{4}\int\dd^4x\sdg \delta G_\mn \delta g^\mn \nonumber\\
&= -\int\dd^4x \sdg h^\mn G[h]_\mn \label{eq:quad-EH}
\end{align}
Here we have used the basic fact that the first variation of the Einstein--Hilbert action yields the Einstein equations,
\begin{equation}
\delta \int\dd^4x\sdg R = \int\dd^4x\sdg G_\mn \delta g^\mn,
\end{equation}
as well as the background Einstein equation $G_\mn=0$ itself. Note that the quadratic action is essentially just a clever way of encoding the different components of the linearized Einstein equations into a single spacetime scalar.\footnote{In a linear theory one can always write down a valid Lagrangian simply by multiplying the equations of motion by the fields.} Finally, we integrate by parts to cast the action in first-order form and obtain the expression \eqref{eq:EHaction2}, which we rewrite here (dropping the overbar for simplicity):
\begin{equation}\label{eq:S-FP}
\delta_2S = \int\dd^4x \sdg\left(-\frac12\nabla_\alpha h_\mn\nabla^\alpha h^\mn + \nabla_\alpha h_\mn \nabla^\nu h^{\mu\alpha} - \nabla_\mu h \nabla_\nu h^\mn  + \frac12(\partial h)^2 \right).
\end{equation}
This is the \emph{Fierz-Pauli} action for a spin-2 field propagating on an arbitrary Ricci-flat background.

\subsection{Schwarzschild perturbations}
\label{sec:BHPT-Sch}

We begin by considering perturbations around a non-rotating black hole. The background metric $g_\mn$ is the Schwarzschild metric \eqref{eq:sch}. Let us decompose $h_\mn$ in a manner that takes maximal advantage of the symmetries of the background. The most important for our purposes is spherical symmetry, which ensures that $h_\mn$ can be decomposed in spherical harmonics. As usual for a linear equation, each multipole decouples from the rest and can be treated on its own. We may further classify modes of $h_\mn$ by their behavior under parity transformations $t\to-t$, decomposing a perturbation into even- and odd-parity modes.\footnote{Even modes are also referred to as ``polar'' and odd modes as ``axial.''} Linearity ensures that even and odd modes do not couple to each other. The quadratic action can therefore be written as
\begin{equation}\label{eq:S-even-odd}
\delta_2S = \displaystyle\sum_{\ell=2}^\infty\displaystyle\sum_{m=-\ell}^\ell \left(S^{\ell m}_\mathrm{even}+S_\mathrm{odd}^{\ell m}\right).
\end{equation}
For Schwarzschild perturbations, we will frequently drop the $\ell m$ labels when there is no chance of ambiguity. Moreover $m$ drops out of all dynamical equations by spherical symmetry, and we can set $m=0$ without loss of generality.\footnote{We saw this when solving the scalar wave equation on Schwarzschild \eqref{eq:KG-sch}, as the radial potential \eqref{eq:V-s0} depends on $\ell$ but not $m$.} Physically this occurs because any $Y_{\ell m}$ can be obtained by suitably rotating $Y_{\ell0}\equiv Y_\ell$. Note that we are not ignoring modes with non-zero $m$, which typically are the dominant modes in gravitational waveforms; rather, on Schwarzschild, we can solve for their behavior using the $m=0$ equations of motion.

The metric perturbation decomposes into
\begin{equation}\label{eq:h-pm}
h_\mn = \sum_{\ell m}\left(h_\mn^\mathrm{even} + h_\mn^\mathrm{odd}\right),
\end{equation}
with\footnote{We follow the definitions of \rcite{Regge:1957td}, although our notation more closely follows that of \rcite{Franciolini:2018uyq}.}
\begin{subequations}\label{eq:h-even-odd}
\begin{align}
h_\mn^\mathrm{even} &= \begin{pmatrix}
f(r)H_0(t,r) & H_1(t,r) & \mathcal H_0(t,r)\partial_\theta & \mathcal H_0(t,r)\partial_\varphi \\
\cdot & \frac1{f(r)}H_2(t,r) & \mathcal H_1(t,r)\partial_\theta & \mathcal H_1(t,r)\partial_\varphi \\
\cdot & \cdot & r^2\left(\mathcal K(t,r) + \mathcal G(t,r) \partial_\theta^2\right) & r^2\mathcal G(t,r)\nabla_\theta\nabla_\varphi\\
\cdot & \cdot & \cdot & r^2\left(\sin^2\theta\mathcal K(t,r) + \mathcal G(t,r) \nabla_\varphi^2\right)
\end{pmatrix}Y_{\ell m}, \\
h_\mn^\mathrm{odd} &= r^2\begin{pmatrix}
0 & 0 & -h_0(t,r)\csc\theta\partial_\varphi & h_0(t,r)\sin\theta\partial_\theta \label{eq:h-odd} \\
\cdot & 0 & -h_1(t,r)\csc\theta\partial_\varphi & h_1(t,r)\sin\theta\partial_\theta \\
\cdot & \cdot & -h_2(t,r)\csc\theta\nabla_\theta\nabla_\varphi & \frac12h_2(t,r) \left(\sin\theta\partial_\theta^2-\csc\theta\nabla_\varphi^2\right) \\
\cdot & \cdot & \cdot & h_2(t,r)\sin\theta\nabla_\theta\nabla_\varphi
\end{pmatrix}Y_{\ell m}.
\end{align}
\end{subequations}

Note that, for later convenience, we have written the odd perturbation with an overall factor of $r^2$ which is not universal in the literature.
The relevant components of the second covariant derivative of $Y_{\ell m}$ on the 2-sphere are
\begin{equation}
\nabla_\theta\nabla_\varphi=\partial_\theta\partial_\varphi-\cot\theta\partial_\varphi,\qquad \nabla_\varphi^2 = \partial_\varphi^2+\sin\theta\cos\theta\partial_\theta.
\end{equation}
The angular dependence of the various components of $h_\mn$ is fixed by spherical symmetry and parity.
Using the parity property of the spherical harmonics,
\begin{equation}
Y_{\ell m}(\pi-\theta,\varphi+\pi) = (-1)^\ell Y_{\ell m}(\theta,\varphi)
\end{equation}
we see that under a parity transformation even modes pick up a factor of $(-1)^\ell$ while odd modes pick up a factor of $(-1)^{\ell+1}$. This demonstrates why it is impossible for the even and odd sectors to communicate with each other at linear order.

At each $(\ell,m)$, $h_\mn$ contains ten fields: $\{H_0,H_1,H_2,\mathcal K,\mathcal G,\mathcal H_0,\mathcal H_1\}$ in the even sector and $\{h_0,h_1,h_2\}$ in the odd sector. The freedom to change coordinates at the linear level, known as \emph{gauge invariance}, allows us to fix four of the ten (three even and one odd). A particularly common choice is \emph{Regge--Wheeler gauge}~\cite{Regge:1957td},
\begin{equation}
\mathcal{H}_0 = \mathcal{H}_1=\mathcal G = h_2 = 0,
\label{eq:RWgaugepart2}
\end{equation}
which we will employ herein. For the gauge transformations and gauge-invariant combinations see \rcite{Martel:2005ir}. Regge--Wheeler gauge is also particularly useful for our purposes as it is one of the gauge choices which can be safely imposed at the level of the action \cite{Motohashi:2016prk,Hui:2020xxx,Solomon:2023ltn}.

Fixing a gauge leaves us with equations of motion for six fields, four in the even sector and two in the odd. The metric perturbation however only contains two genuinely independent dynamical degrees of freedom, corresponding to the two polarizations of the graviton. Indeed, the even and odd modes correspond at infinity to the standard $+$ and $\times$ polarizations, respectively.

To compute the quadratic action, we plug the Regge--Wheeler-gauge metric perturbation \eqref{eq:h-pm} into the action \eqref{eq:S-FP}.\footnote{One may also use the second-order action \eqref{eq:quad-EH}. The relevant components of $G[h]_\mn$ are listed in \rcite{Sarbach:2001qq,Martel:2005ir,Chaverra:2012bh,Solomon:2023ltn}.} Decomposing the action as in \cref{eq:S-even-odd} and using the orthonormality of the spherical harmonics to integrate over the 2-sphere, we find even and odd actions of the form \cite{Hui:2020xxx,Solomon:2023ltn}
\begin{equation}
S_\mathrm{even/odd} = \int\dd t \, \dd r \, \mathcal L_\mathrm{even/odd}
\end{equation}
with Lagrangians
\begin{subequations}\label{eq:L-even-odd}
\begin{align}
\mathcal{L}_\mathrm{even} &= \Delta \mathcal{K}'(\mathcal{K}'-2H_0')-\frac{r^4}\Delta\dot{\mathcal K}(\dot{\mathcal K}+2\dot{H}_2) +4r \dot{H}_1(r\mathcal{K}'-H_2) + \Delta' \mathcal K H_2' + \frac{2\Delta}rH_0'H_2\nonumber\\
&\hphantom{{}=}+ (2r-3\rs)\left(\frac{2r^2}\Delta \dot{H}_1 - H_0'\right) \mathcal K +H_2^2+\ell(\ell+1)\left(H_1^2-H_0H_2-H_0\mathcal K+H_2\mathcal K\right), \label{eq:action-even}\\
\mathcal{L}_\mathrm{odd} &= \ell(\ell+1)\left[r^4(\dot{h}_1 - h_0')^2 + (\ell+2)(\ell-1)\left(\frac{r^4}\Delta h_0^2 - \Delta h_1^2\right)\right].\label{eq:action-odd}
\end{align}
\end{subequations}
Overdots and primes denote $\partial_t$ and $\partial_r$, respectively.
Note that $(\ell+2)(\ell-1)=\ell(\ell+1)-2$.

\subsubsection{Even action}

The even sector contains one independent degree of freedom, but as we have written things it is described by the four fields $\{H_0,H_1,H_2,\mathcal K\}$. Our first task is therefore to eliminate three redundant variables. Two of these are immediately identifiable: up to boundary terms we can write the action without any derivatives on $H_0$ and $H_1$, making it clear they are non-dynamical. $H_0$ appears linearly so is a \emph{Lagrange multiplier}, whose equation of motion enforces a constraint on the fields $H_2$ and $\mathcal K$,
\begin{equation}\label{eq:H0-const}
\partial_r\left[2\Delta\left( \mathcal K' - \frac {H_2} r\right)+(2r-3\rs)\mathcal K\right] - \ell(\ell+1)(H_2+\mathcal K) = 0,
\end{equation}
while $H_1$ appears quadratically, i.e., is \emph{auxiliary}, so it can be solved for algebraically,
\begin{equation}\label{eq:H1-sol}
\ell(\ell+1)H_1 = 2r(r\dot{\mathcal K}'-\dot H_2)+\frac{2r-3\rs}f\dot{\mathcal K}.
\end{equation}
Note that in the static limit ($\partial_t=0$), which is of particular relevance when studying tidal responses, $H_1$ vanishes.

We are now in position to \emph{integrate out} the non-dynamical fields $H_0$ and $H_1$.
To integrate out $H_1$, we need only solve \cref{eq:H1-sol} for $H_1 = H_1(H_2,\mathcal K,\partial_t,\partial_r)$ and plug this directly into $\mathcal{L}_\mathrm{even}$.\footnote{After plugging \cref{eq:H1-sol} into \cref{eq:action-even}, one finds terms with more than two derivatives, namely $\dot{\mathcal K}'^2$ and $\dot{\mathcal K}'\dot H_2$. Such higher-derivative terms are often the hallmark of a fatal ``ghost'' instability \cite{Woodard:2015zca}. In this case they are simply an artifact of our gauge choice; they are not present, for instance, in a gauge where $\mathcal K=0$ and $\mathcal H_1\neq0$ \cite{Hui:2020xxx,Solomon:2023ltn}. In Regge--Wheeler gauge we will compensate for these spurious higher-derivative operators by including derivatives in the field redefinition \eqref{eq:H2-redef}.} Some care must be taken with the constraint \eqref{eq:H0-const}, as we cannot solve it for either $H_2$ or $\mathcal K$. To deal with this we employ a field redefinition,
\begin{equation}\label{eq:H2-redef}
H_2 \equiv \psi + \left(r\partial_r-\frac{\Lambda}{2f}\right)\mathcal K,
\end{equation}
where we have defined the function \cite{Martel:2005ir}
\begin{align}
\Lambda &\equiv \ell(\ell+1)-3f+1\nonumber\\
&= (\ell+2)(\ell-1)+\frac{3\rs}{r}.
\label{eq:LambdaellRW}
\end{align}
Now the constraint \eqref{eq:H0-const} no longer contains any derivatives of $\mathcal K$,
\begin{equation}
-2\partial_r\left(\frac{\Delta\psi}r\right) + \ell(\ell+1)\left(\frac{\Lambda}{2f}\mathcal K - \psi\right) = 0,
\end{equation}
and we can solve for $\mathcal K = \mathcal K(\psi,\psi')$. Plugging this into $\mathcal L_\mathrm{even}$, we find an action for $\psi$ alone,
\begin{align}
\mathcal{L}_\mathrm{even} &= \frac{4(\ell+2)(\ell-1)}{\ell(\ell+1)}\frac\Delta{\Lambda^2}\left[\dot\psi^2 - f^2\psi'^2 - \left(\frac{\ell(\ell+1)\left[(\ell+2)(\ell-1)r-\ell(\ell+1)\rs\right]-2\rs}{r^3\Lambda} +\frac\rs{r^3} \right)\psi^2\right].
\end{align}

Having reduced the action from its dependence on the four fields $\{H_0,H_1,H_2,\mathcal K\}$ to a single one $\{\psi\}$, all that remains is to \emph{canonically normalize} $\psi$, that is, to write the kinetic ($\dot\psi^2$) and gradient ($\psi'^2$) terms of the action in canonical form. There are two levers available to us for this task: we can perform field redefinitions $\psi \to g(r)\Psi_+$ for some function $g(r)$, and we can transform coordinates. First let us observe that the appropriate coordinate change is to the radial ``tortoise'' coordinate $\dd r=f(r)\dd \rst$ (cf. \cref{eq:tortoise}), as this ensures that the kinetic and gradient terms have the same overall factor,
\begin{equation}
\dot\psi^2 - f^2\psi'^2 = \dot\psi^2-(\partial_{\rst}\psi)^2.
\end{equation}
It is important to account for this coordinate transformation in the integration measure,
\begin{equation}
S_\mathrm{even} = \int \dd t \,\dd \rst f \mathcal{L}_\mathrm{even}.
\end{equation}
To determine the required field redefinition we can inspect the kinetic term in $f\mathcal{L}_\mathrm{even}$,
\begin{equation}
\frac{4(\ell+2)(\ell-1)}{\ell(\ell+1)}\frac{r^2f^2}{\Lambda^2}\dot\psi^2 \equiv \frac12\dot\Psi_+^2,
\end{equation}
from which we read off
\begin{equation}
\psi = \sqrt{\frac{\ell(\ell+1)}{2(\ell+2)(\ell-1)}}\frac{\Lambda}{2rf}\Psi_+.
\end{equation}
Finally we obtain the \emph{Zerilli action} for the even sector \cite{Hui:2020xxx},
\begin{equation}\label{eq:S-Zerilli}
\boxed{S_\mathrm{even} = \int \dd t \, \dd \rst \left(-\frac12(\partial_t\Psi_+)^2+\frac12(\partial_{\rst}\Psi_+)^2 - \frac12 V_+\Psi_+^2\right),}
\end{equation}
where the \emph{Zerilli potential} is
\begin{equation}
V_+(r) = \frac{f}{3r^2}\left(\Lambda + \frac{2(\ell+2)^2(\ell-1)^2\left(1+\ell(\ell+1)\right)}{\Lambda^2}\right).
\label{eq:VZ}
\end{equation}
The canonically-normalized field $\Psi_+$ (commonly referred to as a ``master variable'') is known as the \emph{Zerilli variable} \cite{Zerilli:1970se}.\footnote{The standard definition of the Zerilli--Moncrief function $\Psi_\mathrm{even}$ in \rcite{Martel:2005ir} is related to the canonically-normalized field $\Psi_+$ by an $\ell$-dependent constant factor, \[ \Psi_\mathrm{even} = \frac1{\sqrt{2(\ell-1)\ell(\ell+1)(\ell+2)}} \Psi_+.\]} It obeys the Schr\"odinger-like \emph{Zerilli equation},
\begin{equation}\label{eq:Z}
    \left(\partial_t^2-\partial_{\rst}^2\right)\Psi_+ = V_+(r) \Psi_+.
\end{equation}

In passing let us note two simple on-shell relations which are obscured in the quadratic action approach to finding the Zerilli variable and Zerilli potential,
\begin{subequations}
\begin{align}
    H_0 &= H_2,\\
    \partial_a K &= \nabla^b h_{ab}.
\end{align}    
\end{subequations}
The second of these is expressed in terms of the covariant derivative compatible with the metric $\dd s^2_2=-f\dd t^2+f^{-1}\dd r^2$ on the $(t,r)$ subspace.

\subsubsection{Odd action}

The odd sector is simultaneously simpler and more involved than its even counterpart: while we only need to eliminate one non-dynamical field rather than three, neither $h_0$ nor $h_1$ appears without derivatives in \cref{eq:action-odd} and so it is not obvious which degree of freedom \emph{is} non-dynamical. We see that $h_0$ has no time derivatives while $h_1$ has no spatial derivatives. To clarify the structure of the action we package the individual fields $h_0$ and $h_1$ into a 2-vector in the $(t,r)$ subspace,\footnote{It is possible, though not quite as enlightening, to perform the even-sector calculation covariantly in the $(t,r)$ subspace as well \cite{Solomon:2023ltn}.}
\begin{equation}
h_a \dd x^a \equiv h_0\dd t + h_1\dd r,
\end{equation}
so that, defining the usual field strength tensor $F_{ab}=\partial_ah_b-\partial_bh_a$, the odd Lagrangian \eqref{eq:action-odd} takes the compact form \cite{Solomon:2023ltn}
\begin{equation}
\mL_\mathrm{odd} = -\ell(\ell+1)\left(\frac12r^4F_{ab}^2 + (\ell+2)(\ell-1)r^2h_a^2\right).
\end{equation}
Index contractions are performed using the 2-metric $\dd s^2_2=-f\dd t^2+f^{-1}\dd r^2$. We may remove the overall $\ell$-dependent factor by rescaling,
\begin{equation}
\mathfrak{h}_a \equiv \sqrt{2\ell(\ell+1)}h_a,\qquad \mathfrak{F}_{ab}=2\partial_{[a}\mathfrak{h}_{b]}=\sqrt{2\ell(\ell+1)}F_{ab},
\end{equation}
so that
\begin{equation}
\mL_\mathrm{odd} = -\frac14r^4\mathfrak{F}_{ab}^2 - \frac{1}{2}(\ell+2)(\ell-1)r^2\mathfrak{h}_a^2. \label{eq:odd-action-22}
\end{equation}

We see that the odd sector can be concisely described by a two-dimensional vector with radially-dependent kinetic and non-derivative couplings.
To isolate the single independent degree of freedom it is convenient to integrate \emph{in} an auxiliary field $\lambda(t,r)$ via
\begin{align}
\mathcal{L}_\mathrm{odd} &\to \mathcal{L}_\mathrm{odd} +\frac14\left(r^2 \mathfrak{F}_{ab}+\lambda \epsilon_{ab}\right)^2\nonumber\\
&= \frac12\left(r^2\lambda\epsilon^{ab} \mathfrak{F}_{ab} - \lambda^2 - (\ell+2)(\ell-1) r^2\mathfrak{h}^2\right).
\end{align}
Here $\epsilon_{ab}$ is the Levi--Civita tensor on the $(t,r)$ subspace, $\epsilon_{tr}=1=-\epsilon^{tr}$.

The fact that $\mL_\mathrm{odd}$ describes the same dynamics before and after introducing $\lambda$ is guaranteed by our use of a perfect square, since its contributions to the equations of motion will be proportional to $r^2\mathfrak{F}_{ab}+\lambda \epsilon_{ab}$, which vanishes on shell by the $\lambda$ equation of motion. Note that $\mathfrak{F}_{ab}\propto \epsilon_{ab}$, a direct consequence of its antisymmetry and the fact that we are working in two dimensions (recall that in $D$ dimensions any fully antisymmetric, rank-$D$ tensor is a multiple of the Levi--Civita tensor). 

After contracting the $\lambda$ equation of motion with $\epsilon^{ab}$ we find $\lambda = (1/2)r^2\epsilon^{ab} \mathfrak{F}_{ab}$. If we plug this solution for $\lambda$ back into the action, we recover our starting point \eqref{eq:odd-action-22}. This is a useful check that the dynamics are in fact unchanged when we integrate $\lambda$ in.

The upshot of introducing $\lambda$ is that $\mathfrak h_a$ is now auxiliary and can itself be integrated out by algebraically solving its equation of motion,
\begin{equation}
(\ell+2)(\ell+1)r^2\mathfrak h_a = \epsilon_{ab} \partial^b (r^2\lambda). \label{eq:ha-sol}
\end{equation}
Substituting this into the action, and performing a field redefinition to canonically normalize the kinetic term,
\begin{equation}
\lambda = \frac{\sqrt{(\ell+2)(\ell-1)}}{r}\Psi_-,
\end{equation}
the action becomes
\begin{equation}
\mathcal{L}_\mathrm{odd} = -\frac12(\partial\Psi_-)^2 - \frac12\frac{V_-(r)}{f(r)}\Psi_-^2,
\end{equation}
where the master variable $\Psi_-$ is known as the \emph{Regge--Wheeler variable}\footnote{Strictly speaking, $\Psi_-$ is the \emph{Cunningham--Price--Moncrief variable} \cite{Cunningham:1978zfa}, which is a time integral of the variable introduced by Regge and Wheeler. Moreover our canonically-normalized field $\Psi_-$ also contains an $\ell$-dependent factor not present in standard definitions of these variables; comparing to the definition in \rcite{Martel:2005ir},
\[ \Psi_\mathrm{CPM} = \frac{2}{\sqrt{(\ell+2)(\ell-1)}}\Psi_-.\]} and $V_-(r)$ the \emph{Regge--Wheeler potential} \cite{Regge:1957td},\footnote{Note the resemblance between the Regge--Wheeler potential \eqref{eq:VRW} and the scalar field potential \eqref{eq:V-s0}. In fact, both of these cases, as well as the potential experienced by an electromagnetic field, can be written as a single potential for a spin-$s$ field,
\[V_s = \frac f{r^2}\left(\ell(\ell+1)+(1-s^2)\frac\rs r\right).\]
Here $s=0$ for a scalar, $s=\pm1$ for electromagnetism, and $s=\pm2$ for gravity.}
\begin{equation}\label{eq:VRW}
V_-(r)=f\left(\frac{\ell(\ell+1)}{r^2}-\frac{3\rs}{r}\right).
\end{equation}
Putting everything together we obtain the odd-sector \emph{Regge--Wheeler action} \cite{Hui:2020xxx,Solomon:2023ltn},
\begin{equation}\label{eq:S-RW}
\boxed{S_\mathrm{odd} = \int \dd t \,\dd \rst \left(-\frac12(\partial_t\Psi_-)^2+\frac12(\partial_{\rst}\Psi_-)^2 - \frac12 V_-\Psi_+^2\right).}
\end{equation}
Analogously to the even sector, the equation of motion is a Schr\"odinger-like equation called the \emph{Regge--Wheeler equation},
\begin{equation}\label{eq:RW}
    \left(\partial_t^2-\partial_{\rst}^2\right)\Psi_- = V_-'(r) \Psi_+.
\end{equation}

\subsubsection{Static actions}
\label{sec:BHPT-static}

The static Love numbers, i.e., the leading-order conservative tidal response coefficients, are computed for time-independent perturbations, $\partial_th_\mn=0$. The Regge--Wheeler and Zerilli master variables turn out to be slight overkill in this r\'egime and in the latter case even obscure the physics. It is therefore illuminating to analyze the Einstein--Hilbert action in this limit.

The even and odd actions \eqref{eq:L-even-odd} in the static limit are
\begin{subequations}
\begin{align}
\mathcal{L}_\mathrm{even}^\mathrm{static} &= \Delta \mathcal{K}'(\mathcal{K}'-2H_0') + \Delta' \mathcal K H_2' + \frac{2\Delta}rH_0'H_2 +r\left(\Lambda-\ell(\ell+1)\right)H_0' \mathcal K \nonumber\\
&\hphantom{{}=}+H_2^2+\ell(\ell+1)\left(H_1^2-H_0H_2-H_0\mathcal K+H_2\mathcal K\right), \label{eq:action-even-static}\\
\mathcal{L}_\mathrm{odd}^\mathrm{static} &= \ell(\ell+1)\left[r^4h_0'^2 + (\ell+2)(\ell-1)\left(\frac{r^4}\Delta h_0^2 - \Delta h_1^2\right)\right].\label{eq:action-odd-static}
\end{align}
\end{subequations}
The fields $H_1$ and $h_1$ are auxiliary and decouple from the rest of the perturbations, so we set $H_1=h_1=0$.

The static odd sector only contains one variable, $h_0$, whose Lagrangian is quite simple,
\begin{equation}
\boxed{\mathcal{L}_\mathrm{odd}^\mathrm{static} = \ell(\ell+1)r^4\left(h_0'^2 + \frac{(\ell+2)(\ell-1)}\Delta h_0^2\right).}
\end{equation}
The equation of motion is
\begin{equation}
    \label{eq:h0-static}
    h_0''+\frac4rh_0'-\frac{(\ell+2)(\ell-1)}\Delta h_0=0.
\end{equation}
The static Lagrangian can be canonically normalized simply by rescaling,
\begin{equation}\label{eq:h0-can-norm}
h_0 \equiv \frac{\tilde h_0}{\sqrt{2\ell(\ell+1)}r^2}.
\end{equation}
We see that canonical normalization in the static limit gives us a rather different master variable than in the full time-dependent dynamics. The two master variables --- $\tilde h_0$ and $\Psi_-$ --- are related on shell by
transformations of the form
\begin{equation}\label{eq:Darboux-static}
\Psi_- = \left[A(r)\partial_r+B(r)\right]\tilde h_0,\qquad\tilde h_0 = [C(r)\partial_r+D(r)]\Psi_-.
\end{equation}

The even sector takes some more work. We start by making the convenient redefinitions \cite{Combaluzier-Szteinsznaider:2024sgb}
\begin{equation}
\delta = H_2 - H_0,\quad \kappa = \mathcal K - H_0.
\end{equation}
The action is
\begin{align}
\mathcal{L}_\mathrm{even}^\mathrm{static} &= \Delta\left(\kappa'^2-H_0'^2\right) -2\rs H_0\kappa' +\ell(\ell+1)H_0^2 +\delta^2-\delta\left[\Delta'\kappa'+\rs H_0'-(\ell+2)(\ell-1)\kappa\right].
\end{align}
We can solve the $\delta$ equation of motion algebraically and integrate it out,
\begin{align}
\mathcal{L}_\mathrm{even}^\mathrm{static} &= -\frac14\left(\Delta'^2H_0'^2+\rs^2\kappa'^2\right)+ \frac\rs 2\left[\left(\ell(\ell+1)+2\right)\kappa -\Delta'\kappa' \right] H_0' \nonumber\\
&\hphantom{{}=}-\frac{(\ell-1)\ell(\ell+1)(\ell+2)}4\kappa^2 -\ell(\ell+1)H_0^2.
\end{align}
Indeed on shell $\delta=0$, reflecting the fact that the equations of motion impose $H_0=H_2$.
Since there are derivatives on both $H_0$ and $\kappa$, we demix them using a field redefinition,
\begin{equation}
\kappa \equiv Q - \frac{\Delta'}\rs H_0,
\end{equation}
which makes $H_0$ auxiliary,
\begin{align}
\mathcal{L}_\mathrm{even}^\mathrm{static} &= -\left(\frac\rs2Q'+\frac{\ell(\ell+1)}2H_0\right)^2 - \frac{\ell(\ell+1)(\ell+2)(\ell-1)}4\left(Q-\frac{2r}\rs H_0\right)\left(Q-\frac{2r f}{\rs}H_0\right).
\end{align}
Finally we integrate in a field $\lambda$ via a perfect square,
\begin{align}
\mathcal{L}_\mathrm{even}^\mathrm{static} &\to \mathcal{L}_\mathrm{even}^\mathrm{static} + \left[\frac\rs2Q'-\frac{(\ell+2)(\ell-1)}{2\rs}\Delta'Q+\left(\frac{2(\ell+2)(\ell-1)}{\rs^2}\Delta+\frac{\ell(\ell+1)}2\right)H_0-\lambda\right]^2 \nonumber\\
&= \frac{(\ell+2)(\ell-1)}{\rs^4}\left[(\ell+2)(\ell-1)\Delta(2rH-\rs Q)(2rfH-\rs Q)-2\rs^3(\Delta H)'Q+\ell(\ell+1)\rs^2\Delta H^2\right] \nonumber\\
&\hphantom{{}=}-\frac{\ell(\ell+1)\rs^2}{4(\ell+2)(\ell-1)\Delta}\lambda^2,
\label{eq:Levenstatic}
\end{align}
where in the second equality we have demixed $H_0$ and $\lambda$ via another field redefinition,
\begin{equation}
H_0 \equiv H + \frac{\rs^2}{2(\ell+2)(\ell-1)\Delta}\lambda.
\end{equation}
On shell we have $\lambda=0$ and $H=H_0$. Finally we integrate out $Q$ using its equation of motion to obtain
\begin{equation}\label{eq:L-even-static}
\boxed{\mathcal{L}_\mathrm{even}^\mathrm{static} = -\Delta H'^2 - \left(\ell(\ell+1)+\frac{\rs^2}\Delta\right)H^2.}
\end{equation}
As with the odd sector, the canonically-normalized field is a simple rescaling of the relevant metric perturbation living in $h_{t\mu}$, which is related to the Zerilli variable $\Psi_+$, and vice versa, by first-order differential operators of the form \eqref{eq:Darboux-static}.

The remarkably simple Lagrangian \eqref{eq:L-even-static} yields the equation of motion \cite{Riva:2023rcm}
\begin{equation}\label{eq:H0-static}
\Delta H_0'' + \Delta' H_0' - \frac{\rs^2+\ell(\ell+1)\Delta}\Delta H_0 = 0.
\end{equation}
This is rather easier to work with than the $\partial_t=0$ Zerilli equation, because it is a hypergeometric equation which can be solved with hypergeometric functions. By contrast, the Zerilli equation is of the more general Heun type. In the odd sector, both the static and full actions lead to hypergeometric equations of motion. This fact will be important when computing Love numbers in \cref{sec:love-compute}.

\subsubsection{Chandrasekhar duality}
\label{sec:chandra}

The even and odd sectors of $h_\mn$, despite beginning with different descriptions, end up being described by qualitatively similar dynamics: each sector is fully described by a canonically-normalized ``master variable'' $\Psi_\pm$, each of which satisfies a Schr\"odinger-like equation in a potential $V_\pm(r)$. In fact, the only difference between the Zerilli and Regge--Wheeler equations is in the potentials; the even sector has a rather involved potential $V_+(r)$ \eqref{eq:VZ}, while the odd potential $V_-(r)$ \eqref{eq:VRW} is comparatively very simple.

It turns out that there is a single structure underlying both the even and odd dynamics
\cite{1975RSPSA.343..289C,Chandrasekhar:1975zza,Chandrasekhar:1985kt}.\footnote{This structure can also be formulated in the language of supersymmetric quantum mechanics \cite{Cooper:1994eh}.} 
Both potentials can be written in a unified form
\begin{equation}
    V_\pm = W^2\mp \partial_{\rst}W+\beta,
\end{equation}
where $W(r)$ is called the \emph{superpotential}, and $\beta$ is a constant,\footnote{As a result, the Regge--Wheeler and Zerilli potentials differ by a single compact term, $V_+ = V_- - 2\partial_{\rst}^2\ln\Lambda$.}
\begin{subequations}
\begin{align}
    W(r) &= -\left(\frac{3\rs}{r^2}\frac f \Lambda+\sqrt{-\beta}\right)\nonumber\\
    &=\partial_{\rst}\ln\Lambda-\sqrt{-\beta},\\
    \beta &= -\left(\frac{(\ell-1)\ell(\ell+1)(\ell+2)}{6\rs}\right)^2
\end{align}
\end{subequations}
This goes by the name of \emph{Chandrasekhar duality}, and it famously is responsible for a crucial fact of black hole perturbation theory: that even- and odd-parity Schwarzschild quasinormal modes have the same spectrum of frequencies, i.e., they are \emph{isospectral}  \cite{1975RSPSA.343..289C,Chandrasekhar:1985kt,Chandrasekhar:1975zza}.\footnote{See \rcite{Berti:2009kk} for a review. For a generalization to partially-massless fields, see \rcite{Rosen:2020crj}.} 
Given a solution $\Psi_+$ to the Zerilli equation or $\Psi_-$ to the Regge--Wheeler equation, we can construct a solution to the other (in frequency space) via
\begin{equation}\label{eq:sol-chandra}
    \Psi_\pm = \frac1{\beta-\omega^2}\left(\frac{\dd}{\dd \rst}\mp W(r)\right)\Psi_\mp.
\end{equation}
The isospectrality of even and odd modes follows because the duality operator $\partial_\rst\mp W$ preserves the boundary conditions associated to quasinormal mode solutions and does not change the frequency.

Chandrasekhar duality can be extended to an off-shell symmetry (i.e., a symmetry of the action \eqref{eq:L-even-odd}),
\begin{equation}\label{eq:sym-chandra}
    \delta\Psi_\pm = [\partial_\rst\mp W(r)]\Psi_\mp.
\end{equation}
from which we can derive a conserved current via Noether's theorem \cite{Solomon:2023ltn}. Its conservation implies that the even and odd sectors have identical black hole Love numbers \cite{Hui:2020xxx,Solomon:2023ltn}.\footnote{For other compact objects this is typically not true, as the boundary conditions at the surface of the object are generally not invariant under the duality symmetry.}

\subsection{Kerr perturbations}
\label{sec:BHPT-Kerr}

Perturbations of the Kerr spacetime require a rather different treatment than the procedure we used for Schwarzschild perturbations. The key difference is that it is not known how to construct a separable wave equation for the Kerr metric perturbations. Fortunately there does exist a separability structure for Kerr, with the resultant equations of motion acting on certain components of the Weyl tensor rather than the metric tensor. In other words, the tractable study of small deviations from Kerr considers curvature perturbations rather than metric perturbations  \cite{Teukolsky:1972my,Teukolsky:1973ha}. There is a standard procedure to reconstruct the metric perturbations from the linearized Weyl tensor \cite{Pound:2021qin}.
Unlike the Regge--Wheeler and Zerilli equations, it is not known how to derive the master equation for Kerr perturbations by canonically normalizing the linearized Einstein--Hilbert action. 

Rather than the Einstein equations for the metric, we would like a wave-like equation for the curvature. Starting with the Bianchi identity \cite{Misner:1973prb,Wald:1984rg},
\begin{equation}\label{eq:bianchi}
    \nabla_{[\rho}R_{\mn]\ab}=0,
\end{equation}
take the covariant derivative $\nabla^\rho$ to find
\begin{equation} \label{eq:cd-bianchi}
    \Box R_{\mn\ab} + \nabla^\rho\nabla_\mu R_{\nu\rho\ab} + \nabla^\rho\nabla_\nu R_{\rho\mu\ab} = 0.
\end{equation}
Contracting the first two indices of \cref{eq:bianchi} and assuming the vacuum Einstein equation, $R_\mn = 0$, gives the contracted Bianchi identity, $\nabla^\mu R_{\mn\ab}=0$. As a result, the covariant derivatives in the latter two terms of \cref{eq:cd-bianchi} are effectively commutators of covariant derivatives, so we may write those terms sans derivatives but quadratic in the Riemann tensor. The result is the \emph{Penrose wave equation} \cite{Penrose:1960eq},
\begin{equation}\label{eq:penrose-wave}
    \Box R_{\mn\ab} + R_{\mn\rho\sigma}R_\ab{}^{\rho\sigma} + 4R^\rho{}_{\mu\sigma[\alpha}R^\sigma{}_{\beta]\rho\nu} = 0.
\end{equation}
The Teukolsky equation for gravitational perturbations \cite{Teukolsky:1972my,Teukolsky:1973ha}, presented below, follows from projecting the Riemann wave equation along the null tetrad directions and perturbing the Weyl scalars to linear order \cite{Ryan:1974nt,Bini:2002jx}.

To display the Teukolsky equation we will make use of the GHP formalism developed in \cref{sec:GHP}. The Kerr background, being of Petrov type D, is fully encapsulated in the Weyl scalar $\Psi_2$ \eqref{eq:Psi2} and the connection coefficients $(\rho,\tau)$, as well as their primes and complex conjugates (note that $\Psi_2'=\Psi_2$). On the linearized Kerr spacetime, the other Weyl scalars are non-vanishing and of linear order. Recall that a spin-$s$ field is described by a quantity of GHP type $(2s,0)$. This criterion picks out the extreme spin-weight Weyl scalars $\Psi_0$ and $\Psi_4$, which are of type $(4,0)$ and $(-4,0)$, respectively, and so describe gravitational fields of spin $s=\pm2$.

Using the GHP-covariant derivative $\Theta_\mu$ \eqref{eq:Theta-GHP} we can form a wave operator acting on quantities of type $(p,0)$, $\Box_p \equiv \Theta^\mu\Theta_\mu$. It turns out to be more convenient to slightly modify this wave operator, as follows. Recall that $\Theta_\mu$ was defined with the aid of the connection form $\omega$ \eqref{eq:omega-GHP}. We can construct another properly-weighted derivative operator by modifying the connection to $\omega_\mu - B_\mu$, where $B_\mu$ is any unweighted vector. We will choose \cite{Bini:2002jx,Aksteiner:2010rh}
\begin{equation}
    B_\mu \equiv -\rho n_\mu + \tau \bar m_\mu
\end{equation}
to construct the wave operator
\begin{equation}
\teuk_p \equiv (\Theta^\mu+pB^\mu)(\Theta_\mu+pB_\mu)
\end{equation}
which appears in the \emph{Teukolsky master equation} for a spin-$s$ field $\psi^{(s)}$ \cite{Bini:2002jx,Aksteiner:2010rh},\footnote{This equation is typically presented with the sign of the $\Psi_2$ term flipped, cf. \rcite{Bini:2002jx,Aksteiner:2010rh,Araneda:2016iwr}. These references --- and to the best of our knowledge, all other references presenting the form \eqref{eq:teuk} --- work in the mostly-minus metric signature. Changing signatures flips $-4s^2\Psi_2$ to $+4s^2\Psi_2$.}
\begin{equation}\label{eq:teuk}
    \boxed{\left(\teuk_{2s}+4s^2\Psi_2\right)\psi^{(s)} = 0.}
\end{equation}
(We do not reproduce the derivation of this equation, which follows from writing the components of the Bianchi identity and Einstein equations in GHP notation and performing some algebra.)
Here the
spin-$\pm2$
fields $\psi^{(\pm2)}$ are related to the extreme Weyl scalars $(\Psi_0,\Psi_4)$ by a multiplicative factor necessary to achieve separation of variables \cite{Teukolsky:1972my,Teukolsky:1973ha},\footnote{Note that this construction also holds for the other three (linearized) Weyl scalars, albeit with a source term on the right-hand side. Of particular note is the Teukolsky equation for the perturbed part of $\Psi_2$, which is equivalent to the Regge--Wheeler and Zerilli equations when $a=0$ \cite{Aksteiner:2010rh}. The factors in \cref{eq:psi-Psi} are specific to the Kinnersley tetrad.}
\begin{equation}\label{eq:psi-Psi}
    \psi^{(2)} = \Psi_0,\quad \psi^{(-2)} = \zeta^4\Psi_4,
\end{equation}
Remarkably, the Teukolsky master equation also holds for scalar ($s=0$), neutrino ($s=\pm1/2$), and electromagnetic ($s=\pm1$) fields propagating on a type D background \cite{Teukolsky:1972my,Teukolsky:1973ha,Dudley:1977zz}. For a scalar the Teukolsky equation is precisely the Klein--Gordon equation, since $\teuk_0=\Box_0=\Box$. For the electromagnetic field, the fields $\psi^{(\pm1)}$ are built out of the components of the field strength tensor $F_\mn$, analogous to the Weyl scalars for gravitational perturbations, and the Teukolsky equation results from the Bianchi identity for $F_\mn$. For completeness we will leave $s$ general.

Using the Kinnersley tetrad \eqref{eq:kinnersley} and working in Boyer--Lindquist coordinates, the Teukolsky equation \eqref{eq:teuk} can be written explicitly as
\begin{align}\label{eq:teuk-coord}
    &\bigg\{\left[a^2\sin^2\theta-\frac{(r^2+a^2)^2}\Delta\right]\partial_t^2 - \frac{4aMr}\Delta\partial_t\partial_\varphi + \left(\csc^2\theta-\frac{a^2}\Delta\right)\partial_\varphi^2 \nonumber\\
    &\hphantom{\bigg\{}+2s\left[\frac{a(r-M)}\Delta+i\cot\theta\csc\theta\right]\partial_\varphi + 2s\left[\frac{M(r^2-a^2)}{\Delta}-\bar\zeta\right]\partial_t \nonumber\\
    &\hphantom{\bigg\{}+ \Delta^{-s}\partial_r\left(\Delta^{s+1}\partial_r\right) + \frac1{\sin\theta}\partial_\theta\left(\sin\theta\partial_\theta\right) + s\left(1-s\cot^2\theta\right) \bigg\} \psi^{(s)} = 0.
\end{align}
Despite its complicated appearance, this equation, miraculously, is separable: we can rewrite it as
\begin{equation}
    (\mathcal{R}+\mathcal{S})\psi^{(s)} = 0,
\end{equation}
where the radial and angular operators are, after some factorization,
\begin{subequations}
\begin{align}
    \mathcal{R} &=\Delta^{-s}\partial_r\left(\Delta^{s+1}\partial_r\right) + \frac{\mathcal{K}^2-2is(r-M)\mathcal{K}}\Delta + 2a\partial_t\partial_\varphi-4sr\partial_t, \\
    \mathcal{S} &= \frac1{\sin\theta}\partial_\theta\left(\sin\theta\partial_\theta\right) + a^2\sin^2\theta\partial_t^2 -2ias\cos\theta\partial_t- \csc^2\theta\left(i\partial_\varphi-s\cos\theta\right)^2 + s,
\end{align}
\end{subequations}
and we have defined\footnote{Here, the operator $\mathcal{K}$ should not be confused with the scalar metric component $\mathcal{K}$ in \cref{eq:h-even-odd}.}
\begin{equation}\label{eq:K-op}
    \mathcal K \equiv i\left[(r^2+a^2)\partial_t+a\partial_\varphi\right].
\end{equation}
As a result of this separability structure, the spin-$s$ Teukolsky equation admits solutions of the form
\begin{equation}
    \psi^{(s)}_{\ell m \omega} = e^{i(-\omega t+m\varphi)}R^{(s)}_{\ell m \omega}(r)S^{(s)}_{\ell m \omega}(\theta).
\end{equation}
As in the non-rotating case, the simple exponential dependence on $t$ and $\varphi$ is a consequence of the Killing isometries in those directions, or equivalently, the lack of any explicit $t$ or $\varphi$ dependence in \cref{eq:teuk-coord}.
The general solution to the Teukolsky equation is a superposition of such separable solutions,
\begin{equation}
    \psi^{(s)} = \int_{-\infty}^\infty\frac{\dd\omega}{2\pi}\sum_{\ell=|s|}^\infty\sum_{m=-\ell}^\ell \psi^{(s)}_{\ell m \omega}.
\end{equation}

The mode functions $R^{(s)}_{\ell m \omega}(r)$ and $S^{(s)}_{\ell m \omega}(\theta)$ obey the ordinary differential equations
\begin{subequations}
\begin{align}
    \mathcal{R}R^{(s)}_{\ell m \omega}(r) &= A_{\ell m}^{(s)}R^{(s)}_{\ell m \omega}(r),\\
    \mathcal{S}S^{(s)}_{\ell m \omega}(\theta) &= -A_{\ell m}^{(s)}S^{(s)}_{\ell m \omega}(\theta),
\end{align}
\end{subequations}
where $A_{\ell m}^{(s)}$ is a separation constant, and $\mathcal R$ and $\mathcal S$ now act as
\begin{subequations}
\begin{align}
    \mathcal{R} &=\Delta^{-s}\partial_r\left(\Delta^{s+1}\partial_r\right)+\frac{K^2-2is(r-M)K}\Delta+2am\omega+4is\omega r, \\
    \mathcal{S} &= \frac1{\sin\theta}\partial_\theta\left(\sin\theta\partial_\theta\right) -a^2\omega^2\sin^2\theta -2a\omega s \cos\theta - \frac{\left(m+s\cos\theta\right)^2}{\sin^2\theta} + s.
\end{align}
\end{subequations}
Here $K \equiv (r^2+a^2)\omega-am$
is the eigenvalue of $\mathcal K$ \eqref{eq:K-op} for the eigenfunction $e^{i(-\omega t+m\varphi)}$.
It is conventional to define
\begin{equation}
    A_{\ell m}^{(s)} = \Lambda_{\ell m}^{(s)}+2am\omega-a^2\omega^2,
\end{equation}
and shift the constant $a^2\omega^2$ from the radial to the angular equation, so that our decoupled equations take the standard form, 
\begin{subequations}
\begin{align}
    \left[\Delta^{-s}\partial_r\left(\Delta^{s+1}\partial_r\right)+\frac{K^2-2is(r-M)K}\Delta+4is\omega r - \Lambda_{\ell m}^{(s)}\right]R^{(s)}_{\ell m \omega}(r) &= 0,\label{eq:teuk-rad-dontref}\\
    \left[\frac1{\sin\theta}\partial_\theta\left(\sin\theta\partial_\theta\right) +a^2\omega^2\cos^2\theta -2a\omega s \cos\theta - \frac{\left(m+s\cos\theta\right)^2}{\sin^2\theta} + s+A_{\ell m}^{(s)}\right]S^{(s)}_{\ell m \omega}(\theta) &= 0.\label{eq:teuk-ang}
\end{align}
\end{subequations}

In the limit $(a,s)\to0$, the angular equation \eqref{eq:teuk-ang} reduces to the defining equation \eqref{eq:sph-harm-eq} for the (scalar) spherical harmonics $Y_{\ell m}$. The solutions $S^{(s)}_{\ell m \omega}(\theta)$ that are regular at the poles are known as \emph{spin-weighted spheroidal harmonics}; these generalize the usual spherical harmonics to account for the deformations to sphericity caused by the rotation of the object, as well as the internal spin of the field \cite{Teukolsky:1972my,Teukolsky:1973ha}. Note that we have encountered the vector ($s=\pm1$) and tensor ($s=\pm2$) \emph{spherical} harmonics in the angular dependences of the odd and even metric perturbations \eqref{eq:h-even-odd}, respectively, of the Schwarzschild spacetime. We also saw the standard ($s=0$) spheroidal harmonics when studying the scalar wave equation on Kerr in \cref{sec:Kerr-KG}.
The separation constant depends in general on $a\omega$ and must be calculated numerically, although there is an approximate analytic expression in the small-$a\omega$ limit,
\begin{equation}
    A_{\ell m}^{(s)} = \ell(\ell+1)-s(s+1)+\mathcal{O}(a^2\omega^2).
\end{equation}
For a review of the spin-weighted spheroidal harmonics and their properties, see \rcite{Berti:2005gp}.

All told, the dynamics of spin-$s$ perturbations of the Kerr spacetime boil down to the radial Teukolsky equation \eqref{eq:teuk-rad-dontref}, which we repeat here:
\begin{equation}\label{eq:teuk-rad-boxed}
    \boxed{\left[\Delta^{-s}\partial_r\left(\Delta^{s+1}\partial_r\right)+\frac{K^2-2is(r-M)K}\Delta+4is\omega r - \Lambda_{\ell m}^{(s)}\right]R^{(s)}_{\ell m \omega}(r) = 0.}
\end{equation}
This equation will be the main object of study on the ``UV'' side when we compute the Love numbers of Kerr black holes in Section \ref{sec:love-compute}.

\newpage
\section{Neutron stars}
\label{sec:NS}

Neutron stars are fascinating and complex astrophysical objects. Their interiors exhibit forms of matter and physical r\'egimes that are not realized anywhere else in the universe. In their cores, densities exceed those found in atomic nuclei, reaching the most extreme values known in the observable universe. These extreme environments transform neutron stars into cosmic laboratories where the fundamental behavior of matter can be tested at the interface of nuclear physics, particle physics, and astrophysics.
The  description of their interiors is formulated with an equation of state (EoS), which encodes how pressure, density, and temperature are related under such extreme conditions. This relation depends sensitively on the microscopic details and interactions among particles, and dictates the star's observable macroscopic properties, including its mass and radius, the maximum mass it can sustain against collapse, and its tidal response to external gravitational fields. The detection of gravitational waves from merging neutron-star binaries now allows these properties to be measured with increasing precision, offering an unprecedented opportunity to explore the physics of ultra-dense matter beyond the reach of laboratory experiments.

The goal of this section is not to provide a comprehensive review of neutron stars. Numerous detailed reviews and books already exist, covering the full range of topics, from theoretical predictions and microscopic approaches for constructing neutron-star equations of state to observational constraints from both gravitational-wave and X‑ray measurements --- see, e.g.,~\rcite{Lattimer:2012nd,Rezzolla:2018jee,GuerraChaves:2019foa,Burns:2019byj,Chatziioannou:2020pqz,Lattimer:2021emm,Burgio:2021vgk,Yunes:2022ldq,Chatziioannou:2024jsr,Glendenning:1997wn,Haensel:2007yy}. We refer the reader to these  and other references for an in-depth discussion of neutron star properties. Here, our aim is much more modest: we   summarize the basic tools of stellar perturbation theory~\cite{1967ApJ...149..591T,1969ApJ...155..163P,Thorne:1968zz,1970ApJ...159..847C,1973ApJ...181..181I,Thorne:1969rba,Lindblom:1983ps,Detweiler:1985zz,Ferrari2003}, which  provide a straightforward generalization of  the discussion on black hole perturbation theory in \cref{sec:BHPT}. In particular, we  focus on non-rotating objects.

\subsection{Tolman--Oppenheimer--Volkoff   equations}

The metric $\bar{g}_{\mu\nu}$ that describes a static
and spherically-symmetric unperturbed star is given by the line element
\begin{equation}
\dd s^2= \bar{g}_{\mu\nu} \dd x^\mu\dd x^\nu = - e^{\Phi(r)}\dd t^2 + e^{\Psi(r)}\dd r^2 + r^2 \dd \Omega_{S^2}^2 ,
\end{equation}
where $\Phi(r)$ and $\Psi(r)$ are functions of the radial coordinate $r$ only, and $\dd \Omega_{S^2}^2 \equiv \dd\theta^2+ \sin^2\theta \dd \varphi^2$ denotes the line element on the two-sphere. Inside the star, the functions $\Phi(r)$ and $\Psi(r)$ can be determined by solving the equations of hydrostatic equilibrium.  We shall further assume the interior of the  star to be described by the energy-momentum tensor of a perfect fluid,
\begin{equation}
T_{\mu\nu} = (\rho+p)u_\mu u_\nu+ p g_{\mu\nu} ,
\label{eq:Tmustarisotropic}
\end{equation}
where $\rho$ denotes   the energy density, $p$ the pressure, and $u_\mu$ the fluid four-velocity.
At the background level, the unperturbed energy-momentum tensor of the star is denoted by $\bar T_{\mu\nu} = (\bar\rho+\bar p)\bar{u}_\mu \bar{u}_\nu+ \bar{p} \bar{g}_{\mu\nu}$.
For convenience, we will momentarily set $G = 1$ in this section. From the Einstein equations,
\begin{equation}
\Ef_{\mu\nu}\equiv G_{\mu\nu}- 8 \pi T_{\mu\nu} = 0,
\label{eq:EEinsteinEq}
\end{equation}
it follows that, at the background level, the metric $\bar{g}_{\mu\nu}$ and the unperturbed matter variables --- the energy density $\bar \rho$, pressure $\bar p$, and fluid four-velocity $\bar{u}_{\mu}$ --- are related through the Tolman--Oppenheimer--Volkoff (TOV) equations~\cite{Tolman:1939jz,Oppenheimer:1939ne},
\begin{align}
\Mm'(r)&=4\pi r^2\bar \rho(r) \; ,
\label{TOV1}\\
\Phi '(r)&= 2\frac{ \Mm(r)+4 \pi  r^3 \bar p(r)}{r[r-2  \Mm(r)]} \; ,
\label{TOV2}\\
\bar p'(r) &=  -[\bar p(r)+\bar \rho (r)]\frac{\left[\Mm(r)+4 \pi  r^3 \bar p(r)\right]}{r [r-2 \Mm(r)]} \; ,
\label{TOV3}
\end{align}
where we have defined $\Mm(r)$ by
\begin{equation}
e^{-\Psi(r)} \equiv 1-\frac{2\Mm(r)}{r}.
\end{equation}
Since the background is characterized by four quantities, $\{\bar p(r), \bar \rho(r), \Mm(r), \Phi(r)\}$, closing the system requires specifying an EoS. Once the fluid's EoS is provided, the equations can be solved numerically, yielding the radial profiles of pressure and energy density throughout the star.
Outside the star, the metric reduces to the Schwarzschild solution, $e^\Phi = e^{-\Psi}= 1 - \frac{2 M}{r}$, where 
\be
M \equiv \Mm(\Rstar) \;
\ee
denotes the total mass of the star. The stellar radius $\Rstar$ is defined as the location where the pressure vanishes, $\bar p(\Rstar)=0$. 
Furthermore, imposing the normalization condition $u_\mu u^\mu = -1$, the unperturbed fluid four-velocity is
\begin{equation}
\bar{u}^\mu\partial_\mu = e^{-\frac12\Phi(r)}\partial_t.
\end{equation}

\subsection{Perturbation theory of a  star}

Once the background equilibrium configuration of the star is determined, we proceed to study small deformations about it. To this end, we expand both the metric and the matter sector in perturbations as $g_{\mu\nu}=\bar{g}_{\mu\nu}+ \delta g_{\mu\nu}$ and $T_{\mu\nu}=\bar{T}_{\mu\nu}+ \delta T_{\mu\nu}$.

The most general parametrization of the metric fluctuation $\delta g_{\mu\nu}$ was already given in \cref{eq:h-even-odd}. We reproduce it here for completeness:\footnote{Note that we are slightly changing conventions here. In contrast with \cref{eq:h-even-odd}, the metric components in \cref{eq:h-even-odd-NS} are not canonically normalized. This difference in choice is completely immaterial for our purposes, as we always work below at the level of the linearized equations of motion.}
\begin{subequations}\label{eq:h-even-odd-NS}
\begin{align}
\delta g_\mn^\mathrm{even} &= \begin{pmatrix}
e^\Phi H_0(t,r) & H_1(t,r) & \mathcal H_0(t,r)\partial_\theta & \mathcal H_0(t,r)\partial_\varphi \\
\cdot & e^\Psi H_2(t,r) & \mathcal H_1(t,r)\partial_\theta & \mathcal H_1(t,r)\partial_\varphi \\
\cdot & \cdot & r^2\left(\mathcal K(t,r) + \mathcal G(t,r) \partial_\theta^2\right) & r^2\mathcal G(t,r)\nabla_\theta\nabla_\varphi\\
\cdot & \cdot & \cdot & r^2\left(\sin^2\theta\mathcal K(t,r) + \mathcal G(t,r) \nabla_\varphi^2\right)
\end{pmatrix}Y_{\ell m}, \\
\delta g_\mn^\mathrm{odd} &= r^2\begin{pmatrix}
0 & 0 & -h_0(t,r)\csc\theta\partial_\varphi & h_0(t,r)\sin\theta\partial_\theta \\
\cdot & 0 & -h_1(t,r)\csc\theta\partial_\varphi & h_1(t,r)\sin\theta\partial_\theta \\
\cdot & \cdot & -h_2(t,r)\csc\theta\nabla_\theta\nabla_\varphi & \frac12h_2(t,r) \left(\sin\theta\partial_\theta^2-\csc\theta\nabla_\varphi^2\right) \\
\cdot & \cdot & \cdot & h_2(t,r)\sin\theta\nabla_\theta\nabla_\varphi
\end{pmatrix}Y_{\ell m}.
\end{align}
\end{subequations} 
(Note that, compared to \cref{eq:h-even-odd}, the metric perturbation components are defined here with a non-canonical normalization, and the $t$--$t$ and $r$--$r$ components have been rescaled by different prefactors.) 
The fluid four-velocity can be analogously decomposed as~\cite{Kojima:1992ie}
\begin{equation}
    \delta u^\mu = \frac{e^{\Phi/2}}{4\pi(\bar{\rho}+\bar p)}\left[
    \begin{pmatrix}
        \delta u^0(t,r) \\
        e^{-\Psi}R(t,r) \\
        r^{-2} V(t,r)\partial_\theta \\
        r^{-2}\csc^2\theta V(t,r)\partial_\varphi
    \end{pmatrix}+ 
    \begin{pmatrix}
        0 \\
        0 \\
        -r^{-2} U(t,r)\csc\theta\partial_\varphi \\
        r^{-2} U(t,r)\sin\theta\partial_\theta
    \end{pmatrix}\right]Y_{\ellm}
    ,\label{eq:deltau-NS}
\end{equation}
where the two terms in brackets are of even and odd parity, respectively.
The normalization condition for the four-velocity, $u^\mu u_\mu = -1$, determines $\delta u^0$ in terms of the remaining perturbations. To first order in perturbation theory, one obtains\footnote{See \rcite{Pani:2025qxs} for a generalization at second-order in perturbation theory.}
\begin{equation}
\delta u^0 = 2 \pi  e^{-\Phi} (\bar \rho + \bar p) H_0 .
\end{equation}
In the following, we work in Regge--Wheeler gauge~\cite{Regge:1957td}, defined by the conditions $h_2 = \mathcal H_0 = \mathcal H_1 = \mathcal G = 0$.

\subsubsection{Linearized polar equations}

Since the equations of motion for the even- and odd-parity metric components decouple to linear order in perturbation theory, they can be analyzed independently. We begin by focusing on the parity-even (polar) sector~\cite{Chandrasekhar:1991fi,Ipser:1991ind,1991RSPSA.433..423C,Kojima:1992ie,1967ApJ...149..591T,Detweiler:1985zz}.
Because the equation of motion $\Ef_{\mu\nu}$ \eqref{eq:EEinsteinEq} is a rank-two tensor, it can be decomposed into tensor spherical harmonics using the same basis as for the metric perturbations. After projecting out the angular dependence through standard spherical-harmonic identities, and denoting by capital Latin indices $A,B,C,\ldots$ the angular coordinates on the two-dimensional sphere $S^2$, whose line element is $\gamma_{A B} {\rm d}x^A {\rm d}x^B \equiv \dd \Omega_{S^2}^2    = {\rm d}\theta^2 + \sin^2\theta\, {\rm d}\varphi^2$,  the components of $\Ef_{\mu\nu}$ take the form
\begin{align}
    \Ef_{tt}  = & \ \frac{e^{\Phi }}{2 r^2}  \big\{  2 \left[(r-2 \Mm) (\Htwof'-r \, \Kf'')+ \left(5 \Mm+4 \pi  r^3 \bar \rho-3 r\right) \Kf' -8 \pi  r^2 \deltaf  \rho
   \right] \nonumber \\
   &  + \left(\ell(\ell+1)-16 \pi  r^2 \bar\rho +2\right)\Htwof + \left(\ell(\ell+1)-2\right) \Kf \big\} Y_{\ell m}(\theta, \varphi) \;, \label{Ett}\\
\Ef_{tr}  = & \  \frac{1}{2 r^2} \left\{\Honef \ell(\ell+1)+2 r \left[-i \omega  ( \Htwof - r  \Kf') -\frac{i \omega   \left(3 \Mm+4 \pi  r^3 \bar p-r\right) \Kf }{r-2 \Mm}+2 r
   e^{\Phi } \Rf\right] \right\} Y_{\ell m}(\theta, \varphi) \;, \\
    \Ef_{rr}  = & \ \frac{e^{-\Phi }}{2 r (r-2 \Mm)} \big\{ -e^{\Phi } \big[2 (r-2 \Mm) \Hzerf' + 2 \left(1+ 8 \pi  r^2 \bar p \right) \Htwof  + 2\left(\Mm-4 \pi  r^3
   \bar p-r\right) \Kf' +16 \pi  r^2 \deltaf p \nonumber \\
   &  -\ell(\ell+1) \Hzerf +  (\ell(\ell+1)-2 ) \Kf \big] +2 \omega  \big(r^2 \omega 
   \Kf-2 i (r-2 \Mm) \Honef  \big) \big\} Y_{\ell m}(\theta, \varphi) \;, \\
\gamma^{AB} \Ef_{AB}   = & \ \frac{e^{-\Phi }}{2}  \big\{ e^{\Phi } \big[  \ell(\ell+1) \Hzerf - \left(\ell(\ell+1)+32 \pi  r^2 \bar p\right) \Htwof -2  \left(r+\Mm+4 \pi 
   r^3 (2 \bar p- \bar \rho ) \right) \Hzerf' 
   \nonumber \\
&   -2  \left(r-\Mm+4 \pi  r^3 \bar p\right) \Htwof'    
-2 r (r-2 \Mm) (\Hzerf''- \Kf'') +4 
   \left(r -\Mm+2 \pi  r^3 ( \bar p-\bar \rho )\right) \Kf' -32 \pi  r^2 \deltaf p\big] \nonumber \\  
   & +2 r \omega  \left[r \omega 
   \left(\Htwof+\Kf\right)-2 i (r-2 \Mm) \Honef'\right] 
   +4 i \omega  \Honef \left(\Mm+4 \pi  r^3 \bar \rho -r\right)\big\} Y_{\ell m}(\theta, \varphi) \;, \\
    \Ef_{tA} =  & \ \frac{1}{2 r^2} \left\{  r \left[ (r-2 \Mm) \Honef'+4 \pi  r^2 ( \bar p- \bar \rho ) \Honef  +i r \omega  \Kf+4 r e^{\Phi } \Vf\right]+2
    \Mm \Honef +i r^2 \omega  \Htwof \right\} \nabla_A Y_{\ell m}(\theta, \varphi) \;, \\
    \Ef_{rA}  = & \ \frac{1}{2}  \left[ \frac{  (3 \Mm -r +4 \pi  r^3 \bar p) \Hzerf - (\Mm-r+4 \pi  r^3 \bar p) \Htwof  }{r (r-2 \Mm)}+i
   \omega  e^{-\Phi } \Honef  +\Hzerf'-\Kf'\right] \nabla_A Y_{\ell m}(\theta, \varphi) \;, \\
    [\Ef_{AB}]_{\rm{STF}}  = & \ \frac{1}{2} \left(\Hzerf-\Htwof\right)  \left( \nabla_A\nabla_B  -\frac{1}{2} \gamma_{AB} \nabla_C\nabla^C \right) Y_{\ell m}(\theta, \varphi)   \;,
    \label{EAB}
\end{align}
where a prime denotes differentiation with respect to the radial coordinate, $\nabla_A$ is the covariant derivative on $S^2$, and $[{\cdots}]_{\rm STF}$ in \cref{EAB} indicates the symmetric trace-free part. In deriving these expressions, the TOV equations \eqref{TOV1}--\eqref{TOV3} have been used to simplify the results.
The system consists of seven equations for eight perturbation variables: $\Hzerf$, $\Honef$, $\Htwof$, $\Kf$, $\Rf$, $\Vf$, $\deltaf \rho$, and $\deltaf p$. Therefore, it can be  closed upon supplementing it with an EoS. 
Clearly, not all of these variables are dynamical. The step to integrate out the constraint fields and simplify the set of eqs.~\eqref{TOV1}--\eqref{TOV3}, once an EoS is fixed, can be found in, e.g., \rcite{Ferrari2003,Chandrasekhar:1991fi,Ipser:1991ind,1991RSPSA.433..423C,Kojima:1992ie}.

In summary, \cref{EAB} yields the constraint $\Htwof = \Hzerf$. The equations $\Ef_{tr}=0$ and $\Ef_{tA}=0$ can, in general, be used to express $\Rf$ and $\Vf$ in terms of $\Hzerf$, $\Honef$, $\Kf$, and their radial derivatives. Moreover, $\deltaf p$ --- and hence $\deltaf \rho$ through the equation of state --- can be written in terms of the same set of variables by means of $\Ef_{rr}=0$.
The remaining functions, $\Hzerf$, $\Honef$, and $\Kf$, are then determined by solving the reduced system given by $\Ef_{tt}=0$, $\Ef_{rA}=0$, and $\gamma^{AB}\Ef_{AB}=0$.

Note that the system of eqs.~\eqref{Ett}--\eqref{EAB} is exact to linear order in perturbation theory (see \rcite{Pitre:2025qdf,Pani:2025qxs} for an extension to second order). In particular, the system is valid for arbitrary frequency $\omega$. 
To investigate the tidal response in the adiabatic r\'egime, one would like to consider the small-frequency limit of eqs.~\eqref{Ett}--\eqref{EAB}.
However, this limit can be subtle. If one sets $\omega=0$ directly in these equations, the variables  $\Vf$, $\Rf$, $\Honef$  appear to remain unconstrained. As discussed above, the equations $\Ef_{tr}=0$ and $\Ef_{tA}=0$ can be used to express $\Vf$ and $\Rf$ in terms of $\Honef$. However, when $\omega$ vanishes exactly, $\Honef$ drops out of the remaining equations. In particular, one can no longer use, for example $ \Ef_{rA} =0 $ to solve for $\Honef$ in terms of $\Htwof$, $\Hzerf$ and $\Kf$.

Although this does not pose a fundamental problem for the linear static response in the even sector --- where $\Honef$ is not required explicitly to determine  $\Hzerf$~\cite{Hinderer:2007mb} --- it may be problematic at second order~\cite{Pani:2025qxs}. At least formally, this issue can be resolved by working carefully in the adiabatic limit  $\omega\rightarrow0$, rather than setting $\omega=0$  from the outset.\footnote{This subtlety can also be seen in our analysis of vacuum Schwarzschild perturbations in \cref{sec:BHPT-Sch}. For finite $\omega$, $H_1$ is determined by \cref{eq:H1-sol}, and goes to zero in the limit $\omega\to0$. If we instead set $\omega=0$ at the start, $\Honef$ drops out of the action \eqref{eq:action-even} entirely and formally has no equation of motion.}
After solving the constraint equations at finite $\omega$, one finds that the frequency enters only quadratically in the remaining equations for  $\Kf$ and $\Hzerf$ (see, e.g., \rcite{Ipser:1991ind,Kojima:1992ie}). 
Consequently, the solutions at orders $\mathcal{O}(\omega^0)$ and $\mathcal{O}(\omega)$ coincide in the small-$\omega$ limit for these variables.
It then follows that, at order $\mathcal{O}(\omega)$, $\Ef_{rA}=0$  reduces simply to $\omega \Honef=0$, implying $\Honef = \mathcal{O} (\omega)$. Likewise, from $\Ef_{tr}=0$ and $\Ef_{tA}=0$ one finds  $\Vf, \Rf= \mathcal{O} (\omega)$.
Therefore, if one is interested only in the properties of the linearized perturbations at order $\mathcal{O}(\omega^0)$ it is consistent to set  $\Vf = \Rf = \Honef = 0$. As a result, the components $\Ef_{tr} $ and $\Ef_{tA}$ vanish identically at this order.

At leading order in the static limit, the remaining equations simplify significantly. One can  use $\Ef_{rA} = 0$ to solve for $\Kf'$, and  $\gamma^{AB}\Ef_{AB} = 0$ to  solve for $\Kf$ in terms of $\Hzerf$ and $\Hzerf'$. 
In addition, solving $\Ef_{tt}=0$ for $\deltaf \rho$, one finds
\begin{equation}
\deltaf \rho = \frac{1}{2} \frac{\bar\rho'}{\bar p'} 
(\bar p + \bar \rho)\Hzerf \; .
\end{equation}
Finally, after introducing the sound speed
\begin{equation}
c_s^2(r)\equiv\frac{\bar p' (r)}{\bar\rho'(r)}\;,
\end{equation}
such that $\deltaf p =  c_s^2 \deltaf \rho$,\footnote{A major simplification arises in the static limit. Because the perturbations evolve on timescales much longer than those associated with heat transfer, the fluid effectively maintains local thermodynamic equilibrium throughout the process. As a result, the perturbations are fully characterized by a single fluid degree of freedom, leading to the relation $\delta p / p' = \delta \rho / \rho'$, as can be checked explicitly.} one obtains a single second-order differential equation for $\Hzerf$~\cite{Hinderer:2007mb},
\begin{multline}
 0=   \Hzerf''+\Hzerf'\frac{2 \left[r-\Mm+2 \pi  r^3 (\bar p- \bar \rho )\right]}{r (r-2 \Mm)} 
    +\Hzerf \Bigg\{\frac{4 \pi  r (\bar p+ \bar \rho ) }{(r-2 \Mm)c_s^2}  
\\
    -\frac{2 \Mm r \left[- \ell(\ell+1)+52 \pi  r^2 \bar p+20 \pi  r^2 \bar \rho
   \right]+r^2 \ell(\ell+1)+4 \Mm^2+4 \pi  r^4 \left[ \bar p \left(16 \pi  r^2 \bar p-9\right)-5 \bar \rho \right]}{r^2 (r-2
   \Mm)^2}\Bigg\} \;.
   \label{master1}
\end{multline}

\subsubsection{Linearized axial equations}

We now turn to gravito-magnetic --- i.e., parity-odd (axial) --- perturbations of Einstein's equations, working in the Regge--Wheeler gauge ($h_2=0$), for an isotropic perfect-fluid star described by the stress--energy tensor \eqref{eq:Tmustarisotropic}. The odd equations are (see, e.g., \rcite{Kojima:1992ie})
\begin{subequations}
\begin{align}
e^{-\Phi } \dot{h}_0+\left(\frac{2 \mathcal{M}}{r}-1\right) h_1'+ \left(4 \pi  r (\bar \rho -\bar p-\frac{2 \mathcal{M}}{r^2}\right) h_1 &=  0,
\label{eq:oddstar1}
\\
e^{-\Phi } \left(\dot{h}_0' -\frac{2   }{r}\dot{h}_0-  \ddot{h}_1  \right)-\frac{(\ell-1)(\ell+2) }{r^2} h_1 & =0,
\label{eq:oddstar2}
\\
e^{-\Psi} \left(h_0''-\dot{h}_1'\right) -4 \pi  r (\bar p+\bar \rho ) \left(h_0'-\dot{h}_1\right)  -\frac{2 e^{-\Psi} }{r} \dot{h}_1 \nonumber\\
-\frac{\ell(\ell+1) r-4 \mathcal{M}+8 \pi  r^3 (\bar p+\bar \rho )}{r^3} h_0 -4 e^{\Phi } U & =0 . 
\label{eq:oddstar3}
\end{align}
\end{subequations}
It is clear that \cref{eq:oddstar1} can generally be solved for $h_0$ in terms of $h_1$, as long as the perturbations are not exactly static. For non-static perturbations, after  defining the field $\psi$ as
\begin{equation}
h_1 \equiv  e^{\frac{1}{2} (\Psi-\Phi )} r  \psi
\end{equation}
and Fourier transforming in time, one obtains, from \cref{eq:oddstar1}, 
\begin{equation}
h_0 = \frac{i e^{\frac{1}{2} (\Phi-\Psi)} }{\omega } \left(r \psi '+\psi \right).
\label{eq:hopsipsip}
\end{equation}
Plugging this into \cref{eq:oddstar2}, one finds
\begin{equation}
e^{\frac{1}{2} (\Phi-\Psi)}\partial_r \left( e^{\frac{1}{2} (\Phi-\Psi)}  \psi '\right) 
+  \left[\omega ^2- e^\Phi \left(\frac{\ell(\ell+1)}{r^2}-\frac{6 \mathcal{M}}{r^3}+4 \pi  (\bar \rho -\bar p)\right)\right]\psi
 =0,
 \label{eq:psiNSodd}
\end{equation}
which corresponds to the Regge--Wheeler equation describing axial perturbations within the star (see, e.g., \rcite{Kojima:1992ie}). In particular, it recovers the result of \rcite{Damour:2009vw}  upon taking $\omega\rightarrow0$. \Cref{eq:oddstar3} can finally be used to express the fluid perturbation $U$ in terms of the axial metric perturbation. In Fourier space,
\begin{equation}
U = -\frac{4 i \pi  (\bar p+\bar \rho ) e^{-\frac{1 }{2}(\Phi+\Psi)} }{\omega } \left(r \psi '+\psi \right)
= -4 \pi  e^{-\Phi} (\bar p+\bar \rho) h_0.
\label{eq:constraintUodd}
\end{equation}

It has been noted in the literature~\cite{Damour:2009vw,Binnington:2009bb,Favata:2005da,Pani:2015nua,Pani:2018inf} that taking the $\omega \rightarrow 0$ limit in \cref{eq:psiNSodd} yields a result that is \textit{not} equivalent to the one obtained by setting $\omega = 0$ from the outset. Notice that \cref{eq:psiNSodd} at $\omega \neq 0$ can equivalently be written as an equation for $h_0$ by inverting the relation \eqref{eq:hopsipsip}. The resulting expression is somewhat more involved, but it admits a well-defined limit as $\omega \rightarrow 0$. Taking this limit at the very end yields the equation \cite{Landry:2015cva,Pani:2018inf}
\begin{equation}
e^{-\Psi}h_0''
-4 \pi  r (\bar p+\bar \rho ) h_0'
-  \left[\frac{\ell (\ell+1)}{r^2}-\frac{4 \mathcal{M}}{r^3}-8 \pi  (\bar p+\bar \rho)\right]h_0=0.
\label{eq:staticoddcc1}
\end{equation}
This equation can be contrasted with the one obtained by setting $\omega = 0$ directly at the level of the Einstein equations \eqref{eq:oddstar1}--\eqref{eq:oddstar3}. In the strictly static r\'egime, defined by $\dot h_0 = \dot h_1 = 0$ and $U = 0$, one finds from \cref{eq:oddstar2} that $h_1 = 0$, and from \cref{eq:oddstar3} that
\begin{equation}
e^{-\Psi} h_0''
- 4 \pi r (\bar p + \bar \rho) h_0'
- \left[ \frac{\ell (\ell+1)}{r^2} - \frac{4 \mathcal{M}}{r^3} + 8 \pi (\bar p + \bar \rho) \right] h_0 = 0,
\label{eq:staticoddcc2}
\end{equation}
which reproduces the equation derived in \rcite{Binnington:2009bb}, where the axial perturbations of a strictly static fluid were studied.
As noted in \rcite{Landry:2015cva}, \cref{eq:staticoddcc2} is very similar to \cref{eq:staticoddcc1}, the only difference being the opposite sign in front of the $(\bar p + \bar \rho)$ term. 
This discrepancy can be attributed to the apparent loss of a constraint --- specifically, \cref{eq:constraintUodd} --- when $\omega = 0$ is imposed from the outset.  

In summary, the equation derived in \rcite{Damour:2009vw} coincides with that used in \rcite{Landry:2015cva} (\cref{eq:staticoddcc1}) for an \textit{irrotational} fluid. This has been explained by the fact that the zero-frequency limit of the Regge--Wheeler equation enforces the fluid to be irrotational rather than strictly \textit{static}~\cite{Pani:2018inf}. By contrast, the inequivalence between the static-limit master equations obtained in \rcite{Damour:2009vw} (\cref{eq:staticoddcc1}) and \rcite{Binnington:2009bb} (\cref{eq:staticoddcc2}) has been attributed to the fact that the fluid is irrotational in the former case, but strictly static in the latter~\cite{Pani:2018inf}.

This behavior is reminiscent of what occurs, in a different form, in the even-parity sector above, where the strict static limit introduces ambiguities in some of the constraints and appears discontinuous. 
Note that this apparent discontinuity is absent in vacuum general relativity for black hole perturbation equations: in the absence of a matter sector, \cref{eq:staticoddcc1,eq:staticoddcc2} coincide when $\bar p = \bar \rho = 0$.
A detailed discussion of these issues within an EFT  framework for the fluid --- along the lines of modern EFT approaches~\cite{Dubovsky:2005xd,Dubovsky:2011sj} --- remains absent.

We have presented the equations describing a non-rotating star. The study of linear perturbation theory for slowly rotating stars was pioneered in \rcite{1971ApJ166175I,1991RSPSA.433..423C,Kojima:1992ie,1993ApJ...414..247K,1993PThPh..90..977K,Ferrari:2007rc} (see also \rcite{Pani:2013pma} for a review). These equations were later solved to compute the tidal response in \rcite{Pani:2015hfa,Landry:2015zfa}. We will return to the calculation of the tidal Love numbers in \Cref{part:part3Love} below.

\newpage

\addtocontents{toc}{\protect\newpage} %

\part{Love numbers}
\label{part:part3Love}

\epigraphhead[]{
\epigraph{What's Love got to do with it?}{\textsc{Tina Turner}}
}

\section{Gravitational computation of tidal response coefficients}
\label{sec:love-compute}

The Love numbers can be defined within the effective field theory (EFT) or post-Newtonian (PN) frameworks, as discussed in \cref{sec:EFT}. However, in order to compute them, one needs to match to a general relativistic calculation in which the full Einstein equations are solved under the assumption of tidal-field boundary conditions. This section provides explicit results for Love numbers and other response coefficients from gravitational perturbation theory in the linear and nonlinear r\'egimes, covering Schwarzschild and Kerr black holes and neutron stars in four dimensions. 
We further review extensions to charged black holes, higher-dimensional black holes, black branes, black strings, and supergravity black holes, and discuss Love numbers beyond pure general relativity. Results
are summarized in \cref{TableLoveNumbers}.

\begin{table}[p]
\centering
\renewcommand{\arraystretch}{1.3}
\setlength{\tabcolsep}{5pt}
\small
\begin{tabular}{>{\raggedright\arraybackslash}p{4.2cm}
                >{\raggedright\arraybackslash}p{10.8cm}}
\toprule
\textbf{Black hole type} & \textbf{Love number representative references} \\
\midrule
Schwarzschild BHs (4D)
& Fang \&  Lovelace~\cite{Fang:2005qq}; Damour \& Nagar~\cite{Damour:2009vw};
  Binnington \& Poisson \cite{Binnington:2009bb};
Kol \& Smolkin \cite{Kol:2011vg};
Chakrabarti \emph{et al.}~\cite{Chakrabarti:2013lua}; G\"urlebeck~\cite{Gurlebeck:2015xpa};
  Hui \emph{et al.}~\cite{Hui:2020xxx}  \\[2pt]
Kerr BHs (4D)
&  Landry \& Poisson \cite{Landry:2015zfa};  Pani \emph{et al.}~\cite{Pani:2015hfa};  Le Tiec \& Casals~\cite{LeTiec:2020spy}; Le Tiec \emph{et al.}~\cite{LeTiec:2020bos};
  Chia~\cite{Chia:2020yla}; Goldberger \emph{et al.}~\cite{Goldberger:2020fot};
  Charalambous \emph{et al.}~\cite{Charalambous:2021mea} \\[2pt]
Reissner--Nordstr\"{o}m BHs
& Cardoso \emph{et al.}~\cite{Cardoso:2017cfl}; Pere\~{n}iguez \& Cardoso~\cite{Pereniguez:2021xcj};
  Rai \& Santoni~\cite{Rai:2024lho};
  Grilli \emph{et al.}~\cite{Grilli:2024fds}; Gounis \emph{et al.}~\cite{Gounis:2025tmt}
\\[2pt]
Kerr--Newman BHs 
& Pani \emph{et al.}~\cite{Pani:2013wsa};
  Ma \emph{et al.}~\cite{Ma:2024few}; Kehagias \emph{et al.}~\cite{Kehagias:2024yzn} \\[2pt]
Higher-$D$ non-rotating BHs
& Kol \& Smolkin \cite{Kol:2011vg};
  Cardoso \emph{et al.}~\cite{Cardoso:2019vof};
  Chakravarti \emph{et al.}~\cite{Chakravarti:2018vlt}; Hui \emph{et al.}~\cite{Hui:2020xxx};
Charalambous~\cite{Charalambous:2024tdj}; Hadad \emph{et al.}~\cite{Hadad:2024lsf}; Akhtar \emph{et al.}~\cite{Akhtar:2025nmt}; Berens \emph{et al.}~\cite{Berens:2025jfs}; Xia \emph{et al.}~\cite{Xia:2025zfp} \\[2pt]
5D rotating (Myers--Perry)
& Rodriguez \emph{et al.}~\cite{Rodriguez:2023xjd};
  Charalambous \& Ivanov~\cite{Charalambous:2023jgq};
  Glazer \emph{et al.}~\cite{Glazer:2024eyi};
  Berens \emph{et al.}~\cite{Berens:2025jfs} \\[2pt]
Black ring
& Rodriguez \emph{et al.}~\cite{Rodriguez:2023xjd}
\\[2pt]

Large-$D$ BHs
& Glazer \emph{et al.}~\cite{Glazer:2024eyi} \\[2pt]
Ultraspinning Myers--Perry BHs
& Rodriguez \emph{et al.}~\cite{Rodriguez:2023xjd};
  Charalambous \& Ivanov~\cite{Charalambous:2023jgq};
  Glazer \emph{et al.}~\cite{Glazer:2024eyi}
  \\[2pt]
Black strings and $p$-branes
& Rodriguez \emph{et al.}~\cite{Rodriguez:2023xjd}; Charalambous~\cite{Charalambous:2024tdj}; Charalambous \emph{et al.}~\cite{Charalambous:2025ekl};
 Tan~\cite{Tan:2020hog}
  \\[2pt]
(A)dS black holes
& Emparan \emph{et al.}~\cite{Emparan:2017qxd};
  Nair \emph{et al.}~\cite{Nair:2024mya};
  Franzin \emph{et al.}~\cite{Franzin:2024cah};
  Yusmantoro \emph{et al.}~\cite{Yusmantoro:2025ylw} \\[2pt]
Vaidya BHs (dynamical mass) & Capuano \emph{et al.}~\cite{Capuano:2024qhv} \\[2pt]
BTZ (3D)
& Bhatt \& Singha~\cite{Bhatt:2024mvr} \\[2pt]

Magnetic BHs
& Pere\~{n}iguez \& Karnickis~\cite{Pereniguez:2025jxq} \\[2pt]
Fermionic perturbations & Chakraborty \emph{et al.}~\cite{Chakraborty:2025zyb}; Pang \emph{et al.}~\cite{Pang:2025myy}  
\\
Beyond-GR BHs 
& Cardoso \emph{et al.}~\cite{Cardoso:2017cfl}; Cardoso \emph{et al.}~\cite{Cardoso:2018ptl};
   Katagiri \emph{et al.}~\cite{Katagiri:2023umb}; Katagiri \emph{et al.}~\cite{Katagiri:2024fpn};
  Cano~\cite{Cano:2025zyk};
  Garcia-Saenz \& Lin \cite{Garcia-Saenz:2025urd};
  Bhattacharyya \emph{et al.}~\cite{Bhattacharyya:2025slf};  Chakraborty  \emph{et al.}~\cite{Singha:2025xah}; Barbosa \emph{et al.}~\cite{Barbosa:2025uau};
  Barbosa \emph{et al.}~\cite{Barbosa:2026qcv}; Wang \emph{et al.}~\cite{Wang:2026qst} \\[2pt]
  Supergravity BHs
& Cveti\v{c} \emph{et al.}~\cite{Cvetic:2021vxa};
  Cveti\v{c} \emph{et al.}~\cite{Cvetic:2024dvn};
  Cveti\v{c} \emph{et al.}~\cite{Cvetic:2026wht} \\[2pt]
Exotic compact objects/Analog BHs/Non-singular BHs & Mendes \emph{et al.}~\cite{Mendes:2016vdr};  Cardoso \emph{et al.}~\cite{Cardoso:2017cfl}; Maselli \emph{et al.}~\cite{Maselli:2017cmm}; Sennet \emph{et al.}~\cite{Sennett:2017etc};  Cardoso \& Pani~\cite{Cardoso:2019rvt}; Herdeiro \emph{et al.}~\cite{Herdeiro:2020kba}; 
De Luca \emph{et al.}~\cite{DeLuca:2024nih,DeLuca:2025zqr}; Di Russo \emph{et al.}~\cite{DiRusso:2024hmd}; Silvestrini \emph{et al.}~\cite{Silvestrini:2025lbe};   
Wang \emph{et al.}~\cite{Wang:2025oek};
  Coviello \emph{et al.}~\cite{Coviello:2025pla} 
\\ [2pt]
Environmental effects
&  Cardoso \& Duque~\cite{Cardoso:2019upw}; De Luca \& Pani~\cite{DeLuca:2021ite}; De Luca \emph{et al.}~\cite{DeLuca:2022xlz};  Katagiri \emph{et al.}~\cite{Katagiri:2023yzm};  Cannizzaro \emph{et al.}~\cite{Cannizzaro:2024fpz}; Arana \emph{et al.}~\cite{Arana:2024kaz};  
  Chakraborty \emph{et al.}~\cite{Chakraborty:2024gcr};
   De Luca \emph{et al.}~\cite{DeLuca:2024uju}\\[2pt]
Nonlinear Love numbers
& Poisson \& Vlasov~\cite{Poisson:2009qj}; G\"urlebeck~\cite{Gurlebeck:2015xpa}; Poisson \cite{Poisson:2020vap}; De Luca \emph{et al.}~\cite{DeLuca:2023mio};
  Riva \emph{et al.}~\cite{Riva:2023rcm};
  Iteanu \emph{et al.}~\cite{Iteanu:2024dvx};
  Combaluzier-Szteinsznaider \emph{et al.}~\cite{Combaluzier-Szteinsznaider:2024sgb};
  Kehagias \& Riotto \cite{Kehagias:2024rtz};
  Gounis \emph{et al.}~\cite{Gounis:2024hcm};
  Parra-Martinez \& Podo~\cite{Parra-Martinez:2025bcu}  \\ [2pt]
Dynamical tidal response & See Table \ref{Table3} in Section \ref{subsection:DynLN} \\ [2pt]
\bottomrule
\end{tabular}
\caption{Summary of black hole types and representative references for tidal Love number computations in four-dimensional GR, higher dimensions, extended theories of gravity, environmental effects, and nonlinear contributions. The list includes calculations of responses for various fields (e.g., gravitational, scalar). Note that the listed references may contain only partial results and be derived under specific assumptions (such as axisymmetry, test scalar fields, etc.); we refer the reader to the main text for further discussion, and to the original references for a comprehensive account of the assumptions underlying each work.}
\label{TableLoveNumbers}
\end{table}

The term ``compact object'' refers to an object whose radius \(\Rstar\) is comparable to its Schwarzschild radius \(\rs\), i.e., \(\Rstar \sim \mathcal{O}(\rs)\). This condition characterizes the strong-gravity, or UV, r\'egime. When the compactness parameter satisfies
$\Rstar/\rs-1\ll1$, the radial solutions of the perturbation equations~\eqref{eq:teuk-rad-boxed} naturally take a characteristic asymptotic form similar to their Newtonian counterpart \eqref{eq:Uellm-2},
\begin{equation}
\phi \rightarrow \sum_{\ell=2}^\infty  \mathcal{E}_{\ellm} \, r^{-s}\left[ r^\ell\left( 1 + a_1 \frac{\rs}{r} +\dots \right)  +    \frac{\lambda_{\ell}}{r^{\ell+1}}   \left(1 + b_1 \frac{\rs}{r} +\dots\right) \right]  
+ {\cal O}(\dot{{\cal E}},{\cal E}^2).
\label{eq:intro1}
\end{equation}
Here $\phi$ is a generic proxy for a suitably normalized spin-$s$ field,\footnote{For example, the perturbation field $\phi$ can be expressed either in terms of the Newtonian potential $U$ appearing in the metric perturbation 
$g_{tt} = -1 + \frac{2 U}{c^2}$ with $s=0$, cf. \cref{eq:intro-g-ansatz}, or in terms of the Weyl scalars $\Psi_0$ ($s=2$) and $\Psi_4$ ($s=-2$), as used in the Teukolsky formalism. The two descriptions are related because the Weyl scalars are constructed from second derivatives of the metric perturbation. In the weak-field limit, this implies
\[
\Psi_0 \sim \frac{1}{c^2} m^i m^j  \partial_i \partial_j U,\qquad \Psi_4 \sim \frac{1}{c^2} \bar{m}^i \bar{m}^j \partial_i \partial_j U.
\]} and as in \cref{eq:Uellm-2} the coefficients $\mathcal{E}_\ellm$ are constants encoding the external field, $a_i$ and $b_i$ are relativistic corrections determined by the Einstein equations, and we denote by $\lambda_{\ell}$ the ratio of the two falloffs, which can be complex; one would usually identify the real part with the Love numbers and the imaginary part with the dissipative response coefficients, although we remind the reader that this identification should only be made after matching to an EFT or PN calculation.

This makes the computation of Love numbers a direct parallel to the Newtonian case, offering an intuitive bridge between classical and relativistic tidal responses. We look for a solution $\phi$ with the appropriate boundary condition imposed at the surface of the object, and measure the static response from the falloff at infinity, cf. \cref{eq:intro1}. For compact objects the boundary condition is provided by matching to the interior fluid solution at the surface. For black holes the relevant boundary conditions at the horizon are regularity for static perturbations and the ingoing wave condition in the generic dynamical case. In Boyer-Lindquist coordinates 
this corresponds to
\begin{equation}
\phi \sim  \begin{cases}
    \mathrm{const.}\times (r-r_+)^{-i\alpha_+}, & r \to r_+ \\
    c_1\, r^{\ell-s} + c_2\, r^{-\ell-s-1}, & r\to\infty
\end{cases},
\label{eq:BC}
\end{equation}
where
\begin{equation}
    \alpha_+ \equiv \frac\beta2\left(\omega - m \Omega_{\mathrm H}\right) \pm i \frac{s}{2},
\end{equation}
with $\beta = 2\rs r_+/(r_+-r_-)$ the inverse surface gravity of the black hole defined in \cref{eq:inv-hawking-temp}.
The factor in the near-horizon solution can be written
\begin{align}
    (r-r_+)^{-i \alpha_+}& = e^{-i \alpha_+ \log(r-r_+)} \nonumber\\
    &\sim e^{-i (\omega - m \Omega_{\mathrm H}) r_*},
\end{align}
where the tortoise coordinate for Kerr
(satisying $\Delta\,\dd\rst = (r^2 + a^2)\,\dd r$)
is
\begin{equation}
    \rst = r + \frac{r_+^2+a^2}{r_+-r_-}\log\left(\frac r{r_+}-1\right) - \frac{r_-^2+a^2}{r_+-r_-}\log\left(\frac r{r_-}-1\right),
\end{equation}
mapping the exterior black hole region $r\in(r_+,\infty)$ to the full real line $\rst\in(-\infty,+\infty)$. This makes the ingoing/outgoing nature of the field $\phi$ unambiguous: it is just given by the sign of the exponent.

At this point, it is useful to return to our discussion in \cref{sec:Lovecovariantly} on a possible ambiguity in the definition of the response coefficients.
In flat space, the two independent solutions are clearly separated, and no ambiguity arises.
However, once gravitational corrections are included ($a_i,b_i \neq 0$), higher-order terms in the expansion of the tidal solution will appear at the same order in $r$ as terms in the response solution. Concretely, there will be an ambiguity between $a_{i+2\ell+1}$ and $b_i$ (considering $a_0=b_0=1$) as long as $2\ell+1\in\mathbb N$.
Different equivalent ways to deal with this issue without relying on EFT or PN matching have been adopted in the literature.
Convenient approaches include working in general spacetime dimension, where the overlap does not occur for generic values, and only at the end taking the physical limit~\cite{Kol:2011vg,Hui:2020xxx,Hadad:2024lsf} (see \cref{sec:highD});
treating the multipole number as a continuous parameter $\ell\in\mathbb R$ and taking the limit $\ell\in\mathbb N$ at the end~\cite{LeTiec:2020bos,LeTiec:2020spy,Charalambous:2021mea,Rai:2024lho,Mano:1996vt}.
Finally, the ambiguity can also be resolved within effective field theory, where a systematic matching procedure cleanly separates the contributions associated with different EFT operators (see \cref{sec:WEFT}).

Before we proceed with the presentation of Love numbers for individual cases, it is worth clarifying an important subtlety regarding the tidal 
response coefficient $\lambda_{\ell}$, which is generically complex. 
At first glance, one might wonder how $\operatorname{Im}(\lambda_{\ell}) \neq 0$ is consistent with the metric perturbation being real. In the weak field limit, the physical metric perturbation can be obtained by summing over all mode contributions and taking the real part to ensure the metric remains real. Expanding in frequency and angular modes, this can be expressed compactly as
\begin{equation}
h^{\text{phys}}_{tt}
= \sum_{\ell m} \operatorname{Re}\left[
R_{\ell m}(r) S_{\ell m}(\theta) e^{-i\omega t + i m \varphi}
\right],
\end{equation}
where $R_{\ell m}(r)$ are the radial mode functions and $S_{\ell m}(\theta)$ are the (spin-weighted) spheroidal harmonics. The inclusion of the complex conjugate ensures that the perturbation is real-valued, as required for a physical spacetime metric.
By expanding the real part explicitly and keeping only the leading asymptotic terms of \cref{eq:intro1}, with $s=0$ and $\phi = R_{\ell m}(r)$,
the physical static metric perturbation (up to overall real constants) can be written as\footnote{We stress that the asymptotic expansion \eqref{eq:perturbation} in powers of $r$ is valid in the small-$\omega$ limit, or in the near-zone approximation ($r\ll\omega^{-1}$).} 
\begin{equation}
\begin{aligned}
h^{\rm physical}_{tt} 
&= \sum_{\ellm}\mathcal{E}_{\ellm}  S_\ellm(\theta)r^{\ell}
\Bigg[
\left(1 + a_1\frac{r_s}{r} + \cdots\right)\cos(\omega t - m\varphi) \\
&\quad +
\left(\frac{1}{r}\right)^{2\ell+1}
\left(1 + b_1\frac{r_s}{r} + \cdots\right)
\Big(
\operatorname{Re}(\lambda_\ell)\cos(\omega t - m\varphi)
-\operatorname{Im}(\lambda_\ell)\sin(\omega t - m\varphi)
\Big)
\Bigg]\\
& \simeq \sum_{\ellm}\mathcal{E}_{\ellm} S_\ellm(\theta)r^{\ell}
\Bigg[\cos(\omega t - m \varphi) 
+ \left| \lambda_{\ell}(\omega) \right| 
\cos\big(\omega t - m \varphi + \delta_{\ell}(\omega)\big) 
\left(\frac{1}{r}\right)^{2\ell+1}\Bigg],
\end{aligned}
\label{eq:perturbation}
\end{equation}
where $|\lambda_{\ell}|$ is the magnitude and $\delta_\ell$ is the phase of $\lambda_\ell$.\footnote{For a complex number $\lambda = a + i b$, we can write it in exponential (polar) form $\lambda = |\lambda| \, e^{i \delta}$, where the magnitude and phase are defined as $|\lambda| = \sqrt{a^2 + b^2}$ and the phase is the angle with respect to the real axis, $\delta = \arctan(b/a)$ taken in $(-\pi,\,\pi]$. Note also that \[
\operatorname{Re}(\lambda_\ell)\cos\theta - \operatorname{Im}(\lambda_\ell)\sin\theta
= |\lambda_\ell| \left(\cos\delta_\ell \cos\theta - \sin\delta_\ell \sin\theta\right)
= |\lambda_\ell| \cos(\theta + \delta_\ell).
\]}  We see that $\operatorname{Im}(\lambda_{\ell})$ when $\omega \ne 0$ produces a time delay in additon to an azimuthal phase lag in $\varphi$ between the tidal forcing and the metric response. The factor $\left(1/r\right)^{2\ell+1}$ acts as a multipole filter at large $r$. To be more precise, considering the $\ell = 2$ quadrupole which falls as $\left(1/r\right)^{5}$ to dominate at large $r$,
 the $\ell = 3$ octupole falls as $\left(1/r\right)^{7}$ and becomes negligible. Higher $\ell$ are exponentially irrelevant at large
separations. The observable metric at large $r$ is therefore dominated by the lowest
allowed $\ell$, which for a symmetric tidal environment is $\ell = 2$ --- this is why
gravitational wave models need only a handful of multipoles.

The complexity of $\lambda_{\ell}(\omega)$ is therefore 
a feature of the mode decomposition, not of the metric itself, and its real 
and imaginary parts carry distinct physical information. The reality condition 
on the metric translates into the following constraint on the tidal response 
coefficient
\begin{equation}
\lambda_{\ell\, -m}(-\omega) = \lambda_{\ell m}^*(\omega)
\end{equation}
which can be verified directly.

For $\omega=0$ the $\operatorname{Im}(\lambda_{\ell})$ term in \cref{eq:perturbation} produces an azimuthal phase lag
$\sim \sin(m\varphi)$ --- the metric response is rotated in $\varphi$ relative to 
the tidal source by $\pi/(2m)$. This is the static manifestation of frame-dragging: even 
though the external perturbation is time-independent in the inertial frame, 
the black hole's rotation twists the metric response azimuthally.  This effect has a concrete observational consequence in binary systems: when a small body orbits a rotating black hole, the tidal distortion of the larger body would be slightly twisted relative to the orbital plane. This contributes a measurable phase shift in the gravitational waveform that, in principle, allows measurement of the imaginary tidal Love numbers --- a direct probe of horizon dissipation that LISA-band sources may constrain.

\subsection{Static tidal response}
\label{sec:love-static}

Our aim in this section is to begin with the simplest setups and progressively introduce various bells and whistles. We will start off by describing the tidal responses for Schwarzschild and Kerr black holes in a purely static (time-independent) external environment in linear perturbation theory, showing that the Love numbers vanish for all four-dimensional black holes. We expand the discussion to include dynamical tidal responses in \cref{subsection:DynLN} and nonlinear tidal responses in \cref{sec:love-nonlin}

\subsubsection{Schwarzschild black holes}
\label{sec:love-static-sch}
Starting with the Schwarzschild solution as our background, we study linear perturbations describing the black hole in an external environment, encoding its Love numbers. This simplified setting allows us to clearly introduce the relevant framework and boundary conditions before moving on to the Kerr background, where the radial equations are more involved and the extraction of the corresponding Love numbers is correspondingly more subtle.

As a first step, we focus on the massless scalar, whose radial equation is well known and provides a simple and transparent illustration of the method.
While gravitational perturbations are intrinsically spin-2, the spin-0 case retains the essential radial structure of the gravitational perturbation equations and therefore serves as a useful “toy model” for understanding how an external tidal field elicits a response from a black hole. 
We employ this toy model for several reasons. First, it is technically simpler and allows us to present the method in a transparent way. Second, the gravitational problem is so closely analogous that most of the main qualitative conclusions still hold. Third, the spin-0 equation of motion is directly shared by a number of gravitational master variables of interest, including the Kaluza--Klein field $\phi=\ln(-g_{tt})$ in four dimensions \cite{Combaluzier-Szteinsznaider:2024sgb} (cf. \cref{sec:love-nonlin}) and the master field governing tensor-mode metric perturbations in higher spacetime dimensions 
$D>4 $ \cite{Kodama:2003jz,Ishibashi:2003ap} (cf. \cref{sec:statichighD}).

\paragraph{Scalar field}

Recall that the Klein--Gordon equation $\Box\phi=0$ expanded in spherical harmonics is \eqref{eq:sch-KG}
\begin{equation}
\partial_r\left(\Delta\partial_r\phi\right)+\frac{\omega^2r^4}{\Delta}\phi -\ell(\ell+1)\phi = 0.
\end{equation}
We remind the reader of the definitions $\Delta=r(r-\rs)=r^2f(r)$ and $f(r)=1-r/\rs$, where $\rs=2GM$ is the Schwarzschild radius.
To compute the static tidal response we may ignore time derivatives by setting $\omega=0$ (cf. \cref{sec:EFT}),
\begin{equation}\label{eq:kg-static}
\partial_r\left(\Delta\partial_r\phi\right)-\ell(\ell+1)\phi = 0.
\end{equation}
It is convenient to redefine the radial variable as
\begin{equation}
x \equiv \frac{\Delta'}\rs = \frac{2r}\rs-1,
\end{equation}
after which \cref{eq:kg-static} reveals itself to be Legendre's differential equation,
\begin{equation}\label{eq:phi-rad}
\partial_x \left[(1-x^2)\partial_x\phi\right]+\ell(\ell+1)\phi=0.
\end{equation}
This equation and its solutions are discussed in detail in \ref{app:legendre}. The solutions are the Legendre functions of the first and second kind,
\begin{equation}\label{eq:phi-sol-sch}
\phi = c_1P_\ell(x)+c_2Q_\ell(x).
\end{equation}
Asymptotically at infinity and the horizon they behave like
\begin{subequations}\label{eq:legendre-asymptotics}
\begin{align}
P_\ell(x) &\longrightarrow \begin{cases}
\frac{(2\ell)!}{\ell!^2}\left(\frac r\rs\right)^\ell,&r\to\infty \\
1,&r\to\rs
\end{cases}, \\
Q_\ell(x)&\longrightarrow \begin{cases}
\frac{\ell!(\ell+1)!}{(2\ell+2)!}\left(\frac \rs r\right)^{\ell+1},&r\to\infty \\
-\frac12\ln\left(\frac r\rs-1\right) - H_\ell,&r\to\rs
\end{cases},
\end{align}
\end{subequations}
where $H_n=\sum_{k=1}^nk^{-1}$ are the harmonic numbers. These values of $r$ have straightforward asymptotics because (along with $r=0$) they are two of the three regular singular points of \cref{eq:phi-rad}.

To fix the solution \eqref{eq:phi-sol-sch} we must impose two boundary conditions. One is the overall amplitude at infinity $\mathcal{E}_{\ell m}$, cf. \cref{eq:intro1}, which does not affect the computation of the tidal response $\lambda_\ell$. The other boundary condition is typically set at the surface of the object, which for a black hole we take to be the horizon. For a compact object such as a white dwarf or neutron star, this boundary condition is determined by the object's internal fluid dynamics, and is generally an admixture of the $P_\ell(x)$ and $Q_\ell(x)$ solutions.
For a black hole, we note that \(Q_\ell(x)\) diverges at the horizon. Imposing regularity of the solution, as required by the boundary condition~\eqref{eq:BC}, selects the regular branch and therefore sets 
\begin{equation}
c_2 = 0.
\end{equation}

To calculate the Love number, we need to follow the solution $\phi(x)=c_1P_\ell(x)$ from the surface of the object to infinity, where (cf. \cref{eq:intro1}) we read off the response coefficient from the coefficient in front of the decaying mode. From \cref{eq:legendre-asymptotics} we see that the growing solution (corresponding to the external tidal field) at infinity lives in $P_\ell(x)$, while the decaying solution (the induced response) is in $Q_\ell(x)$. Indeed, for integer values of $\ell$ the Legendre functions of the first kind $P_\ell(x)$ are more commonly known as Legendre polynomials because they are just polynomials in $x$, lacking any $1/r$ tails. As a result we find that the static, scalar Love numbers of Schwarzschild black holes vanish,
\begin{equation}
\lambda^{s=0}_\mathrm{Sch,BH} = 0.
\end{equation}
Around a compact object the static response generally does not vanish, as there is typically no reason to expect the internal dynamics of the star to yield zero contribution from $Q_\ell(x)$ at the surface. Following the same logic as for the black hole, the tidal response is $c_2/c_1$ (which can be determined just from the matching at the surface).
This contrast underscores a fundamental difference between black holes and ordinary celestial objects: whereas planets and stars respond to tidal forces with measurable deformations, Schwarzschild black holes are uniquely insensitive to such external influences. This feature has significant implications for gravitational-wave physics and black hole modeling.

\paragraph{Gravitational perturbations}

The generalization to gravitational perturbations is obtained by solving the Einstein equations in the static limit for linearized metric fluctuations. The calculation of the relativistic Love numbers of Schwarzschild black holes in general relativity was pioneered independently by Damour \& Nagar~\cite{Damour:2009vw} and Binnington \& Poisson \cite{Binnington:2009bb}.

As we saw in \cref{sec:BHPT-static}, the metric perturbations $h_0$ and $H_0$ are (up to rescalings) the natural master variables in the static limit~\cite{Fang:2005qq,Hinderer:2007mb,Damour:2009vw,Binnington:2009bb}.
Their static equations of motion are \cref{eq:h0-static,eq:H0-static},
\begin{align}
\Delta H_0'' + \Delta' H_0' - \frac{\rs^2+\ell(\ell+1)\Delta}\Delta H_0 &= 0\label{eqH0_lin} ,\\
h_0''+\frac4rh_0'-\frac{(\ell+2)(\ell-1)}\Delta h_0=0,\label{eqh0_lin}
\end{align}
where primes denote $\partial_r$.

Both of these are hypergeometric differential equations, which we discuss in detail in \ref{app:hypergeo}. Like the Legendre equation \eqref{eq:phi-rad} for the scalar field, which is itself a special case of the hypergeometric class, it possesses three regular singular points located at $r=(0,\rs,\infty)$ (indeed the hypergeometric equation is the general second-order ordinary differential equation with three regular and no irregular singularities). The odd equation \eqref{eqh0_lin} is already in hypergeometric form \eqref{eq:hypergeo} while the even equation \eqref{eqH0_lin} takes this form acting on $\Delta H_0$. The parameters in each case are
\begin{subequations}
\begin{align}
    (\mathfrak a,\mathfrak b,\mathfrak c)_\mathrm{even} &= (\ell+3,2-\ell,3),\\
    (\mathfrak a,\mathfrak b,\mathfrak c)_\mathrm{odd} &= (\ell+2,1-\ell,4),
\end{align}    
\end{subequations}
although for the even equation it is simpler to notice that (after changing to the variable $x=\Delta'/\rs$ but without rescaling $H_0$) it takes the form of the associated Legendre equation \eqref{eq:legendre-assoc} with $m=2$. This puts the (even-parity) spin-2 field $H_0$ on the same footing as the spin-0 field $\phi$, as both radial profiles satisfy the associated Legendre equation with $m=|s|$.
Using the asymptotics of the hypergeometric function listed in \ref{app:hypergeo-sols}, we find the horizon-regular solutions,
\begin{subequations}
\label{eq:H0h0linsol}
\begin{align}
    \Hf_0 (r) &  \propto P^2_{\ell}(\Delta'/\rs) , 
    \label{H0linsol} \\
    h_0 (r) & \propto {}_2F_1(1-\ell,\ell+2;4;r/\rs) ,
\label{h0linsol}
\end{align}
\end{subequations}
where $P_\ell^m$ is the associated Legendre polynomial and $_2F_1$ is the hypergeometric function.
and $\mathcal{E}^{(E/B)}_\ellm$ are constants corresponding to the asymptotic tidal field amplitudes, cf.~\cref{eq:intro-g-ansatz}.
Because the parameter $\mathfrak a$ in both hypergeometric equations is a negative integer, the hypergeometric series \eqref{eq:hypergeo-series} truncates to a finite polynomial \eqref{eq:hypergeo-series-poly}. As in the spin-0 case, our horizon-regular solutions are therefore pure polynomials in $r$, and we again infer the vanishing of the static tidal coefficients (see \cref{eq:intro-g-ansatz}):
\begin{equation}
k_{\ell}^{(\mathrm{E})}=k_{\ell}^{(\mathrm{B})}=0 .
\end{equation}
The gauge-invariant Love numbers, determined, for instance, by matching with the EFT as described  in \cref{sec:EFTapproach} (see, e.g., \cref{eq:matchingcEl}),
therefore vanish identically for Schwarzschild black holes. 

The gravitational computation can also be performed with the various master equations we have found, namely the Zerilli \eqref{eq:Z}, Regge--Wheeler \eqref{eq:RW}, and Teukolsky equations \eqref{eq:teuk-rad-boxed} (the last of which we will study on Kerr in the next section). Of these, only the Zerilli equation is not hypergeometric in the static limit (it is of the more general Heun type), and it can be made hypergeometric using the Chandrasekhar transformation \eqref{eq:sol-chandra}. All of these computations again yield vanishing Love numbers \cite{Kol:2011vg,Hui:2020xxx,Chia:2020yla}, as we would expect since the differential operators relating these quantities to the metric perturbations do not change the Love number boundary conditions.

\subsubsection{Kerr black holes}

Black hole rotation enriches the tidal response with new features, and a thorough understanding of the corresponding static Love numbers is essential to quantify how angular momentum modifies the deformability of the horizon and its induced multipole moments.
As summarized in \cref{sec:includingspin}, the coupling between the object's angular momentum and the external tidal field introduces new families of rotational tidal response coefficients.\footnote{The geometry of a tidally deformed black hole in the small-rotation limit and at higher orders  was derived in \rcite{Yunes:2005ve,OSullivan:2014ywd,Poisson:2014gka}.} Tidal deformations of slowly-spinning objects were studied in \rcite{Poisson:2014gka,Landry:2015zfa,Pani:2015hfa} (see also \rcite{Pani:2015nua,Landry:2017piv,Gagnon-Bischoff:2017tnz}), which found that the Love numbers of a black hole vanish through quadratic order in spin in the axisymmetric sector.
These results were placed on a firmer basis more recently by \rcite{LeTiec:2020spy,LeTiec:2020bos}, who computed the tidal response of a Kerr black hole in full generality and showed that the static Love numbers (i.e., the conservative tidal response coefficients) vanish for all multipole moments. As remarked in \rcite{Chia:2020yla} and independently studied in \rcite{Goldberger:2020fot,Charalambous:2021mea} in the language of point-particle EFT, the non-vanishing imaginary terms in the induced falloffs found in \rcite{LeTiec:2020spy,LeTiec:2020bos} correspond to a dissipative tidal response, often referred to as tidal heating~\cite{Hartle:1973zz,Hughes:2001jr,Poisson:2009qj}.
The vanishing of the conservative response was later interpreted from the perspective of hidden symmetries of the Kerr perturbations in \rcite{Hui:2021vcv,Hui:2022vbh,Charalambous:2021kcz,Charalambous:2022rre}, which we review in more detail in \cref{sec:sym}.

In the following, we briefly review the computation of the static response coefficients for spin-$s$ perturbations of Kerr black holes in general relativity.

The radial Teukolsky equation \eqref{eq:teuk-rad-boxed} in the static limit is (setting $G=1$)
\begin{equation}
   \left[\Delta^{-s}\partial_r\left(\Delta^{s+1}\partial_r\right)+\frac{a^2m^2+2isam(r-M)}\Delta - \ell(\ell+1)+s(s+1)\right]R^{(s)}_{\ell m \omega}(r) = 0,
   \label{eq:TeuAlls}
\end{equation}
where
\begin{equation}
\Delta = r(r-2M)+a^2 = (r-r_+)(r-r_-),
\qquad
r_\pm = M \pm \sqrt{M^2-a^2}.
\end{equation}
Here $a$ is the Kerr spin parameter (with dimensions of length), satisfying $0<a<M$ for a non-extremal black hole.
For compactness we write $\phi\equiv R^{(s)}_{\ell m \omega}(r)$ in this section.

\Cref{eq:TeuAlls} exhibits regular singular points at $r=r_\pm$, corresponding to the outer and inner horizons, and at $r=\infty$. Near the horizons the solution behaves as a power law,
\begin{equation}
\phi \sim (r-r_\pm)^{\alpha_\pm},
\end{equation}
with indicial exponents
\begin{equation}
\alpha_\pm = -\frac{s}{2} \pm i m \gamma,
\qquad
\gamma = \frac{a}{r_+ - r_-}.
\end{equation}
At spatial infinity one finds solutions going as
\begin{equation}
\phi(r) \sim r^{\ell-s} \quad \text{or} \quad r^{-(\ell+1+s)},
\end{equation}
corresponding respectively to the applied tidal field and the induced multipole response.

Because the equation possesses exactly three regular singular points, it can be mapped to the Gauss hypergeometric equation. To achieve this, introduce the dimensionless coordinate
\begin{equation}
z = \frac{r-r_+}{r-r_-},
\label{eq:coord}
\end{equation}
so that the outer horizon is at $z=0$, the inner horizon at $z=\infty$, and spatial infinity at $z=1$. We further redefine 
\begin{equation}
\phi(r) = (r-r_-)^{p}(r-r_+)^{q} w(z),
\label{eq:phisol}
\end{equation}
with exponents chosen as
\begin{equation}
p = -\frac{s}{2}-i m \gamma,
\qquad
q = -\frac{s}{2}+i m \gamma.
\end{equation}
In terms of $z$, \cref{eq:TeuAlls} takes the hypergeometric form \eqref{eq:hypergeo} with parameters
\begin{subequations}
\begin{align}
\mathfrak{a} &= 1+\ell -s, \\
\mathfrak{b} &= 1+\ell + 2 i m \gamma, \\
\mathfrak{c} &= 1-s +2 i m \gamma .
\end{align}
\label{eq:kerr-static-hypergeo-param}
\end{subequations}
As in the Schwarzschild case, the parameters are degenerate,
\begin{equation}
\mathfrak{c}-\mathfrak{a}-\mathfrak{b}
= -2\ell-1 \in \mathbb{Z}_{\le 0},
\end{equation}
which implies that $\mathfrak{c}-\mathfrak{a}-\mathfrak{b}$ is a negative integer. 
The general solution is the linear combination
\begin{align}
w(z) = {} & d_1\, {}_2F_1(\mathfrak{a},\mathfrak{b};\mathfrak{c};z)
+ d_2\, z^{1-\mathfrak{c}}
\, {}_2F_1(\mathfrak{a}-\mathfrak{c}+1,\mathfrak{b}-\mathfrak{c}+1;2-\mathfrak{c};z),
\label{eq:wsol}
\end{align}
with arbitrary constants $d_1,d_2$.
We impose boundary conditions at the horizon and at infinity. Regularity on the future horizon $r=r_+$ requires that the solution be analytic at $z=0$, which selects the branch proportional to ${}_2F_1(\mathfrak{a},\mathfrak{b};\mathfrak{c};z)$, i.e., $d_2=0$.
The full solution is then obtained by analytically continuing this branch to $z\to 1$ (i.e.,\ $r\to\infty$), where it becomes a linear combination of the two independent asymptotic behaviors,
\begin{align}
{}_2F_1(\mathfrak{a},\mathfrak{b};\mathfrak{c};z)
= {} & \frac{\Gamma(\mathfrak{c})\Gamma(\mathfrak{c}-\mathfrak{a}-\mathfrak{b})}{\Gamma(\mathfrak{c}-\mathfrak{a})\Gamma(\mathfrak{c}-\mathfrak{b})}
\, {}_2F_1(\mathfrak{a},\mathfrak{b};\mathfrak{a}+\mathfrak{b}-\mathfrak{c}+1;1-z) \\
& + (1-z)^{\mathfrak{c}-\mathfrak{a}-\mathfrak{b}}
\frac{\Gamma(\mathfrak{c})\Gamma(\mathfrak{a}+\mathfrak{b}-\mathfrak{c})}{\Gamma(\mathfrak{a})\Gamma(\mathfrak{b})}
\, {}_2F_1(\mathfrak{c}-\mathfrak{a},\mathfrak{c}-\mathfrak{b};\mathfrak{c}-\mathfrak{a}-\mathfrak{b}+1;1-z).
\label{eq:ACsol}
\end{align}
Substituting the solution \eqref{eq:wsol} with $d_2=0$ into \cref{eq:phisol} and using the asymptotic expansion \eqref{eq:ACsol}, we obtain the large-$r$ behavior
\begin{equation}
\phi(r) \sim r^{\ell-s}
\left(c_1 + c_2 r^{-(2\ell+1)}\right),
\label{eq:phiKerr0}
\end{equation}
which allows the static tidal response to be computed as
\begin{align}
\lambda_\ell^{\rm Kerr}
&= \frac{c_2}{c_1}\nonumber\\
&= \frac{
\Gamma(\mathfrak{a})
\Gamma(\mathfrak{b})
\Gamma(\mathfrak{c}-\mathfrak{a}-\mathfrak{b})
}{
\Gamma(\mathfrak{c}-\mathfrak{a})
\Gamma(\mathfrak{c}-\mathfrak{b})
\Gamma(\mathfrak{a}+\mathfrak{b}-\mathfrak{c})
}(r_+-r_-)^{2\ell+1}.
\label{eq:lambdaKerrs}
\end{align}
Substituting the parameters \eqref{eq:kerr-static-hypergeo-param} and simplifying yields\footnote{Although the individual gamma functions on the first line diverge for negative integer arguments, the poles cancel exactly between the numerator and denominator, leading to the finite final result. We have written the second line as it usually appears in the literature, but note that the product can be expressed using the Pochhammer symbol, $(a)_n = \Gamma(a+n)/\Gamma(a)$, as
\[ \prod_{n=1}^{\ell} \left(n^2 + 4 m^2 \gamma^2\right)=(1+2im\gamma)_\ell(1-2im\gamma)_\ell.\]
}
\begin{align}
\lambda_\ell^{\rm Kerr} &=
\frac{
\Gamma(-2\ell-1)\,
\Gamma(1+\ell-s)\,
\Gamma(1+\ell+2 i m \gamma)
}{
\Gamma(-\ell-s)\,
\Gamma(-\ell+2 i m \gamma)\,
\Gamma(2\ell+1)
}(r_+-r_-)^{2\ell+1}\nonumber\\
&= (-1)^{s+1} i m \gamma  \frac{(\ell + s)!  (\ell - s)!}{(2\ell + 1)! \, (2\ell)!}(r_+-r_-)^{2\ell+1}
\prod_{n=1}^{\ell} \left(n^2 + 4 m^2 \gamma^2\right).
\label{eq:lambdaKerrell}
\end{align}
This is imaginary, establishing that the static tidal response of Kerr black holes is purely dissipative in nature \cite{LeTiec:2020spy,LeTiec:2020bos}.

Several key features follow immediately. First, despite the presence of rotation, all static electric- and magnetic-type Love numbers vanish identically for non-extremal Kerr black holes in classical vacuum GR. This result holds for arbitrary spin $a<M$ and for all multipoles $\ell\ge2$. Physically, imposing regularity at the event horizon removes the real part of the decaying (response) mode and leaves only the externally applied tidal field and an imaginary (dissipative) response. Consequently, Kerr black holes do not develop induced multipole moments under static tidal deformations, reinforcing their no-hair-like character and sharply distinguishing them from material compact objects such as neutron stars. Second, the coefficients \eqref{eq:lambdaKerrell} are purely imaginary, depend on the azimuthal number $m$ while coupling to the Kerr black hole spin through the product $m \gamma$ and only involve odd powers of $m\gamma$. Third, the large-$\ell$ asymptotic behavior of \cref{eq:lambdaKerrell} takes the simple form, for  a generic black hole spin $a$, 
\begin{equation}
\lambda^{\rm Kerr}_{\ell}\sim (-1)^{s+1} \frac{\ell}{2^{4\ell+2}}  i \sinh(2\pi m \gamma)
\quad \text{as } \ell \to \infty \, ,
\label{eq:4.35}
\end{equation}
demonstrating that the linear response of the black hole is exponentially suppressed at large multipolar order. Fourth, analytic continuation shows that in the Schwarzschild limit $(a\to0)$ one recovers
\begin{equation}
\lim_{a\rightarrow 0}\lambda^{\rm Kerr}_{\ell} =\lambda_\ell^\mathrm{Sch}=0.
\end{equation}

\begin{figure}[h!]
  \centering
  \includegraphics[width=7.5cm]{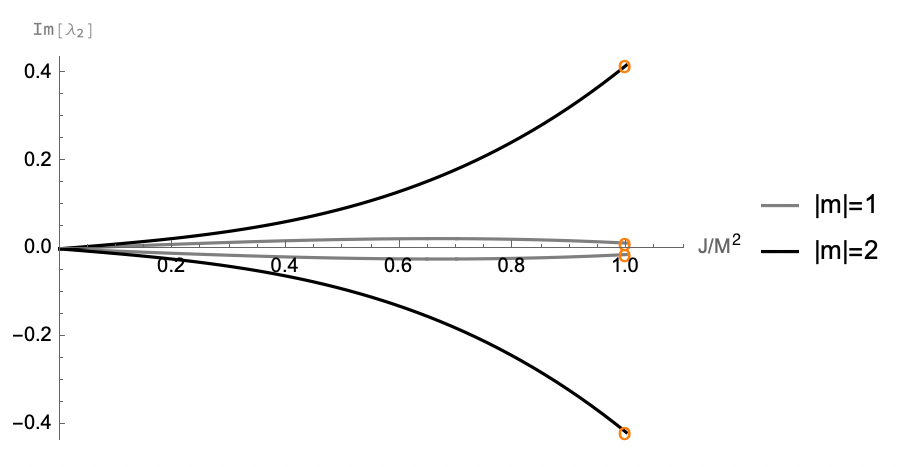}
  \includegraphics[width=7.5cm]{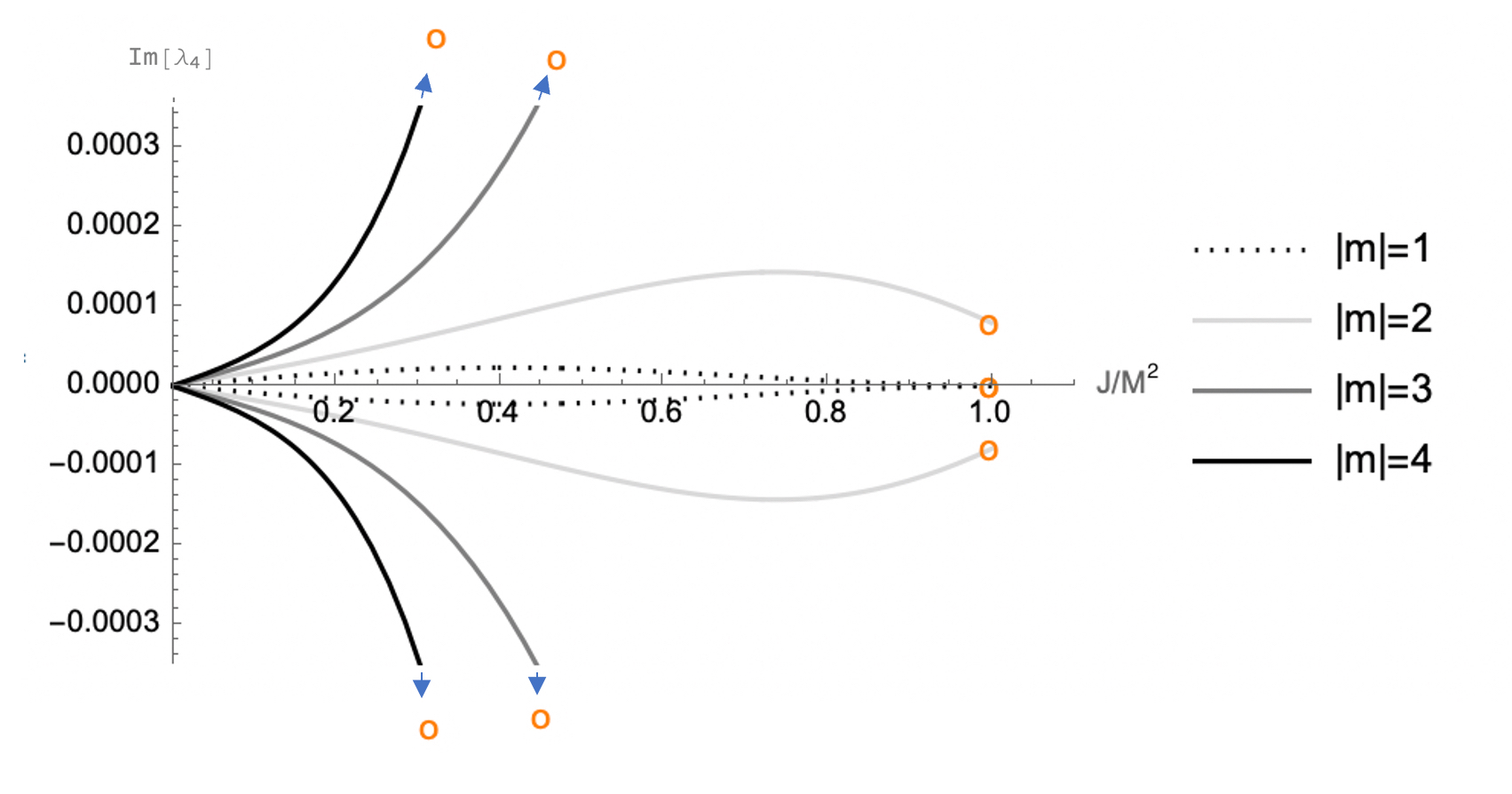}
  \caption{Static dissipative response coefficients \eqref{eq:lambdaKerrell} for Kerr black holes
  $k_\ell^\mathrm{Kerr} =\rs^{-2\ell-1}\lambda_{\ell}^\mathrm{Kerr}$
  for gravitational fields $s=2$ as a function of the dimensionless angular momentum $J/M^2$ for $\ell=2,4$. The endpoints of these curves correspond to extremal black holes, which are maximally rotating and have zero temperature.}
  \label{fig:love_number_l2}
\end{figure}

The dimensionless static dissipative coefficients
$k_\ell^\mathrm{Kerr} = \rs^{-2\ell-1}\lambda^{\rm Kerr}_{\ell}$
are shown in \cref{fig:love_number_l2} as a function of the dimensionless spin $J/M^2$, for gravitational ($s=2$) perturbations at $\ell=2$ and $\ell=4$. All curves vanish at $\chi=0$ (Schwarzschild) and grow monotonically with spin, reaching either the maximum at extremality or at an intermediate point before decreasing to a finite, non-zero value in the extremal limit $a/M \to 1$. This endpoint behavior, corresponding to a maximally rotating Kerr black hole, was first noted in \rcite{LeTiec:2020spy,LeTiec:2020bos} and has recently been computed explicitly using the Leaver--MST method \cite{Perry:2023wmm}.

 The full general-relativistic solution \eqref{eq:phiKerr0} with response coefficients \eqref{eq:lambdaKerrs} for a Kerr black hole can be  mapped onto the language of point-particle EFT, following the matching procedure reviewed in \cref{sec:matching}. This has been done explicitly in \rcite{Goldberger:2020fot,Charalambous:2021mea}. In the notation of \cref{eqlambdas} (following \rcite{Saketh:2023bul}), one finds~\cite{Goldberger:2020fot,Charalambous:2021mea,Saketh:2023bul}
\begin{equation}
\Lambda^{(\mathrm{E},\mathrm{B})}_{\omega^0,S^0} = \Lambda^{(\mathrm{E},\mathrm{B})}_{\omega^0,S^2} = \Lambda^{(\mathrm{E},\mathrm{B})}_{\omega^0,S^4} = 0 ,
\end{equation}
for the static Love numbers, and (in units of $-M(GM)^4$)
\begin{equation}
H^{(\mathrm{E},\mathrm{B})}_{\omega^0,S^1} = -\frac{8}{45}(\chi+3\chi^3) ,
\qquad
H^{(\mathrm{E},\mathrm{B})}_{\omega^0,S^3} = \frac{2}{3}\chi^3 ,
\label{eq:HHomega0S13}
\end{equation}
for the static dissipation numbers.
Here $\chi$ is the dimensionless spin parameter defined by $J\equiv \chi GM^2= \sqrt{(1/2)S^{ij}S_{ij}}$, with $J$ the spin magnitude and $S_{ij}$ the spin tensor~\cite{Saketh:2023bul}.

The computation of static tidal Love numbers has been extended to fermionic perturbations in \rcite{Chakraborty:2025zyb}. This work highlights a distinction between the (static) bosonic and fermionic solutions to the Teukolsky equation, suggesting that fermionic Love numbers of a Kerr black hole are non-zero, while their corresponding dissipation numbers vanish, contrary to the integer-spin results. A later extension to Reissner--Nordstr\"om black holes is discussed in \rcite{Pang:2025myy}.

\subsection{Dynamical tidal response}
\label{subsection:DynLN}

The tidal Love numbers discussed in the preceding sections are defined in the  static limit $\omega = 0$, characterizing the adiabatic deformation of a compact  object in a time-independent external field. In a realistic binary, however, the  tidal field varies on the orbital timescale, and one must consider the  \textit{dynamical} tidal response at finite driving frequency $\omega$. A comprehensive, if not exhaustive, classification of the literature on 
dynamical tidal deformations --- spanning approximately fifteen years of 
developments --- is provided in \cref{Table3}.

\begin{table}[h!]
\centering
\renewcommand{\arraystretch}{1.3}
\setlength{\tabcolsep}{5pt}
\small
\begin{tabular}{>{\raggedright\arraybackslash}p{4.2cm}
                >{\raggedright\arraybackslash}p{10.8cm}}
\toprule
\normalfont\bfseries Dynamical tides & \bfseries References \\
\midrule

Foundational / EFT
& Goldberger \& Rothstein \emph{et al.}~\cite{Goldberger:2004jt, Goldberger:2005cd};  Chakrabarti \emph{et al.}~\cite{Chakrabarti:2013lua}${}^\dagger$;  Porto~\cite{Porto:2007qi}; Endlich \& Penco~\cite{Endlich:2015mke}; Goldberger \emph{et al.}~\cite{Goldberger:2020fot} \\[6pt]

Waveform modelling
& Hinderer \emph{et al.}~\cite{Hinderer:2016eia}${}^\dagger$; Steinhoff \emph{et al.}~\cite{Steinhoff:2016rfi}${}^\dagger$; Schmidt \& Hinderer~\cite{Schmidt:2019wrl}${}^\dagger$; Pratten \emph{et al.}~\cite{Pratten:2021pro}${}^\dagger$;  Saketh \emph{et al.}~\cite{Saketh:2022xjb}; Pitre \& Poisson \cite{Pitre:2023xsr,Pitre:2025qdf}${}^\dagger$; Jakobsen \emph{et al.}~\cite{ Jakobsen:2023pvx}${}^\dagger$;  Mandal \emph{et al.}~\cite{Mandal:2023hqa,Mandal:2023lgy}${}^\dagger$; Hegade \emph{et al.}~\cite{HegadeKR:2024agt}${}^\dagger$; Yu \emph{et al.}~\cite{Yu:2024uxt}${}^\dagger$; Chakraborty \emph{et al.}~\cite{Chakraborty:2025wvs} \\[6pt]

BH perturbation theory / PN framework
& Poisson \cite{Poisson:2020vap}; Pitre \& Poisson \cite{Pitre:2023xsr}${}^\dagger$; Katagiri \emph{et al.}~\cite{Katagiri:2024wbg,Katagiri:2024fpn}; Chakraborty \emph{et al.}~\cite{ Chakraborty:2023zed, Chakraborty:2025wvs}; Bhatt \emph{et al.}~\cite{Bhatt:2024rpx}; Andersson \emph{et al.}~\cite{Andersson:2025iyd}${}^\dagger$; Hegade \emph{et al.}~\cite{HegadeKR:2024agt,HegadeKR:2024slr,HegadeKR:2025qwj,HegadeKR:2026iou}${}^\dagger$ \\[6pt]

Scattering amplitudes /  
Raman scattering
& Saketh \emph{et al.}~\cite{Saketh:2023bul};  Ivanov \emph{et al.}~\cite{Ivanov:2024sds, Ivanov:2025ozg, Ivanov:2026icp, Ivanov:2022qqt}; Saketh \emph{et al.}~\cite{Saketh:2024juq}${}^\dagger$; Caron-Huot  \emph{et al.}~\cite{ Caron-Huot:2025tlq}; Correia \& Isabella \cite{Correia:2024jgr}; Correia \emph{et al.}~\cite{Correia:2025enx}; Jarequi \emph{et al.}~\cite{Jarequi:2026cyp}${}^\dagger$ \\[6pt]

EFT   
off-shell matching
&  Chakrabarti \emph{et al.}~\cite{Chakrabarti:2013lua}${}^\dagger$; Combaluzier--Szteinsznaider \emph{et al.}~\cite{Combaluzier--Szteinsznaider:2025eoc}
\\[6pt]

Dissipation, tidal heating ($\omega\neq0$)
& Goldberger \emph{et al.}~\cite{Goldberger:2005cd,Goldberger:2020fot}; Chia~\cite{Chia:2020yla}; Starobinsky  \& Churilov~\cite{Starobinskil:1974nkd}; Page \cite{Page:1976df}; Chia \emph{et al.}~\cite{Chia:2024bwc};  Saketh \emph{et al.}~\cite{ Saketh:2022xjb}; Bhatt  \emph{et al.}~\cite{  Bhatt:2024yyz}; %
Hegade \emph{et al.}~\cite{HegadeKR:2024slr,HegadeKR:2026iou}${}^\dagger$;
Kobayashi \emph{et al.}~\cite{Kobayashi:2025swn} \\ [6pt]

Dynamical Love numbers  (RG running)
& Chakrabarti \emph{et al.}~\cite{Chakrabarti:2013lua}${}^*$;  Charalambous  \emph{et al.}~\cite{Charalambous:2021mea}; Saketh \emph{et al.}~\cite{Saketh:2023bul}; Perry \& Rodriguez \cite{Perry:2023wmm}; Ivanov \emph{et al.}~\cite{Ivanov:2024sds,Ivanov:2026icp}${}^*$; Katagiri \emph{et al.}~\cite{Katagiri:2024wbg}; De Luca \emph{et al.}~\cite{DeLuca:2024ufn}; Chakraborty \emph{et al.}~\cite{Chakraborty:2025wvs}; Caron-Huot \emph{et al.}~\cite{Caron-Huot:2025tlq}${}^*$; Combaluzier--Szteinsznaider \emph{et al.}~\cite{ Combaluzier--Szteinsznaider:2025eoc}${}^*$; Kobayashi \emph{et al.}~\cite{Kobayashi:2025vgl}${}^*$
\\[6pt]

Spinning objects
& Landry \& Poisson \cite{Landry:2015snx}${}^\dagger$; Steinhoff \emph{et al.}~\cite{Steinhoff:2021dsn}${}^\dagger$; Gupta  \emph{et al.}~\cite{Gupta:2020lnv, Gupta:2023oyy}${}^\dagger$
\\[6pt]

Shell EFT
& Kosmopoulos \emph{et al.}~\cite{Kosmopoulos:2025vgl}${}^*$; Bhattacharyya \emph{et al.}~\cite{Bhattacharyya:2026itm}${}^*$ \\[6pt]

Neutron star dynamical tides (normal-mode description) 
& Hinderer \emph{et al.}~\cite{Hinderer:2016eia}${}^\dagger$; Steinhoff \emph{et al.}~\cite{Steinhoff:2016rfi}${}^\dagger$; Andersson \& Pnigouras~\cite{Andersson:2019dwg}${}^\dagger$; Passamonti \emph{et al.}~\cite{Passamonti:2020fur,Passamonti:2022yqp}${}^\dagger$;  Steinhoff \emph{et al.}~\cite{Steinhoff:2021dsn}${}^\dagger$; Yu \& Lau~\cite{Yu:2025ptm}${}^\dagger$;  Ghosh \emph{et al.}~\cite{Ghosh:2025glz}${}^\dagger$;  Pereira \emph{et al.}~\cite{Pereira:2025xsi}${}^\dagger$;  Pnigouras \emph{et al.}~\cite{Pnigouras:2025muo}${}^\dagger$
\\[6pt]

Symmetries and dualities
& Hui \emph{et al.}~\cite{Hui:2022vbh}; Charalambous \emph{et al.}~\cite{Charalambous:2021kcz}; Ivanov \& Zhou~\cite{ Ivanov:2022hlo}; Combaluzier--Szteinsznaider \emph{et al.}~\cite{Combaluzier--Szteinsznaider:2025eoc}; Chandrasekhar \emph{et al.}~\cite{Chandrasekhar:1975zza, Chandrasekhar:1985kt}; Berti \emph{et al.}~\cite{ Berti:2009kk}; Solomon~\cite{Solomon:2023ltn} %
 \\[6pt]

Teukolsky equation
& Teukolsky \cite{Teukolsky:1973ha}
\\[6pt]

MST formalism
& Mano \emph{et al.}~\cite{Mano:1996vt}; Sasaki \& Tagoshi~\cite{Sasaki:2003xr} \\

\bottomrule
\end{tabular}

\caption{A comprehensive (though not exhaustive) classification of references on  dynamical tidal deformations of black holes and neutron stars (for the  latter, see \cref{sec:dynamicaltidesNS} for a more complete list). 
We introduce the symbol ${}^\dagger$ to highlight works with a substantial focus on neutron stars, as opposed to black holes (which do not carry a dagger). Moreover, in the context of dynamical Love numbers, the asterisk ${}^*$ denotes references in which a prescription is introduced to compute finite contributions to the EFT response coefficients, in addition to their logarithmic running. 
We do not distinguish between works that compute effects at linear versus higher order in $\omega$, nor between those that study scalar versus gravitational dynamical effects; we  refer the reader to the individual references for  a precise account of the underlying assumptions and approximations. We also include in the table more general or classical references that provide useful context for the works listed.
}
\label{Table3}
\end{table}

The  frequency-dependent response is defined schematically through\footnote{See~\cref{sec:dissipation} for a more precise definition at the level of the point-particle EFT~\cite{Goldberger:2004jt,Goldberger:2005cd}.}
\begin{equation}
    Q_{\ell m}(\omega) = \lambda_\ell(\omega)\,\mathcal{E}_{\ell m}(\omega),
\label{eq:dynamicall}
\end{equation}
where $\mathcal{E}_{\ell m}(\omega)$ is the Fourier-transformed external tidal multipole. The response function $\lambda_\ell(\omega)$ encodes two physically distinct  effects: its real part describes a conservative (elastic) deformation, and its  imaginary part describes irreversible energy absorption. 
Assuming that the internal dynamics occur on a timescale much shorter than that of the external probing field ($\omega^{-1}$), \cref{eq:dynamicall} can be expanded in powers of $\omega$. For black holes, we will assume $\omega \rs \ll 1$, i.e., we work with frequencies smaller than the leading quasinormal-mode frequencies. For neutron stars, we will assume that the internal timescale is set by the hydrodynamical timescale of the star.

The study of horizon absorption, tidal torquing, tidal heating, greybody factors, and superradiance is an old subject~\cite{Starobinskil:1974nkd,Starobinskii:1973vzb,Maldacena:1996ix}. More recently, a formalism to compute the horizon absorption of a rotating black hole induced by an external gravitational perturbation --- expanded in powers of the ratio of the black hole mass to the radius of curvature of the external universe --- was developed in \rcite{Poisson:2004cw} and later applied in \rcite{Chatziioannou:2012gq,Chatziioannou:2016kem} to evaluate tidal heating and torquing of a Kerr black hole at subleading order.
Time-dependent tidal effects were instead reformulated and pioneered in the modern language of EFT in \rcite{Goldberger:2005cd,Chakrabarti:2013lua}, and the study of time-dependent responses and dynamical Love numbers has since developed rapidly.
This progress has been driven both by their intrinsic theoretical interest as probes of near-horizon physics and by their role in gravitational waveform modeling~\cite{Hinderer:2016eia,Steinhoff:2016rfi,Poisson:2020vap,Charalambous:2021mea,Pratten:2021pro,Saketh:2022xjb,Pitre:2023xsr,Chakraborty:2023zed,Perry:2023wmm,Saketh:2023bul,Jakobsen:2023pvx,Mandal:2023hqa,Ivanov:2024sds,Katagiri:2024wbg,HegadeKR:2024agt,Yu:2024uxt,DeLuca:2024ufn,Katagiri:2024fpn,Bhatt:2024rpx,Saketh:2024juq,Chakraborty:2025wvs,Pitre:2025qdf,Kobayashi:2025swn}. In particular, it is well known that, in contrast to the static Love numbers (obtained from the $\omega \rightarrow 0$ limit of $\lambda_\ell(\omega)$ in \cref{eq:dynamicall}), frequency-dependent effects in $\lambda_\ell(\omega)$ are non-vanishing for black holes.\footnote{A note of caution: some works in the literature refer to dynamical Love numbers as any frequency correction appearing in the expansion~\eqref{eq:dynamicall}, without distinguishing between time-reversal-invariant and time-reversal-breaking contributions. Here we follow the most common convention and use the term \emph{dynamical Love numbers} only for the conservative tidal response at non-zero $\omega$, reserving the term \emph{tidal heating} for the dissipative contribution (see also the terminology section in the Introduction).} For example, a black hole absorbs radiation through its horizon, which at linear order in frequency corresponds to a positive imaginary part of the response coefficients~\cite{Chia:2020yla,Charalambous:2021mea,Chia:2024bwc}.

A number of works have addressed the computation of the dynamical tidal response of black holes and compact objects from complementary perspectives. These approaches include the Mano--Suzuki--Takasugi (MST) formalism~\cite{Mano:1996vt,Sasaki:2003xr}, which expresses the solution as a series of special functions truncated at a chosen order in the small-frequency limit~\cite{Kobayashi:2025vgl}; black hole perturbation theory (e.g., \cite{Poisson:2020vap,Pitre:2023xsr,Katagiri:2024wbg,Chakraborty:2025wvs}); on-shell scattering amplitudes and gravitational Raman scattering~\cite{Saketh:2023bul,Ivanov:2024sds,Saketh:2024juq,Caron-Huot:2025tlq,Ivanov:2025ozg,Ivanov:2026icp} (see also \rcite{Ivanov:2022qqt,Correia:2025enx,Correia:2025dyy,Markovic:2025kvr}); worldline EFT matching from one-point functions~\cite{Chakrabarti:2013lua,Combaluzier--Szteinsznaider:2025eoc,Kobayashi:2025vgl}; and shell EFT~\cite{Kosmopoulos:2025vgl,Bhattacharyya:2026itm}. 

At second order in $\omega$, it is well understood that dynamical Love numbers, unlike their static counterparts, exhibit a logarithmic dependence on distance. This behavior is interpreted as a classical example of renormalization-group running in the point-particle EFT description of the compact object~\cite{Chakrabarti:2013lua,Charalambous:2021mea,Saketh:2023bul,Ivanov:2024sds,Katagiri:2024wbg,Chakraborty:2025wvs,Caron-Huot:2025tlq}. Although the coefficient of the logarithm has been computed using the methods above, as well as symmetry-based arguments~\cite{Perry:2023wmm}, most results available until recently were incomplete. A full computation of the renormalized dynamical Love numbers cannot be performed at tree level but requires evaluating certain classical higher-loop diagrams in the point-particle EFT after subtracting ultraviolet divergences.

For a test scalar field, some complete results have been obtained from the perspective of scattering amplitudes \cite{Ivanov:2024sds,Caron-Huot:2025tlq,Ivanov:2026icp}. In particular, \rcite{Caron-Huot:2025tlq} introduced an efficient method based on a Born-series expansion \cite{Correia:2024jgr} to resum gravitational nonlinearities and perform matching at very high order in $GM$ for gravitational perturbations. This approach was subsequently applied in \rcite{Combaluzier--Szteinsznaider:2025eoc} to obtain the first complete calculation of the renormalized dynamical tidal Love numbers of Schwarzschild black holes in general relativity, including scheme-dependent finite terms. Independently, the same problem was investigated in \rcite{Kobayashi:2025vgl} using the MST formalism (see also \rcite{Chakrabarti:2013lua}).
Taking the $\omega \rightarrow 0$ limit, these results  also provide a firm basis for the computation of EFT graviton amplitudes at $\mathcal{O}((GM)^5)$, confirming the vanishing of the static Love number couplings in the point-particle EFT~\cite{Goldberger:2004jt,Goldberger:2005cd}.

\subsubsection{Schwarzschild black holes}

The dynamical tidal response has two physically distinct components. The \textit{dissipative} part --- the imaginary component of $\lambda_\ell(\omega)$ --- encodes irreversible energy absorption through the horizon and has been known since the foundational work of Starobinsky and Page~\cite{Starobinskil:1974nkd,Page:1976df}. See~\rcite{Goldberger:2005cd,Chakrabarti:2013lua,Goldberger:2019sya,Goldberger:2020fot,Goldberger:2020wbx,Charalambous:2021mea,Creci:2021rkz,Saketh:2023bul,Chia:2024bwc} for recent results on horizon absorption and linear-in-$\omega$ tidal effects within the EFT framework.

The \textit{conservative} (real) part describes a reversible elastic deformation and has only been understood more recently, as discussed above. 
Dynamical Love numbers are generically non-zero and exhibit logarithmic running with scale, with two sources of running identified: a universal gravitational dressing and a self-induced running set by lower-order tidal Wilson coefficients.
In the case of a free scalar theory, complete results --- including both finite scheme-dependent contributions and logarithmic running at second order in $\omega$ --- have been obtained in \rcite{Ivanov:2024sds,Caron-Huot:2025tlq,Ivanov:2025ozg,Ivanov:2026icp} using on-shell scattering amplitudes. The complete dynamical gravitational response of Schwarzschild black holes was subsequently computed within the point-particle EFT framework in \rcite{Combaluzier--Szteinsznaider:2025eoc} by matching the renormalized one-point function in the EFT to the classical field profile in GR.

In \rcite{Combaluzier--Szteinsznaider:2025eoc} (see also \rcite{Caron-Huot:2025tlq}), the computation proceeds by solving the Regge--Wheeler (odd-parity) and Zerilli (even-parity) equations perturbatively in the small-frequency limit $\rs\omega \ll 1$. After resumming gravitational nonlinearities, one finds for the Fourier transform of the response function $K(\tau)$ in \cref{eq:KpmGR} an expression of the schematic form
\begin{equation}
    K_\ell(\omega) = c_\ell^{(0)} 
    + i\omega   c_\ell^{(1)} 
    + \omega^2 c_\ell^{(2)} + \cdots,
\end{equation}
where the static coefficient  vanishes, $c_\ell^{(0)} = 0$, and the leading 
dynamical contribution enters at first order in $\omega$ through the 
dissipative channel. 
For example, for the quadrupole $\ell=2$ in the even sector of perturbations, one finds
\begin{equation}
    \frac{45}{2\pi \Mp^2}K^{(\mathrm{E})}_{\ell=2}(\omega) = i\omega \rs^6-\omega^2\rs^7\left(-\frac{797}{1260}+\log(\mu \rs)\right) + \mathcal{O}(\omega^3),
    \label{eq:elecl0}
\end{equation}
where $\mu$ is the renormalization scale. Note that the coefficient of the linear term in $\omega$ and the coefficient of the logarithmic running have long been known~\cite{Goldberger:2005cd,Chakrabarti:2013lua}.
An expression for the non-logarithmic coefficient at second order in $\omega$ was derived in \rcite{Chakrabarti:2013lua} by solving the Regge--Wheeler equation~\eqref{eq:RW} with a point-source term, using the MST method~\cite{Mano:1996mf,Leaver:1986vnb,Sasaki:2003xr} and an analytic continuation in $\ell$ in the limit $\delta_\ell \equiv \ell-2 \to 0$ (see also \rcite{Kobayashi:2025vgl} for a more recent treatment). 
However, it was only very recently that the finite terms at order $\omega^2$ were robustly obtained within a dimensional-regularization scheme \cite{Combaluzier--Szteinsznaider:2025eoc}, thanks to a resummation of gravitational nonlinearities in the worldline EFT \cite{Caron-Huot:2025tlq}.

The logarithmic term in \cref{eq:elecl0} signals that the dynamical tidal operators mix under renormalization-group flow. This contrasts sharply with the static sector, where the vanishing of $c_\ell^{(0)}$ is exact and scheme independent, protected by symmetries (see \cref{sec:sym} for a summary). A notable structural feature of \rcite{Combaluzier--Szteinsznaider:2025eoc} is that the even- and odd-parity dynamical Love numbers become equal in an appropriate renormalization scheme, reflecting the Chandrasekhar duality between the Regge--Wheeler and Zerilli equations~\cite{Chandrasekhar:1975zza,Chandrasekhar:1985kt,Berti:2009kk,Solomon:2023ltn}, as discussed in \cref{sec:chandra}.

The scheme dependence of $K_\ell(\omega)$ is not a pathology but a standard feature of EFT. Physical observables, such as the gravitational-wave phase, are scheme-independent combinations of Wilson coefficients, even though the latter may depend on the scale at which the system is probed. The physical content of the dynamical Love numbers therefore resides in the logarithmic running %
and in observable waveform corrections, which enter the gravitational-wave phase formally at 8PN order.

A noteworthy property to mention is that, for Schwarzschild black holes, explicit calculations show that the imaginary part of $K_\ell(\omega)$ at linear order in $\omega$ (for generic $\ell$) is directly related to the beta function of dynamical Love numbers (e.g., \cite{Goldberger:2005cd,Chakrabarti:2013lua,Combaluzier--Szteinsznaider:2025eoc,Chakraborty:2025wvs}),
\begin{equation}
\omega \rs \operatorname{Im} K_\ell(\omega) = -   {\mu}\frac{\dd}{\dd\mu} K_\ell(\omega) + \mathcal{O}(\omega^3).
\end{equation}
At first glance, this connection may seem surprising. The long-distance logarithmic running of the conservative response is sourced entirely by the Schwarzschild potential and is therefore universal, whereas the dissipative response is a finite-size effect specific to black holes. At subleading orders, this relation becomes more intricate~\cite{Caron-Huot:2025tlq}. An alternative perspective is that the discontinuity of the conservative part of $K$ is proportional to the dissipative part, with proportionality factor $-\beta \omega/2$, where $\beta$ denotes the inverse Hawking temperature \eqref{eq:inv-hawking-temp}. Remarkably, this relation appears to extend to the most transcendental components of the response kernel at higher orders in $\omega$ in the examples studied in \rcite{Caron-Huot:2025tlq}. All this is suggestive of a fluctuation--dissipation-type correspondence. Nonetheless, a fully fundamental understanding of this connection remains elusive.

\subsubsection{Kerr black holes}

While the static tidal Love numbers of black holes vanish 
identically~\cite{Fang:2005qq,Damour:2009vw,Binnington:2009bb,
LeTiec:2020spy,LeTiec:2020bos,Chia:2020yla,Charalambous:2021mea,
Charalambous:2021kcz}, the tidal response of a Kerr black hole 
becomes non-trivial when the external tidal field varies in time. 
As in the Schwarschild case, the dynamical response has two qualitatively distinct 
contributions: a \emph{conservative} piece $\kappa_{\ell m}$, encoded by the real 
part of the frequency-dependent tidal coefficient and characterizing 
genuine tidal deformation of the geometry, and a \emph{dissipative} 
piece $\nu_{\ell m}$, encoded by the imaginary part and characterizing the 
irreversible exchange of energy and angular momentum through 
the event horizon \cite{Chia:2020yla,Charalambous:2021mea,
LeTiec:2020spy},
\begin{align}
\lambda_{\ell}(\omega) 
= \kappa_{\ell m} + i\nu_{\ell m}.
\end{align}

For Kerr black holes, the two sectors are 
intertwined in a way that has no analogue in the non-rotating 
case: the broken spherical symmetry mixes multipoles of different 
parity~\cite{Saketh:2023bul}, and the horizon angular velocity 
$\Omega_{\mathrm H}$ \eqref{eq:ang-vel-hor} shifts the effective tidal frequency from $\omega$ 
to the superradiant frequency $\omega - m\Omega_{\mathrm H}$, so that 
even a \emph{static} external tidal field ($\omega = 0$) drives 
a non-zero dissipative response proportional to $m\Omega_{\mathrm H}$, 
which changes sign in the superradiant 
r\'egime~\cite{Hartle:1973zz,Teukolsky:1974yv,Chia:2020yla}. 
Understanding this dynamical response is essential both 
conceptually --- as a window into the near-horizon structure 
and symmetries of the Kerr 
geometry~\cite{Charalambous:2021mea,Charalambous:2021kcz,Hui:2022vbh} --- and practically, since the dynamical 
tidal coefficients of spinning black holes enter binary 
gravitational waveforms at lower effective post-Newtonian 
order than their Schwarzschild counterparts, owing to the 
$\omega \to \omega - m\Omega_{\mathrm H}$ 
shift~\cite{Saketh:2023bul,Chia:2024bwc,Perry:2023wmm,
Bhatt:2024yyz}.
We now turn to a more detailed, historical review.

\paragraph{Origins: 1972--1974}

Hawking and Hartle~\cite{Hawking:1972hy} showed that the area of the event horizon of a rotating black hole increases in 
the presence of a non-axisymmetric or time-dependent perturbation, and that the 
coupling between the black hole's rotation and an orbiting companion becomes 
significant precisely when the orbital angular velocity approaches $\Omega_{\mathrm H}$. 
The vanishing of the torque at $\omega = m\Omega_{\mathrm H}$ is the first physical 
instance of this combination in tidal physics.

Hartle~\cite{Hartle:1973zz} subsequently made this quantitative in the 
slow-rotation limit, computing the rate at which a stationary external matter 
distribution decreases the angular momentum of a Kerr black hole via tidal 
friction. The result is proportional to $\omega - m\Omega_{\mathrm H}$: when the tidal 
field co-rotates with the horizon ($\omega = m\Omega_{\mathrm H}$), there is no net 
torque, while for $\omega < m\Omega_{\mathrm H}$ the black hole loses angular momentum 
to the orbit; this is the tidal superradiance r\'egime. This established the 
combination $\omega-m\Omega_\mathrm{H}$ as the fundamental measure of how far a tidal perturbation is from 
co-rotating with the horizon.

Independently, in the wave-scattering context, Starobinsky~\cite{Starobinskii:1973vzb} 
and Starobinsky and Churilov~\cite{Starobinskil:1974nkd} showed that the 
amplification coefficient of electromagnetic and gravitational waves scattered 
by a Kerr black hole is governed by the sign of $\omega - m\Omega_{\mathrm H}$: waves 
with $\omega < m\Omega_{\mathrm H}$ are superradiantly amplified, extracting rotational 
energy from the black hole, while waves with $\omega > m\Omega_{\mathrm H}$ are absorbed. 
This is the superradiance condition, independent of the spin of the wave field.

The definitive early calculation of absorption cross sections using the 
Teukolsky equation --- which explicitly expresses all horizon fluxes in terms 
of $\omega - m\Omega_{\mathrm H}$ --- was given originally by Starobinsky and Churilov \cite{Starobinskil:1974nkd} and later by Teukolsky and 
Press~\cite{Teukolsky:1974yv} in their systematic study of the interaction of 
the Kerr black hole with gravitational and electromagnetic radiation. The low-frequency expansion of the absorption probability, at leading order in $\omega$, takes the form
\begin{equation}
\Gamma_{\ell m}(\omega) \propto \omega^{2\ell+1}\, \beta (\omega - m\Omega_{\mathrm H})
\prod_{n=1}^{\ell} \left[n^2 + \beta^2(\omega - m\Omega_{\mathrm H})^2\right]
+ \mathcal{O}(\omega^{2\ell+2}),
\end{equation}
where $\beta=2\rs r_+/(r_+-r_-)$ is the inverse surface gravity, cf. \cref{eq:inv-hawking-temp}.
This expression reproduces the vanishing of the absorption at 
$\omega = m\Omega_{\mathrm H}$ and the superradiant r\'egime, and it forms the basis for the 
modern tidal dissipation numbers. Higher-order terms in $\omega$ modify the detailed 
frequency dependence but do not affect the location of the superradiant zero.

\paragraph{Tidal heating and binary dynamics: 1990s--2000s}

The relevance of $\omega - m\Omega_{\mathrm H}$ for binary black hole dynamics was 
further developed by Poisson and Sasaki~\cite{Poisson:1994yf}, who computed 
gravitational-wave absorption into a Kerr black hole from a particle in 
circular orbit in the post-Newtonian framework, identifying the horizon flux 
as proportional to $\omega_{\rm orb} - m\Omega_{\mathrm H}$ at each multipole order. 
Hughes~\cite{Hughes:2001jr} extended this to generic Kerr orbits and showed 
that tidal heating --- the rate at which the black hole gains mass and loses 
angular momentum due to the tidal field of the companion --- changes sign 
precisely at $\omega_{\rm orb} = m\Omega_{\mathrm H}$, so that rapidly spinning black 
holes in the superradiant r\'egime are tidally \emph{accelerated} rather than 
braked.

In the context of wave scattering, the absorption cross section of the radial problem for Kerr black holes was computed in \rcite{Maldacena:1997ih}, where it was shown to be proportional to 
\begin{equation}
    \sigma \propto \omega - m\Omega_{\mathrm H},
\end{equation}
explicitly highlighting the crucial role of the combination $\omega - m \Omega_{\mathrm H}$ in determining the flux through the horizon.

\paragraph{Modern Love number era: 2020--present}

The connection between the combination $\omega - m\Omega_{\mathrm H}$ and the modern decomposition into conservative and dissipative tidal Love numbers was clarified in a series of works in 2020--2021. In particular, Chia~\cite{Chia:2020yla} derived the expression for the dynamical tidal coefficients of Kerr black holes at finite frequency, within a suitably defined near zone,
\begin{equation}
\lambda^{\rm Kerr}_{\ell}(\omega) = \frac{i}2
\frac{(\ell-2)!\,(\ell+2)!}{(2\ell)!\,(2\ell+1)!} \beta
(\omega - m \Omega_{\mathrm H})
\prod_{n=1}^{\ell} \left[ n^2 + \beta^2 (\omega - m \Omega_{\mathrm H})^2 \right],
\label{eq:LoveDiss}
\end{equation}
which is valid for arbitrary spin and multipole order.
We stress that \cref{eq:LoveDiss} is not exact to all orders in $\omega$ appearing in the product; it is valid only within the near-zone approximation. Specifically, it correctly captures the dissipative response to linear order in $\omega$~\cite{Chia:2020yla,Charalambous:2021mea,Hui:2022vbh}.

The real part of \cref{eq:LoveDiss} vanishes identically, implying that the conservative (static) Love numbers of rotating black holes are zero. In contrast, the non-vanishing imaginary part corresponds to non-trivial dissipative response numbers. This generalizes the Schwarzschild result to the rotating case, as the prefactor $\beta(\omega-m\Omega_\mathrm{H})$
implies that for rotating black holes, tidal dissipation persists even in the static limit $\omega = 0$. This reflects the fact that the intrinsic rotation of the black hole induces a relative time dependence with respect to an otherwise static external tidal field.\footnote{Equivalently, in the co-rotating frame of the Kerr black hole, a static tidal field is perceived as oscillating with frequency $-m\Omega_{\mathrm H}$, thereby generating dissipation.}

The EFT description of tidal interactions provides a complementary perspective. In the worldline EFT framework developed by Goldberger and Rothstein \cite{Goldberger:2005cd,Goldberger:2020fot}, dissipative effects are encoded in the coupling constants of an in-in effective action (see \cref{sec:dissipation}). Porto \cite{Porto:2007qi,Porto:2016pyg} extended this framework to spinning compact objects, while Goldberger, Li, and Rothstein \cite{Goldberger:2020fot} showed that horizon absorption for rotating black holes is governed precisely by the combination $\omega - m\Omega_{\mathrm H}$. Within this framework, even a static external field ($\omega = 0$) induces a non-zero dissipative response, as it corresponds to an effective frequency $-m\Omega_{\mathrm H}$ in the horizon frame.

A notable feature of \cref{eq:LoveDiss} is that the prefactor $\beta(\omega-m\Omega_\mathrm{H})$ is not simply proportional to $\omega$, which itself scales as the third power of the orbital velocity in a binary system. As a consequence, tidal dissipation for rotating black holes is enhanced by $1.5$ post-Newtonian (PN) orders relative to the Schwarzschild case~\cite{Poisson:1994yf,Tagoshi:1997jy}. Furthermore, the sign of $\beta(\omega-m\Omega_\mathrm{H})$ can be either positive or negative, reflecting the presence of superradiance: the dissipative response changes sign when $\omega < m\Omega_{\mathrm H}$.

\Cref{eq:LoveDiss} generalizes the static result in \cref{eq:lambdaKerrell} to finite frequency. Whereas earlier works interpreted these coefficients as non-zero Love numbers, it has now become more standard to refer to them as dissipative tidal response coefficients, since they are purely imaginary and do not correspond to conservative effects.
At leading order, the tidal dissipation numbers take the form
\begin{equation}
\nu_{\ell m} = r_s\,(\omega - m\Omega_{\mathrm H})\,
\frac{(\ell+s)!\,(\ell-s)!\,(\ell!)^2}{(2\ell)!\,(2\ell+1)!}
+ \mathcal{O}(\omega\Omega_{\mathrm H},\,\omega^2,\,\Omega_{\mathrm H}^2),
\end{equation}
making explicit that the dissipative response is governed entirely by the combination $\omega - m\Omega_{\mathrm H}$.
Further EFT matching calculations~\cite{Goldberger:2020fot,Charalambous:2021mea} confirmed the non-vanishing dissipative coefficients and revealed an underlying near-zone $\mathrm{SL}(2,\mathbb{R})$ symmetry structure \cite{Charalambous:2021kcz,Hui:2022vbh,Perry:2023wmm} (see \cref{sec:sym}).  This symmetry explains both the vanishing of conservative Love numbers and the universal linear dependence of the dissipative response on $\omega - m\Omega_{\mathrm H}$.

As a concrete example, the leading near-zone quadrupolar dissipation coefficient ($s=-2$) takes the form
\begin{equation}
\lambda_2(\omega)
= \frac{16i}{45}\,M^5\,(\omega - m\Omega_{\mathrm H})
\bigl[(\omega - m\Omega_{\mathrm H})^2 + 4\Omega_{\mathrm H}^2\bigr],
\label{eq:lambda2kerr}
\end{equation}
which reduces in the Schwarzschild limit ($a \to 0$) to
\begin{equation}
\lambda_2(\omega) \simeq \frac{16 i}{45} M^5 \omega^5,
\end{equation}
consistent with the low-frequency absorption scaling $\sigma_{\rm abs} \sim (M\omega)^{2\ell+1}$~\cite{Page:1976df}.

These results collectively demonstrate that the dissipative tidal response vanishes at $\omega = m\Omega_{\mathrm H}$ and changes sign in the superradiant r\'egime, while the conservative Love numbers remain identically zero at all multipole orders.

Finally, we note that numerical values of dynamical tidal coefficients reported in the literature depend on normalization conventions.\footnote{For instance, Chia~\cite{Chia:2020yla} defines dimensionful tidal deformabilities via induced multipole moments, while EFT approaches (e.g., \rcite{Charalambous:2021mea}) extract Wilson coefficients, and near-horizon analyses~\cite{Perry:2023wmm} define dimensionless response functions.} These definitions differ by factors of $M^{2\ell+1}$, combinatorial coefficients, and scheme-dependent logarithmic terms associated with renormalization-group running~\cite{Saketh:2023bul,Perry:2023wmm}. Nevertheless, all approaches agree on physical observables such as absorption cross sections and gravitational-wave phase shifts. In particular, the dissipative Love numbers universally scale as
\[
\nu_{\ell m} \propto \omega - m\Omega_{\mathrm H}
\]
at leading order.

In \rcite{Perry:2023wmm} the Love numbers in the
extremal limit were determined explicitly using the Leaver--MST method.
Expanding
order by order in frequency,
\begin{equation}
    k_{\ell m}(\omega) = \sum_{n=0}^{\infty}
    \left(\kappa_\ellm^{(n)} + i\nu_\ellm^{(n)}\right)\omega^n,
\end{equation}
the leading coefficients are
\begin{equation}
    \kappa^{(0)}_\ellm = 0, \qquad
    \nu^{(0)}_\ellm = -\frac{(-1)^s\,m^{2\ell+1}\,(\ell-s)!\,(\ell+s)!}
                      {2\,(2\ell)!\,(2\ell+1)!},
    \label{eq:PR_static}
\end{equation}
so the static conservative Love number vanishes, $\kappa^{(0)}=0$,
while the static dissipative coefficient $\nu^{(0)}$ is finite and
non-zero --- consistent with the results for non-extremal Kerr, where
the static conservative Love numbers also vanish~\cite{LeTiec:2020bos,Chia:2020yla,Charalambous:2021kcz,LeTiec:2020spy}. The first subleading
coefficients are
\begin{align}
    \kappa^{(1)}_\ellm &= -\nu^{(0)}\left(2\sum_{j=0}^{s-1}
    \frac{1}{\ell+j+1-s} + 2\log r\right), \\
    \nu^{(1)}_\ellm    &= -\nu^{(0)}\left(\frac{1+2\ell}{m}
    + \frac{m s(1+2\ell)(s-2\ell(\ell+1))}{2\ell^2(\ell+1)^2}\right).
\end{align}
Three features of this result are noteworthy. First, all dynamical
corrections are proportional to the zeroth-order dissipative
coefficient $\nu^{(0)}_\ellm$; this proportionality was previously observed
for non-extremal Kerr black holes in \rcite{Perry:2023wmm}. Second, extremal Kerr black holes
dissipate energy despite having zero Hawking temperature: $\nu^{(0)}_\ellm
\neq 0$ even at $\chi = 1$, so extremality does not eliminate the
dissipative tidal response. This is in contrast to the conservative sector,
where $\kappa^{(0)}_\ellm = 0$ holds for both extremal and non-extremal values of the spin.
Third, the logarithmic term in $\kappa^{(1)}_\ellm$ signals a classical
renormalization-group running of the Love number under changes of the
EFT matching scale $r$~\cite{Charalambous:2021mea,Ivanov:2022hlo},
an ambiguity that can be resolved by fixing a suitable renormalization scheme~\cite{Kol:2011vg,Charalambous:2021mea,LeTiec:2020spy,LeTiec:2020bos,Ivanov:2024sds, Caron-Huot:2025tlq, Combaluzier--Szteinsznaider:2025eoc, Ivanov:2025ozg, Ivanov:2026icp}.

\subsection{Nonlinear tidal response}
\label{sec:love-nonlin}

As a coalescing binary approaches merger, its dynamics becomes highly non-trivial. In particular, the adiabatic approximation breaks down and the nonlinearities of general relativity become increasingly important.
Closely related to dynamical Love numbers are nonlinear tidal distortions. Both effects can yield comparably significant contributions to the gravitational waveform, formally entering the dynamics at 8PN order.

Nonlinear Love number couplings of black holes have been computed explicitly in a number of cases~\cite{Poisson:2009qj,Gurlebeck:2015xpa,Poisson:2020vap,DeLuca:2023mio,Riva:2023rcm,Iteanu:2024dvx,Combaluzier-Szteinsznaider:2024sgb,Kehagias:2024rtz,Gounis:2024hcm,Parra-Martinez:2025bcu} (see \rcite{Pitre:2025qdf,Pani:2025qxs} for recent studies of relativistic nonlinear tides in neutron stars, and \cref{sec:neutronstarnonlinear} for a broader discussion).
In particular, \rcite{DeLuca:2023mio} considers a toy scalar-field example with self-interactions. 
\rcite{Poisson:2020vap,Riva:2023rcm,Iteanu:2024dvx} examine the quadratic gravitational tidal deformation of non-rotating compact bodies following different approaches:  \rcite{Poisson:2020vap} uses a post-Newtonian definition for the tidally induced multipole moments, focusing on the even-parity sector of perturbations, while \rcite{Riva:2023rcm,Iteanu:2024dvx} define the Love number couplings as in \cref{eq:E2B2-0nonlinear} at the level of the point-particle EFT and compute them by performing an explicit matching with the full general-relativistic solution. In particular, these references show that, in the case of four-dimensional general relativity, the quadratic Love numbers of Schwarzschild black holes vanish identically for generic multipolar tidal perturbations. 
This result has been extended to higher nonlinear order in perturbation theory by~\rcite{Gurlebeck:2015xpa,Combaluzier-Szteinsznaider:2024sgb,Kehagias:2024rtz,Gounis:2024hcm,Parra-Martinez:2025bcu}, which provide  evidence that the nonlinear Love numbers of Schwarzschild black holes vanish to all orders in perturbation theory. 

In the following, we first review the calculation at second order in perturbation theory, showing that the quadratic Love numbers vanish for Schwarzschild black holes in general relativity. We then discuss the extension of this result to all orders in perturbation theory.

\subsubsection{Second-order perturbation theory}

As we have seen  in \Cref{Part:BHPT}, black hole perturbation theory has a long and well-established history, originating with the seminal work of Regge and Wheeler in the late 1950s on linearized perturbations of Schwarzschild spacetime~\cite{Regge:1957td}. In more recent years, this framework has been systematically extended beyond the linear r\'egime, with significant progress achieved in the development of second-order perturbation theory~\cite{Nicasio:1998aj,Gleiser:1998rw,Brizuela:2006ne,Brizuela:2007zza,Nakano:2007cj,Brizuela:2009qd}. See, e.g., \rcite{Lagos:2022otp,Mitman:2022qdl,Cheung:2022rbm,Bourg:2024jme,Bucciotti:2024zyp,Bucciotti:2024jrv,Lagos:2024ekd,Bucciotti:2025rxa,Bourg:2025lpd}  for recent advances and applications in the study of the quasinormal mode spectrum of black holes.

In the context of gravitational tidal deformations, the study of nonlinear tides and the explicit calculation of higher-order Love numbers were pioneered in the PN framework in \rcite{Poisson:2009qj,Poisson:2020vap} and in the EFT framework in \rcite{Riva:2023rcm,Iteanu:2024dvx,Combaluzier-Szteinsznaider:2024sgb}. In this section, we review the second-order equations and their solutions in the static tidal limit. We mostly follow the notation and conventions of \rcite{Riva:2023rcm,Iteanu:2024dvx}.

The general idea is straightforward: we start from the general parametrization~\eqref{eq:h-even-odd} for the metric perturbation, substitute it into the Einstein equations, and expand to second order in perturbation theory. We will take the following expansion for the metric tensor,
\begin{equation}
     h_{\mn} = {}^{(1)}\! h_{\mn} + \frac{1}{\Mp} \, {}^{(2)}\! h_{\mn} +\dots ,
\end{equation}
where the superscript counts the perturbative order. The natural expansion parameter is $1/\Mp$; alternatively, one can introduce a bookkeeping parameter multiplying the metric perturbation to track the perturbative order.

Fixing a gauge simplifies both the derivation of the perturbation equations and the structure of their solutions. A natural and convenient choice, well-adapted to the symmetries of the problem, is the Regge--Wheeler defined in \cref{eq:RWgaugepart2}, which remains valid at higher orders~\cite{Nicasio:1998aj,Gleiser:1998rw,Brizuela:2006ne,Brizuela:2007zza,Nakano:2007cj,Brizuela:2009qd}.\footnote{This choice completely eliminates the gauge freedom at all orders, apart from  the lowest multipoles $\ell = 0$ and $\ell = 1$. See~\rcite{Iteanu:2024dvx} for a discussion.}
Following essentially the same steps as discussed in \cref{sec:BHPT} (see~\rcite{Riva:2023rcm,Iteanu:2024dvx} for details), one obtains the following set of second-order static (i.e., $\omega=0$) equations for the  even and odd metric components $H_0$ and $h_0$:
\begin{align}
\Delta {}^{(2)}\! H_0'' + \Delta' {}^{(2)}\! H_0' - \frac{\rs^2+\ell(\ell+1)\Delta}\Delta {}^{(2)}\! H_0 &= 
\sum_{\ell_1, m_1, \ell_2, m_2} {\cal S}_{H_0 \ \ell \ell_1  \ell_2}^{m  m_1 m_2} (r)
\label{eqH0_quad} ,\\
{}^{(2)}\! h_0''+\frac4r {}^{(2)}\! h_0'-\frac{(\ell+2)(\ell-1)}\Delta {}^{(2)}\! h_0 & = \sum_{\ell_1, m_1, \ell_2, m_2} {\cal S}_{h_0 \ \ell \ell_1 \ell_2}^{ m m_1 m_2}  (r)  ,
\label{eqh0_quad}
\end{align}
where ${\cal S}_{H_0 \ \ell \ell_1  \ell_2}^{m  m_1 m_2}$ and ${\cal S}_{h_0 \ \ell \ell_1 \ell_2}^{ m m_1 m_2}$ are source terms that depend exclusively on the linear-order fields ${}^{(1)}\! H_0$ and ${}^{(1)}\! h_0$, once all linear constraints variables are integrated out. Here $\ell_1$ and $\ell_2$  denote the angular momentum quantum numbers of two arbitrary linear modes, while $\ell$ denotes the multipole  of the resulting second-order sourced mode.
Note that starting from second-order, the axial and polar sectors couple to each other: the sources on the right-hand sides mix ${}^{(1)}\! H_0$ and ${}^{(1)}\! h_0$, i.e., the odd sector can now source the even sector and vice versa. The explicit expressions of the sources are not particularly illuminating for our purposes and we omit them here (they can be found in \rcite{Iteanu:2024dvx}). In what follows, we will just briefly comment on the main points. 

First, note that the right-hand sides involve sums over the harmonics of the linear solutions. In particular, the spherical harmonic coefficients enforce the standard angular momentum selection rules. The allowed combinations of linear-mode couplings for a given parity of the sum $\ell + \ell_1 + \ell_2$ are summarized in \cref{table:SR}.
\begin{table}[h!]
\centering
\renewcommand{\arraystretch}{1.6}  %
\setlength{\tabcolsep}{12pt}       %
\begin{tabular}{|>{\columncolor{gray!20}}c||c|c|}
\hline
\rowcolor{gray!30} $\ell + \ell_1 + \ell_2:$ even & ${}^{(1)}\!H_0^{\ell_2 m_2}$ & ${}^{(1)}\!h_0^{\ell_2 m_2}$ \\ [0.5ex]
\hline \hline
${}^{(1)}\!H_0^{\ell_1 m_1}$ & ${}^{(2)}\!H_0^{\ell m}$ & ${}^{(2)}\!h_0^{\ell m}$ \\
\hline
${}^{(1)}\!h_0^{\ell_1 m_1}$ & ${}^{(2)}\!h_0^{\ell m}$ & ${}^{(2)}\!H_0^{\ell m}$ \\
\hline
\end{tabular}
\quad
\begin{tabular}{|>{\columncolor{gray!20}}c||c|c|}
\hline
\rowcolor{gray!30} $\ell + \ell_1 + \ell_2:$ odd & ${}^{(1)}\!H_0^{\ell_2 m_2}$ & ${}^{(1)}\!h_0^{\ell_2 m_2}$ \\ [0.5ex]
\hline \hline
${}^{(1)}\!h_0^{\ell_1 m_1}$ & ${}^{(2)}\!h_0^{\ell m}$ & ${}^{(2)}\!H_0^{\ell m}$ \\
\hline
${}^{(1)}\!h_0^{\ell_1 m_1}$ & ${}^{(2)}\!H_0^{\ell m}$ & ${}^{(2)}\!h_0^{\ell m}$ \\
\hline
\end{tabular}
\vspace{0.5em}
\caption{Table adapted from \rcite{Iteanu:2024dvx}. Selection rules for the quadratic sources are shown for even (left) and odd (right) values of $\ell + \ell_1 + \ell_2$. In the superscripts of each field, we specify the perturbative order and restore the $\ell m$ labels for clarity.}
\label{table:SR}
\end{table}

Once the sources ${\cal S}_{H_0 \ \ell \ell_1  \ell_2}^{m  m_1 m_2}$ and ${\cal S}_{h_0 \ \ell \ell_1 \ell_2}^{ m m_1 m_2}$ are evaluated on the linearized solutions~\eqref{eq:H0h0linsol}, \cref{eqH0_quad,eqh0_quad} become inhomogeneous differential equations, which can be solved using the standard Green’s function method. As an illustration, we consider the case $\ell = \ell_1 = \ell_2 = 2$ with generic magnetic quantum numbers. Imposing regular boundary conditions at the horizon $r = \rs$ at second order, one finds:
\begin{align}
    {}^{(2)}\! H_0^{2 m} & =  \sum_{m_1,m_2} {}^0\! I_{222}^{m m_1 m_2} \frac14 \left[ \bigg. \mathcal{E}^{(\mathrm{B})}_{2 m_1} \mathcal{E}^{(\mathrm{B})}_{2 m_2} \frac{3}{2} \frac{r}{\rs}\bigg(12 \frac{r^3}{\rs^3} - 14 \frac{r^2}{\rs^2} + \frac{r}{\rs} - 1\bigg) 
-\mathcal{E}^{(\mathrm{E})}_{ 2 m_1 } \mathcal{E}^{(\mathrm{E})}_{2 m_2 }   \frac{r^2}{\rs^2} \bigg(2 \frac{r^2}{\rs^2} + \frac{r}{\rs}-3 \bigg)    \right],
\label{eq:H02nnl}
\\
{}^{(2)}\!h_0^{2 m} & = - \sum_{m_1,m_2} {}^0\! I_{222}^{m m_1 m_2} \mathcal{E}^{(\mathrm{B})}_{2 m_1} \mathcal{E}^{(\mathrm{E})}_{2 m_2} \frac{1}{2} \frac{r^3}{\rs^2} \bigg(\frac{r^2}{\rs^2} + \frac{r}{\rs} - 2\bigg) ,
\label{eq:h02nnl}
\end{align}
where we  introduced the quantity
\begin{equation}
{}^{0}\!I_{\ell \ell_1  \ell_2}^{m  m_1 m_2} = \int {\rm d} \Omega  \; \bar{Y}_{\ellm} (\theta, \varphi) Y_{\ell_1m_1} (\theta, \varphi) Y_{\ell_2m_2}   (\theta, \varphi) ,
\end{equation}
and where $\mathcal{E}^{(\mathrm{E})}_\ellm$ and $\mathcal{E}^{(\mathrm{B})}_\ellm$ are the amplitude of the external (linear) gravito-electric and gravito-magnetic tidal fields, respectively.

One crucial aspect worth emphasizing is that the quadrupole solutions  \eqref{eq:H02nnl}--\eqref{eq:h02nnl} are finite polynomials involving only  positive powers of $r$~\cite{Taylor:2008xy,Poisson:2009qj,Riva:2023rcm,Iteanu:2024dvx}. This property has been verified to hold for generic  multiplets $\ell \ell_1 \ell_2$ in \rcite{Iteanu:2024dvx} and, from a technical standpoint, follows  from two key ingredients: \textit{(i)} the structure of the homogeneous  equation (a degenerate hypergeometric equation), which is known to exhibit  symmetry constraints (see \cref{sec:sym}); and \textit{(ii)} the  special form of the source terms. Taken together, these features ensure that  a generic second-order static solution for gravitational perturbations (in the Regge--Wheeler gauge) around a Schwarzschild background reduces to a finite sum of positive powers  of $r$ (see also \rcite{DeLuca:2023mio} for a detailed discussion in the case 
of scalar fields).

The solutions \eqref{eq:H02nnl}--\eqref{eq:h02nnl} should be compared with  the general form of a second-order metric perturbation (see, e.g.,  \cref{eq:gttPN}~\cite{Poisson:2020vap,Pitre:2025qdf}). After being careful with the gauge  fixing, this comparison implies that $p_2 = 0$. In particular, all terms in the second line of \cref{eq:gttPN}, with the exception of the leading $r^4$ term --- corresponding to the quadratic correction to the quadrupole tidal field --- vanish for a Schwarzschild black hole in general relativity.

Similarly, the solutions \eqref{eq:H02nnl}--\eqref{eq:h02nnl}, as well as  their counterparts for generic $\ell$, can be matched onto the point-particle  EFT. A detailed analysis was carried out in  \rcite{Riva:2023rcm,Iteanu:2024dvx}, showing that all quadratic Love number  couplings in the higher-order action \eqref{eq:E2B2-0cubic} (and likewise at  higher multipoles) vanish for generic multiplets, i.e.,
\begin{equation}
c_{\ell\ell_1\ell_2}^{(\mathrm{E})} 
    = c_{\ell\ell_1\ell_2}^{(\mathrm{EB^2})} = 0 \, .
\end{equation}
These results extend the well-known vanishing of linear Love numbers for  Schwarzschild black holes in general relativity to second order in the tidal 
deformation.

This result strengthens the naturalness puzzle originally recognized at linear order for black hole Love numbers~\cite{Rothstein:2014sra,Porto:2016zng}. The persistence of vanishing Love numbers at nonlinear orders, combined with the special structure of the source terms and of the resulting second-order solutions, suggests that a reorganization of perturbation theory may be possible, enabling a more efficient computation of higher-order responses and the identification of hidden symmetry structures. In \cref{sec:nonlineartheory} we review how to resum the perturbative expansion for a special subset of tidal deformations, while in \cref{sec:sym} we focus on the discussion of symmetries.

\subsubsection{Nonlinear theory}
\label{sec:nonlineartheory}

It is well known that analytic solutions of Einstein's equations only exist in very special circumstances, typically when there is some degree of symmetry or when linearizing about a known background. In more general settings, the highly nonlinear Einstein equations can only be solved numerically. Nevertheless, the relative simplicity of the Love number setup --- most notably the fact that in the adiabatic limit the perturbed metric maintains a timelike Killing vector --- permits some statements to be made about tidal responses in fully nonlinear general relativity.

\paragraph{Kaluza--Klein dimensional reduction}

A useful tool to study Love numbers in a linear or nonlinear setting is \emph{dimensional reduction}, which we perform using a \emph{Kaluza--Klein decomposition}. This approach leverages the existence of a Killing field to covariantly decompose the full metric tensor into a more manageable set of variables.

Consider a standard $3+1$ splitting,
\begin{equation}\label{eq:metric-3-1}
    \dd s^2 = -e^{-\phi}(\dd t+A)^2+e^\phi\dd s_3^2,
\end{equation}
where $A=A_i\dd x^i$ and $\dd s_3^2=g_{3,\Ij}\dd x^i\dd x^j$. When $t$ is the coordinate adapted to a timelike Killing vector, $\xi^\mu\partial_\mu=\partial_t$, the fields $\phi$, $A_i$, and $g_{3,\Ij}$ by construction only depend on $x^i$, so we are left with a theory of a scalar, vector, and tensor in $D=3$.
The Einstein--Hilbert action decomposes as
\begin{equation}
    \sdg R = \sqrt{g_3}\left(R_3-\frac12(\partial\phi)^2+\frac14e^{-2\phi}F_\Ij F^\Ij\right),
\label{eq:dimredEH}
\end{equation}
where we raise and lower indices with $g_{3,\Ij}$ and its inverse, and $F=\dd A$ is a field strength tensor. Varying with respect to $A_i$ we find
\begin{equation}\label{eq:A-eom}
    \nabla^i(e^{-2\phi}F_\Ij)=0.
\end{equation}
This equation implies the existence of a dual scalar $\chi$ satisfying\footnote{This is easiest to see in the language of differential forms, which we review in \ref{app:diff-forms}. The $A$ equation of motion \eqref{eq:A-eom} reads $\dd(e^{-2\phi}\star_3\! F)=0$,
so by the Poincar\'e lemma there must exist a potential $\chi$ solving $F = e^{2\phi}\star_3\!\dd\chi$, which in coordinates is \cref{eq:A-dual}.}
\begin{equation}\label{eq:A-dual}
    F_\Ij = e^{2\phi}\epsilon_{ijk}\partial^k\chi.
\end{equation}
This reflects the well-known fact that a massless 3-vector is dual to a scalar. Substituting this back into the action we have
\begin{equation}
    \sdg R = \sqrt{g_3}\left(R_3-\frac12(\partial\phi)^2-\frac12e^{2\phi}(\partial\chi)^2\right).\label{eq:KK-3D}
\end{equation}
The 3-metric is non-dynamical, as in standard 3D general relativity, leaving the gravitational field encoded in the \emph{Kaluza--Klein scalars} $(\phi,\chi)$, which form an $\mathrm{SL}(2,\mathbb R)$/SO(2) sigma model. They may be interpreted as encoding the electric and magnetic degrees of freedom, respectively. For instance, when this ansatz is applied to the perturbed Schwarzschild metric \eqref{eq:h-even-odd}, $\phi$ turns out to only contain even modes and $\chi$ only odd modes. The $\mathrm{SL}(2,\mathbb R)$ symmetry group that emerges in the reduction from four dimensions to three is known as the \emph{Ehlers group} \cite{Ehlers:1957zz,Maison:2000fj,Mars:2001gd}.

Even though we performed the decomposition \eqref{eq:metric-3-1} in a coordinate system, everything we have done is fully covariant when we have a timelike Killing vector to work with; for instance, instead of thinking of $e^{-\phi}=-g_{tt}$ as a metric component, we can think of it as the norm of the Killing vector, $e^{-\phi}=-\xi^\mu\xi_\mu$.

If the metric is static rather than stationary, i.e., $A_i=0$, then the equation of motion for $\phi$ is linear,
\begin{equation}
    \Box_3\phi=0.
\end{equation}
It is largely this fact --- that the nonlinear Einstein equations for a static spacetime can be encoded in a single linear equation --- which can be used to make significant progress in computing Love numbers beyond perturbation theory \cite{Gurlebeck:2015xpa,Combaluzier-Szteinsznaider:2024sgb,Kehagias:2024rtz}, though see \rcite{Gounis:2024hcm,Parra-Martinez:2025bcu} for computations of nonlinear Love numbers in a stationary spacetime, i.e., with $A_i$ turned on.

It is frequently the case that there is a second Killing vector which commutes with $\xi^\mu$. In the black hole setting this is typically the azimuthal isometry $\eta^\mu\partial_\mu=\partial_\varphi$, in which case the spacetime is \emph{axisymmetric}. We can reduce the 3D theory \eqref{eq:KK-3D} to 2D along this direction by parametrizing the 3-metric as
\begin{equation}
    \dd s_3^2 = e^{2\gamma}\dd s_2^2 +\rho^2(\dd\varphi+b)^2,
\end{equation}
in terms of which the action is
\begin{equation}
    \sdg R = \sqrt{g_2}\rho\left(R_2 -\frac14\rho^2e^{-\gamma}(\dd b)^2 +\frac2\rho\partial\rho\cdot\partial\gamma-\frac12(\partial\phi)^2-\frac12e^{2\phi}(\partial\chi)^2\right).\label{eq:KK-2D-1}
\end{equation}
There are several simplifications available. The field $b$ can be set to zero by its equation of motion. Any 2D metric is conformally flat, so in a suitable coordinate system we can absorb the conformal factor into $\gamma$ and set $g_{2,IJ}=\delta_{IJ}$. We can identify such a set of coordinates by varying with respect to $\gamma$, for which the Euler--Lagrange equation is
\begin{equation}
    g^{IJ}_2\nabla_I\nabla_J\rho=\delta^{IJ}\partial_I\partial_J\rho=0.
\end{equation}
As long as $\rho$ is not a constant, it is a harmonic function on $\mathbb R^2$ which we can use as a coordinate. The natural choice for the second spatial coordinate is its dual $z$, defined by
\begin{equation}\label{eq:z-dual}
    \partial_Iz =- \epsilon_{IJ}\partial^J\rho\quad\Longleftrightarrow\quad\dd z = -\star_2\!\dd\rho.
\end{equation}
In these coordinates, known as \emph{Weyl canonical coordinates}, the 2-metric is $\dd s_2^2=\dd\rho^2+\dd z^2$.

Interpreting $\rho$ as a coordinate rather than a field, the three fields left in the picture are $(\phi,\chi,\gamma)$. The conformal factor $\gamma$ is non-dynamical; it obeys a constraint equation (namely the 2D Einstein equation) and can be obtained by line integration once a solution for $(\phi,\chi)$ is in hand. Combining $\phi$ and $\chi$ into the Ernst potential $\mathcal E$,
\begin{equation}
    \mathcal E \equiv e^{-\phi}+i\chi,
\label{eq:mathcalEphichi}
\end{equation}
the equation of motion for the sigma model is the \emph{Ernst equation},
\begin{equation}
    \partial^I(\rho\partial_I\mathcal E) = \frac\rho{\operatorname{Re}\mathcal E}(\partial\mathcal E)^2\quad\Longleftrightarrow\quad\dd(\rho\star_2\!\dd\mathcal E) = \frac{\rho}{\operatorname{Re}\mathcal E}\dd\mathcal{E}\wedge\star_2\dd\mathcal{E}.
\end{equation}
When the metric is static ($\chi=0$), this simplifies to
\begin{equation}
\left(\partial_\rho^2+\frac1\rho\partial_\rho+\partial_z^2\right)\phi\quad\Longleftrightarrow\quad\dd(\rho\star_2\!\dd\phi)=0.\label{eq:phi-weyl}
\end{equation}

From the coordinate expression we see that the equation of motion \eqref{eq:phi-weyl} is nothing other than the three-dimensional Laplace equation for a flat space in cylindrical coordinates, with $\rho$ and $z$ the radial and height coordinates, respectively. The resulting 4D metric, which is the general solution to Einstein's equations for a static and axisymmetric spacetime, is known as the \emph{Weyl metric} \cite{Stephani:2003tm},
\begin{equation}
    \dd s^2_\mathrm{Weyl} = -e^{-\phi(\rho,z)}\dd t^2 + e^{\phi(\rho,z)}\left[e^{2\gamma(\rho,z)}(\dd \rho^2+\dd z^2)+\rho^2\dd\varphi^2\right].\label{eq:weyl}
\end{equation}
The general \emph{stationary} and axisymmetric spacetime is of the same form with $\dd t\to\dd t+A_\varphi(\rho,z)\dd \varphi$.

The reduction to two dimensions expands the Ehlers group into the infinite-dimensional \emph{Geroch group} \cite{Geroch:1970nt,Breitenlohner:1986um,Maison:2000fj,Lu:2007zv,Lu:2007jc,Maison:1978es,Schwarz:1995td,Schwarz:1995af}. This occurs because there are now two Ehlers groups available, corresponding to the two possible orders in which we can perform the dimensional reduction from 4D to 2D. The intertwining of these groups leads to the Geroch symmetry. We comment on the relation of these symmetries to Love numbers in \cref{sec:sym-nonlin}.

\paragraph{Love numbers of distorted Schwarzschild black holes}

A Schwarzschild black hole placed in a static external environment, or \emph{distorted black hole} \cite{Geroch:1982bv}, is a Weyl metric \eqref{eq:weyl} with
\begin{equation}
    \phi = \phi_\mathrm{Sch} + \hat\phi,\quad\gamma = \gamma_\mathrm{Sch} + \hat\gamma,\label{eq:distorted-BH}
\end{equation}
where the Schwarzschild fields are
\begin{subequations}\label{eq:coord-weyl-sch}
\begin{align}
    e^{-\phi_\mathrm{Sch}} &= f(r),\\
    e^{-2\gamma_\mathrm{Sch}} &= 1+\frac{\rs^2\sin^2\theta}{4\Delta},
\end{align}
\end{subequations}
and the Weyl canonical coordinates $(\rho,z)$ are related to the Schwarzschild coordinates $(r,\theta)$ by
\begin{subequations}
\begin{align}
    \rho&=\sqrt\Delta\sin\theta,\\
    z &= \frac12\Delta'\cos\theta.
\end{align}    
\end{subequations}
We emphasize that the distortions $\hat\phi$ and $\hat\gamma$ are not necessarily small. We will ignore $\hat\gamma$ herein as it is non-dynamical. The main constraint on $\hat\phi$ is that it must be regular on the horizon, which ensures that the distorted black hole has the same distributional singularity as its undistorted counterpart \cite{Geroch:1982bv,Combaluzier-Szteinsznaider:2024sgb}.

Because the equation of motion \eqref{eq:phi-weyl} is linear in $\phi$, it is also satisfied by $\hat\phi$. In Schwarzschild coordinates, this reads
\begin{equation}\label{eq:eom-deformaton}
\partial_r(\Delta\partial_r\hat\phi)+\frac1{\sin\theta}\partial_\theta(\sin\theta\partial_\theta\hat\phi)=0.
\end{equation}
This is precisely the equation of motion \eqref{eq:KG-sch} for a static, azimuthally-symmetric massless scalar field $\phi(r,\theta)$, which we studied in \cref{sec:love-static-sch}. Since $\hat\phi$ obeys the same boundary condition as the scalar (regularity at the horizon), the same analysis applies, and we conclude that the distorted potential lacks a tidal tail in $r$ \cite{Combaluzier-Szteinsznaider:2024sgb,Kehagias:2024rtz},
\begin{equation}
    \hat\phi(r,\theta)=\sum_{\ell=2}^\infty c_\ell P_\ell(\Delta'/\rs)P_\ell(\cos\theta),
\label{eq:hatphinonlinear}
\end{equation}
with $c_\ell$ constants. The absence of tidal tails for $\hat\phi$ was first noted by G\"urlebeck, who demonstrated using source integrals that the Weyl multipole moments of distorted Schwarzschild black holes vanish \cite{Gurlebeck:2015xpa}. A careful matching calculation shows that this implies the vanishing of an infinite subset of nonlinear static response coefficients in the EFT \cite{Combaluzier-Szteinsznaider:2024sgb}.\footnote{Recall the discussion in \cref{sec:nonlinearities}. At sufficiently high perturbative order, multiple independent couplings can appear in the EFT for a fixed multiplet of angular momentum quantum numbers. In such cases, performing the matching with an axisymmetric solution such as \cref{eq:hatphinonlinear} is not sufficient to determine all the free couplings. At most, one can constrain one coupling per multiplet. This explains why, for axisymmetric distortions of Schwarzschild black holes, only a \textit{subset} of the EFT couplings can be determined through matching. See appendix~D of \rcite{Combaluzier-Szteinsznaider:2024sgb} for an explicit example.}

An analogous computation for the distorted Kerr black hole \cite{Tomimatsu:1984sx,BRETON19977} was performed in \rcite{Gounis:2024hcm}, which found a vanishing quadrupolar tidal tail under the assumption of an axisymmetric tidal distortion. A full computation of the nonlinear stationary and axisymmetric Kerr Love numbers, entailing an extension to all multipoles and an EFT matching calculation, has not yet been performed.

A generalization of these analyses beyond axisymmetry has been obtained in~\rcite{Parra-Martinez:2025bcu} for Schwarzschild black holes. In that work, the full non-axisymmetric solution to the nonlinear perturbation equations is not computed explicitly, but a symmetry argument is provided establishing the absence of a decaying falloff for both even and odd nonlinear Love numbers of four-dimensional Schwarzschild black holes to all orders in perturbation theory.  We will return to the analysis of \rcite{Parra-Martinez:2025bcu} in \cref{sec:sym-nonlin}.

In closing let us remark on Poisson's computation of vanishing tidally-induced static mass multipole moments in the post-Newtonian approach (cf.~\cref{sec:post-newtonian}) \cite{Poisson:2021yau}. This is a rather involved calculation, so to make progress it was performed in an unphysical, highly specialized setting: an equilibrium state consisting of a black hole with (arbitrarily small but non-vanishing) electric charge balanced out by a charged point particle. It is argued that this result captures the tidal responses of uncharged black holes in generic environments due to 1) the universality of the relationship between the tidal tensor and the mass multipole moments, and 2) the expectation of continuity in the limit of vanishing charge. The result holds to all orders in the post-Newtonian expansion, but is derived at linear order in metric perturbations. As such, it is complementary to the other nonlinear results discussed in this subsection, which define Love numbers through the point-particle EFT and work to all orders in perturbation theory.

\subsection{Love numbers in higher dimensions}
\label{sec:highD}

In spacetime dimensions $D>4$, black holes can exhibit richer structures and novel physical phenomena, including classical instabilities such as the Gregory--Laflamme instability \cite{Gregory:1993vy}, non-spherical horizon topologies \cite{Emparan:2001wn}, and hydrodynamic-like behavior in the long-wavelength limit \cite{Hubeny:2011hd,Bhattacharyya:2007vjd}.\footnote{See \rcite{Emparan:2008eg} for a review of black holes in higher dimensions.}
Extending the study of Love numbers to higher-dimensional spacetimes opens new avenues for exploring gravitational dynamics in theories beyond four-dimensional general relativity, including string theory, supergravity, and holography. Moreover, investigations in higher dimensions can shed light on the properties of compact objects in four-dimensional gravity by revealing structural features and symmetries of the gravitational response.

In \cref{sec:love-static} we uncovered a striking property of black holes in four-dimensional general relativity: their static Love numbers vanish identically. As we will see, this is a very special feature of $D=4$; in higher dimensions,
the static response coefficients are generically non-zero and exhibit a richer structure that depends on the spacetime dimension and multipole order \cite{Kol:2011vg,Cardoso:2019vof,Hui:2020xxx}.

In this subsection, we review the current literature and highlight recent progress. We begin with the static Love numbers for non-rotating Schwarzschild--Tangherlini black holes and subsequently extend the discussion to rotating cases and extended black objects. Finally we emphasize the existence of ultra-spinning r\'egimes in $D \ge 6$ and in the large-$D$ limit, an area of study which has attracted considerable attention in recent years.

\subsubsection{Definition}

As in four dimensions, Love numbers in $D$ spacetime dimensions describe the multipolar response of a compact object to an external tidal perturbation, generalizing the classical notion to higher-dimensional geometry. In this context, one typically considers the massless Klein--Gordon equation in $D$ spacetime dimensions,\footnote{This is because gravitational perturbation theory of higher-dimensional black objects is rather less developed and more involved than its 4D counterpart, often lacking separable equations and a Teukolsky-like formalism. When it is possible to study gravitational perturbations in this subsection, we do so.} often in the presence of a black hole or black string or black brane background, and studies the linearized response to a small external perturbation. The field profile behaves asymptotically as
\begin{equation}
\Phi(r) \overset{r\to\infty}{\longrightarrow} c_1 r^{-\ell} + c_2 r^{-(\ell + D - 3)},
\end{equation}
where $c_1$ is the amplitude of the applied external field and $c_2$ represents the induced response. 
The Love number is defined (up to normalization) as the ratio,
\begin{equation}
\lambda_\ell \sim \frac{c_2}{c_1}.
\end{equation}
Essentially, the only modification from 4D involves replacing the radial decay of the potential $1/r$ with its $D$-dimensional form, $1/r^{D-3}$. %

In previous analyses restricted to four-dimensional spacetime, an ambiguity was present in the definition of the response coefficient.
In higher dimensions, this ambiguity does not always arise~\cite{Kol:2011vg}.
However, from the perspective of effective field theory, there is an additional normalization that must be taken into account when matching to the mass multipole at the origin. Following \rcite{Kol:2011vg}, to match the definitions with the induced mass multipole, the physical response coefficient is proportional to 
\begin{equation}
    \hat{\lambda} = \mathcal{N}\lambda ,
\end{equation}
where the normalization constant is 
\begin{equation}
\mathcal{N} = \frac{4\pi^{(D-3)/2}}{2^{\ell}\,\Gamma\!\left(\frac{D-3}{2}+\ell\right)} .
\end{equation}

\subsubsection{Non-rotating higher-dimensional black holes} \label{sec:statichighD}
We begin with the static Love numbers for non-rotating Schwarzschild--Tangherlini black holes. This is the unique spherically-symmetric black hole in $D$ dimensions, and is a straightforward generalization of the 4D Schwarzschild black hole,
\begin{equation}
    \dd s^2 = -f(r)\dd t^2 + \frac1{f(r)}\dd r^2+r^2\dd\Omega^2_{S^{D-2}},\quad f(r) = 1-\left(\frac\rs r\right)^{D-3},\label{eq:S-T-metric}
\end{equation}
where $\dd\Omega^2_{S^{D-2}}$ is the metric on the unit $(D-2)$-sphere.

The gravitational perturbation theory for Schwarzschild--Tangherlini, generalizing the Regge--Wheeler--Zerilli formalism discussed in \cref{sec:BHPT-Sch}, was developed by Kodama and Ishibashi \cite{Kodama:2003jz,Ishibashi:2003ap}, and is largely analogous to its four-dimensional counterpart. The split into even- and odd-parity sectors is replaced by a decomposition into modes that transform as scalars, vectors, and tensors under rotations. The scalar and vector modes generalize the even- and odd-parity sectors, respectively, while the tensor mode is only present in higher dimensions. The three sectors are decoupled at linear order due to their different transformation properties. The hyperspherical symmetry of the metric \eqref{eq:S-T-metric} allows each sector to be expanded in hyperspherical harmonic modes which also decouple in the linear theory. Each sector is encoded in a master variable, generalizing the Regge--Wheeler and Zerilli variables, which satisfies a Schr\"odinger-type equation. The equation obeyed by the tensor mode is equivalent to the Klein--Gordon equation for a massless scalar on the same background \cite{Kodama:2003jz,Ishibashi:2003ap,Hui:2020xxx}.

The static responses of Schwarzschild--Tangherlini black holes were first computed in \rcite{Kol:2011vg,Hui:2020xxx} (see also \rcite{Hadad:2024lsf,Ivanov:2026icp} for subsequent work).
The static master equations for the vector and tensor sectors are hypergeometric, while the scalar master equation is of Heun type but can be made hypergeometric by means of a field redefinition involving the master variable and its first derivative, in complete analogy to $D=4$.\footnote{The analogy breaks down somewhat in that the transformation which makes the scalar equation hypergeometric is not related to an electric--magnetic duality transformation, which does not exist in $D\neq4$. This is in contrast to $D=4$ where it is the Chandrasekhar duality that sends the Zerilli equation to a hypergeometric equation, which so happens to be the Regge--Wheeler equation for the odd sector, as discussed in \cref{sec:chandra} \cite{Kodama:2003jz,Ishibashi:2003ap,Hui:2020xxx}.}
The main new feature compared to Schwarzschild can be seen from the Klein--Gordon equation (which is equivalent to the master equation for the gravitational tensor perturbation),
\begin{equation}
    \frac1{r^D}\partial_r(r^{D-2}f(r)\partial_r\phi) + \nabla^2_{S^{D-2}}\phi = 0.
\end{equation}
The hyperspherical harmonics are eigenfunctions of $\nabla^2_{S^{D-2}}$ with eigenvalue $-\ell(\ell+D-3)$. Defining $y\equiv(r/\rs)^{D-3}$ and $\Delta=y(y-1)$, the equation for a single multipole is
\begin{equation}
    \partial_y(\Delta\partial_y\phi_\ell)-\hat\ell(\hat\ell+1)\phi_\ell=0,
\end{equation}
where we have defined
\begin{equation}
    \hat\ell \equiv \frac\ell{D-3}.
\end{equation}
This is of precisely the same form as its 4D counterpart \eqref{eq:kg-static}, with the replacements $r/\rs\to y$ and $\ell\to\hat\ell$. The key difference in higher dimensions is that $\hat\ell$ is not necessarily an integer. Indeed we will see that when $\ell$ is an integer multiple of $D-3$ (i.e., $\hat\ell\in\mathbb N$), the Love numbers vanish, just as in 4D, while they display new phenomenology for other multipoles.

We review the Love number computations for the scalar, vector, and tensor sectors of gravitational perturbation theory on Schwarzschild--Tangherlini.\footnote{The response coefficients for the massless scalar ($s=0$) are the same as for tensor gravitational perturbations, while the electromagnetic ($s=1$) responses can be found in \rcite{Hui:2020xxx,Ivanov:2026icp}.} In each case the response coefficient is obtained by solving a hypergeometric equation, imposing regular boundary conditions at the horizon, and reading off the coefficient of the decaying term at infinity. The asymptotics can change if some of the parameters of the hypergeometric equation are integers, depending on the values of $\ell$ and $D$.

\paragraph{Tensor modes}

We begin with the tensor gravitational perturbations; these results also apply to a massless scalar field, since it satisfies an equivalent equation of motion. The asymptotics in this case depend on whether $\hat\ell$ is an integer, half-integer, or neither. The response coefficients are \cite{Kol:2011vg,Hui:2020xxx}
\be
\lambda_{\rm T} = 
\begin{cases}
\frac{\Gamma(-2\hat{\ell}-1) \Gamma(\hat{\ell}+1)^2}{\Gamma(-\hat{\ell})^2\Gamma(2\hat{\ell}+1)}=\frac{(2\hat{\ell}+1)\Gamma(\hat{\ell}+1)^4}{2\pi \Gamma(2\hat{\ell}+2)^2} \tan(\pi \hat{\ell})
&\qquad\text{for generic $\hat \ell$},\\
\frac{(-1)^{2\hat{\ell}}(D-3) \Gamma(\hat{\ell}+1)^2}{(2\hat{\ell})!(2\hat{\ell}+1)!\Gamma(-\hat{\ell})^2}  \log{\left(\frac{r_0}{r}\right)}
 & \qquad\text{for half-integer $\hat\ell$},\\
0&\qquad\text{for integer $\hat\ell$}.
\end{cases}
\label{eq:spin2summaryRW}
\ee
The vanishing of $\lambda_\mathrm{T}$ for integer $\hat\ell$ generalizes the spin-0 result derived in \cref{sec:love-static-sch} in $D=4$, where $\hat\ell=\ell$ is always an integer. The half-integer $\hat\ell$ case displays a classical running, meaning the induced response depends on the distance at which it is measured, in analogy to running in quantum field theory. The generic result $\lambda_\mathrm{T}\propto\tan(\pi\hat\ell)$ contains both of these special cases, vanishing for integer $\hat\ell$ and diverging for half-integer $\hat\ell$; the logarithmic running results from cancelling divergences \cite{Kol:2011vg}.

\paragraph{Vector modes}

The vector-type (odd parity) perturbations  are given by \cite{Hui:2020xxx,Hadad:2024lsf}
\be
\lambda_{\rm RW} = 
\begin{cases}
(2\hat \ell +1)
 \frac{\Gamma(\hat \ell +2 + \frac{1}{D-3})^2\Gamma(\hat \ell - \frac{1}{D-3})^2}{ \Gamma(2\hat \ell +2)^2  } \frac{\sin[\pi (\hat \ell + \frac{1}{D-3}) ] \sin[\pi (\hat \ell - \frac{1}{D-3}) ]}{\pi \sin (2\pi \hat \ell)} &\qquad\text{for generic $\hat \ell,D$},\\
 \frac{ (-1)^{2\hat{\ell}}(D-3) \Gamma(\hat{\ell}-\tfrac{1}{D-3})\Gamma(\hat{\ell}+2+\tfrac{1}{D-3})  }{(2\hat \ell+1)!(2\hat \ell)!\Gamma(-\hat{L}-1-\tfrac{1}{D-3})\Gamma(-\hat{\ell}+1+\tfrac{1}{D-3})}\log\left(\frac{r_0}{r}\right)
 & \qquad\text{for integer $\mathfrak c$},\\
0&\qquad\text{for integer $\mathfrak a$ or $\mathfrak b$},
\end{cases}
\ee
where the hypergeometric parameters $\mathfrak{a,b,c}$ are
\be
\mathfrak a=\hat \ell - \frac{1}{D-3} ,
\qquad
\mathfrak b= \hat \ell +2 + \frac{1}{D-3}   ,
\qquad
\mathfrak c= 2  \hat \ell +2  .
\label{abcoddspin2}
\ee
Again, note the appearance of a logarithmic running term in the case where $\mathfrak c$ is integral and $\mathfrak a$ and $\mathfrak{b}$ are not. This corresponds to both $\hat\ell$ integer in $D\geq5$ and $\hat\ell$ half-integer in $D>5$. The case where the Love numbers vanish, with $\mathfrak{a}$ and $\mathfrak{b}$ integral, corresponds to all multipoles in $D=4$, half-integer $\hat\ell$ in $D=5$, and $\ell = n(D-3)\pm1$ with $n\in\mathbb N$ in $D>5$. This includes the 4D odd-parity gravitational perturbations studied in \cref{sec:love-static-sch}.

\paragraph{Scalar modes}

In the scalar-type (even parity) sector, the generic Love number is given by \cite{Kol:2011vg,Hui:2020xxx}
\be
\lambda_{\rm Z}
	= -\frac{1}{4^{2\hat \ell+1}} \frac{(\ell+D-3) (\ell+D-2)^2 }{\ell (\ell-1)^2  } \frac{\Gamma(\hat \ell)\Gamma (\hat{\ell}+2) }{\Gamma (\hat{\ell}+\frac{1}{2}) \Gamma (\hat{\ell}+\frac{3}{2}) } {\tan} (\pi\hat \ell).  %
\label{eq:zerillilovesummary}
\ee
In $D=4$ we recover the even-parity result of \cref{sec:love-static-sch}.

\subsubsection{Rotating five-dimensional black holes}
\label{sec:5DMP}
Rotating spherical black holes in $D>4$ are described by the Myers--Perry solution \cite{Myers:1986un,Myers:2011yc,Emparan:2008eg}.
Love numbers for 5D Myers--Perry black holes were independently derived analytically in \rcite{Rodriguez:2023xjd,Charalambous:2023jgq}. The problem is made tractable by two conditions that reduce the five-dimensional problem to a concrete equation that can be solved exactly. The separability property for massless scalar field perturbations on Kerr (cf. \cref{sec:Kerr-KG}) famously persists to Myers--Perry black holes due to hidden symmetries generated by one or more Killing tensors, generalizing the construction discussed in \cref{sec:Kerr-syms} \cite{Frolov:2006dqt,Frolov:2017kze}. In addition, the radial equation, which originally has five regular singular points, two of which (corresponding to the inner and outer horizons) are degenerate, can be simplified to a hypergeometric equation with a change of variable $r \rightarrow r^2$. This reduces the number of regular singular points is reduced to three, ensuring that the radial equation can be expressed in terms of hypergeometric functions.

The static response for a massless scalar on 5D Myers--Perry is \cite{Rodriguez:2023xjd,Charalambous:2023jgq}
\begin{equation}
\lambda_\ell = 2(-1)^{\ell} \frac{(-\hat\ell+2i\tilde m_L)_{\ell+1}(-\hat\ell+2i\tilde m_R)_{\ell+1}}{\ell!(\ell+1)!}\ln\left(\frac{r_0}{r}\right),\label{eq:LoveMP}
\end{equation}
where $(x)_n=\Gamma(x+n)/\Gamma(x)$ denotes the Pochhammer symbol and $\tilde{m}_{L,R}$ encode the left- and 
right-moving thermal parameters of the black hole.  The rescaled parameters are defined as
\begin{equation}
\tilde m_L = \frac{a-b}{r_+ +r_-}(m_\phi-  m_\psi),\quad\tilde m_R = \frac{a+b}{r_+ - r_-}(m_\phi+  m_\psi).
\end{equation}
In these expressions, the rotational parameters $a$ and $b$ are related to the physical angular momenta 
$J_{\phi}$ and $J_{\psi}$ of the five-dimensional Myers--Perry black hole. 
The quantities $m_{\phi}$ and $m_{\psi}$ denote the corresponding eigenvalues on the hypersphere. In the $(\tilde m_L,\tilde m_R)\to0$ limit, the Pochhammer symbols take the values
\begin{equation}
(-\hat\ell)_{\ell+1} = \frac{\Gamma(\hat\ell+1)}{\Gamma(-\hat\ell)} = \begin{cases}0, & \text{$\hat\ell$ integer}, \\ (-1)^{(\ell+1)/2}\frac{\ell!!^2}{2^{\ell+1}}, & \text{$\hat\ell$ half-integer}.
\end{cases}
\end{equation}

Note that the logarithmic behavior in $\lambda_\ell$ persists for rotating black holes and is confirmed as a characteristic feature of static deformations in higher spacetime dimensions. In the EFT language, this corresponds to a logarithmic running of the tidal coupling, with $r$ playing the role of the EFT renormalization scale. Different choices of $r$ correspond to different renormalization schemes, and the physical content resides in the coefficient of the logarithm rather than in the scale itself.
Notably, the Love numbers can take negative values for certain spin configurations,  and their magnitude and sign evolve non-trivially as the spin parameter increases.  This change in behavior --- from positive to negative values across the spin parameter space,  as noticed in \rcite{Rodriguez:2023xjd} --- is not merely a quantitative shift but signals a qualitative  change in the tidal response of the black hole.  The spin-induced sign change thus serves as an  indicator that the physics governing tidal deformability is truly different  across different r\'egimes of the rotation parameter, and warrants further investigation.

\subsubsection{Five-dimensional black rings}

One of the distinctive features of higher-dimensional general relativity is that black hole solutions exhibit qualitatively new properties absent in four dimensions. In particular, non-spherical horizons are possible, and the standard notions of black hole uniqueness no longer apply. An example of such a solution is the black ring~\cite{Emparan:2006mm}. Unlike the five-dimensional Myers--Perry black hole, whose horizon is a three-sphere $S^3$, the black ring describes a rotating, ring-shaped object with an event horizon of topology $S^1 \times S^2$.
For spherical black holes, the perturbation equations are separable in the radial and angular directions; however, this is not generally the case for black rings. While separability can occur in spacetimes admitting a Killing tensor, the black ring geometry does not, in general, possess this property~\cite{Durkee:2010ea}. Two notable exceptions are the \textit{static limit} ($\omega = 0$)~\cite{Cardoso:2005sj} and the infinite-radius limit ($R \to \infty$), which yields a boosted black string~\cite{Rodriguez:2023xjd}.

In the case of a black ring, one finds the static response coefficients \cite{Rodriguez:2023xjd}
\begin{equation}
\lambda^{\rm BR}_{\ell \in \mathbb{N}} = (-1)^{\ell+1} \frac{\Gamma(\ell+1)^2\Gamma(\ell - 2i {\cal W}+1)}{2  \Gamma(2\ell+1)  \Gamma(2\ell+2) \Gamma(-\ell -2i {\cal W})}  ,
\label{lovenumbersbr5d}
\end{equation}
 with
\begin{equation}
 \mathcal{W}=\frac{\nu  r_0 \sinh \sigma}{1-\frac{r_0^2}{R^2}},
\end{equation}
where $r_0$  characterizes  the thickness or the radius of the $S^2$ at the horizon, while $R$ determines the ring's overall radius. The factor $\cosh^2 \sigma$ quantifies the ring's rotational motion and can be approximated by the local boost velocity $v = \tanh \sigma$.
 
As in the four-dimensional black hole case, the response coefficients are purely imaginary, implying that the static Love numbers vanish for the black ring for all values of $\ell$ and $\nu$. It is instructive to consider the limit $ \mathcal{W} \to 0$ and compare the resulting expression with the induced response of a Schwarzschild black hole. In this limit, the static response coefficients vanish, consistent with the fact that, for $\mathcal{W}= 0$, formally the perturbation equation reduces to the equation of a massless scalar field in a four-dimensional Schwarzschild spacetime.

\subsubsection{Black branes and Kaluza--Klein excitations}

Tidal deformations of extended black objects and their Kaluza--Klein (KK) excitations provide
a natural bridge between four-dimensional observations and higher-dimensional theories of
gravity. The study of Love numbers in these settings reveals both new vanishing mechanisms
and genuinely non-zero responses, enriching the picture established for isolated spherical black
holes.

\paragraph{Kaluza--Klein excitations and flat extra dimensions}
An important early question was whether gravitational-wave observations could constrain
scenarios with flat large extra dimensions, such as the Arkani-Hamed--Dimopoulos--Dvali
(ADD) model. \rcite{Cardoso:2019vof} addressed this directly, showing that
physically motivated setups in this model are essentially unconstrained by current
gravitational-wave data. The key results are twofold: dynamical processes that do not excite
the KK modes produce a gravitational-wave signal identical to that of four-dimensional vacuum
general relativity; and any excitation of KK modes is highly suppressed relative to the
dominant quadrupolar term, appearing only at post-Newtonian order $\sim 10^{11}$ given
existing constraints on the size of extra dimensions and the masses of the observed binary
components.

\paragraph{Black $p$-branes}
A comprehensive treatment of tidal Love numbers for extended black objects was given
in \rcite{Charalambous:2024gpf,Charalambous:2025ekl} (see also \rcite{Rodriguez:2023xjd}), which computed scalar static response coefficients for
non-dilatonic black $p$-brane solutions in higher-dimensional supergravity. The key findings
are:
\begin{enumerate}[(i)]
    \item The Love numbers exhibit a fine-tuning behavior analogous to that of
    higher-dimensional black holes, with near-exact cancellations that are not accidental
    but are instead enforced by hidden near-zone symmetries acting on the perturbation
    equations (rather than on the background geometry itself);

    \item The algebraic structure of the Love symmetry depends on the brane dimension: for
    point-like ($p=0$) charged branes it is the familiar $\mathrm{SL}(2,\mathbb{R})$, while
    for string-like ($p=1$) branes it extends to
    $\mathrm{SL}(2,\mathbb{R})\times\mathrm{SL}(2,\mathbb{R})$; and

    \item In the near-extremal finite-temperature limit, these Love symmetries
    \emph{geometrize}: they reduce to the isometries of the near-horizon
    Schwarzschild--AdS$_{p+2}$ metric, and further to pure AdS$_{p+2}$ isometries in the
    zero-temperature extremal limit.
\end{enumerate}
This geometrization process provides a unified perspective on why Love numbers are finely
tuned across a broad class of extended black objects, connecting the abstract algebraic
explanation to the concrete near-horizon geometry. The symmetry arguments are discussed in more detail (primarily in the 4D setting) in \cref{sec:sym}.

Taken together, these works demonstrate that the vanishing (or fine-tuning) of tidal Love
numbers in higher-dimensional and extended-object settings is not accidental but is controlled
by a hierarchy of hidden symmetries whose algebraic structure reflects the near-horizon
geometry of the background. Open questions include understanding Love number observability via
KK mode excitation in next-generation detectors, and establishing a full EFT dictionary for
$p$-brane tidal operators.

\subsubsection{Large-\texorpdfstring{$D$}{D} black holes}
\label{sec:LargeD_Love}

The large-$D$ limit of general relativity, where $D$ is the number
of spacetime dimensions, provides a powerful analytical framework for
studying black hole dynamics~\cite{Emparan:2020inr,Emparan:2013moa}.
In this limit, the gravitational field becomes strongly localized near
the horizon, and the perturbation equations simplify considerably,
allowing for systematic $1/D$ expansions of physical observables
including tidal Love numbers. We now present the matching of the
large-$D$ tidal response coefficients to the point-particle EFT,
following the gravitational computation of Glazer \emph{et
al.}~\cite{Glazer:2024eyi}.

We consider a $(2N+3)$-dimensional Myers--Perry black hole with
a single spin parameter $a$, and work with the rescaled dimensionless
quantities
\begin{equation}
    \hat{a} \equiv \frac{a}{\rs}, \qquad
    \hat{\ell} \equiv \frac{\ell}{2N}, \qquad
    \hat{m} \equiv \frac{m}{2N}, \qquad
    \hat{\omega} \equiv \frac{\omega}{2N},
    \label{eq:LargeD_rescaled}
\end{equation}
where $\rs$ is the horizon radius parameter. The dynamics of black
holes simplify greatly in the large-$D$
limit~\cite{Emparan:2020inr,Emparan:2013moa}. In the present context,
this manifests as the radial perturbation equation reducing to
hypergeometric form in the limit of large $N$. Specifically, taking
$N\to\infty$ while keeping $\hat{\ell}$, $\hat{m}$, and $\hat{\omega}$
fixed, the radial coordinate satisfies
\begin{equation}
    \rho^{1/N} \to 1.
    \label{eq:LargeD_rho}
\end{equation}
For fixed $N$ this approximation is valid in the region $\log\rho \ll
N$, but since we are interested only in the leading-order behavior we
take the strict $N\to\infty$ limit, where it holds to arbitrarily
large radius. In this sense the perturbation equation reduces
\emph{exactly} to a solvable hypergeometric form.

In the infinite-$N$ limit, the radial equation of motion becomes
\begin{equation}
    \partial_\rho\!\left(\rho\Delta_0\partial_\rho R_{\omega\ell m}
    \right) +\left[
    \frac{\left(\hat{\omega} \rs - \hat{a}\hat{m}\right)^2}{\Delta_0}
    - \hat{\ell}(\hat{\ell}+1)
    + \rs^2\hat{\omega}^2
    \right] R_{\omega\ell m} = 0,
    \label{eq:LargeD_EOM}
\end{equation}
where we have approximated $\hat{\Delta}$ by
\begin{equation}
    \Delta_0 \equiv \rho - 1 + \hat{a}^2.
    \label{eq:Delta0}
\end{equation}
This equation can be recast in a more familiar form by introducing
the parameters
\begin{align}
    \beta &\equiv \frac{2\rs}{\sqrt{1-\hat{a}^2}}, \qquad
    \Omega_{\mathrm H} \equiv \frac{\hat{a}}{\rs}, \qquad
    \nu(\nu+1) \equiv \hat{\ell}(\hat{\ell}+1) - \hat{\omega}^2 \rs^2,
    \label{eq:LargeD_params1}\\[4pt]
    U &\equiv \frac{\beta\hat{m}\Omega_{\mathrm H}}{2}, \qquad
    P^2 \equiv \frac{\hat{a}\hat{m} - \rs\hat{\omega}^2}{1-\hat{a}^2}
    = \left(U - \frac{\beta\hat{\omega}}{2}\right)^2,
    \label{eq:LargeD_params2}
\end{align}
where $\Omega_{\mathrm H}$ is the angular
velocity at the horizon and $\beta$ is the inverse surface gravity, cf. \cref{eq:ang-vel-hor,eq:inv-hawking-temp} in $D=4$. With these definitions,
\cref{eq:LargeD_EOM} takes exactly the same structural form as
the four-dimensional Schwarzschild radial equation near the horizon,
with $\nu$ playing the role of the effective angular momentum and $U$
encoding the rotational coupling. This structural similarity underlies
the close relationship between the large-$D$ and four-dimensional
results noted below.

Comparing the near- and far-zone solutions, one finds that all
response coefficients share a common real prefactor,
\begin{equation}
    \lambda^{\mathrm{LD}}_{\ell m} \equiv r_h^{2\ell+2N}
    \frac{2^{2-\ell}\pi^{N+1}
    \left|\Gamma\!\left(\hat{\ell}+1+iU\right)\right|^4}
    {\pi\Gamma(N+\ell)
    \Gamma\!\left(\tfrac{\ell+m}{2}+1\right)
    \Gamma\!\left(\tfrac{\ell-m}{2}+1\right)
    \Gamma(2\hat{\ell}+1)\Gamma(2\hat{\ell}+2)},
    \label{eq:lambdaLD}
\end{equation}
which it is convenient to define as a separate parameter. Here $r_h$
is the horizon radius, $N$ counts the number of transverse dimensions,
$\hat{\ell}$ is the rescaled angular momentum \eqref{eq:LargeD_rescaled},
$U$ is the dimensionless rotation parameter \eqref{eq:LargeD_params2},
and $m$ is the azimuthal quantum number.

\paragraph{Conservative response.}
Matching the conservative sector of the large-$D$ tidal response to
the EFT gives
\begin{align}
    \lambda^{(0)}_{\ell m} &= \lambda^{\mathrm{LD}}_{\ell m}
    \left[\csc(2\pi\hat{\ell})
    - \cot(2\pi\hat{\ell})\cosh(2\pi U)\right],
    \label{eq:kLD0} \\[4pt]
    \lambda^{(1)}_{\ell m} &= \frac{\lambda^{\mathrm{LD}}_{\ell m}}{2N}
    \frac{\beta}{\rs}
    \left[\pi\cot(2\pi\hat{\ell})\sinh(2\pi U)
    + \operatorname{Im}(\Delta\psi)
    \left(\cosh(2\pi U)\cot(2\pi\hat{\ell})
    - \csc(2\pi\hat{\ell})\right)\right],
    \label{eq:kLD1}
\end{align}
where $\Delta\psi$ denotes the logarithmic correction arising from the
Mano--Suzuki--Takasugi (MST) formalism~\cite{Mano:1996vt}.

These responses share several properties with the finite-$D$ cases
discussed earlier. Much like the Myers--Perry case
(cf. \cref{sec:5DMP}), when $U \neq 0$ the Love number
$\lambda^{(0)}_{\ell m}$ does not vanish for any $\ell$. The response
diverges for integer values of $\hat{\ell}$, corresponding to a
classical running of the coupling, in direct analogy with the
logarithmic running observed for Schwarzschild black holes in four
dimensions~\cite{Ivanov:2024sds,Combaluzier-Szteinsznaider:2024sgb,
Kobayashi:2025vgl}. As a consistency check, one can verify that both
expressions reduce to the four-dimensional Schwarzschild results in
the $U \to 0$ limit.

The conservative responses display a curious feature: if one
artificially sets $N=1$ and identifies the two 5D Myers--Perry spin
parameters with $U$ (i.e., $U_{\mathrm{MP}} = V_{\mathrm{MP}} =
U_{\mathrm{LD}}$), then the responses agree exactly with the 5D
Myers--Perry result of \rcite{Rodriguez:2023xjd,
Charalambous:2023jgq}. Whether there is a deeper physical origin for
this coincidence remains an open question.

\paragraph{Dissipative response}
Turning to the dissipative sector, the matching procedure yields
\begin{align}
    \nu^{(0)} &= -\lambda^{\mathrm{LD}}_{\ell m}\sinh(2\pi U),
    \label{eq:nuLD0}\\[4pt]
    \nu^{(1)} &= \frac{\lambda^{\mathrm{LD}}_{\ell m}}{2N}
    \frac{\beta}{\rs}
    \left[\pi\cosh(2\pi U)
    + \operatorname{Im}(\Delta\psi)\sinh(2\pi U)\right].
    \label{eq:nuLD1}
\end{align}
As in the conservative sector, these quantities reduce to the
Schwarzschild dissipative responses in the $U\to 0$ limit, providing
a non-trivial consistency check of the large-$D$ expansion.
Interestingly, the leading $1/D$ results accurately reproduce the
finite-$D$ Schwarzschild results once appropriate overall factors in
the definition of $\hat{\ell}$ are scaled out, suggesting that the
large-$D$ expansion captures the essential structure of the tidal
response at finite $D$ already at leading order.

The dissipative coefficients $\nu^{(0)}$ and $\nu^{(1)}$ encode
tidal heating of the large-$D$ black hole, in direct analogy with
the absorption cross sections computed in the Starobinsky--Churilov
and Page analyses for four-dimensional Kerr black holes
\cite{Starobinskil:1974nkd,Page:1976df}. For $U=0$ (non-rotating
limit), $\nu^{(0)}$ vanishes at leading order, consistent with the
vanishing of the dissipative tidal coefficient  for Schwarzschild black holes.

\subsection{Love numbers beyond vacuum general relativity}
\label{sec:beyondGR}

In this subsection we study tidal responses in the presence of some source for the Einstein equations, which may be due to external fields, a cosmological constant, or a modification of gravity. In particular, we discuss black holes with electromagnetic charge, supergravity black holes, asymptotically (A)dS black holes, and black holes in theories beyond general relativity.

\subsubsection{Charged black holes}

Extending the study of tidal Love numbers to charged black holes provides new insights 
into how electromagnetic interactions modify spacetime polarizability. Several recent works 
have investigated this extension across four and higher dimensions, exploring how charge 
influences the static response and the emergence of new polarization modes~\cite{Cardoso:2017cfl,Pereniguez:2021xcj,Berens:2022ebl,Rai:2024lho,Pani:2013wsa,Ma:2024few, Xia:2025zfp, Pereniguez:2025jxq, Barbosa:2026qcv}.  %

When charge is introduced --- as in the Reissner--Nordstr\"{o}m or Kerr--Newman families --- the 
coupling between the gravitational and electromagnetic sectors enriches the structure of the 
perturbations. Collectively, these works reveal that:
\begin{enumerate}[(i)]
    \item Charge --- whether residing in the black hole itself or in the perturbing field --- can 
    give rise to qualitatively new static responses, yielding non-zero Love numbers and 
    opening  up additional polarization channels;

    \item Symmetry structures --- such as ladder symmetries and the near-zone $\mathrm{SL}(2,\mathbb{R})$ --- account for many of the vanishing results and, once combined with EFT matching, help clarify which couplings are physically meaningful, as opposed to being scheme- or definition-dependent; and

    \item Dimensionality and extremality play a decisive role: responses typically scale with 
    temperature and vanish in the extremal limit, while higher dimensions allow for non-zero 
    tensorial Love numbers.
\end{enumerate}
Note that the symmetry structures generalize those we will discuss in more detail for Schwarzschild and Kerr in \cref{sec:sym}.
We now elaborate on the main findings, highlighting the key results.

\paragraph{Reissner--Nordstr\"{o}m black hole families} In the static limit, perturbations of Reissner--Nordstr\"{o}m black holes are governed by a single, decoupled hypergeometric master equation \cite{Rai:2024lho}, that unifies scalar, electromagnetic, and gravitational modes via a generalized Darboux-type field redefinition \cite{Chandrasekhar:1985kt,Glampedakis:2017rar}. 
The Love numbers were computed in closed form and argued to vanish in all cases in \rcite{Rai:2024lho}, which generalized a previous result by \rcite{Cardoso:2017cfl} (see \rcite{Berens:2022ebl} for a scalar-field analysis). The analysis further uncovered a potential ambiguity in defining mixed tidal response coefficients and demonstrated that all Love number couplings vanish in four dimensions within the point-particle EFT. This outcome is rooted in a hidden set of ladder symmetries that dictate the form of the static solutions and their tidal behavior~\cite{Berens:2022ebl,Rai:2024lho}.

Employing an analogy with the higher-dimensional neutral case, the definitions of tidal Love 
numbers can be extended to charged backgrounds. \rcite{Pereniguez:2021xcj} examined 
the impact of electric charge on the static tidal response of 
Reissner--Nordstr\"{o}m--Tangherlini black holes in arbitrary spacetime dimensions 
($D \geq 4$). The tensor tidal Love numbers were found to scale with the black hole temperature 
as $T_{\mathrm{H}}^{2\ell+1}$, and hence vanish in the extremal limit. In contrast, the electric charge 
$Q$ activates new polarization modes in the vector sector that are absent for neutral black 
holes. In four dimensions, both the Love numbers and magnetic susceptibilities vanish for all 
subextremal values of the charge.

Departing from this pattern, it was recently shown~\cite{Pereniguez:2025jxq}  
that magnetically charged black holes acquire genuinely non-zero Love numbers when perturbed by 
a charged scalar field, providing one of the clearest counterexamples to the general expectation 
of vanishing Love numbers in four-dimensional general relativity. Complementing this from an EFT 
perspective, Barbosa, Fichet, and de Souza~\cite{Barbosa:2026qcv} demonstrated that loop 
corrections generate running Love numbers for charged black holes, which are non-zero and flow 
logarithmically with scale, reflecting the renormalization group structure of the point-particle 
EFT.
Other studies, focusing in charged binary systems, first construct a stationary gravitational tide acting on a dyonic Reissner–Nordström black hole, analyze the motion of a test particle, finding that tidal corrections are suppressed as the black hole’s charge increases but remain finite in the extremal limit \cite{Grilli:2024fds}.

\paragraph{Kerr--Newman black hole families}
In the context of non-supersymmetric solutions, \rcite{Ma:2024few} computed the Love numbers for a charged scalar field on the Kerr--Newman background. While the Love numbers vanish for neutral scalars, \rcite{Ma:2024few} claims that for charged scalars the coefficients are inversely proportional to the scalar charge, revealing a novel effect of electromagnetic interactions on the tidal response. For static 
perturbations ($\omega = 0$), the radial equation is solved exactly using hypergeometric 
functions. The presence of scalar charge modifies the effective multipole index, avoiding the 
ambiguities encountered in the neutral case. For dynamical perturbations ($\omega \neq 0$), the 
near-zone displays an $\mathrm{SL}(2,\mathbb{R})$ symmetry, defining a pseudo-static frequency 
r\'egime with a critical frequency $\omega_{\rm cr}$ at which the Love number vanishes, recovering 
the neutral scalar case. The study further extends to higher-dimensional 
Reissner--Nordstr\"{o}m black holes, showing that exact solutions for charged scalars exist in 
four and five dimensions with Love numbers vanishing in the extremal limit; additionally, massive 
scalar perturbations in extremal backgrounds can be solved via double confluent Heun functions. 
Complementary results for the Kerr--Newman family can be found in \rcite{Charalambous:2022rre,Charalambous:2024gpf}. For rotating Kerr--Newman black holes, the Love symmetry algebra extends to an 
infinite-dimensional $\mathrm{SL}(2,\mathbb{R})\ltimes\hat{U}(1)_{\mathcal{V}}$ 
structure containing  physically distinct $\mathrm{SL}(2,\mathbb{R})$ subalgebras, 
whose extremal limit recovers the Killing vectors of the AdS$_2$ throat.
Note that, while analytic closed-form results can be obtained for the tidal response of a test scalar field on Kerr--Newman spacetime~\cite{Ma:2024few,Cvetic:2021vxa}, a generalization to gravitational and electromagnetic responses is currently available only in the small black-hole-spin limit~\cite{Pani:2013wsa}. A complete analysis to all orders in the black-hole spin is still missing, mainly due to technical difficulties related to the lack of separability of the perturbation equations~\cite{Chandrasekhar:1985kt} (see also \rcite{Mark:2014aja,Dias:2015wqa,Dias:2022oqm}).

\subsubsection{Supergravity black holes}

A particularly rich arena for the study of tidal Love numbers is provided by the STU family of black holes, solutions of four-dimensional $\mathcal{N}=2$ gauged supergravity coupled to three vector multiplets. This class is especially useful because it admits separation of variables for the scalar wave equation and exhibits hidden conformal symmetry, and encompasses several physically important solutions as special cases, including the Kerr--Newman metric \cite{Kerr-NewmanSol}, its Kaluza--Klein analogues \cite{Larsen:1999pp}, and the Kerr--Sen black hole of string theory \cite{Sen:1992ua} --- the latter having been widely used as a black hole foil in the gravitational-wave literature. The tidal response is characterized by dimensionless Love numbers, in direct analogy with the Newtonian framework. Within the STU family, these investigations reveal that the vanishing of static Love 
numbers persists for the rotating charged solutions in four dimensions, reflecting the 
emergence of a near-zone $\mathrm{SL}(2,\mathbb{R})$ Love symmetry. The generality of 
the STU framework --- bridging Kerr--Newman, Kaluza--Klein, and string-theoretic black 
holes within a single solution space --- makes it an ideal setting in which to disentangle 
the roles of rotation, charge, and supersymmetry in controlling the tidal response. A key development 
in this direction is the explicit embedding of the effective Kerr spacetimes --- those 
pertinent to the vanishing of static Love numbers, soft hair descriptions, and 
low-frequency scalar-Kerr scattering amplitudes --- as solutions within 
$\mathcal{N}=2$ supergravity~\cite{Cvetic:2024dvn}.

While significant progress has been made in this area, several open questions remain. These 
include establishing a precise EFT mapping between full general relativity calculations and 
point-particle Wilson coefficients in the presence of charge and sector mixing, understanding 
the role of dissipation and dynamical (frequency-dependent) responses, and exploring the 
observational prospects for detecting charge-induced tidal effects in gravitational-wave signals.

\subsubsection{(A)dS black holes}

The vanishing of tidal Love numbers is a feature tied specifically to asymptotically-flat 
black holes in four-dimensional general relativity. The introduction of a cosmological 
constant breaks the boundary conditions underlying this result, and Love numbers generically become 
non-zero. This setting is of independent theoretical interest because of the role tidal 
response coefficients play in gauge/gravity duality: via AdS/CFT, the Love numbers of AdS 
black objects map onto linear-response polarization coefficients of the dual strongly-coupled 
plasma, directly related to two-point functions of the stress-energy tensor.

\paragraph{(A)dS$_4$ black holes}
In asymptotically (A)dS$_4$ spacetimes, the tidal Love numbers of both static and rotating 
black holes have been computed in \rcite{Nair:2024mya, Yusmantoro:2025ylw}. 
\rcite{Nair:2024mya} showed that asymptotically de~Sitter black holes acquire 
non-zero tidal Love numbers, establishing that the cosmological constant generically breaks 
the vanishing that holds in the flat case. Building on this, \rcite{Yusmantoro:2025ylw} 
performed a systematic study of scalar tidal perturbations for Schwarzschild-AdS, Kerr-AdS, 
Reissner--Nordstr\"{o}m-AdS, and Kerr--Newman-AdS black holes, finding that the tidal Love 
numbers are always non-zero in the presence of a cosmological constant for all these 
families. In contrast, tidal dissipation can vanish for certain values of the scalar field 
frequency and black hole parameters, showing a non-trivial interplay between conservative 
and dissipative response in the AdS setting.

\paragraph{AdS black branes}
The holographic interpretation of tidal Love numbers was developed in \rcite{Emparan:2017qxd}, 
which used AdS/CFT to study how a strongly-coupled plasma polarizes when placed in a curved 
geometry. In the gravitational dual, this polarization corresponds to the tidal deformation 
coefficients of an AdS black brane. The paper computed both the gravitational Love numbers 
of the brane and the coefficients of static electric polarization of the plasma, establishing 
a concrete dictionary between bulk tidal response and boundary transport data. This work 
highlighted that the Love numbers of AdS black branes are generically non-zero, in contrast 
to their asymptotically flat counterparts, and gave a physical interpretation to this 
non-vanishing in terms of the geometric polarizability of the dual fluid.

\paragraph{BTZ black holes}
The three-dimensional Ba\~{n}ados--Teitelboim--Zanelli (BTZ) black hole provides a 
particularly tractable example for exploring Love numbers in AdS, owing to the exact 
solvability of the perturbation equations in terms of hypergeometric functions. 
\rcite{Bhatt:2024mvr} studied the scalar tidal response of the rotating BTZ 
black hole and found that the real part of the tidal response function is non-zero, 
establishing that the rotating BTZ black hole possesses non-zero tidal Love numbers. 
Additionally, the tidal response exhibits scale-dependent (log-running) behavior. A 
separate analysis of the extremal rotating BTZ black hole found qualitative similarities 
with the non-extremal case. The paper also outlined a procedure to extend these results to 
the charged rotating BTZ black hole, demonstrating that the presence of an AdS cosmological 
constant robustly produces non-zero tidal responses across the BTZ family.

These results establish that the cosmological constant acts as a universal 
source of non-zero tidal deformability for black holes, removing the fine-tuning responsible 
for the vanishing of Love numbers in asymptotically flat four-dimensional gravity. The 
AdS/CFT interpretation of these results is relevant beyond purely the gravitational context, 
connecting black hole tidal response to measurable properties of strongly-coupled field 
theories.

\subsubsection{Beyond general relativity}
\label{sec:love-beyond-gr}

Since any non-zero measurement of tidal Love numbers from a binary black hole signal would 
constitute an unambiguous signature of physics beyond vacuum gneral realtivity, there has been sustained 
effort to compute and characterize tidal responses in modified gravity theories, 
quantum-corrected spacetimes, and non-vacuum environments. The results fall naturally into 
four categories.

\paragraph{Modified gravity and higher-derivative extensions}
A systematic program to compute Love numbers in EFT extensions of GR has been developed 
through the Regge--Wheeler--Zerilli approach in \rcite{Cardoso:2018ptl, DeLuca:2022tkm, 
Katagiri:2024fpn, Barbosa:2025uau,Barura:2024uog,Wang:2026qst},
which established that higher-curvature operators generically produce non-zero, logarithmically running Love numbers, whose beta functions encode the theory-dependent deformation of the geometry. 
A unified treatment of the perturbations   using the modified Teukolsky equation  --- previously introduced independently in \rcite{Li:2022pcy,Hussain:2022ins} (see also \rcite{Cano:2023tmv,Cano:2023jbk,Wagle:2023fwl,Cano:2024ezp,Guo:2024bqe} for  related subsequent studies) ---  was achieved in \rcite{Cano:2025zyk}, which describes the full set of electric, magnetic,  and parity-violating tidal responses through a single theory-dependent coefficient,  identifying a new ``mixing'' Love number arising in parity-breaking theories. Complementing this, \rcite{Garcia-Saenz:2025urd} showed that the logarithmic Love 
number is directly calculable from the Fuchsian theory of differential equations in a 
model-independent way, 
establishing general results on its sign and 
magnitude across broad classes of modified gravity theories, and demonstrating that 
perturbativity is essential: explicit black hole solutions beyond GR with exactly zero 
running exist once this assumption is relaxed. 
A parametrized formalism for extracting Love numbers across a general class of spherically symmetric and stationary Konoplya--Rezzolla--Zhidenko black holes was developed in \rcite{Sharma:2025xii}, which employed a ladder symmetry (cf. \cref{sec:sym-ladder}) to show that, in the case of a test scalar field, any deviation from a ladder-symmetric effective background produces non-zero static scalar Love numbers, establishing ladder symmetry as both a necessary and sufficient condition for vanishing.
A particularly instructive example within higher-curvature (but second-order) extensions of GR is Lovelock gravity, where black hole Love numbers are generically non-zero and directly probe the additional curvature couplings in the action \cite{Singha:2025xah}. Love numbers from an EFT description of black hole perturbations in  theories featuring a scalar field with a timelike background profile have been computed in \rcite{Barura:2024uog}.\footnote{See \rcite{Khoury:2022zor,Mukohyama:2022enj,Mukohyama:2022skk} for an explicit construction of the EFT. See, instead, \rcite{Franciolini:2018uyq,Hui:2021cpm,Mukohyama:2025jzk} for an EFT description of beyond-GR theories in the presence of an additional scalar degree of freedom with a radially-dependent  profile.}

\paragraph{Quantum gravity and regular black holes}
Quantum corrections to the spacetime geometry break the conditions underlying the 
vanishing of Love numbers in classical GR, generically producing non-zero tidal 
responses that serve as potential signatures of quantum hair. This has been studied 
in two complementary approaches: loop-quantized black holes~\cite{Motaharfar:2025typ,
Motaharfar:2025ihv,Brustein:2020tpg}, where the Love numbers are non-zero and Planck-scale suppressed 
for astrophysical masses but grow in significance near the Planck scale; and 
semiclassical QFT~\cite{Kim:2020dif}, where a sum-rule argument based on the Hawking 
radiation spectrum yields a finite non-zero quantum correction to the classically 
vanishing Love number of the Schwarzschild black hole. We note that the sign of the 
tidal response in these works depends sensitively on conventions, and care should be 
taken when comparing results across the literature. A unified analytic study of Love numbers for regular black holes --- covering the 
Bardeen model, sub-Planckian curvature black holes, and asymptotically safe gravity --- was 
carried out in \rcite{Wang:2025oek}, finding that Love numbers for all three classes are 
generically non-zero, exhibit strong model and mode dependence, and develop logarithmic 
scale dependence resembling renormalization-group running in quantum field theory. Quantum corrections --- whether from loop quantum gravity~\cite{Motaharfar:2025typ,
Motaharfar:2025ihv} or semiclassical QFT~\cite{Kim:2020dif} --- generically seem to render 
the Love numbers of Schwarzschild black holes non-zero, opening a potential 
observational window into quantum gravity.

\paragraph{Exotic compact objects}  
As discussed above, four-dimensional asymptotically flat black holes in general relativity have vanishing Love numbers. Therefore, a hypothetical measurement of a non-zero Love number for a dark compact object with mass above the neutron-star range would likely imply one of the following:  the object is not a black hole but rather an exotic compact object (ECO)~\cite{Mendes:2016vdr,Cardoso:2017cfl,Maselli:2017cmm,Sennett:2017etc,Cardoso:2019rvt,Herdeiro:2020kba};  general relativity receives corrections at scales accessible to gravitational-wave interferometers through tidal effects~\cite{Endlich:2017tqa,Cardoso:2018ptl,Franciolini:2018uyq,Noller:2019chl,Hui:2021cpm}; or  the black hole does not reside in a vacuum region \cite{Baumann:2018vus,Cardoso:2019upw,DeLuca:2021ite,DeLuca:2022xlz,Katagiri:2023yzm,Cannizzaro:2024fpz,Arana:2024kaz,Chakraborty:2024gcr}.
ECOs are hypothetical alternatives to black holes (see \rcite{Cardoso:2019rvt} for an extensive review). Theorized examples include boson stars, gravastars, and wormholes, which have been generically shown to have non-zero tidal Love numbers~\cite{Cardoso:2017cfl}, in sharp contrast to the vanishing result for black holes in GR. This makes Love numbers a powerful 
observational discriminator: a measurement of non-zero tidal deformability 
in a compact binary would be a smoking-gun signal of new physics beyond 
classical GR~\cite{Cardoso:2017cfl,Cardoso:2019rvt}. The Love numbers of 
ECOs depend sensitively on the object's internal structure and 
compactness~\cite{Silvestrini:2025lbe, Mendes:2016vdr,Maselli:2017cmm}, and can in principle 
encode signatures of new fields or quantum corrections near the would-be 
horizon \cite{Herdeiro:2020kba,Sennett:2017etc}. 
See \rcite{Chia:2023tle} for first results on matched-filtering searches for compact binaries involving exotic objects with large tidal deformabilities using LVK gravitational-wave  data.

At the theoretical level, ECOs offer a useful testing ground for studying various aspects of the fundamental properties of gravity. For instance, in the notation of \cref{eq:intro-g-ansatz}, the quadrupolar Love numbers of the following categories of ECOs have been computed in \rcite{Cardoso:2017cfl} in the large compactness limit $\xi\equiv\Rstar/\rs\rightarrow1$:
\begin{align}
\text{Wormhole:} &  \qquad\qquad k_{\ell=2}^{(\mathrm{E})}  \approx k_{\ell=2}^{(\mathrm{B})}  \approx \frac{4}{15\log\xi} ,
   \hspace{3cm}      \,  \\
\text{Perfect mirror:} &   \qquad\qquad k_{\ell=2}^{(\mathrm{E})}  \approx k_{\ell=2}^{(\mathrm{B})} \approx \frac{8}{15\log\xi}, 
    \\
\text{Gravastar:} &  \qquad\qquad  k_{\ell=2}^{(\mathrm{E})}  \approx k_{\ell=2}^{(\mathrm{B})} \approx \frac{16}{45\log\xi},
\end{align}
and similarly for higher $\ell$. From these results, one sees that the Love numbers vanish in the black hole limit (as expected), and they do so in a way that preserves the even-odd duality --- a clear remnant of the Chandrasekhar symmetry in general relativity (cf. \cref{sec:chandra}) \cite{1975RSPSA.343..289C,Chandrasekhar:1975zza,Chandrasekhar:1985kt,Solomon:2023ltn}. This suggests that an explanation in terms of (weakly broken) symmetries should exist.

\paragraph{Environmental effects}
Non-vacuum environments surrounding black holes --- including accretion disks, boson clouds, 
and dark matter halos --- can  induce non-zero tidal Love numbers, and can mimic or mask signatures 
of modified gravity. 
{The specific case of tidal responses of black holes surrounded by ultralight scalar fields has been studied by \rcite{DeLuca:2021ite,DeLuca:2022xlz} for scalar- or vector-type perturbations, and by \rcite{Arana:2024kaz} for gravitational tides in the Newtonian limit.\footnote{These systems are usually referred to as \textit{gravitational atoms}~\cite{Arvanitaki:2010sy,Baumann:2019eav} because, in the non-relativistic limit, they are described by the Schr\"odinger equation. They are hypothesized to form either through accretion (see, e.g., \rcite{Barranco:2012qs,Hui:2019aqm,Clough:2019jpm,Cardoso:2022vpj,Cardoso:2022nzc}) or through superradiant instabilities~\cite{Arvanitaki:2009fg,Brito:2014wla,East:2017ovw,Hui:2022sri} (see \rcite{Brito:2015oca} for a review).} }
\rcite{Cannizzaro:2024fpz} studied black holes surrounded 
by thin accretion disks, showing that the environmentally-induced Love numbers can be 
large enough to obscure any modified-gravity tidal signal and to intrinsically limit tidal 
tests of exotic compact objects, while also finding that next-generation detectors such as 
LISA and the Einstein Telescope could measure these environmental parameters with high 
precision. Environmental Love numbers induced by surrounding matter distributions 
were further analyzed in \rcite{DeLuca:2024uju}, establishing scaling relations between 
tidal deformability and environmental mass that provide a model-agnostic framework for

The pattern that emerges is unambiguous: vanishing is fragile. Zero Love numbers survive only in 
the exact vacuum GR black hole, protected by a hidden symmetry (cf. \cref{sec:sym}) that departures 
from that idealization tend to immediately destroy. In this sense, a future non-zero 
measurement would be less a detection of new physics than a confirmation that real 
astrophysical black holes are not fully the classical vacuum solutions 
of textbook GR.

\subsection{Love numbers of neutron stars}
\label{sec:NS_Love}

During the final stages of a neutron-star binary inspiral, the stars experience tidal distortions caused by their mutual gravitational interaction while the orbit shrinks through gravitational-wave emission. These tidal effects leave characteristic signatures in the gravitational-wave signal. It has long been appreciated~\cite{Flanagan:2007ix,Damour:2012yf,Favata:2013rwa} that observing such signatures can provide constraints on the equation of state (EoS) of ultra-dense nuclear matter, a property that remains poorly understood and impossible to test in terrestrial labs~\cite{Ozel:2016oaf,Oertel:2016bki,Baym:2017whm}. An attempt to extract the tidal deformability of a neutron star was carried out for the gravitational-wave event GW170817~\cite{LIGOScientific:2017vwq,LIGOScientific:2018cki,De:2018uhw,Narikawa:2021pak}, and the inferred upper limit supports relatively soft equations of state that predict comparatively compact neutron stars~\cite{Landry:2020vaw}.\footnote{The first measurement of $\lambda_2$ from GW170817 placed the constraint $\tilde{\Lambda} \leq 800$ at the $90\%$ credible level~\cite{LIGOScientific:2018cki,Raithel:2018ncd} (see these references for the notation), inaugurating the era of multimessenger neutron-star EoS inference~\cite{Chatziioannou:2020pqz,Yunes:2022ldq,Chatziioannou:2024jsr}.}
An overview of these developments can be found in \rcite{Lattimer:2012nd,Rezzolla:2018jee,GuerraChaves:2019foa,Burns:2019byj,Chatziioannou:2020pqz,Lattimer:2021emm,Burgio:2021vgk,Yunes:2022ldq,Chatziioannou:2024jsr,Glendenning:1997wn,Haensel:2007yy}, while additional discussions on the potential for future observational constraints are presented in \rcite{Pacilio:2021jmq} and subsequent studies. (See also \rcite{ET:2025xjr} for an analysis of the Einstein Telescope's ability to detect the tidal Love numbers of neutron stars.)

\subsubsection{Static tides}

The tidal effects that are most directly measurable with current gravitational-wave observations correspond to the r\'egime of static (or adiabatic) tides, in which the temporal variation of the external tidal field can be neglected. To be precise, for neutron stars the adiabatic r\'egime is defined as an expansion in the ratio of the internal hydrodynamical timescale to the external orbital timescale.
The study of the static Love numbers of neutron stars was pioneered in \rcite{Flanagan:2007ix,Hinderer:2007mb}, where the dimensionless static quadrupolar (electric) tidal Love number $k_2$ was first computed in full general relativity and shown to enter the gravitational-wave phase at fifth post-Newtonian (5PN) order, as had long been recognized~\cite{Damour:1984rbx}. Solving the $H_0$ equation of motion \eqref{eq:H0-static} for $\ell=2$ %
yields~\cite{Hinderer:2007mb} (momentarily setting $G=1$)
\begin{equation}
H_0 =
c_1 \left(\frac{r}{M}\right)^2 \left(1-\frac{2M}{r}\right)
+ 3c_2 \left(\frac{r}{M}\right)^2
\left[
1-\frac{2M}{r}
- \frac{M(M-r)(2M^2+6Mr-3r^2)}{r^2(2M-r)^2}
+ \frac{3}{2}\log\!\left(\frac{r}{r-2M}\right)
\right].
\end{equation}
The integration constants can be rewritten in terms of the external field amplitude $\mathcal{E}_2$ and the induced falloff $\lambda_2$ via
\begin{equation}
c_1 = \frac{1}{M^2} \mathcal{E}_2,
\qquad
c_2 = \frac{15}{8}\frac{1}{M^3}\lambda_2\mathcal{E}_2.
\end{equation}
In terms of the dimensionless coefficient $k_2$, which is related to the dimensionful coupling $\lambda_2$ via $\lambda_2 = \frac{2}{3} k_2 \Rstar^5$, where $\Rstar$ is the stellar radius, one finds~\cite{Hinderer:2007mb}
\begin{equation}
\begin{split}
k_2 & =
\frac{8C^5}{5}(1-2C)^2[2+2C(y-1)-y] \times
\\
& \quad \times
\Bigg\{
2C(6-3y+3C(5y-8))
+4C^3[13-11y+C(3y-2)+2C^2(1+y)]
\\
&\quad \quad\quad
+3(1-2C)^2[2-y+2C(y-1)]\log(1-2C)
\Bigg\}^{-1},
\end{split}
\label{eq:k2Hinderer}
\end{equation}
where $C$ is the compactness, defined by\footnote{Note that formally the compactness $C$ is a number between zero and $1/2$ (when $\Rstar$ equals the Schwarzschild radius). However, more stringent bounds on $C$ exist, derived under relatively mild assumptions about the background solution or physical considerations about the matter sector. One notable example is the Buchdahl limit~\cite{Buchdahl:1959zz}, $C < 4/9$, valid for static, spherically symmetric, non-vacuum spacetimes, with roots tracing back to Schwarzschild~\cite{1916skpa.conf..424S} (see also~\rcite{Bondi:1964zz,Wald:1984rg}). Further bounds exist that employ microphysical EoS models; see, e.g.,~\rcite{1984ApJ...278..364L,Haensel:1989mvc,Koranda:1996jm,Lattimer:2015nhk}.} 
\begin{equation}
C\equiv \frac{M}{\Rstar} ,
\end{equation}
and $y$ is the  logarithmic derivative at the stellar surface, 
\be
y \equiv   \frac{ \Rstar H_0'(\Rstar)}{H_0(\Rstar)} 
\;. 
\label{eq:ydef}
\ee
Once the expression for $k_2$ is obtained, the next step is to solve the Einstein equations inside the object, imposing regular boundary conditions at the origin. Matching and continuity conditions at the surface then completely determine the variable $y$, and thus the asymptotic falloff of the metric perturbation and the Love numbers~\cite{Damour:2009vw,Binnington:2009bb}. A generalization of \cref{eq:k2Hinderer} to higher multipole Love numbers was obtained in \rcite{Damour:2009vw,Binnington:2009bb}. Note that for generic EoS, closed-form expressions for the Love numbers are generally not available. Numerical results for various choices of the EoS can be found in \rcite{Hinderer:2007mb,Damour:2009vw,Binnington:2009bb}. Some plots for more realistic EoS can be found in \rcite{Hinderer:2009ca}; see also \cref{fig:MR} below.

\begin{figure*}[t]
\begin{center}
\includegraphics[width=0.435\textwidth]{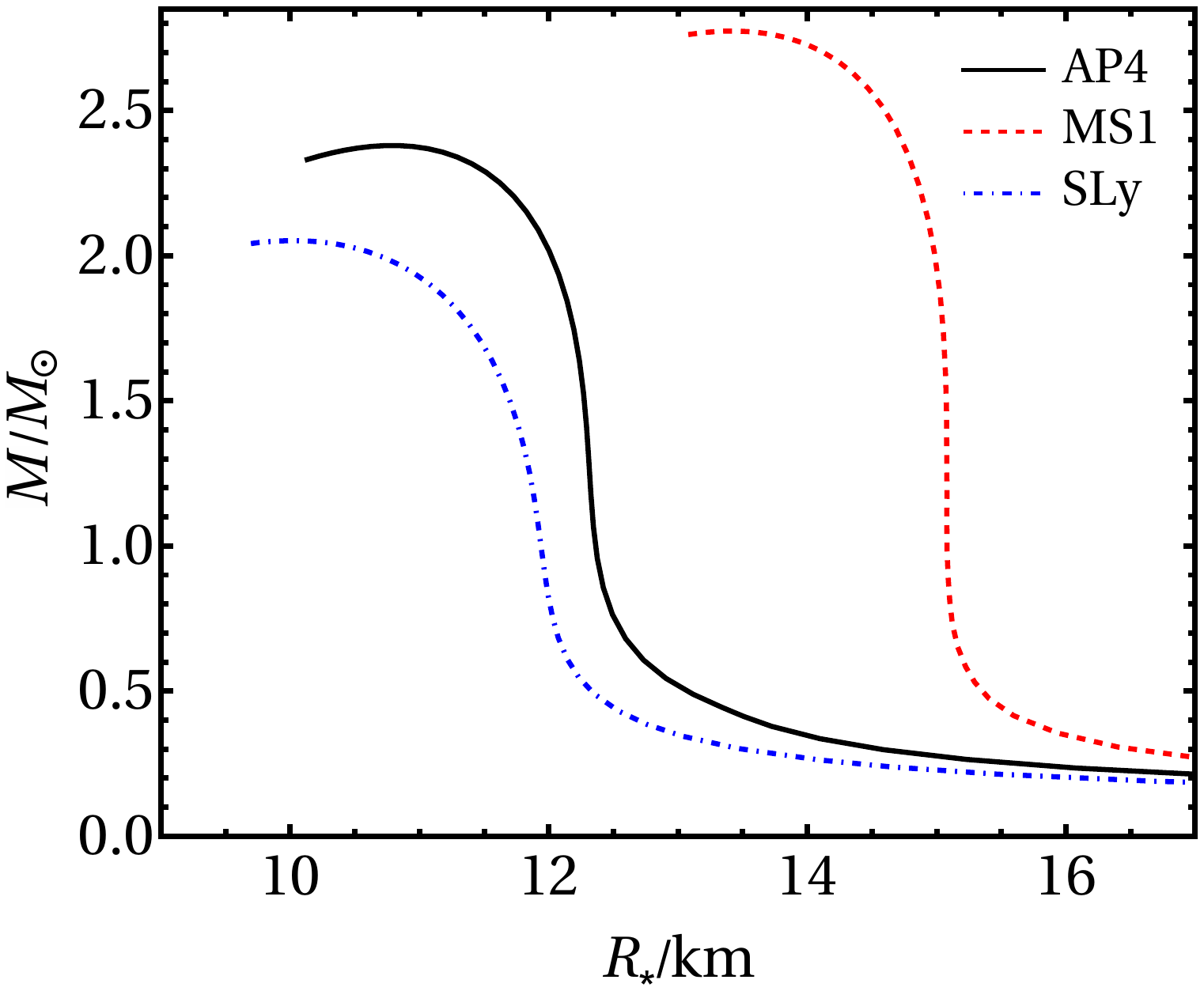}
\hspace{1cm}
\includegraphics[width=0.435\textwidth]{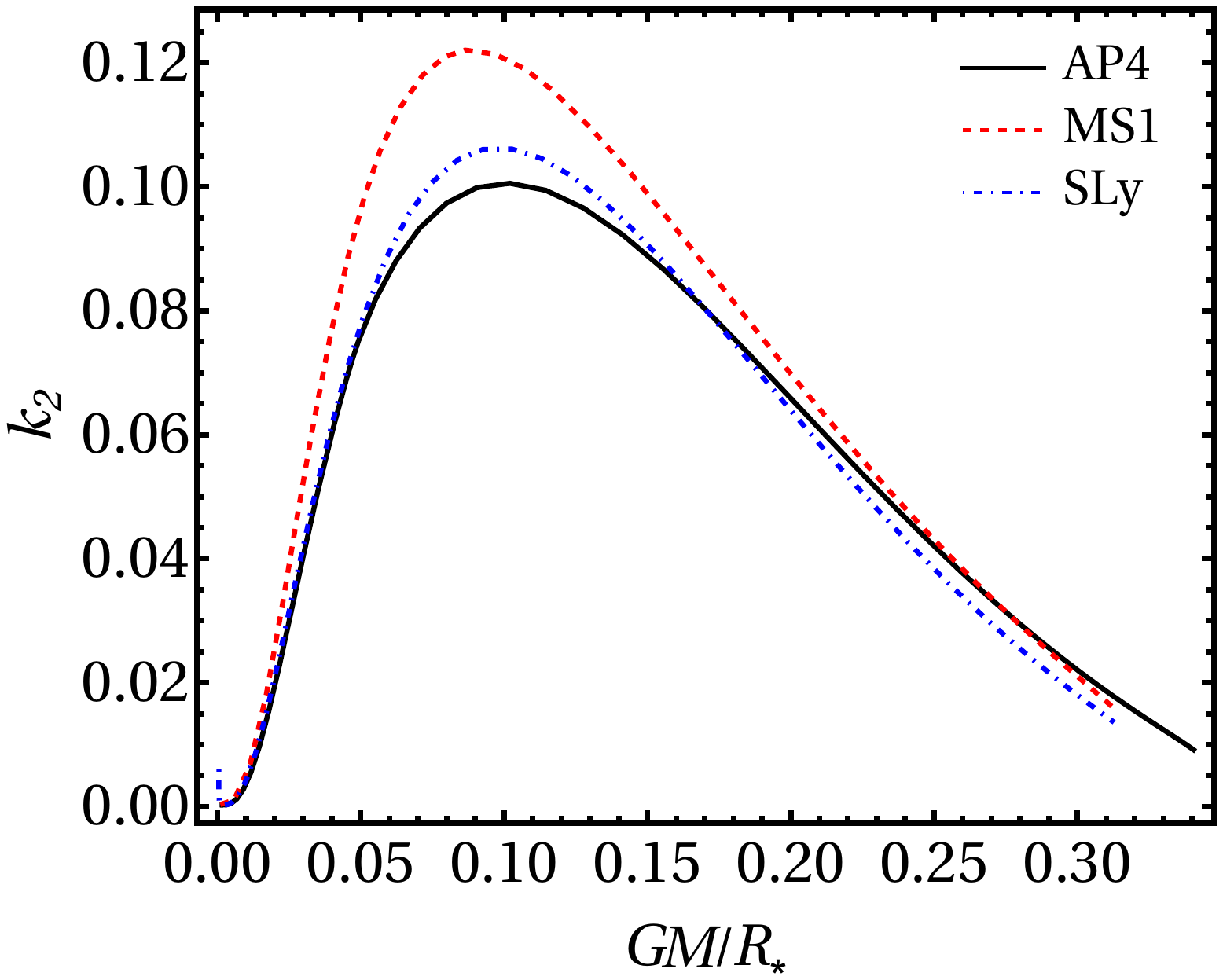}
\end{center}
\caption{Figure adapted from \rcite{Pani:2025qxs}. Examples of tabulated, nuclear-physics-based EoS --- specifically APR~\cite{Akmal:1998cf}, MS1~\cite{Mueller:1996pm}, and SLy4~\cite{Douchin:2001sv} --- are shown (see \rcite{Pani:2025qxs} and references therein for details).
The left panel displays the relation between the mass parameter $M \equiv \mathcal{M}(\Rstar)$ and the stellar radius $\Rstar$ for the neutron-star EoS considered. The right panel shows the corresponding (linear) tidal Love number as a function of the compactness $C = GM/\Rstar$. See \rcite{Hinderer:2009ca} for earlier results.}
\label{fig:MR}
\end{figure*}

For both simple polytropic and more realistic EoS, the quantity $y$ typically varies only slowly as the compactness changes within a physically relevant range. Performing a Taylor expansion of $k_2$ in the limit of small compactness, while keeping $y$ fixed, gives
\begin{equation}
k_2 = -\frac{y-2}{2 (y+3)} + \frac{5 C \bigl(y(y+2)-6\bigr)}{2 (y+3)^2} + \mathcal{O}(C^2),
\end{equation}
i.e.,  $k_2 = \mathcal{O}(1)$ as $C \to 0$. In particular, the first term in the expansion  in $M/\Rstar$ reproduces the Newtonian
result~\cite{Hinderer:2007mb}. For typical realistic EoS (see, e.g., \rcite{Pani:2025qxs} and references therein), one finds that $y \to 2$ in the small-compactness limit, so that $k_2 \sim \mathcal{O}(C)$.

As summarized in \cref{sec:NS}, magnetic (axial) Love numbers were computed in \rcite{Damour:2009vw,Binnington:2009bb,Landry:2015cva} and later revisited in \rcite{Pani:2018inf}. The computation of Love numbers for (slowly) rotating stars was carried out in \rcite{Pani:2015hfa,Pani:2015nua,Landry:2015zfa,Gagnon-Bischoff:2017tnz,Poisson:2020mdi,Castro:2021wyc}.

Although the static response provides the dominant contribution to the tidal deformability of neutron stars within the adiabatic approximation to a binary inspiral, future observations may become sensitive to subleading tidal effects~\cite{ET:2025xjr}. These include dynamical tides, which account for the time dependence of the external tidal field, and nonlinear tidal responses, which arise from higher-order couplings in the stellar deformation. Probing these r\'egimes would offer additional insight into the internal structure and composition of neutron stars. This prospect has recently motivated substantial theoretical and phenomenological work, including the modeling of neutron stars subject to more general tidal perturbations, the incorporation of such effects into gravitational-waveform models, and studies assessing their detectability and astrophysical implications. 
Below we survey some of these aspects.

\subsubsection{Dynamical tides}
\label{sec:dynamicaltidesNS}

In the context of dynamical tides, previous works in the literature (see \cref{Table3} for a summary) can be broadly distinguished according to the following criteria: whether they adopt a Newtonian approximation~\cite{Lai:1993di,1994ApJ...426..688R,Kokkotas:1995xe,Chakrabarti:2013xza,Passamonti:2020fur,Passamonti:2022yqp,Yu:2024uxt,Pnigouras:2025muo} or a fully general relativistic framework~\cite{Chakrabarti:2013lua,Hinderer:2016eia,Steinhoff:2016rfi,Pitre:2023xsr,Saketh:2024juq,HegadeKR:2025qwj,HegadeKR:2026iou,Jarequi:2026cyp}; how they parametrize the stellar interior (using normal modes~\cite{Hinderer:2016eia,Steinhoff:2016rfi,Schmidt:2019wrl,Andersson:2019dwg,Gupta:2020lnv,Steinhoff:2021dsn,Gupta:2023oyy,HegadeKR:2024agt,HegadeKR:2025qwj,HegadeKR:2026iou} or a mode-independent description~\cite{Chakrabarti:2013lua,Pitre:2023xsr,Saketh:2024juq,Jarequi:2026cyp}); and how they define the induced response at the level of the exterior solution (effective field theory~\cite{Chakrabarti:2013lua,Steinhoff:2016rfi,Jakobsen:2023pvx,Mandal:2023hqa,Mandal:2023lgy,Saketh:2024juq,Jarequi:2026cyp} versus post-Newtonian approaches~\cite{Poisson:2020vap,Pitre:2023xsr,HegadeKR:2024agt,HegadeKR:2025qwj}).

A substantial body of work computes dynamical Love numbers within a Newtonian-inspired fluid description, in which the tidal deformation is expressed as a spectral expansion over the normal modes of oscillation of the neutron star. In this approach, the frequency-domain Love number is given by (see, e.g., \rcite{Pitre:2023xsr}; see also \rcite{Poisson_Will_2014} for an introduction to the tidal deformation
of fluid bodies)
\begin{equation}
{k}_\ell(\omega)
=
\frac{2\pi \ell^{2}}{2\ell+1}
\sum_{n}
\frac{GM/\Rstar^{3}}{\omega_{n\ell}^{2}} \mathcal{A}_{n\ell}
\frac{\mathcal{I}_{\ell n}^2}{\mathcal{N}_{n\ell}},
\label{eq:freq_love}
\end{equation}
where $\omega_{n\ell}$ is the frequency of the $n\ell m$ mode,
\begin{equation}
\mathcal{A}_{n\ell} \equiv \left(1 - \frac{\omega^{2}}{\omega_{n\ell}^{2}}\right)^{-1}
\end{equation}
is the amplification factor,\footnote{In the context of a binary inspiral, taking a representative tidal frequency of $500~\mathrm{Hz}$ and a typical $f$-mode frequency of $1000~\mathrm{Hz}$, the amplification factor evaluates to $\mathcal{A} \simeq 1.19$, corresponding to an approximately $20\%$ increase in the tidal deformation.} $\mathcal{I}_{n\ell}$ is an overlap integral that depends on the mode solutions for the star and is sensitive to the internal EoS,  and $\mathcal{N}_{n\ell}$ is a (positive) normalization factor.
In principle, the mode expansion \eqref{eq:freq_love} involves a sum over an infinite set of stellar oscillation modes, but in practice it can typically be truncated to a finite number of terms with little loss of accuracy. In the most extreme simplification, only the $n=0$ term is retained. This corresponds to the fundamental mode, or $f$-mode, whose eigenfunctions have the smallest number of radial nodes and yield the largest overlap integrals~\cite{Pitre:2023xsr}. As a result, the $f$-mode generally provides the dominant contribution to the frequency-domain Love numbers, and the $f$-mode approximation often offers a good description of the tidal response.\footnote{Note that, in units of $(\Rstar^3/GM)^{1/2}$, the $f$-mode frequency is order unity (see e.g.,~\rcite{Pitre:2023xsr} for some explicit numbers in polytropic stellar models). The truncation to the $f$-mode becomes exact in the incompressible stellar model with constant density ($\rho=\mathrm{const}$).} In this approximation, the static and dynamical Love numbers are related by (in the notation of \rcite{Pitre:2023xsr}):
\begin{equation}
\ddot{k}_\ell(\omega) \simeq \frac{GM}{\Rstar^{3}} \frac{1}{\omega_{0\ell}^{2}}\, k_\ell .
\end{equation}

Within a normal-mode description, the impact of dynamical tides on gravitational-wave observations has been  studied in \rcite{Hinderer:2016eia,Steinhoff:2016rfi,Yu:2025ptm,Ghosh:2025glz,Pereira:2025xsi}. Models describing neutron stars undergoing dynamical tidal deformations have been developed in \rcite{Steinhoff:2021dsn,Andersson:2019dwg,Passamonti:2020fur,Passamonti:2022yqp,Pnigouras:2025muo}. Gravitational-waveform models incorporating these effects have been presented in \rcite{Schmidt:2019wrl,Mandal:2023lgy,Mandal:2023hqa}, while analyses of the prospects for future detections can be found in \rcite{Andersson:2017iav,Williams:2022vct,Pratten:2021pro}.
Direct constraints on the fundamental $f$-mode frequencies from GW170817 were derived in \rcite{Pratten:2019sed} using an inspiral gravitational-wave phase model that explicitly incorporates the $f$-mode frequency.
In addition, the description of dynamical tides has been extended to include the gravitomagnetic sector of the tidal interaction \cite{Flanagan:2006sb,Poisson:2020eki,Gupta:2020lnv,Gupta:2023oyy}, and related effects have been considered in studies of the neutron-star $p$-$g$ mode instability~\cite{Weinberg:2015pxa}.
An analysis of dynamical tides in a gravitational-wave-driven coalescing binary involving a rapidly spinning neutron star is done in \rcite{Yu:2024uxt}.

In contrast to a normal-mode description of tidal deformation, in a general relativistic setup, a ``mode-less'' approach has been employed  to determine the frequency-dependent tidal response of a compact star without relying on Newtonian analogies in~\rcite{Chakrabarti:2013lua,Pitre:2023xsr,Andersson:2025iyd}.
\rcite{Chakrabarti:2013lua} defines the tidal response coefficients of non-rotating neutron stars using the effective field theory approach of \rcite{Goldberger:2004jt} to gravitational interactions in classical general relativity  (see also \rcite{Chakrabarti:2013xza} for a Newtonian application), while solving  the perturbation equations  numerically in the stellar interior for a simple equation of state. For this setup, it is found that the response function is well approximated by an $f$-mode description~\cite{Chakrabarti:2013lua,Steinhoff:2016rfi}.
A post-Newtonian framework to define the tidally induced multipole moments is instead adopted in \rcite{Pitre:2023xsr}, where the Einstein equations of the relativistic star are solved in a small-frequency expansion. In particular, for certain choices of EoS, this formulation of dynamical tides is shown to accurately reproduce the $f$-mode truncation of the stellar mode expansion, assuming negligible contributions from $g$-modes.
To address the potential breakdown of this expansion during the late inspiral due to low-frequency $g$-modes, the relativistic framework of \rcite{Pitre:2023xsr} was extended in \rcite{HegadeKR:2024agt,HegadeKR:2024slr,HegadeKR:2025qwj}. Building on definitions introduced in \rcite{Ripley:2023qxo}, these works implement a resummation scheme in the frequency domain, while treating the exterior problem analogously to \rcite{Pitre:2023xsr}.
In particular, in \rcite{HegadeKR:2024agt}, resonant effects due to the $f$- and $g$-modes were included, and the dissipative tidal deformability arising from bulk and shear viscous dissipation --- assuming a simple viscous profile for both bulk and shear viscosity --- was computed. 
A study of 1PN corrections to the orbital dynamics of binaries with dissipative tidal interactions was presented in \rcite{HegadeKR:2024slr},\footnote{ See \rcite{Damour:1990pi,Damour:1991yw,Damour:1992qi,Racine:2004xg,Racine:2004hs,Vines:2010ca} for foundational works in the PN literature on this topic.} improving on the heuristic results previously obtained in \rcite{Ripley:2023lsq}; in \rcite{HegadeKR:2024slr}, the resulting waveform model was applied to constrain the individual dissipative tidal deformabilities of each binary component responsible for the GW170817 event using observational data.
The framework of relativistic tides of \rcite{HegadeKR:2024agt,Poisson:2020vap,Pitre:2023xsr} was extended in \rcite{HegadeKR:2026iou} to include frequency-dependent conservative and dissipative tidal responses, incorporating input from microphysical models of neutron star matter.

Relativistic dynamical tidal  effects of non-spinning neutron stars were also studied in \rcite{Saketh:2024juq,Jarequi:2026cyp} within the framework of the worldline EFT, where the dissipative coefficients were obtained via scattering-amplitude matching with relativistic stellar perturbation theory. In particular, \rcite{Saketh:2024juq} computed the electric quadrupolar Love number in the static limit ($\omega=0$) and the leading dissipation number at linear order in $\omega$, both to all orders in compactness for various realistic equations of state. That reference also provided an estimate of the gravitational-wave dephasing accumulated during inspiral in the LVK band due to tidal heating. The results of \rcite{Saketh:2024juq} were then extended to NNLO in \rcite{Jarequi:2026cyp}, i.e., through second order in $\omega$ in the conservative sector, although at $\mathcal{O}(\omega^2)$ only the logarithmic running of the dynamical Love numbers was obtained.

For a modern EFT treatment based on a covariant fluid effective action for perfect-fluid neutron stars, see \rcite{Martinez-Rodriguez:2026omk}.

\subsubsection{Nonlinear tides and mode coupling}
\label{sec:neutronstarnonlinear}

Similarly to the case of the dynamical tidal response, studies of nonlinear tidal deformability of neutron stars in the literature can broadly be divided into Newtonian-inspired analyses and fully general-relativistic calculations. \rcite{Yu:2022fzw} originally employed a Newtonian description of the stellar fluid and gravitational field, representing the tidal deformation in terms of the star's normal modes of oscillation and analyzing the nonlinear driving and coupling among these modes. Within this framework, the authors demonstrated that the dynamical-tide r\'egime is strongly affected by nonlinear aspects of the tidal dynamics. In particular, they showed that nonlinearities can induce a phase shift in the gravitational-wave signal corresponding to a correction of order 10\% -- 20\% relative to predictions obtained within a linearized treatment. The theoretical framework underlying these calculations is described in \rcite{2012ApJ...751..136W}, itself based on the formalism developed in \rcite{Schenk:2001zm}.
Other scenarios in which nonlinear tidal effects have been investigated include the instability arising from nonlinear couplings between the $p$- and $g$-modes of a neutron star \cite{Weinberg:2013pbi,Weinberg:2015pxa}. Possible observational signatures of this mechanism in gravitational-wave signals have been discussed in \rcite{Essick:2016tkn,Essick:2018wvj,LIGOScientific:2018ehx,Reyes:2018bee}. Another example is resonant locking between a $g$-mode frequency and the tidal driving frequency \cite{Kwon:2024zyg}.
Nonlinear relativistic corrections to neutron-star mode-tide couplings have also been explored, for instance in \rcite{Nouri:2021mvb}. 

By contrast, fully relativistic analyses of nonlinear tidal effects that do not rely on a mode decomposition of the deformation have only recently appeared in \rcite{Pitre:2025qdf,Pani:2025qxs}. These works solve the full Einstein equations both inside and outside the compact object, assuming a perfect-fluid stellar model, and compute the leading quadrupolar nonlinear Love number in the even-parity sector.
\rcite{Pitre:2025qdf} focuses primarily on polytropic stellar models within a post-Newtonian framework (see \cref{sec:post-newtonian} and \rcite{Poisson:2020vap}) and shows that the nonlinear tidal coefficient $p_2$ (see \cref{eq:gttPN}) can reduce the characteristic frequency parameter by up to $\sim15$\%  compared with estimates based on a purely linear description of the tidal deformation, effectively triggering a resonance at an earlier frequency during the inspiral. \rcite{Pani:2025qxs}, on the other hand, adopts the effective field theory framework developed in \rcite{Goldberger:2004jt,Goldberger:2005cd}, deriving nonlinear Love numbers by matching the relativistic solution to the point-particle EFT description (see \cref{sec:nonlinearities}) for a range of realistic, nuclear-physics-motivated equations of state. In particular, it is shown in \rcite{Pani:2025qxs} that quadratic Love numbers can reach values as large as $\sim10$\% of the linear Love numbers during the late inspiral phase.
It is worth recalling that quadratic Love numbers --- much like the dynamical tidal response --- receive an additional enhancement in the small-compactness limit. As a consequence, despite formally entering the gravitational-wave phase only at 8PN order, their contribution may nevertheless be non-negligible and has been argued in those references to be a necessary ingredient for future high-precision gravitational-wave modeling.

\begin{figure*}[t]
\begin{center}
\includegraphics[width=0.5\textwidth]{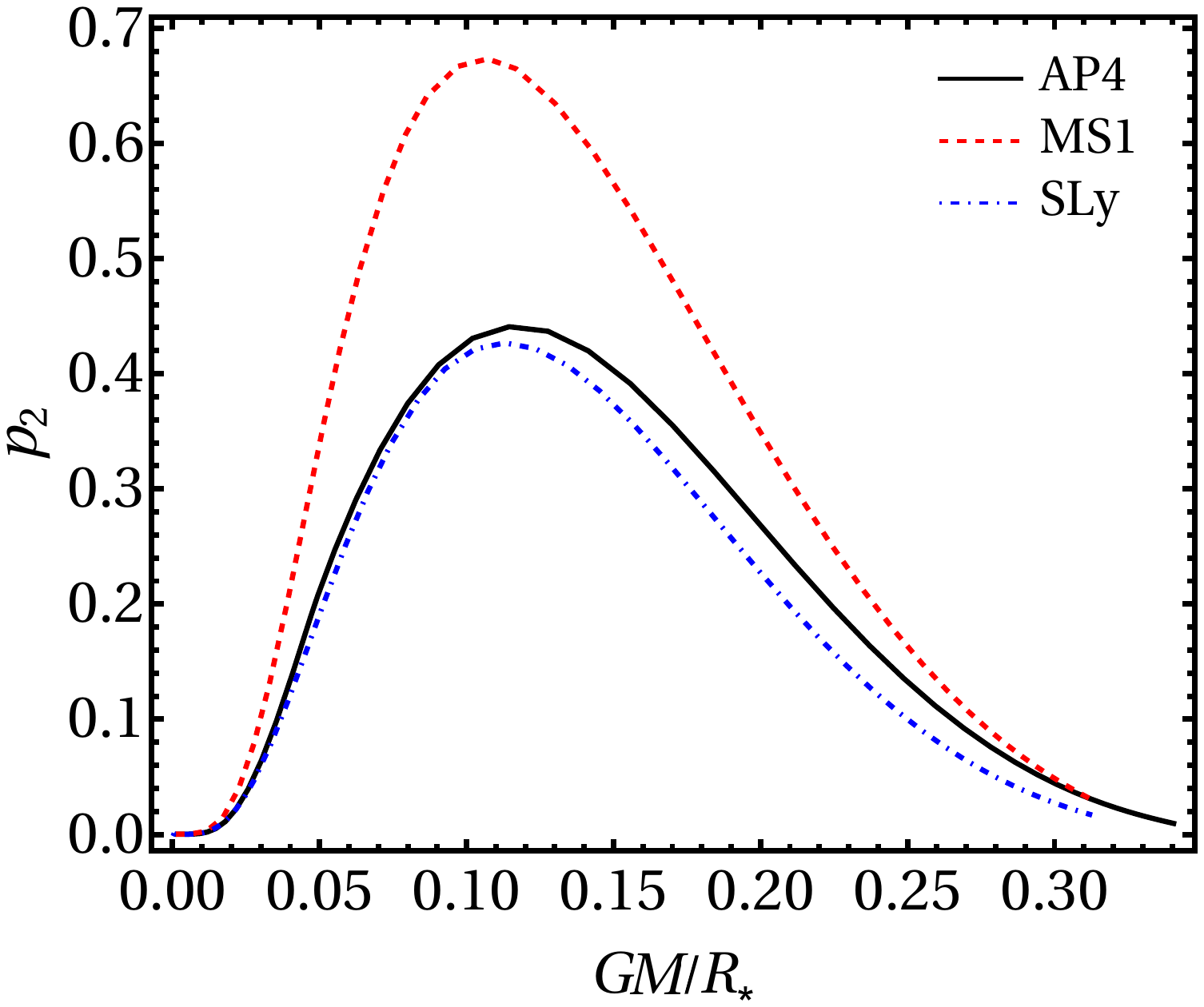}
\end{center}
\caption{Figure adapted from \rcite{Pani:2025qxs}. The dimensionless coefficient $p_2$, capturing the leading quadratic tidal deformation as a function of compactness for different EoS, is shown.}
\label{fig:p2-222}
\end{figure*}

In summary, the procedure used to solve the nonlinear Einstein equations in the static limit and compute the nonlinear (static) Love numbers closely follows that for black holes~\cite{Riva:2023rcm,Iteanu:2024dvx,Combaluzier-Szteinsznaider:2024sgb}. As an example, at second order in perturbation theory the equation of motion \eqref{master1} for $H_0$ reads~\cite{Pani:2025qxs}
\begin{multline}
    {}^{(2)}\! H_0 \Bigg[\frac{4 \pi  r (\bar p+ \bar \rho ) }{(r-2 \Mm) c_s^2}-\frac{2 \Mm \left(-r \ell(\ell+1)+52 \pi  r^3 \bar p+20 \pi  r^3 \bar \rho
   \right)+r^2 \ell(\ell+1)+4 \Mm^2+4 \pi  r^4 \left[ \bar p \left(16 \pi  r^2 \bar p-9\right)-5 \bar \rho \right]}{r^2 (r-2
   \Mm)^2}\Bigg]
   \\
   +{}^{(2)}\! H_0'\frac{2 \left[r-\Mm+2 \pi  r^3 (\bar p- \bar \rho )\right]}{r (r-2 \Mm)}+ {}^{(2)}\! H_0'' =S_{H_0} \;,
   \label{master2}
\end{multline}
where ${}^{(2)}\! H_0$ is a second-order field and  $S_{H_0}$ is a source term, evaluated on the static linear solution for $H_0$, which we shall denote by ${}^{(1)}\! H_0$. In particular, focusing on the modes $\ell_1 = \ell_2 = \ell = 2$ and $m_1 = m_2 = m = 0$, the source is~\cite{Pani:2025qxs}
\begin{multline}
   - \left(\tfrac{1}{7}\sqrt{\tfrac{5}{\pi}}\right)^{-1} S_{H_0} =    ({}^{(1)}\! H_0')^2 + {}^{(1)}\! H_0  {}^{(1)}\! H_0' \frac{3   \left(\Mm
   +4 \pi  r^3 \bar  p\right)}{r(r-2
   \Mm)} + ({}^{(1)}\! H_0)^2 \Bigg[ \frac{\pi  r (\bar p+ \bar \rho )}{c_s^4(r-2\Mm)} \left( 1 + (c_s^2)'\frac{r(r-2\Mm)}{\Mm + 4\pi r^3 \bar p} \right) 
   \\ 
     +\frac{6 \pi  r  (\bar  p+ \bar  \rho ) }{(r-2
   \Mm) c_s^2}+\frac{  r \Mm \left[ 2 \pi  r^2 (71 \bar  p+15 \bar  \rho )+3\right]-7
   \Mm^2+\pi  r^4 \left[  \bar p \left(176 \pi  r^2 \bar  p-27\right)-15 \bar  \rho \right] +3 r^2 }{r^2 (r-2
   \Mm)^2} \Bigg]
      \;.
      \label{source}
\end{multline}
\Cref{master2} can be solved in perturbation theory. Outside the star, the most general solution can be obtained using the Green's function method and admits an analytic closed-form expression. Inside the star, the equation can be solved numerically for a given equation of state. After imposing a tidal-field boundary condition at infinity and regularity at the center of the star, the coefficient of the induced falloff can be extracted from the large-distance expansion. In the small-compactness limit, the falloff coefficient $p_2$ (see \cref{eq:gttPN}) can be expressed as~\cite{Pani:2025qxs}
\begin{equation}
    p_2 = \frac{50 z}{7 (y+3)^3} + \frac{25 C [ y (y+3)-2 z(5 y+24)]}{7 (y+3)^4}+ \mathcal{O}(C^2),
\end{equation}
in terms of the logarithmic derivative $y$ defined in \cref{eq:ydef} and
\begin{equation}
z \equiv \frac{ 7 \rs}{2  \sqrt{\frac{5}{\pi}} {}^{(1)}\! H_0(\Rstar)}\left( \frac{{}^{(2)}\! H_0(\Rstar)}{{}^{(1)}\! H_0(\Rstar)} \right)' \;.
\end{equation}
A plot of $p_2$ for some realistic equations of state is shown in \cref{fig:p2-222}.

\subsubsection{Universal I-Love-Q relations and beyond}

Special empirical relations among various couplings, hinting at universality properties largely insensitive to the internal structure and the matter equation of state, have been identified for neutron stars. A notable example is provided by the $I$-\emph{Love}-$Q$ universal relations discovered by Yagi and Yunes~\cite{Yagi:2013awa} (see \rcite{Yagi:2016bkt} for a review), which connect the moment of inertia $I$, the quadrupolar tidal Love number, and the quadrupole moment $Q$ in a manner that is insensitive to the EoS at the few-percent level. These relations hold for neutron stars across a wide range of realistic nuclear EoSs, significantly enhancing the utility of tidal measurements for neutron-star science~\cite{Yagi:2013bca,Yagi:2013baa,Yagi:2016bkt,Gagnon-Bischoff:2017tnz,Abdelsalhin:2018reg,Kunz:2022wnj}. 

The $I$-Love-$Q$ relations have since been extended in several directions: to higher multipole orders~\cite{Yagi:2013sva,Pappas:2013naa}; to differentially rotating stars~\cite{Bretz:2015rna}; to no-hair-like universal relations for neutron stars on spin-induced multipole moments~\cite{Stein:2013ofa,Yagi:2014bxa,Chatziioannou:2014tha}; to include anisotropic pressure~\cite{Yagi:2015hda,Guedes:2025gqi} (see in particular \rcite{Guedes:2025gqi} for a study of universal relations between the Love number and $f$-mode oscillations in anisotropic neutron stars, as well as preliminary constraints on the anisotropy parameter from GW170817);  to combinations of the individual tidal deformabilities in gravitational-wave models of binary systems~\cite{Yagi:2015pkc,Yagi:2016qmr}; to Love-$C$ relations (between the tidal deformability and the compactness)~\cite{Yagi:2016bkt,Maselli:2013mva,Silva:2020acr,Saffer:2021gak,Dong:2024opz,Lowrey:2024anh} (see e.g.~\rcite{Zhao:2018nyf,Jiang:2019vmf,Jiang:2020uvb} for some model-related analyses);  to the dynamical tidal sector~\cite{Saes:2025jvr}; to nonlinear order~\cite{Pani:2025qxs}; to alternative theories of gravity~\cite{Gupta:2017vsl,Saffer:2021gak,Ajith:2022uaw,Vylet:2023pkp}; and to exotic compact objects~\cite{Pani:2015tga}. These extended universal relations strengthen the case that the tidal response of neutron stars can be characterized by a small number of bulk parameters, largely independent of poorly known microphysics, and provide robust observational handles on fundamental physics. 

Despite their utility, a fundamental explanation of this universality in terms of (weakly broken) symmetries remains elusive. (See \rcite{Yagi:2014qua} for an earlier attempt, which however remained limited to an empirical isodensity-contour approximation for the star, and \rcite{Yagi:2016ejg} for a toy model of an incompressible star with anisotropic pressure that approaches the black hole limit arbitrarily closely.)
It would be interesting to investigate whether the different classes of universality relations summarized above are connected to the hidden symmetries which control black hole numbers, as we review in \cref{sec:sym}.

Overall, neutron-star Love numbers encode rich information about the internal structure, microphysical composition, and dynamics of compact stars. Their precise extraction from gravitational-wave observations is one of the primary scientific goals of current and future detector networks~\cite{AmaroSeoaneEtAl2017,Burns:2019byj,Yunes:2022ldq,Chatziioannou:2024jsr,ET:2025xjr}.

\newpage
\section{Symmetries}
\label{sec:sym}

\epigraphhead[]{
\epigraph{Love is not Love
    
    Which alters when it alteration finds.}{\textsc{Shakespeare}}
}

\subsection{Love and naturalness}
\label{sec:naturalness}

We have seen that the Love numbers of \emph{all} black holes in four dimensional general relativity vanish. We have also identified the Love numbers as Wilson coefficients of the point-particle effective field theory discussed in \cref{sec:EFT}. Together these facts reveal a naturalness problem \cite{Rothstein:2014sra,Porto:2016pyg,Porto:2016zng}.

According to standard EFT lore, an effective action should contain every term allowed by its symmetries. On ``naturalness'' grounds one generally expects the dimensionless Wilson coefficients accompanying these terms to be $\sim\mathcal{O}(1)$. When this expectation is violated by a Wilson coefficient being tiny or even vanishing, we call it a fine tuning problem. The three major, open naturalness problems in high-energy physics --- the cosmological constant, electroweak hierarchy, and strong CP problems --- are classic examples of this notion of naturalness \cite{Craig:2022eqo}. In the case of the cosmological constant problem \cite{Weinberg:1988cp,Weinberg:2000yb,Burgess:2013ara,Padilla:2015aaa}, the Wilson coefficient in question is the cosmological constant itself, multiplying the unique zero-derivative term $\sdg$ that is built out of the metric $g_\mn$ and transforms properly under diffeomorphisms, and which on naturalness grounds would be expected to be comparable to the cutoff scale of the EFT of gravity, likely somewhere between the eV scale and the Planck scale, in contrast to its much smaller observed value (which would correspond approximately to an meV cutoff). The electroweak hierarchy problem is the smallness of the Higgs mass relative to the Standard Model cutoff, which is somewhere between the TeV scale and the Planck scale. For the strong CP problem \cite{Dine:2000cj,Peccei:2006as,Hook:2018dlk} the unnaturally small parameter is the $\theta$ angle of quantum chromodynamics.
The vanishing of black hole Love numbers poses a similar problem, although this is a classical ``fine tuning'' rather than a quantum one. 

The problem is not simply the physicist's aesthetic preference for order-unity Wilson coefficients. It is that a small Wilson coefficient may receive large corrections from other sources. In the case of naturalness problems in quantum field theory, for instance, each loop order introduces new large corrections and one must perform the tuning anew \cite{Padilla:2015aaa}. This occurs because there is no symmetry to ``protect'' these terms from such large corrections.

Concretely, from the point-particle EFT perspective there is no
obstruction to writing down non-zero tidal operators --- they are
perfectly consistent with all the (apparent) symmetries of the low-energy
effective theory, namely diffeomorphism invariance and the isometries
of the background. Yet their Wilson coefficients are observed (and can
be verified by explicit calculation) to vanish exactly, to all
multipole orders and for all black holes in four-dimensional general
relativity, across both the electric and magnetic sectors
\cite{Binnington:2009bb,Damour:2009va,Hui:2021vcv,Charalambous:2021kcz}.
This exact vanishing
is not an artifact of any particular gauge or matching scheme
\cite{Gralla:2017djj,Ivanov:2022hlo}. From the EFT standpoint, such
an exact cancellation with no apparent symmetry reason is precisely
what one means by a fine-tuning. The resolution, as we discuss in the
rest of this section, lies in the existence of novel symmetries of general
relativity that are not manifest in the low-energy point-particle
description, but which nonetheless are present and enforce the vanishing of all tidal
Wilson coefficients.

In the literature, the various approaches for the resolution of this puzzle can be broadly divided into three categories: the first is based on a set of symmetry generators --- originally proposed in \rcite{Hui:2021vcv,Hui:2022vbh} and later generalized to nonlinear order in \rcite{Combaluzier-Szteinsznaider:2024sgb,Kehagias:2024rtz} --- dubbed ``\textit{ladder symmetries}'' for the way they act on the space of solutions decomposed in spherical harmonics (see also, e.g.,~\rcite{Katagiri:2022vyz,Lupsasca:2025pnt} for related, albeit distinct, constructions); the second is usually referred to as \textit{``Love symmetry''}~\cite{Charalambous:2021kcz,Charalambous:2022rre} and is based on a hidden $\mathfrak{sl}(2,\mathbb R)$ algebra in a suitably defined near-zone approximation of the perturbation equations; and the third combines the $\mathrm{SL}(2,\mathbb R)$ Ehlers symmetry for GR solutions admitting a timelike Killing vector (cf. \cref{sec:nonlineartheory}) with a \textit{spurion argument}~\cite{Kol:2011vg,Parra-Martinez:2025bcu}.

In this section, we discuss these symmetries and their subsequently proposed variants and generalizations~\cite{BenAchour:2022uqo,Katagiri:2022vyz,Berens:2022ebl,Charalambous:2023jgq,Atkins:2023axs,Sharma:2024hlz,Charalambous:2024tdj,Rai:2024lho,Gounis:2024hcm,BeltranJimenez:2024zmd,Charalambous:2025ekl,Lupsasca:2025pnt,Berens:2025jfs,Sharma:2025xii,DeLuca:2025zqr,Cvetic:2026wht,Kumar:2026mpv}, organizing the presentation according to whether they operate in the exact static limit or at non-zero frequency. Further details are summarized in \cref{tab:KEG_summary}.

\begin{table}[h!]
\centering
\renewcommand{\arraystretch}{1.3}
\setlength{\tabcolsep}{5pt}
\footnotesize
\begin{tabular}{>{\raggedright\arraybackslash}p{4.0cm}
                >{\raggedright\arraybackslash}p{4.0cm}
                >{\raggedright\arraybackslash}p{4.0cm}
                >{\raggedright\arraybackslash}p{3.0cm}}
\toprule
{}
& \textbf{Static ($\omega=0$)} & \textbf{Small frequencies ($\omega r\ll 1$)} & \textbf{Dynamical (finite $\omega$)} \\
\midrule
Metric perturbation theory ($\delta g_\mn\ll g_\mn$) & Ladder symmetries \cite{Hui:2021vcv} & Love symmetry \cite{Charalambous:2021kcz,Charalambous:2022rre,Charalambous:2024gpf}, near-zone geometry \cite{Hui:2022vbh} & ? \\[2pt]
Nonlinear general relativity ($\delta g_\mn\sim g_\mn$) & Nonlinear ladders/Geroch \cite{Combaluzier-Szteinsznaider:2024sgb}, spurions \cite{Parra-Martinez:2025bcu} & ? & ?
\\
\bottomrule
\end{tabular}
\caption{Symmetries that control tidal response coefficients and underlie the vanishing of black hole Love numbers, organized by whether they are formulated in linear or nonlinear general relativity and to what extent they apply to time-dependent perturbations. 
The question marks refer to open questions or aspects that have not yet been addressed in the literature in relation to tidal responses and Love numbers. For instance, a complete understanding of the symmetries at nonlinear order in the small-frequency limit is still lacking (see \rcite{Parra-Martinez:2025bcu} for a preliminary discussion). The last column of the table is more speculative, and refers to the open possibility of studying Love number symmetries --- both linear and nonlinear --- at larger values of the frequency. We return to this speculative discussion in \cref{sec:sym-BHPT}.
}
\label{tab:sym_summary}
\end{table}

\subsection{Symmetries of the static sector}

Being the Wilson coefficients of static operators, the natural setting for the computation of Love numbers is the static sector of a field theory, i.e., its restriction to solutions of the equations of motion that do not depend on time, or equivalently have zero frequency. For massless field theories on four-dimensional black hole backgrounds, these static sectors possess rich symmetry structures that protect the vanishing of black hole Love numbers. We focus on these static symmetries in this subsection, in both linear perturbation theory and nonlinear general relativity.

\subsubsection{Linear theory: ladder symmetries}
\label{sec:sym-ladder}

Linear gravitational perturbations of Kerr black holes are described by the Teukolsky equation \eqref{eq:teuk-rad-boxed}. For non-spinning black holes, we can equivalently use the Regge--Wheeler \eqref{eq:RW} and Zerilli \eqref{eq:Z} equations, or even (for static perturbations) \cref{eq:h0-static,eq:H0-static} for the metric perturbations. The profiles of massless scalar, neutrino, and electromagnetic fields can also be computed using the Teukolsky equation. All of these equations possess enhanced symmetry structures in the static limit $\partial_t=\omega=0$, which we group under the general heading of \emph{ladder symmetries}.

\paragraph{Scalar ladders on Schwarzschild}

Let us illustrate these symmetries using the simplest example: a static, massless scalar field on Schwarzschild \cite{Hui:2021vcv}. As we have seen throughout \cref{sec:love-compute}, the spin-0 case models the qualitative features and in some cases exactly mimics the behavior of gravitational perturbations.

Expanding a time-independent scalar field profile in spherical harmonics, as discussed in \cref{sec:Sch-KG}, the Klein--Gordon equation for a given multipole $\phi_\ell(r)$ is
\begin{equation}\label{eq:kg-zerofreq}
\partial_r\left(\Delta\partial_r\phi_\ell\right)=\ell(\ell+1)\phi_\ell ,
\end{equation}
where $\Delta=r(r-\rs)$.
Of course this is also the Teukolsky master equation \eqref{eq:teuk-rad-boxed} setting $a=\omega=s=0$.
We define a ``Hamiltonian''
\begin{equation}
    H_\ell = -\Delta\partial_r(\Delta\partial_r)+\ell(\ell+1)\Delta,
\end{equation}
such that the equation of motion \eqref{eq:kg-zerofreq} is $H_\ell\phi_\ell=0$. This Hamiltonian admits raising and lowering operators,
\begin{subequations}\label{eq:ladders0}
\begin{align}
    D_\ell^+ &= -\Delta\partial_r-\frac{\ell+1}2\Delta',\label{eq:ladder-plus}\\
    D_\ell^- &= \Delta \partial_r-\frac\ell2\Delta',
\end{align}
\end{subequations}
in the sense that $D_\ell^\pm\phi_\ell$ solves \cref{eq:kg-zerofreq} with $\ell\to\ell\pm1$ \cite{Hui:2021vcv}. This can be seen by explicit calculation of the ``commutation'' relations
\begin{equation}\label{eq:laddercomm}
    H_{\ell\pm1}D_\ell^\pm=D_\ell^\pm H_\ell.
\end{equation}
They also satisfy convenient ``intertwining'' relations,
\begin{subequations}\label{eq:intertwining2}
\begin{align}
    H_\ell &= D^-_{\ell+1}D^+_\ell - \frac{(\ell+1)^2}4\rs^2\label{eq:intertwin2a} \\
    &= D^+_{\ell-1}D^-_\ell - \frac{\ell^2}4\rs^2, \label{eq:intertwin2b}.
\end{align}
\end{subequations}
which implies
\begin{equation}\label{eq:intertwining1}
    D^-_{\ell+1}D^+_\ell - D^+_{\ell-1}D^-_\ell = \frac{2\ell+1}4\rs^2.
\end{equation}
In analogy to the ladder operators in quantum mechanics, the $D_\ell^\pm$ operators allow one to construct a solution to \cref{eq:kg-zerofreq} with multipolar order $\ell\pm1$ given a solution at level $\ell$.

The ladder operators provide an explanation for the vanishing of black hole Love numbers. Recall that, viewed as a statement about the ordinary differential equation \eqref{eq:kg-zerofreq}, the vanishing of the scalar Love numbers corresponds to the fact that the solution with regular asymptotics at the horizon corresponds solely to a growing mode at infinity. In fact the horizon-regular solution is a pure polynomial in $r$, schematically $\phi_\ell \sim 1+r+\cdots+r^\ell$. While \cref{eq:kg-zerofreq} admits solutions with powers of $1/r$ in the $r\to\infty$ limit, these terms turn out to only be present if the solution also blows up at $r=\rs$.

We can use the ladder operators to explicitly construct such solutions. Consider the lowest ``rung'' of the ladder, i.e., the mode with the lowest allowed value of $\ell$, namely $\ell=|s|=0$. Its equation of motion is
\begin{equation}
    \partial_r(\Delta\partial_r\phi_0) = 0.\label{eq:kg-ell0}
\end{equation}
Note that the zero-derivative term in the general equation \eqref{eq:kg-zerofreq} is absent in this special case, so \cref{eq:kg-ell0} has an additional shift symmetry not present for $\ell>0$. As a result, we can integrate \cref{eq:kg-ell0} directly, $\Delta\partial_r\phi_0=\mathrm{const.}$, and integrate again to find the linearly-independent solutions $\phi_0^\mathrm{reg}=1$ and $\phi_0^\mathrm{irreg}=\ln f(r)$. The latter is divergent at $r=\rs$, so imposing regularity at the horizon selects the constant solution.\footnote{The fact that the  purely decaying solution is divergent at the horizon is consistent with the absence of linear (perturbative) hair for black holes~\cite{Israel:1967wq,Carter:1968rr,Carter:1971zc,Wald:1971iw,Hartle:1971qq,Bekenstein:1971hc,Bekenstein:1972ky,Fackerell:1972hg,Price:1972pw,Bekenstein:1995un,Hui:2012qt,Graham:2014mda,Herdeiro:2015waa,Capuano:2023yyh}.}
Starting from $\phi_0^\mathrm{reg}=1$ we can build the regular solution at $\ell=1$ using the raising operator,
\begin{equation}
    \phi_1^\mathrm{reg} = D_0^+1 = -\frac{\Delta'}2 = -r+\frac12\rs.
\end{equation}
This is indeed a horizon-regular solution to \cref{eq:kg-zerofreq} for $\ell=1$. Similarly a divergent solution at $\ell=1$ is $D_0^+\ln f$. From the form of the operators $D_\ell^\pm$ it is apparent that they preserve the boundary conditions relevant for computing Love numbers, namely (ir)regularity at the horizon and the growing or decaying nature at infinity. As a result we can continue this process and build horizon-regular solutions at arbitrary $\ell$ by ``climbing the ladder,''
\begin{equation}\label{eq:phireg}
    \phi^\mathrm{reg}_\ell = D_{\ell-1}^+\cdots D_0^+ 1.
\end{equation}
It is not difficult to see from the form \eqref{eq:ladder-plus} of $D_\ell^+$ that this solution is an order-$\ell$ polynomial in $r$; we conclude that the scalar Love numbers, which are the coefficients of terms that decay at large $r$, vanish for all $\ell$, as we have already seen in \cref{sec:love-static-sch}.\footnote{The full solutions to \cref{eq:kg-zerofreq} are simply Legendre functions $P_\ell(x)$ and $Q_\ell(x)$, where $x=\Delta'/\rs = 2r/\rs-1=r/M-1$, as discussed in \cref{sec:love-static-sch}. At the horizon ($x=1$) the Legendre function of the second kind, $Q_\ell(x)$, is divergent, so the horizon-regular solutions are proportional to the Legendre functions of the first kind, $P_\ell(x)$, which are finite polynomials as advertised, cf. \ref{app:legendre}. The growing mode \eqref{eq:phireg} obtained from acting $\ell$ raising operators on 1 is
\begin{equation}
    \phi^\mathrm{reg}_\ell = \left(-\frac\rs2\right)^\ell \ell! P_\ell\left(\frac{\Delta'}\rs\right).
\end{equation}}

We can also phrase this argument in the more familiar language of off-shell symmetries and Noether charges. At $\ell=0$ the Klein--Gordon equation \eqref{eq:kg-ell0} (and the associated action) is invariant under constant shifts of $\phi_0$. One can use this fact to deduce the existence of the constant solution, which we have seen is regular at the horizon. The solution $\phi_0\sim\ln f(r)$, which blows up at the horizon, spontaneously breaks the shift symmetry. We can express this symmetry with a linear operator,\footnote{That this (linear) symmetry is equivalent to the (nonlinear) shift symmetry can be seen from the fact that the conserved quantity associated to the former is the square of the conserved quantity for the latter. Each symmetry operator can in turn be reconstructed from its conserved charge via the Poisson bracket; see \cref{foot:Poisson-bracket} for more detail.}
\begin{equation}
\delta_0 \equiv \Delta \partial_r = D_0^-,    
\end{equation}
and the fact that it is a symmetry corresponds to the operator identity $[\delta_0,H_0]=0$. To see this note that, per \cref{eq:intertwin2b}, $H_0 = D^+_{-1}D^-_0=-\delta_0^2$.

At higher values of $\ell$ we do not have an obvious shift symmetry, but we can nevertheless construct a symmetry operator using the ladder operators: acting on a field $\phi_\ell$, we lower to $\ell=0$, apply the shift symmetry $\delta_0$, and then raise back to $\ell$,
\begin{align}\label{eq:ladder-horizontal}
    \delta_\ell &= D^+_{\ell-1}\delta_{\ell-1}D_\ell^-\nonumber\\
    &= D^+_{\ell-1}\cdots D^+_0 D_0^- D_1^-\cdots D_\ell^-.
\end{align}
The ``horizontal'' ladder operator $\delta_\ell$ (so named in contrast to the ``vertical'' operators $D^\pm_\ell$) commutes with the Hamiltonian $H_\ell$, so it sends solutions to solutions at level $\ell$. This is most readily proven by induction. Assuming that $[\delta_{\ell-1},H_{\ell-1}]=0$, we have
\begin{align}
    [\delta_\ell,H_\ell] &= [D^+_{\ell-1}\delta_{\ell-1}D^-_\ell,H_\ell] \nonumber\\
    &= D^+_{\ell-1}\delta_{\ell-1}D^-_\ell H_\ell - H_\ell D^+_{\ell-1}\delta_{\ell-1}D^-_\ell \nonumber\\
    &= D^+_{\ell-1}\delta_{\ell-1}H_{\ell-1}D^-_\ell - D^+_{\ell-1}H_{\ell-1}\delta_{\ell-1}D^-_\ell \nonumber\\
    &= D^+_{\ell-1}[\delta_{\ell-1},H_{\ell-1}]D^-_\ell = 0.
\end{align}
In going to the third line we used \cref{eq:laddercomm}.

Let us note the interesting fact that, analogously to $\ell=0$, each $\delta_\ell$ is a symmetry of the horizon-regular solution and is broken by the horizon-divergent solution. We see that the vanishing or non-vanishing of the Love numbers can be equivalently phrased as a statement about \emph{spontaneous symmetry breaking}: an object with non-vanishing Love numbers, such as a neutron star, spontaneously breaks the horizontal ladder symmetry. We will see below that the ladder symmetries forbid Love number couplings in the point-particle EFT; the fact that they appear in the EFT for a neutron star can be tied to the star's explicitly breaking this symmetry, so that it is either absent or nonlinearly realized in the EFT.

The horizontal ladders are genuine symmetries of the action, and therefore come with a corresponding conserved charge, which we will see is closely related to the Love numbers. Performing a multipole expansion in the Klein--Gordon action and ignoring time derivatives, we have
\begin{equation}
    S = -\frac12\int\dd^4x\sdg(\partial\phi)^2 = \sum_\ell\int\dd t\, S^\mathrm{static}_\ell,
\end{equation}
where
\begin{align}
    S_\ell^\mathrm{static} &\equiv \frac12\int \dd r \,\phi_\ell\left[\partial_r(\Delta\partial_r)-\ell(\ell+1)\right]\phi_\ell \nonumber\\
    &= -\frac12\int\dd y\,\phi_\ell H_\ell\phi_\ell.
\end{align}
Here we have introduced the radial coordinate $y\equiv \rs^{-1} \ln f(r)$, such that $\partial_y = \Delta\partial_r$ is the derivative operator appearing in $D^\pm_\ell$. At $\ell=0$ the variation of the action under the horizontal ladder symmetry $\delta_0$ is a total derivative,
\begin{align}
    \delta_0S_0^\mathrm{static} &= -\int\dd y \,\partial_y\phi_0\,\partial_y^2\phi_0 \nonumber\\
    &= -\frac12\,\int\dd y\partial_y(\partial_y\phi_0)^2.
\end{align}
To obtain the conserved charge we follow the canonical Noether procedure to find
\begin{equation}
    J_0 = (\Delta\partial_r\phi_0)^2 = (D_0^-\phi_0)^2,
\end{equation}
which satisfies $\partial_rJ_0=0$ on shell. The invariance of $S^\mathrm{static}_\ell$ under $\delta_\ell$ for $\ell>0$ follows by induction. Assuming $\delta_{\ell-1}$ is a symmetry of $S_{\ell-1}^\mathrm{static}$ and using the fact that the adjoints of $D_\ell^\pm$ are (ignoring boundary terms)
\begin{equation}
    \int \dd y\, A D_\ell^\pm B = \int \dd y \,B D^\mp_{\ell\pm1}A,
\end{equation}
we find that $\delta_\ell$ is a symmetry of $S_\ell^\mathrm{static}$
with corresponding conserved charge
\begin{equation}
    J_\ell = (D_0^-\cdots D_\ell^-\phi_\ell)^2.
\end{equation}

To interpret this charge, we evaluate it on a (generic) solution constructed using the raising operators,
\begin{align} \label{eq:Jell}
    \sqrt{J_\ell} &= D_0^-\cdots D_\ell^- D^+_{\ell-1}\cdots D_0^+\phi_0 \nonumber\\
    &= (\ell!)^2\left(\frac{\rs^2}4\right)^\ell \Delta\partial_r\phi_0,
\end{align}
where in going to the second line we have repeatedly used the intertwining relation \eqref{eq:intertwin2a}. We immediately see that $J_\ell$ vanishes when evaluated on the horizon-regular solution \eqref{eq:phireg} built from a constant $\phi_0$, and is non-vanishing for the divergent solution built from $\phi_0\propto\ln f(r)$. By the same logic we see that the solution with growing asymptotics at infinity produces $J_\ell=0$, so that a non-vanishing charge corresponds to a non-vanishing tidal response, $J_\ell\propto\lambda_\ell^2$, as indeed is apparent from \cref{eq:Jell}.

It is interesting to note a relation between the Noether charge $J_\ell$ and another conserved quantity, the Wronskian (with respect to $y$) of a given solution $\phi_\ell$ with the horizon-regular solution \eqref{eq:phireg},
\begin{equation}
    W[\phi_\ell^\mathrm{reg},\phi_\ell] \equiv \phi_\ell^\mathrm{reg}\partial_y\phi_\ell - \phi_\ell\partial_y\phi_\ell^\mathrm{reg}.
\end{equation}
On shell (i.e., when the equation of motion $H_\ell\phi_\ell=0$ is satisfied), the two conserved quantities are related very simply,
\begin{equation}
    \sqrt{J_\ell} = W[\phi_\ell^\mathrm{reg},\phi_\ell].
\end{equation}
This is remarkable because as constructed the two quantites have very different numbers of derivatives ($\ell$ for $\sqrt{J_\ell}$ and one for $W$). Related to this is the fact that the horizontal operator $\delta_\ell$ \eqref{eq:ladder-horizontal}, which as written has $(2\ell+1)$ derivatives acting on $\phi_\ell$, can be reduced using the equations of motion to an equivalent operator that is first order in derivatives \cite{Combaluzier-Szteinsznaider:2024sgb}.\footnote{By ``equivalent'' we mean that the variation of the action under the difference between $\delta_\ell$ and their corresponding first-order operators is a total derivative.}

This equivalence is apparent if one uses Poisson brackets to generate symmetries from their associated conserved quantities \cite{BenAchour:2022uqo}:\footnote{Recall that a symmetry $\delta\phi$ can be constructed from its conserved charge $Q$ by $\delta\phi=\{Q,\phi\}$, where the Poisson bracket is
\[ \{A,B\}=\frac{\delta A}{\delta\phi}\frac{\delta B}{\delta\Delta\partial_r\phi}-\frac{\delta A}{\delta\Delta\partial_r\phi}\frac{\delta B}{\delta\phi}.\]
\label{foot:Poisson-bracket}} the symmetry constructed this way from the conservation of the squared Wronskian is the reduced version of the horizontal ladder, while the symmetry constructed from the conservation of the Wronskian is simply a shift by a growing-mode solution, which is clearly a symmetry (and indeed is a symmetry of any linear theory).

\paragraph{Gravitational ladders on Kerr}

The ladder structure we have discussed for spin-0 fields on Schwarzschild generalizes straightforwardly to gravitational (and indeed spin-$s$) fields on Kerr \cite{Hui:2021vcv}. These are described by the Teukolsky master equation \eqref{eq:teuk-rad-boxed}, which for static perturbations takes the form \eqref{eq:TeuAlls},
\begin{equation}\label{eq:teuk-static}
    \left[\Delta^{-s}\partial_r\left(\Delta^{s+1}\partial_r\right)+\frac{a^2m^2+2iams(r-M)}\Delta - \ell(\ell+1)+s(s+1)\right]R^{(s)}_{\ell m}(r) = 0.
\end{equation}
To elucidate the ladder structure it is convenient to make the field redefinition
\begin{equation}
    R^{(s)}_{\ell m} \equiv \Delta^{-s}e^{iq\ln\left(\frac{z-z_k}{z}\right)}\psi^{(s)}_{\ell m},
\end{equation}
where we have defined
\begin{equation}
    z\equiv r-r_-,\quad z_k\equiv r_+-r_- = z(r_+),\quad q\equiv\frac{am}{z_k}.
\end{equation}
The field $\psi^{(s)}_{\ell m}(z)$ has the same asymptotic properties we have found useful: at infinity there are growing and decaying solutions going as $z^{\ell+s}$ and $z^{-\ell+s-1}$, respectively, and at the horizon there is a constant solution and a divergent solution going as $e^{(s-2iq)\ln\left(\frac{z-z_k}{z}\right)}$.

The raising and lowering operators are \cite{Hui:2021vcv}
\begin{subequations}\label{eq:ladders-kerr}
\begin{align}
    D_{\ell s}^+ &= -\Delta\partial_z-\frac{\ell-s+1}{2(\ell+1)}\left[(\ell+1)\partial_z\Delta+2iqz_k\right],\\
    D_{\ell s}^- &= \Delta \partial_z-\frac{\ell+s}{2\ell}\left(\ell\partial_z\Delta-2iqz_k\right).
\end{align}
\end{subequations}
The horizontal symmetry structure can be derived analogously to the $s=a=0$ case discussed above, assuming $s<0$, by starting at $\ell=|s|$, where \cref{eq:teuk-static} possesses the accidental symmetry
\begin{equation}
    \delta_{|s|} = \Delta^{-s+1}e^{2iq\ln\left(\frac{z-z_k}{z}\right)}\partial_z.
\end{equation}

Remarkably, the Teukolsky equation \eqref{eq:teuk-static} also has a ladder structure which raises and lowers a field's \emph{spin} $s$ rather than its angular momentum quantum number $\ell$. These \emph{spin ladders} are given by
\begin{subequations}\label{eq:ladders-spin}
\begin{align}
    E_s^+ &= \Delta\partial_z-s\partial_z\Delta+2iqz_k,\\
    E^- &= \partial_z.
\end{align}
\end{subequations}
In other words, given a solution $\psi^{(s)}_{\ell m}$, we can construct another solution $\psi^{(s\pm 1)}_{\ell m} = E_s^\pm\psi^{(s)}_{\ell m}$. This enables one to compute the vanishing of electromagnetic and gravitational Love numbers around Kerr from the vanishing of the scalar Love numbers, as well as the construction of horizontal ladders for $s>0$ starting from the $s<0$ ladders discussed above. The spin ladders interpolate, for static perturbations, the well-known Teukolsky--Starobinsky identities which relate (time-dependent) solutions at spins $\pm|s|$ \cite{Press:1973zz,Starobinskil:1974nkd}.

For Schwarzschild perturbations, we may also obtain ladder structures within metric perturbation theory. The static Regge--Wheeler equation \eqref{eq:RW} admits raising and lowering operators of the form
\begin{subequations}\label{eq:ladders-RW}
\begin{align}
    D_\ell^+ &= -\Delta\partial_r+\frac{\ell^2+3}{2(\ell+1)}\rs-\ell r,\\
    D_\ell^- &= \Delta \partial_r +\frac{\ell^2(\rs-2r)-2\ell(r-\rs)+4\rs}{2\ell}.
\end{align}
\end{subequations}
By contrast, identifying a similar structure for the Zerilli equation \eqref{eq:Z} is less straightforward. One approach is to use the Chandrasekhar duality that is, raising and lowering operators for the Zerilli equation can be obtained by sandwiching a Regge--Wheeler ladder operator between two copies of the Chandrasekhar duality operator \eqref{eq:sym-chandra}, i.e., $(\partial_\rst-W)D_\ell^\pm(\partial_\rst+W)$, with $D_\ell^\pm$ the Regge--Wheeler ladders.
Further computational details for the Teukolsky, spin, and Regge--Wheeler ladders may be found in \rcite{Hui:2021vcv}.
There also exist ladder operators acting directly on metric perturbations in the Kaluza--Klein decomposition discussed in \cref{sec:nonlineartheory} \cite{Berens:2025okm}.\footnote{Ladder operators for \cref{eq:h0-static,eq:H0-static} acting on the metric perturbations $H_0$ and $h_0$ have also recently been derived in unpublished work by Pradhan, Katagiri, and Yagi (private communication).}

Generalizations of the ladder operators to charged Reissner--Nordstr\"om black holeve has been explored in \rcite{Berens:2022ebl} for a test scalar field and in \rcite{Rai:2024lho} for gravito-electromagnetic perturbations.

\paragraph{Geometric origin of ladder symmetries}

Mathematically the ladder symmetries are a consequence of the \emph{hypergeometric} form of the equations describing static perturbations. As discussed in more detail in \ref{app:hypergeo}, the hypergeometric functions (which solve the hypergeometric equation) are subject to well-known recurrence relations which may be used to derive the ladder operators \cite{Hui:2021vcv}.\footnote{Relatedly, a connection between ladder generators and Darboux transformations was emphasized in \rcite{DeLuca:2025zqr}, specifically in the context of analog black holes. In this work, the Darboux transformation, rather than relating different parity sectors, is used to connect different $\ell$ levels within each sector. See also \rcite{Ghosh:2026vig} for a recent exploration of the conditions under which general second-order physical ODEs admit ladder operators.} This also explains why there is not a nice ladder story for the static Zerilli equation, as this is of the Heun type rather than hypergeometric (although, as we have seen, the static Zerilli equation is special in that it can be made hypergeometric through the Darboux transformation \eqref{eq:sol-chandra}).

At a physical level, the ladder symmetries have a geometric origin \cite{Hui:2021vcv,Berens:2025okm}. Let us begin as usual with the massless, static scalar on Schwarzschild. It lives on an effective 3D metric
\begin{equation}\label{eq:geff-scalar}
    g^\eff_{ij} = f(r)g^\mathrm{Sch}_{ij} = \diag(1,\Delta,\Delta\sin^2\theta),
\end{equation}
in the sense that it is minimally coupled to that metric,
\begin{equation}
    \sqrt{-\det g^\mathrm{Sch}_\mn}g^\Ij_\mathrm{Sch}  \equiv \sqrt{-\det g^\eff_\Ij}g^\Ij_\eff.
\end{equation}
Performing the Weyl transformation
\begin{equation}
    \tilde g_\Ij \equiv \Omega^2g^\eff_\Ij,\quad \tilde\phi\equiv\Omega^{-1/2}\phi,
\end{equation}
with conformal factor
\begin{equation}
    \Omega \equiv \frac{L^2}\Delta
\end{equation}
and $L$ an arbitrary length scale, the action becomes
\begin{align}
    S &= -\frac12\int\dd t\int\dd^3x\left(\tilde g^\Ij\partial_i\tilde\phi\partial_j\tilde\phi-\frac{\rs^2}{4L^4}\tilde\phi^2\right),
\end{align}
which is the action for a \emph{massive} scalar on the background $\tilde g_\Ij$. The remarkable thing that has happened here is that the $r$-dependent potential which we expect to appear in front of $\tilde\phi^2$ after the Weyl transformation turns out to be a constant mass term. As a result, $\tilde\phi$ inherits the isometries of the background $\tilde g_\Ij$. This space is nothing other than Euclidean AdS$_3$, as can be seen by calculating its scalar curvature,
\begin{equation}
    \tilde R = -\frac{3\rs^2}{2L^4}.
\end{equation}
Since this is a maximally-symmetric spacetime in three dimensions, it has six Killing isometries $\xi^i$, which act on the effective metric as conformal Killing vectors. These induce symmetry transformations\footnote{This is despite the field $\phi$ not actually being conformally coupled. The underlying reason is that these conformal Killing vectors obey the \emph{melodic condition} $\Box\nabla_i\xi^i=0$ \cite{Berens:2022ebl}.} of the form
\begin{equation}
    \delta\phi = \left(\xi^i\partial_i+ \frac16\nabla_i\xi^i\right)\phi,
\end{equation}
with $\nabla_i$ the covariant derivative associated to $g^\eff_\Ij$.

The full symmetry group is $\mathrm{SL}(2,\mathbb R)$. The symmetry of particular interest to us is generated by the conformal Killing vector
\begin{equation}\label{eq:sch-eff-CKV}
    \xi^i\partial_i = \Delta\cos\theta\partial_r+\frac12\Delta'\sin\theta\partial_\theta,
\end{equation}
which descends from the $r$-$\theta$ ``boost'' isometry of the conformally-related Euclidean AdS$_3$, and causes the scalar to transform as
\begin{equation}\label{eq:phi-ladder-CKV}
    \delta\phi = \Delta\cos\theta\partial_r\phi + \frac12\Delta'\partial_\theta(\sin\theta\phi).
\end{equation}
The ladder operators $D^\pm_\ell$ \eqref{eq:ladders0} emerge from the action of this transformation on a single multipole,
\begin{equation}
    \delta\left(\phi_\ell(r)Y_{\ell}(\theta)\right) = -\frac{\ell+1}{2\ell+1}D^+_\ell\phi_\ell(r) Y_{\ell+1}(\theta) + \frac{\ell}{2\ell+1}D^-_\ell\phi_\ell(r)Y_{\ell-1}(\theta).
\end{equation}
Since this is a symmetry, the object on the right-hand side must be a solution of the equation of motion, and based on the angular dependence we deduce it is a sum of solutions with $\ell\pm1$. We see that the ladder operators have a geometric origin: they arise from the conformal Killing isometries of the effective metric seen by a static field.

The geometric interpretation applies straightforwardly to electromagnetic and gravitational perturbations, in the sense that suitable field variables see an effective metric that is conformally related to $g_\Ij^\eff$ \cite{Berens:2025okm}.\footnote{See also \rcite{Katagiri:2022vyz}, which identified AdS$_2$ conformal isometries underlying the vanishing of spin-$s$ Love numbers on Schwarzschild and Kerr.} The even-parity gravitational perturbation, encoded in the Kaluza--Klein scalar $\phi\equiv-\ln (-g_{tt})$, sees the same effective metric \eqref{eq:geff-scalar} as the spin-0 field, while the odd-parity mode, encoded in the other Kaluza--Klein scalar $\chi$, obtained by dualizing the 3-vector $e^{-\phi}g_{ti}$ as in \cref{eq:A-dual}, sees an effective metric related to \cref{eq:geff-scalar} by a conformal factor $r^8/\Delta^4$.
We conclude that both Kaluza--Klein scalars are invariant under \cref{eq:phi-ladder-CKV}.

We are now in a position to address the naturalness problem
discussed in \cref{sec:naturalness}. It is resolved by the flat-space (IR) limit of \cref{eq:phi-ladder-CKV},
\begin{equation}\label{eq:flat-space-love-sym}
\delta\phi = r^2\cos\theta\partial_r\phi + r\partial_\theta(\sin\theta\phi).
\end{equation}
This is a special conformal transformation in the $x^3$ direction for a field with conformal weight $1/2$, as we can see by transforming to Cartesian coordinates:
\begin{equation}
    \delta\phi = c_i\left(x^i-\vec{x}^2\partial^i+2x^i\vec{x}\cdot\vec\partial\right)\phi
\end{equation}
with $c_i=(0,0,1)$. Acting on (the Kaluza--Klein components of) a spin-$s$ field, the kinetic term (cf. \cref{eq:EHaction2} in the gravitational case) is invariant under \cref{eq:flat-space-love-sym} while the finite-size couplings (cf. \cref{eq:E2B2-0}) break it, forbidding them from appearing in the EFT \cite{Hui:2021vcv,Berens:2025okm}. 

In $D>4$ a similar story holds, with the vertical ladder operators raising and lowering in integer steps of $\hat\ell\equiv\ell/(D-3)$ \cite{Hui:2021vcv,Berens:2025jfs}. This provides a symmetry explanation for the observation that the Love numbers of higher-dimensional Schwarzschild--Tangherlini black holes only vanish when $\hat\ell\in\mathbb{N}$ \cite{Kol:2011vg,Hui:2020xxx}, as discussed in \cref{sec:statichighD}, since only these solutions can be lowered to the bottom rung of the ladder.

\subsubsection{Nonlinear gravity}
\label{sec:sym-nonlin}

\paragraph{Ladders and the Geroch group}

We have seen in \cref{sec:nonlineartheory} that for deformed Schwarzschild black holes \eqref{eq:distorted-BH}, the Einstein equation \eqref{eq:eom-deformaton} for the deformed potential $\hat\phi$ is of the same form as the static Klein--Gordon equation. It follows that there exists an analogous ladder structure \cite{Combaluzier-Szteinsznaider:2024sgb}. The scalar ladder operators \eqref{eq:ladders0} serve to raise and lower the angular momentum quantum number of individual modes $\hat\phi_\ell$, and can be used to construct the horizontal symmetries \eqref{eq:ladder-horizontal}. It is worth emphasizing that the coordinates $(r,\theta)$ are now \emph{defined} by \cref{eq:coord-weyl-sch} as functions of the Weyl canonical coordinates $(\rho,z)$, which are spacetime scalars: $\rho=\sqrt{-\xi^2\eta^2}$ is built from the norms of the timelike and azimuthal Killing vectors, and its dual $z$ has the covariant definition \eqref{eq:z-dual}. In particular, when comparing to a perturbed Schwarzschild metric, e.g., in Regge--Wheeler gauge, the coordinates $(r,\theta)$ which appear in the ladder operators for $\hat\phi$ are generally not the same as the Schwarzschild coordinates $(r,\theta)$ we encountered in \cref{sec:BHPT}; see App.~C of \rcite{Combaluzier-Szteinsznaider:2024sgb} for further details.

In Weyl canonical coordinates it is easy to write the generators of the full $\mathrm{SL}(2,\mathbb R)$ symmetry group of \cref{eq:phi-weyl}:
\begin{equation}
    P=\partial_z,\quad D=-\left(\frac12+\rho\partial_\rho+z\partial_z\right),\quad K=2\rho z\partial_\rho+(z^2-\rho^2)\partial_z+z,
\end{equation}
which form the $\mathfrak{sl}(2,\mathbb R)$ algebra
\begin{equation}
    [D,P]=P,\quad[D,K]=-K,\quad[P,K]=-2D.
\end{equation}
The transformation \eqref{eq:phi-ladder-CKV} corresponds to the combination $K-(\rs^2/4)P$.\footnote{The symmetry algebra of \cref{eq:kg-zerofreq} for the test scalar is given by the same expressions, using \cref{eq:coord-weyl-sch} to translate from Weyl to Schwarzschild coordinates.}

The dimensional reduction procedure outlined in \cref{sec:nonlineartheory} generically unveils symmetry structures that are hidden in the higher-dimensional theory \cite{Maison:2000fj}. In the case of the reduction from four dimensions to two, this hidden symmetry is described by the infinite-dimensional \emph{Geroch group} \cite{Ehlers:1957zz,Geroch:1970nt,Breitenlohner:1986um,Maison:2000fj,Lu:2007zv,Lu:2007jc,Maison:1978es,Schwarz:1995td,Schwarz:1995af}. The Geroch symmetry turns out to be intimately related to the ladder symmetries outlined for $\hat\phi$ \cite{Combaluzier-Szteinsznaider:2024sgb}. At the infinitesimal level, the Geroch group (restricted to the static case $g_{t\varphi}=0$) acts as \cite{Lu:2007zv,Lu:2007jc}
\begin{equation}
\delta \hat\phi =\frac{w}{\sqrt{\rho^2+(w-z)^2}}, \label{eq:perry-sym-static}
\end{equation}
where $w$ is a constant \emph{spectral parameter}. Its presence reflects the infinite-dimensional nature of the Geroch group: we can decompose the symmetry operator $\delta\equiv\sum_{n\geq 0}w^{-n}\delta_{(n)}$ into an infinite number of symmetries $\delta_{(n)}$. The function on the right-hand side of \cref{eq:perry-sym-static} is a generating function for growing-mode solutions,
\begin{equation}
    \frac{w}{\sqrt{\rho^2+(w-z)^2}} = \sum_{n\geq0}w^{-n}\mathcal{R}^nP_n(\cos\vartheta),
\end{equation}
where we have introduced polar coordinates $(\rho,z)=(\mathcal R\sin\vartheta,\mathcal R\cos\vartheta)$. In these coordinates the mode solutions to \cref{eq:eom-deformaton} are $\mathcal R^\ell\cos\vartheta$ and $\mathcal R^{-\ell-1}\cos\vartheta$; they are linear combinations of the modes we computed in $(r,\theta)$ coordinates in 
\cref{eq:hatphinonlinear}, and share the same interpretation as being tidal and response pieces, respectively. We see that each symmetry operator $\delta_{(n)}$ shifts $\hat\phi$ by the level-$n$ growing mode in polar coordinates,
\begin{equation}
    \delta_{(n)}\hat\phi = \mathcal{R}^nP_n(\cos\vartheta).
\end{equation}
Applied to a single mode $\hat\phi=\hat\phi_k(\mathcal R)P_k(\cos\vartheta)$, this symmetry acts on the radial piece as
\begin{equation}\label{eq:rad-geroch}
    \delta_{(n)}\hat\phi_k = \mathcal{R}^n\delta_{kn}.
\end{equation}
The associated conserved quantity is a Wronskian,
\begin{equation}
\tilde Q_n \equiv \mathcal R^2\left(\mathcal R^n\partial_{\mathcal R}\hat\phi-\hat\phi\partial_{\mathcal R}\mathcal R^n\right).
\end{equation}
As we saw in the spin-0 case, the conservation of the Wronskian and its square generate via the Poisson bracket, respectively, the shift-by-a-solution symmetry \eqref{eq:rad-geroch} and a linear symmetry which is equivalent to the horizontal ladder symmetry.

\paragraph{Spurionic argument}

A different symmetry perspective on the vanishing of the nonlinear Love numbers has been advocated in \rcite{Parra-Martinez:2025bcu}, building upon an earlier observation made in \rcite{Kol:2011vg}. As we now review, the argument is based on a symmetry rooted in the dimensional reduction discussed above and the $\mathrm{SL}(2,\mathbb{R})$ Ehlers group relating gravitational solutions with a timelike Killing vector~\cite{Harrison:1968wue,Ehlers:1959aug,Ernst:1967wx,Geroch:1970nt}, supplemented by a spurion transformation. When combined with the spurion and suitable assumptions about the boundary conditions at the horizon, the symmetry rules out both even and odd nonlinear Love numbers of non-rotating black holes in four-dimensional general relativity, and establishes a non-renormalization theorem forbidding renormalization group running in the static sector of the point-particle EFT. The argument of \rcite{Kol:2011vg,Parra-Martinez:2025bcu} admits an extension to higher spacetime dimensions, recovering the results of previous explicit calculations~\cite{Kol:2011vg,Hui:2020xxx,Hadad:2024lsf}.

The starting point is the dimensionally reduced action \eqref{eq:dimredEH}. Introducing a complex variable $w$, defined such that~\cite{Parra-Martinez:2025bcu}\footnote{To recover the notation of~\rcite{Parra-Martinez:2025bcu}, one should replace $\chi\mapsto a$, $\phi\mapsto -2\phi$, and $\mathcal{E}\mapsto i \bar{z}$.}
\begin{equation}
    w\equiv \frac{\bar{\mathcal{E}}-1}{\bar{\mathcal{E}}+1},
\label{eq:defwSM}
\end{equation}
where $\mathcal{E}$ is defined in \cref{eq:mathcalEphichi},
the Einstein--Hilbert action takes the form 
\begin{equation}
    \sdg R = \sqrt{g_3}\left(R_3-
    2 \frac{\partial_i w \partial^i \bar{w}}{(1- \bar{w}w )^2}
    \right).
\label{eq:dimredEHPMP}
\end{equation}
As discussed above, the action for $w$ forms an $\mathrm{SL}(2,\mathbb R)$/SO(2) sigma model.  In the  parametrization \eqref{eq:defwSM}, the unbroken SO(2) symmetry group
corresponds to a rotation,
\begin{equation}
    w\rightarrow e^{i \theta}w,
\label{eq:wrotation}
\end{equation}
with $0\leq \theta < 2\pi$.
In particular, for $\theta =\pi$, the rotation \eqref{eq:wrotation} reduces to a $\mathbb{Z}_2$ symmetry, sending
\begin{equation}
  \mathbb{Z}_2 :
  \qquad w\rightarrow -w .
\label{eq:Z2actionw}
\end{equation}

The Minkowski solution corresponds to choosing $w=0$ and $g_{3 , ij}=\delta_{ij}$, which leaves the SO(2) symmetry of \cref{eq:wrotation} unbroken. The Schwarzschild background, by contrast, induces a spontaneous breaking of  the SO(2) symmetry. As argued in \rcite{Parra-Martinez:2025bcu}, it is convenient to express the Schwarzschild line element in isotropic coordinates as
\begin{equation}
    \dd s^2 =-\left(\frac{1-X}{1+X} \right)^2\dd t^2 +(1+X)^4 \delta_{ij} \dd x^i\dd x^j ,
\end{equation}
with 
\begin{equation}
    X\equiv \frac{\rs}{4r} = \frac{GM}{2r} , 
    \qquad
    r=\sqrt{\delta_{ij}x^i x^j},
\end{equation}
which corresponds to setting
\begin{equation}
    w^{\mathrm{Sch}} = -\frac{2X}{1+X^2},
    \qquad
    g_{3,ij}^{\mathrm{Sch}}= (1-X^2)^2\delta_{ij}.
\label{eq:Schwwg3}
\end{equation}

The solution \eqref{eq:Schwwg3} breaks the $\mathbb{Z}_2$ symmetry. 
However, \rcite{Kol:2011vg,Parra-Martinez:2025bcu} observed that it can be restored  as a spurionic symmetry by assigning a transformation rule to the mass  $M$ (or $\rs$), namely
\begin{equation}
\rs \to -\rs \, .
\label{eq:spurionMmM}
\end{equation}
The combined action of the $\mathbb{Z}_2$ transformation \eqref{eq:Z2actionw} and this spurionic  symmetry leaves the Schwarzschild solution invariant.\footnote{The argument was first introduced by \rcite{Kol:2011vg} for the $\phi$ sector of the perturbations and was later generalized to all (even and odd) perturbations in \rcite{Parra-Martinez:2025bcu}, where the connection to the vanishing of Love numbers was made more systematic.}

The extension of the argument to perturbations of Schwarzschild proceeds as follows. Before splitting into background and perturbations,  the Einstein--Hilbert action is separately invariant under the  $\mathrm{SL}(2,\mathbb{R})$ generators and under the spurion transformation  (since $\rs$ does not appear explicitly in \cref{eq:dimredEHPMP}, the term  $\sdg R$ is trivially invariant under the transformation \eqref{eq:spurionMmM}). After  decomposing the fields into background plus perturbations, we have seen  that the background \eqref{eq:Schwwg3} is invariant under the combined action of  \cref{eq:Z2actionw,eq:spurionMmM}. Therefore, in order to  preserve the overall invariance of \cref{eq:dimredEHPMP}, the action for the perturbations must also  be invariant under this combined transformation. 

The conclusion of \rcite{Parra-Martinez:2025bcu} is thus that the action  for perturbations around a Schwarzschild black hole inherits a symmetry  given by the combination of the $\mathbb{Z}_2$ transformation  \eqref{eq:Z2actionw} and the spurion \eqref{eq:spurionMmM}. This statement 
holds fully nonlinearly.

Since this is an exact invariance of the action --- and hence of the equations of motion --- any breaking of the symmetry in the solution must be attributed to the boundary conditions. 
Given that the perturbation equations are second order, only two boundary conditions are required. 
The first corresponds to imposing an $r^\ell$ tidal field profile at infinity, which can always be endowed with a transformation rule consistent with the symmetries. Whether the resulting solution breaks the symmetry is therefore determined by the second boundary condition, imposed at the surface of the object. 

In the case of a Schwarzschild black hole, the relevant radius is $\rs$, and one imposes that no scalar quantity constructed from the perturbative solution diverges at this location. The argument of \rcite{Parra-Martinez:2025bcu} is that regularity at the horizon selects the solution with the same parity as the mode growing as $r^\ell$ at large $r$. Since this mode has definite parity under the spurion transformation, the solution must asymptotically take the form
\begin{equation}
    w \sim r^{\ell} \left( 1 + \left(\frac{\rs}{r}\right)^2 
    + \left(\frac{\rs}{r}\right)^4 + \cdots \right),
\end{equation}
where only non-negative powers of $\rs/r$ appear in  parentheses. 
In conclusion,  the perturbation $w$ cannot develop a $1/r^{\ell+1}$ falloff. This statement holds (nonlinearly)  beyond axisymmetry for the perturbations,  thereby extending the conclusions of \rcite{Combaluzier-Szteinsznaider:2024sgb,Kehagias:2024rtz}.

When combined with the point-particle EFT, the spurion argument provides a stringent power-counting scheme that allows one to rule out a large class of classical worldline loop diagrams, establishing a  non-renormalization theorem in the EFT. Schematically, the argument proceeds as follows (see \rcite{Parra-Martinez:2025bcu} for details).

Let us return to the EFT action~\eqref{eq:S-EFT} for non-rotating objects.  As argued above, the Einstein--Hilbert term \eqref{eq:S-EH} is invariant  under the symmetry given by \cref{eq:Z2actionw,eq:spurionMmM}. The point-particle term \eqref{eq:PP} is what is required to  reconstruct, perturbatively in $\rs/r$, the full nonlinear background  solution. In the parametrization \eqref{eq:metric-3-1}, it takes the form
\begin{equation}
    S_{\mathrm{p.p.}} \supseteq \frac{M}{2} \int \dd \tau \, \phi = -M \int \dd \tau \, \mathrm{Re }\, w +\dots \, ,
\label{eq:sppspurion}
\end{equation}
where we display only the term linear in the field. This is the only vertex needed for the background reconstruction (see \cref{sec:ADMmassex}) and the only one that contributes non-trivially to the classical loop diagrams of interest, potentially renormalizing the static Love number couplings.  Since the Schwarzschild background is invariant under the combined $\mathbb{Z}_2$ and spurionic transformation, so is the  term in \cref{eq:sppspurion}, as can be readily verified~\cite{Parra-Martinez:2025bcu}.

The couplings of the higher-order operators $S_\mathrm{ho}$ (see \cref{eq:S-ho} for the quadratic operators, and \cref{eq:E2B2-0cubic,eq:E2B2-0nonlinear} for the cubic and higher-order terms), when determined via matching to the full theory, generally receive two types of contributions. The first is a universal correction arising from gravitational nonlinearities, which can be represented by classical worldline loop diagrams obtained from multiple insertions of the mass coupling \eqref{eq:sppspurion}. The second arises from correlation functions involving the energy-momentum tensor of additional matter fields in the full theory (e.g., the energy-momentum 
tensor of a star).

As expected from simple power counting, these couplings enter the EFT (nonlinear) response for the multipole-$\ell$ field $w$ (in $D=4$ dimensions) schematically as
\begin{equation}
    w^{\mathrm{resp}}_\ell 
    \propto \left( \prod_{i=1}^n w_{\ell_i}^{\mathrm{tidal}} \right) 
    c_{\ell\ell_1\cdots\ell_n}\,\frac{1}{r^{\ell+1}} \, ,
\end{equation}
for a given set of $n$ external tidal source amplitudes $w_{\ell_i}^{\mathrm{tidal}}$. Compared to the overall $r^\ell$ tidal profile, the induced response therefore falls off as $r^{-2\ell-1}$. 
Restoring dimensions, if $\rs$ is the only length scale in the problem --- as in the case of Schwarzschild black holes --- the Love number couplings must be odd functions of $\rs$. Hence, $c_{\ell\ell_1\cdots\ell_n}$ must vanish by the spurionic argument, in agreement with the discussion above. 

From a purely EFT perspective, this does not allow one to determine the actual values of the Love number couplings for generic objects, such as neutron stars.\footnote{Unless additional assumptions about boundary  conditions are made, which cannot be inferred from the EFT alone.} 
However, it does constrain the universal gravitational contributions to these couplings. In particular, it rules out all loop diagrams that scale as $(\rs/r)^{2\ell+1}$ relative to the external tidal field and that arise  purely from vertices of the Einstein--Hilbert action (irrespective of  whether the compact object is a black hole or not). 
In particular, this implies that the couplings are not logarithmically  renormalized within the EFT~\cite{Parra-Martinez:2025bcu}.\footnote{We emphasize that this statement applies only to diagrams constructed from graviton vertices in $S_{\mathrm{EH}}$, which we called ``universal'' since they do not depend on the nature of the compact object. Diagrams involving vertices arising from the matter action, for instance, can yield non-vanishing contributions and are not excluded, unless additional assumptions about the boundary conditions at the horizon are made.}

\subsection{Symmetries of dynamical perturbations}

We have seen that the vanishing of the static black hole Love number couplings has its origin in hidden symmetries of time-independent perturbations. It is likely that this is just a limit of a larger, more general symmetry structure, which has yet to be fully unveiled. Nevertheless some progress has been made for sufficiently small frequencies, which we review in this subsection.

\subsubsection{The near zone approximation}

The dynamical symmetries discovered so far have been defined in the \emph{near zone},
\begin{equation}\label{eq:near-zone-range}
    \rs \leq r \ll \omega^{-1},
\end{equation}
where $r$ is the distance from the object and $\omega$ is the characteristic frequency of the perturbation (e.g., the orbital frequency in a binary). This is the region where the gravitational field of the body dominates over radiation, and any external perturbation --- such as the tidal field from a companion --- can be treated as \emph{effectively instantaneous}. In this limit, retardation effects are negligible, and the tidal response can be computed in a quasi-static approximation.

The near-zone approximation is crucial for defining tidal Love numbers beyond the static limit: it allows one to separate the elastic response of the body from radiation-reaction effects, and to encode this response in effective parameters (Wilson coefficients) in a point-particle effective action. 
Physically, the near zone can be thought of as the region where the object ``feels'' the tidal field but has not yet radiated it away, analogous to the dominance of electrostatic fields at distances much smaller than the wavelength of emitted radiation in electromagnetism. (The \emph{far zone}, by contrast, is the region where gravitational radiation dominates and propagation effects cannot be neglected.)

\subsubsection{The Schwarzschild near zone and Love symmetry}

The near-zone approximation is
\begin{equation}\label{eq:near-zone-approx}
    (\omega^2r^2,\omega^2\rs^2)\ll1.
\end{equation}
There is a natural way to implement this in the non-spinning case ($a=0$). Considering for simplicity a scalar field ($s=0$), we can approximate the kinetic term of the wave equation \eqref{eq:sch-KG}
as \cite{Starobinskii:1973vzb,Starobinskil:1974nkd,Page:1976df,Maldacena:1997ih,Bertini:2011ga}
\begin{align}\label{eq:near-zone-sch}
    \frac{\omega^2r^4}\Delta &= \omega^2r^2+\omega^2r\rs+\omega^2\rs^2+\frac{\omega^2\rs^3}r+\frac{\omega^2\rs^4}\Delta \nonumber\\
    &\approx \frac{\omega^2\rs^4}\Delta.
\end{align}
The first four terms in the expansion are all of order $\omega^2r^2\ll1$, while the final term is relatively enhanced near the horizon due to the factor of $r-\rs$ in the denominator. The near-zone scalar equation of motion,
\begin{equation}
    \partial_r\left(\Delta\partial_r\phi\right)+\frac{\omega^2\rs^4}{\Delta}\phi - \ell(\ell+1)\phi = 0,\label{eq:near-zone-sch-eom}
\end{equation}
shares a crucial property with its static counterpart \eqref{eq:kg-zerofreq}: they have the same singularity structure. For the full equation \eqref{eq:sch-KG}, the singular point at infinity is irregular, so the equation is of Heun type. The approximation \eqref{eq:near-zone-sch} makes this singularity regular, as it is in the static case, so that \cref{eq:near-zone-sch-eom} is hypergeometric. It follows that the near zone physics possesses much of the symmetry structure we discussed for static configurations, including ladder operators and an effective metric \cite{Hui:2022vbh}. In what follows we treat the near-zone equation \eqref{eq:near-zone-sch-eom} (and its generalizations to $a,s\neq0$) as exact. When the near-zone conditions \eqref{eq:near-zone-approx} hold, solutions to this equation will coincide with solutions to the full time-dependent equation within the region \eqref{eq:near-zone-range} up to $\mathcal{O}(\omega^2r^2)$.

The near-zone effective metric following from \cref{eq:near-zone-sch-eom},
\begin{equation}\label{eq:metric-NZ}
    \dd s^2_\mathrm{NZ}=-\frac\Delta{\rs^2}\dd t^2+\frac{\rs^2}\Delta\dd r^2 + \rs^2\dd\Omega_2^2,
\end{equation}
is $\mathrm{AdS}_2\times S^2$, which has six Killing vectors. This effective metric has two other very useful properties \cite{Hui:2022vbh}. It is conformally flat, so possesses the maximal number (fifteen) of Killing and conformal Killing vectors. One of the CKVs is \cref{eq:sch-eff-CKV}, from which the ladder structure for the static sector discussed in \cref{sec:sym-ladder} descends. The other useful property is that it has vanishing scalar curvature; a massless scalar is therefore conformally coupled. The ladder symmetry arguments of the previous subsection then imply the vanishing of scalar Love numbers in the near-zone approximation.

Because the near-zone metric \eqref{eq:metric-NZ} is conformally flat, the full symmetry algebra of the near-zone equation \eqref{eq:near-zone-sch-eom} is the conformal algebra $\mathfrak{so}(4,2)$, possessing an $\mathfrak{so}(3,1)$ subalgebra and four $\mathfrak{sl}(2,\mathbb R)$ subalgebras \cite{Hui:2022vbh}. One of these $\mathfrak{sl}(2,\mathbb R)$ subalgebras, first identified in \rcite{Bertini:2011ga}, is of particular interest. The generators are
\begin{subequations}\label{eq:sch-love-sym}
    \begin{align}
        L_0&=-2\rs\partial_t,\\
        L_\pm&=e^{\pm t/2\rs}\left(2\rs\partial_r\sqrt{\Delta}\partial_t\mp\sqrt\Delta\partial_r\right),
    \end{align}
\end{subequations}
which satisfy the commutation relations
\begin{equation}\label{eq:sch-sl2}
    [L_0,L_\pm] = \mp L_\pm,\quad [L_+,L_-] = 2L_0.
\end{equation}
The quadratic Casimir operator of this algebra matches the radial part of the Klein--Gordon operator in the near-zone approximation,
\begin{align}
    \mathcal C_2 &= L_0^2 - \frac12\left(L_+L_-+L_-L_+\right) \nonumber\\
    &= \partial_r(\Delta\partial_r)-\frac{\rs^4}\Delta\partial_t^2,
\end{align}
that is, \cref{eq:near-zone-sch} can be written as an eigenvalue equation for $\mathcal C_2$,
\begin{equation}
    \mathcal C_2 \phi = \ell(\ell+1)\phi.
\end{equation}

The generators \eqref{eq:sch-love-sym} are the $a=0$ limit of the \emph{Love symmetry}, which forms the basis for a representation-theoretic argument for the vanishing of black hole Love numbers \cite{Charalambous:2021kcz,Charalambous:2022rre,Charalambous:2024gpf}. The key observation is that static scalar field profiles belong to a \emph{finite-dimensional} representation of $\mathrm{SL}(2,\mathbb R)$. The highest-weight vector $v_{-\ell,0}$, which satisfies
\begin{equation}
    L_+ v_{-\ell,0}=0,\quad L_0 v_{-\ell,0} = -\ell v_{-\ell,0}
\end{equation}
and is given explicitly by
\begin{equation}
    v_{-\ell,0} = e^{\frac{\ell}{2\rs}t}\Delta^{\ell/2},
\end{equation}
satisfies the near-zone equation of motion \eqref{eq:near-zone-sch-eom} and is regular at the horizon. This is not the low-frequency limit of a solution to the full equation of motion, due to its time dependence: the frequency $\rs\omega=i\ell/2\sim\mathcal{O}(1)$ takes it outside the r\'egime of validity \eqref{eq:near-zone-approx} of the near-zone approximation.\footnote{The use of high-frequency modes to ascertain properties of low-frequency modes has led to the intriguing suggestion that the Love symmetry is an example of UV/IR mixing \cite{Charalambous:2021kcz,Charalambous:2022rre}.} It can however be used to construct a valid near-zone solution by lowering $\ell$ times with the operator $L_-$,
\begin{equation}
    v_{-\ell,\ell} = L_-^\ell v_{-\ell,0} = v_{-\ell,\ell} (r).
\end{equation}
This solution is static, as each application of $L_-$ lowers the frequency by $i/2\rs$, and it is regular at the horizon because $L_-$ and $v_{-\ell,0}$ are. Having identified the horizon-regular static solution, it remains to show that it is a finite polynomial in $r$. This is done by raising back to $v_{-\ell,0}$ and using the fact that it is annihilated by $L_+$, i.e.,
\begin{align}
    0 &= L_+^{\ell+1}v_{-\ell,\ell} (r) \nonumber\\
    &= \left(-e^{t/2\rs}\sqrt\Delta\right)^{\ell+1}\partial_r^{\ell+1}v_{-\ell,\ell}(r),
\end{align}
which implies that $v_{-\ell,\ell}(r)$ is indeed a degree-$\ell$ polynomial and the scalar Love numbers vanish. The Love symmetry can be straightforwardly generalized to spin-$s$ fields, and the vanishing of electromagnetic and gravitational Love numbers follows in full analogy to the scalar case \cite{Charalambous:2021kcz,Charalambous:2022rre}.

\subsubsection{Kerr near zones}

\epigraph{Sometimes Science is more\\Art than Science.}{\textsc{Rick Sanchez}}

For spinning black holes, there is more freedom in taking the near-zone %
limit \eqref{eq:near-zone-approx} in the Teukolsky equation \eqref{eq:teuk-rad-boxed}, depending on which subleading terms one chooses to keep, and many proposals exist in the literature. In general there are infinitely many ways to perform a near-zone truncation of the radial equation while still accurately capturing the properties of the full solution in the region \eqref{eq:near-zone-range}; we will study choices for which the truncated equation has particularly nice properties, such as enhanced symmetries. Writing \cref{eq:teuk-rad-boxed} and its approximations in the form
\begin{equation}\label{eq:NZ-eom}
    \left[\Delta^{-s}\partial_r(\Delta^s\partial_r) + V(r)\right]R^{(s)}_{\ell m \omega}(r) = \ell(\ell+1)R^{(s)}_{\ell m \omega}(r),
\end{equation}
choosing a near zone amounts to specifying a function $V(r)$. There are two direct generalizations of the Schwarzschild near zone we have just discussed: the \emph{Starobinsky near zone} \cite{Starobinskii:1973vzb,Starobinskil:1974nkd,Page:1976df,Maldacena:1997ih,Bertini:2011ga},
\begin{equation}\label{eq:NZ-starobinski}
V(r)=\frac{\rs^2r_+^2}\Delta(\omega-m\Omega_{\mathrm H})^2 - i s \rs r_+\frac{\Delta'}\Delta(\omega-m\Omega_{\mathrm H}),
\end{equation}
and the \emph{Love near zone} \cite{Charalambous:2021kcz,Charalambous:2022rre}
\begin{equation}\label{eq:NZ-sergei}
V(r)=\frac{\rs^2r_+^2}\Delta\left((\omega-m\Omega_{\mathrm H})^2 - 4\omega\Omega_{\mathrm H}m\frac{r-r_+}{r_+-r_-}-\frac{2is\omega}\beta\right)+iasm\frac{\Delta'}\Delta,
\end{equation}
where we remind the reader that the angular velocity at the outer horizon $\Omega_{\mathrm H}$ \eqref{eq:ang-vel-hor} and the inverse Hawking temperature $\beta=1/(2\pi T_\mathrm{H})$ \eqref{eq:inv-hawking-temp} are
\begin{equation}
    \Omega_{\mathrm H}=\frac a{\rs r_+},\qquad \beta = \frac{2\rs r_+}{r_+-r_-}.
\end{equation}
Each of these near zone equations reduces to \cref{eq:near-zone-sch-eom} in the limit $a=s=0$. More general near zones which have a singular Schwarzschild limit are also quite interesting, as the $\mathrm{SL}(2, \mathbb{R})$ symmetry present in the Starobinsky and Love near zones can be enlarged to $\mathrm{SL}(2, \mathbb{R})\times\mathrm{SL}(2, \mathbb{R})$. This observation has led to the proposal that non-extremal black holes are dual to a 2D CFT \cite{Castro:2010fd,Lowe:2011aa}.\footnote{It is interesting to note the connection of the Love number symmetries to the fascinating Kerr/CFT proposal. Originally conjectured for (near-)extremal black holes \cite{Guica:2008mu}, the discovery of the $\mathrm{SL}(2, \mathbb{R})\times\mathrm{SL}(2, \mathbb{R})$ symmetry of the scalar wave equation away from extremality has been interpreted as evidence that non-extremal black holes also admit a 2D dual CFT interpretation \cite{Castro:2010fd}.}
Of this class, we consider in particular the near zone approximation \cite{Castro:2010fd,Rodriguez:2023xjd}
\begin{equation}\label{eq:NZ-Kerr-CFT}
V(r)=\frac{\left(\rs r_+\omega-\frac i2s(r_+-r_-)-am\right)^2}{(r-r_+)(r_+-r_-)}+\frac{\left(\rs r_-\omega+\frac i2s(r_+-r_-)-am\right)^2}{(r-r_-)(r_+-r_-)}.
\end{equation}
For $s=0$, all of these choices are cases of the one-parameter family of near-zone approximations of \rcite{Lowe:2011aa}. This does not exhaust the possibilities; see for instance \rcite{Gray:2024qys}, which proposed a systematic construction of globally well-defined near-zone generators that results in a one-parameter extension of the (scalar) Love symmetry \eqref{eq:Love-generators}.

Each of these near zones has a different symmetry structure, leading to complementary viewpoints on tidal responses. We review each in turn.

\paragraph{Starobinsky near zone}
In the Starobinsky near zone \eqref{eq:NZ-starobinski}, the effective metric for the massless scalar is \cite{Hui:2022vbh}
\begin{equation}\label{eq:NZ-metric-kerr}
    \dd s^2_\mathrm{NZ} = -\frac{\Delta-a^2\sin^2\theta}{\rs r_+}\dd t^2-2a\sin^2\theta\,\dd t\,\dd\varphi + \frac{\rs r_+}\Delta\dd r^2+\rs r_+\dd \Omega_2^2.
\end{equation}
A change of angular coordinate $\varphi'=\varphi-(a/\rs r_+)t$ removes the cross terms and makes this look very similar to its Schwarzschild counterpart \eqref{eq:metric-NZ},
\begin{equation}
    \dd s^2_\mathrm{NZ} = -\frac\Delta{\rs r_+}\dd t^2+ \frac{\rs r_+}\Delta\dd r^2+\rs r_+\dd {\Omega'_2}^2.
\end{equation}
Indeed, away from the extremal limit ($a\neq M$)\footnote{In the extremal limit, the $(t,r)$ subspace is AdS$_2$ in Poincar\'e coordinates with radial coordinate $r-r_-$. The extremal near-zone metric coincides, up to an identification of constants, with the near-horizon limit of the extremal Reissner--Nordstr\"om metric \cite{Hui:2022vbh}.} we can rescale the time and radial coordinates,
\begin{equation}
    t'=\frac{r_+-r_-}{\sqrt{\rs r_+}}t,\qquad r'=\frac{\sqrt{\rs r_+}}{r_+-r_-}(r-r_-),
\end{equation}
to write the effective metric as
\begin{equation}
    \dd s^2_\mathrm{NZ} = -\frac{\tilde\Delta}{\rs r_+}\dd {t'}^2 + \frac{\rs r_+}{\tilde\Delta}\dd{r'}^2+\rs r_+\dd {\Omega'_2}^2,
\end{equation}
where we have defined
\begin{equation}
    \tilde\Delta\equiv r'(r'-\sqrt{\rs r_+})=\frac{\rs r_+}{(r_+-r_-)^2}\Delta.
\end{equation}
In these coordinates we see that the Kerr near-zone effective metric is the same as the Schwarzschild near-zone metric \eqref{eq:metric-NZ}, up to the replacement $\rs\to\sqrt{\rs r_+}$.

We remind the reader that this spacetime is $\mathrm{AdS}_2\times S^2$, is conformally flat, and has vanishing scalar curvature; its symmetry group is again $\mathrm{SO}(4,2)$.\footnote{For the precise form of the fifteen generators of this group, see \rcite{Hui:2022vbh}.} This is the largest symmetry group of the three near zone approximations we discuss. It is also distinguished by having non-trivial symmetries that act on the static ($\omega=0$) sector, including the special conformal generator \eqref{eq:phi-ladder-CKV} that contains the ladder operators. It is therefore in the Starobinsky near zone that the ladder symmetry story of \cref{sec:sym-ladder} generalizes neatly to dynamical perturbations. The conserved charge evaluated on the solution with ingoing boundary conditions at the horizon is \cite{Hui:2022vbh}
\begin{equation}\label{eq:resp-coeff-staro}
    \sqrt{J_\ell} = -\frac i2\frac{(\ell!)^2}{(2\ell)!(2\ell+1)!}(r_+-r_-)^{2\ell+1}\beta(m\Omega_\mathrm{H}-\omega)\prod_{n=1}^\ell\left[n^2+\beta^2(m\Omega_\mathrm{H}-\omega)^2\right].
\end{equation}
This is purely imaginary and agrees, up to a constant factor, with the near-zone dynamical dissipative response coefficients \eqref{eq:LoveDiss} \cite{Wong:2019yoc,Chia:2020yla}, and in the limit $\omega=0$ with the Kerr static dissipative coefficients \eqref{eq:lambdaKerrell}.
An interesting feature of the Starobinsky near zone is that the response coefficients \eqref{eq:resp-coeff-staro} are imaginary (i.e., the Love numbers vanish) for \emph{all} (i.e., finite) frequencies, although of course these only accurately capture the actual response when the near-zone approximation \eqref{eq:near-zone-approx} holds \cite{Charalambous:2022rre}.

The representation-theoretic arguments associated to the Love symmetry can be applied in the Starobinsky near zone. The generators of the relevant $\mathrm{SL}(2,\mathbb R)$, which generalize \cref{eq:sch-love-sym} to the Starobinsky near zone and account for internal spin, are \cite{Hui:2022vbh,Charalambous:2022rre}
\begin{subequations}\label{eq:staro-love-sym}
    \begin{align}
        L_0&=-\beta(\partial_t+\Omega_\mathrm{H}\partial_\varphi),\\
        L_\pm&=e^{\pm t/\beta}\left(\beta\partial_r\sqrt{\Delta}(\partial_t+\Omega_\mathrm{H}\partial_\varphi)\mp\sqrt\Delta\partial_r \mp s\frac{r-r_\mp}{\sqrt\Delta}\right).
    \end{align}
\end{subequations}
The representation-theoretic argument outlined above now applies to modes at the locking frequency $\omega=m\Omega_\mathrm H$ \cite{Charalambous:2022rre}. This is complementary to the ladder symmetry arguments, which apply most naturally to static ($\omega=0$) perturbations. As in the Schwarzschild case, the representation-theoretic argument uses an $\mathfrak{sl}(2,\mathbb R)$ subalgebra that does not contain the CKV \eqref{eq:phi-ladder-CKV} which generates the ladder symmetries. It is interesting that the $\mathrm{SO}(4,2)$ symmetry of the Starobinsky near zone is large enough to accommodate these complementary symmetry perspectives on the properties of tidal response coefficients.

\paragraph{Love near zone}

The near zone \eqref{eq:NZ-sergei} is the setting for the $\mathrm{SL}(2,\mathbb R)$ Love symmetry, with generators \cite{Charalambous:2021kcz,Charalambous:2022rre}
\begin{subequations}\label{eq:Love-generators}
\begin{align}
    L_0 &= -\beta\partial_t+s,\\
    L_\pm &= e^{\pm t/\beta}\left(\mp\sqrt\Delta\partial_r+\beta\partial_r\sqrt\Delta\partial_t+\frac a{\sqrt\Delta}\partial_\varphi-s(1\pm 1)\partial_r\sqrt\Delta\right).
\end{align}
\end{subequations}
As in the non-spinning case, the equation of motion (\cref{eq:NZ-eom} with the potential \eqref{eq:NZ-sergei}) can be rewritten in terms of the quadratic Casimir of this algebra,
\begin{align}
    \mathcal{C}_2 \phi &= \left[L_0^2-\frac12\left(L_+L_-+L_-L_+\right)\right]\phi \nonumber\\
    &= \left[\Delta^{-s}\partial_r(\Delta^s\partial_r) +\frac{\rs^2r_+^2}\Delta\left((\omega-m\Omega_{\mathrm H})^2 - 4\omega\Omega_{\mathrm H}m\frac{r-r_+}{r_+-r_-}-\frac{2is\omega}\beta\right)+iasm\frac{\Delta'}\Delta\right]\phi \nonumber\\
    &= \ell(\ell+1)\phi.
\end{align}
The vanishing of spin-$s$ Love numbers follows from a representation-theoretic argument analogous to the spin-0 Schwarzschild case discussed above: the horizon-regular solutions form a finite-dimensional representation of the Love symmetry, and the \emph{static} horizon-regular solution is a descendent of the highest-weight state with weight $-\ell$.

It is worth comparing this to the situation in the Starobinsky near zone. The analogue of the Love generators \eqref{eq:Love-generators} in Starobinsky can only be used to show that the static and axisymmetric ($\omega=m=0$) response coefficients vanish, while in the Love near zone they control the Love numbers and dissipative response coefficients for all $(\ell,m)$.
Instead, the symmetry controlling tidal response coefficients in the Starobinsky near zone is the special conformal symmetry that is also present in the static sector. By contrast, in the Love near zone, there is no symmetry that acts only on static modes.
We note that the $\mathrm{SL}(2,\mathbb R)$ symmetries of the Starobinsky and Love near zones can be combined into a single, infinite-dimensional $\mathrm{SL}(2,\mathbb R)\ltimes\mathrm{U}(1)$ symmetry \cite{Charalambous:2022rre}.

\paragraph{Kerr/CFT near zone}

The near zone \eqref{eq:NZ-Kerr-CFT} was used to study dynamical tidal responses in \rcite{Perry:2023wmm}. An advantage of this near zone in contrast to the Starobinsky and Love near zones is that it is able to derive the logarithmic running of these response coefficients and other $\mathcal{O}(a\omega)$ terms. The $s=0$ wave equation can again be written in terms of the Casimir element of an $\mathrm{SL}(2,\mathbb R)$ group,
\begin{align}
    \mathcal{C}_2 \phi &= \left[H_0^2-\frac12\left(H_+H_-+H_-H_+\right)\right]\phi \nonumber\\
    &= \left[\partial_r \Delta \partial_r 
- \frac{(\rs r_+ \partial_t + a \partial_\phi)^2}{(r - r_+)(r_+ - r_-)}
+ \frac{(\rs r_- \partial_t + a \partial_\phi)^2}{(r - r_-)(r_+ - r_-)}\right]\phi \nonumber\\
    &= \ell(\ell+1)\phi.
\end{align}
where the generators of the associated algebra are locally defined as\footnote{A disadvantage of these generators is that they are not globally well-defined, as they are not periodic under $\varphi\to\varphi+2\pi$. This is in contrast to the generators of the Starobinsky and Love near zones, which \emph{are} globally defined, cf. \cref{eq:staro-love-sym,eq:Love-generators}.}
\begin{align}
H_0 &= i \left(
\frac{1}{2\pi T_R} \partial_\varphi
+ \frac{T_L}{T_R}\rs \partial_t
\right), \\
H_\pm &= i e^{\mp 2\pi T_R \varphi} \left(
\pm \sqrt\Delta \partial_r
+ \frac{1}{2\pi T_R} \frac{r - \frac\rs2}{\sqrt{\Delta}} \partial_\varphi
+ \frac{T_L}{T_R} \frac{\rs r - 2a^2}{\sqrt{\Delta}} \partial_t
\right).
\end{align}
Here the parameters 
\begin{equation}
T_L \equiv \frac{r_+ + r_-}{4 \pi a}    , \qquad T_R \equiv \frac{r_+ - r_-}{4 \pi a}
\end{equation}
define the right- and left-moving temperatures of the dual two-dimensional dual thermal conformal field theory description.
We have already mentioned that the full symmetry of the wave equation in this near zone is $\mathrm{SL}(2, \mathbb{R})\times\mathrm{SL}(2, \mathbb{R})$. This is manifested in the existence of a second set of $\mathfrak{sl}(2,\mathbb R)$ generators $(\bar H_0,\bar H_\pm)$ whose Casimir matches that of the first $\mathfrak{sl}(2,\mathbb R)$ \cite{Castro:2010fd}.

For recent work further exploring the symmetry structure underlying the vanishing of Kerr Love numbers from a representation-theoretic perspective, see \rcite{Lupsasca:2025pnt,Sharma:2025xii}.

\subsubsection{Subtracted/effective geometries}

The symmetry structures appearing in the various near zone approximations we have discussed can all be understood in the context of \emph{subtracted geometries} \cite{Cvetic:2011dn,Cvetic:2011hp,Charalambous:2022rre}. Let us write the Kerr metric \eqref{eq:Kerr} as
\begin{equation}\label{eq:subtracted-metric}
\dd s^2 = -\frac G{\sqrt{\Delta_0}}(\dd t+\mathcal A)^2+\sqrt{\Delta_0}\left(\frac{\dd r^2}\Delta+\dd\theta^2+\frac \Delta G\sin^2\theta\dd\varphi^2\right),    
\end{equation}
with
\begin{equation}
    \Delta_0 = \Sigma^2,\qquad\mathcal A = \frac{a\rs r \sin^2\theta}G\dd\varphi,\qquad G=\Delta-a^2\sin^2\theta.
\end{equation}
The key observation of \rcite{Cvetic:2011dn,Cvetic:2011hp} is that the functions $\Delta_0$ and $\mathcal A$ can be modified without changing the black hole's causal structure and thermodynamics.\footnote{The function $G$ can also be modified, although it is unchanged from its Kerr value for the subtracted geometries we consider. This is reminiscent of the off-shell nature of the Killing tensor symmetry on Kerr, as we discussed in \cref{sec:Kerr-syms}.} This suggests that the modifications correspond purely to the \emph{external environment} of the black hole, while leaving its internal structure untouched. Such a modified geometry is known as a \emph{subtracted} or \emph{effective geometry}.

We associate a subtracted geometry to a given near zone if the Klein--Gordon equation in that geometry reproduces the spin-0 equation of motion in the near zone. We have already encountered such a geometry in the context of the effective metric \eqref{eq:NZ-metric-kerr} of the Starobinsky near zone \eqref{eq:NZ-starobinski} and its Schwarzschild counterpart \eqref{eq:metric-NZ}. In the notation of \cref{eq:subtracted-metric}, this corresponds to
\begin{equation}
    \Delta_0 = \rs^2r_+^2,\qquad\mathcal{A} = \frac{a\sin^2\theta}G\rs r_+\dd\varphi.
\end{equation}
We remind the reader that this geometry is highly symmetric: it is a conformally-flat $\mathrm{AdS}_2\times S^2$, with symmetry group $\mathrm{SO}(4,2)$. Similarly, the Love near zone \eqref{eq:NZ-sergei} is associated to a subtracted geometry with
\begin{equation}
    \Delta_0 = \rs^2r_+^2(1+\beta^2\Omega_{\mathrm H}^2\sin^2\theta),\quad \mathcal A = \frac{a\sin^2\theta}G\left(\rs r_++\beta(r-r_+)\right)\dd\varphi.
\end{equation}
The $s=0$ generators \eqref{eq:Love-generators} of the Love symmetry are related to the Killing isometries of this subtracted geometry \cite{Charalambous:2022rre}.\footnote{For the Starobinsky near zone, the $s$-dependent pieces in the generators \eqref{eq:staro-love-sym} can also be derived using a suitable spin-weighted modification of the GHP Lie derivative, although this procedure does not correctly reproduce the $s$-dependent pieces of the generators \eqref{eq:Love-generators} in the Love near zone \cite{Ludwig_2000,Ludwig:2001hx,Charalambous:2022rre}.},
Finally, the Kerr/CFT near zone \eqref{eq:NZ-Kerr-CFT} has the subtracted geometry
\begin{equation}
    \Delta_0 = \rs^2 (\rs r - a^2 \cos^2 \theta ) ,
\end{equation}
with $\mathcal A$ unchanged from Kerr.

In each of these cases, the Klein--Gordon equation exhibits at least one $\mathrm{SL}(2, \mathbb{R})$ hidden symmetry in the near-region approximation; the symmetry groups for each are summarized in \cref{tab:KEG_summary}. 
While these geometries share the thermodynamic properties of Kerr black holes, they differ in their hidden symmetries and in their ability to capture dynamical versus static tidal responses. 

\begin{table}[h!]
\centering
\renewcommand{\arraystretch}{1.3}
\setlength{\tabcolsep}{5pt}
\footnotesize
\begin{tabular}{>{\raggedright\arraybackslash}p{4.0cm}
                >{\raggedright\arraybackslash}p{4.0cm}
                >{\raggedright\arraybackslash}p{7.0cm}}
\toprule
\textbf{Subtracted geometry} & \textbf{Hidden symmetry} & \textbf{References} \\
\midrule
Love geometry
& $\mathrm{SL}(2,\mathbb{R})$
& Charalambous \emph{et al.}~\cite{Charalambous:2021kcz,Charalambous:2022rre};
  Charalambous~\cite{Charalambous:2024gpf} \\[2pt]
Starobinsky geometry
& $\mathrm{SO}(4,2)$
& Hui \emph{et al.}~\cite{Hui:2022vbh} \\[2pt]
Kerr/CFT geometry
& $\mathrm{SL}(2,\mathbb{R}) \times \mathrm{SL}(2,\mathbb{R})$
& Perry \emph{et al.}~\cite{Perry:2023wmm} \\
\bottomrule
\end{tabular}
\caption{Summary of near-zone effective geometries. Each geometry
preserves the internal near-horizon structure and thermodynamics of
Kerr black holes, while modifying the asymptotics and realizing different hidden
symmetries in the massless wave equation.}
\label{tab:KEG_summary}
\end{table}

\subsection{A broader view: symmetries of black hole perturbation theory}
\label{sec:sym-BHPT}

Having presented the reader a panoply of symmetries purporting to answer the naturalness puzzle posed in \cref{sec:naturalness}, it is not unreasonable to still wonder: \emph{why} do the Love numbers of black holes vanish? We close this section with a speculative discussion as to what the ``bigger picture'' might be, in the hopes that we will inspire readers to further deepen our fundamental understanding of general relativity.

The theoretical setting for all of our investigations has been black hole perturbation theory. What are the symmetries of black hole perturbation theory? This is such a broad question that it is difficult (perhaps impossible) to even define precisely, but as we have seen throughout this section, this breadth goes hand in hand with a rich space of possible answers. We have encountered symmetries of linear and nonlinear perturbations; field theoretic symmetries and geometric symmetries; symmetry explanations for the vanishing of black hole Love numbers based on conventional off-shell symmetries and conserved quantities, but also explanations based on representation-theoretic arguments and others invoking selection rules.

And for all of this, we have focused exclusively on symmetry structures arising in the static and low-frequency limits, since this is the r\'egime of greatest relevance for the tidal responses which are the subject of this review. But there is no reason to think that these symmetries must stop there. Indeed we have already seen how symmetries of the strictly static sector can be enlarged to near-zone symmetries acting on modes with small but non-vanishing frequency.

It is natural to ask: do these symmetries in turn fit into an even larger structure, with a broader r\'egime of applicability? We already know of symmetries of black hole perturbations which act at arbitrary frequency, including the Chandrasekhar duality between even- and odd-parity modes discussed in \cref{sec:chandra}, and the recently-discovered $\mathrm{SL}(2,\mathbb R)$ invariances of the photon ring \cite{Hadar:2022xag,Kapec:2022dvc,Chen:2022fpl}. These act naturally on quasinormal modes, which are the paradigmatic example of black hole perturbations beyond the low-frequency limit.
Since the (retarded) Green's function whose non-analyticities dictate the quasinormal frequencies and tails at large $\omega$ is the same object that encodes information about static response at low $\omega$,
it is tempting to conjecture that some or all of these symmetries in different frequency r\'egimes might have a deeper relation than is apparent at present, and to wonder what more fundamental structure might underlie them.

In the strictest sense, general relativity has only a single symmetry, namely diffeomorphism invariance. If one broadens the list to include asymptotic symmetries, which act non-trivially at, e.g., the horizon or infinity, then many more arise \cite{Bondi:1962px,Sachs:1962wk,Newman:1966ub,Barnich:2010ojg,Barnich:2011mi,Strominger:2013jfa,He:2014laa,Campiglia:2014yka,Flanagan:2015pxa,Donnay:2015abr,Hawking:2016msc,Hawking:2016sgy,Donnay:2016ejv,Strominger:2017zoo,Alessio:2017lps,Compere:2018aar,Haco:2018ske,Chandrasekaran:2018aop,Donnay:2019jiz,Henneaux:2019yax,Flanagan:2019vbl,Campiglia:2020qvc,Pasterski:2021raf,Pasterski:2021rjz,Raclariu:2021zjz,Freidel:2021fxf,Ciambelli:2022vot}. The solutions of general relativity can themselves be symmetric. Indeed we have seen (cf. \cref{sec:Kerr-syms}) that the Schwarzschild and Kerr metrics have a rather rich symmetry structure, including both conventional Killing isometries and the ``hidden symmetries'' described by higher-order Killing tensors. These symmetries in turn leave imprints on fluctuations about these backgrounds. This happens in the standard way, with the Killing vectors and tensor leading to conservation of energy, angular momentum, mass, and the Carter constant, which in turn are responsible for the rather useful property that the spin-$s$ wave equation on Kerr is separable. But we have also seen (cf. \cref{sec:nonlineartheory,sec:sym-nonlin}) that the presence of Killing isometries leads to the more intricate symmetries of Ehlers and Geroch, which act on the space of solutions to Einstein's equations possessing those isometries.

One distinction worth making is between symmetries of the theory and symmetries of the phase space, i.e., the space of solutions. The fact that we have been looking at symmetries of low-frequency perturbations would seem to place us firmly in the latter category. But this line of reasoning remains in need of further development. In particular, it would be very interesting to understand the symmetries we have discussed in the language of the covariant phase space formalism \cite{Lee:1990nz,Wald:1993nt,Iyer:1994ys,Wald:1999wa,Compere:2018aar,Harlow:2019yfa}.

Another potential connection which has yet to be fully fleshed out is between the symmetries underlying Love numbers and asymptotic symmetries, such as the Bondi--Metzner--Sachs (BMS) symmetry \cite{Bondi:1962px,Sachs:1962wk,Newman:1966ub} and its extensions (e.g., \cite{Flanagan:2019vbl,Freidel:2021fxf,Ciambelli:2022vot}). The asymptotic symmetries of the near-horizon region are well-studied \cite{Hotta:2000gx,Afshar:2015wjm,Donnay:2015abr,Hawking:2016msc,Hawking:2016sgy,Donnay:2016ejv,Haco:2018ske,Donnay:2019jiz}. 
An early but inconclusive attempt by Penna to connect these symmetries to Love numbers identified a possible role played by the Carroll symmetry, which is a contraction of BMS \cite{Penna:2018gfx}. The study of asymptotic symmetries has attracted a significant amount of research in recent years, much like Love numbers and their symmetries have during a similar period. It would be of great interest to see concrete connections between these two highly active and promising areas of research.

For any symmetry of black hole perturbation theory, it is worth asking: what property of the black hole metric is responsible? Around what other backgrounds of GR will similar symmetry structures exist? Some analogues in the perturbation theory of the maximally symmetric backgrounds --- Minkowski space and (anti-)de Sitter space --- are already known. Linearized gravitational duality was first shown to be an off-shell symmetry for perturbations of Minkowski \cite{Henneaux:2004jw}, and soon thereafter extended to dS \cite{Julia:2005ze} and later AdS \cite{Hortner:2019iip}. The more recent addition of Schwarzschild to this list \cite{Solomon:2023ltn}, by extending the Chandrasekhar duality (which reduces to gravitational duality in the flat-space limit) to an off-shell invariance of the linearized action \eqref{eq:L-even-odd}, shows at least that maximal symmetry is not a requirement. In the case of duality, it is reasonable to speculate that the timelike Killing vector is responsible, since the Ehlers symmetry is closely related to gravitational duality, though this is not known for sure at present.

The ladder symmetries also have an analogue in de Sitter space, which were discovered in \rcite{Compton:2020cjx} as an explanation for the transparency of de Sitter in odd dimensions \cite{Lagogiannis:2011st}.\footnote{Historically, \rcite{Compton:2020cjx} was a direct inspiration for the formulation of the static ladder symmetries in \rcite{Hui:2021vcv}.} The analogy is in fact exact: as we have seen, the ladder operators emerge from the EAdS$_3$ effective geometry of static black hole perturbations, and the (3D) de Sitter ladder operators are related by analytic continuation. How far does this analogy extend? Can the symmetries of black hole perturbation theory teach us, for example, something about cosmological perturbations?

Finally, a prominent theme in both the symmetries underlying Love numbers and other famous black hole perturbation symmetries is conformal invariance. The ladder symmetries are a consequence of the special conformal transformation \eqref{eq:flat-space-love-sym}; see also \rcite{Katagiri:2022vyz}, which used conformal symmetry to present a ladder argument for vanishing Love numbers. Conformal symmetry is at the heart of the Kerr/CFT proposal \cite{Castro:2010fd,Aggarwal:2019iay}. It is also responsible for symmetries of the photon ring: defining a ``near-ring'' region analogously to the near zone region, the massless wave equation again acquires a conformal invariance that acts on quasinormal modes \cite{Hadar:2022xag,Kapec:2022dvc}. It seems likely that conformal symmetry still has more to teach us about the physics of black holes.

\newpage

\part*{Conclusions and outlook}
\addcontentsline{toc}{part}{Conclusions and outlook}
\label{sec:conclusions}

\epigraphhead[]{
\epigraph{Stop! In the name of Love.}{\textsc{The Supremes}}
}

Tidal Love numbers provide a precise and remarkably sensitive probe of gravitational structure. In classical ($D=4$) vacuum general relativity, the static Love numbers of black holes vanish identically. This result --- established through a wide variety of methodologies, including the Regge--Wheeler--Zerilli formalism, the Teukolsky equation, monodromy methods,  effective field theory (EFT) matching, and the post-Newtonian approach --- reflects a hidden structural property of the Kerr black hole and its deformations. In the presence of a quasi-static external tidal field, ingoing boundary conditions at the horizon rule out any conservative tidal response, leaving at most a dissipative contribution  associated with absorption.
In this precise sense, vacuum black holes in general relativity appear to behave as rigid  objects under static tidal deformations, as seen by a distant observer.

From the EFT viewpoint, tidal Love numbers are Wilson coefficients in the worldline action describing compact objects at long distances. They encode the finite-size response of the object after short-distance degrees of freedom have been integrated out. In this language, the vanishing of black hole Love numbers corresponds to the absence of certain operators in the low-energy effective action, a feature that can be traced back to the special properties and symmetries of black hole solutions in general relativity, as reviewed in \cref{sec:sym}.
Remarkably, recent developments have shown that (static) Love numbers vanish  beyond the r\'egime of linear response, providing evidence that black holes  in general relativity do not develop tidally-induced multipole moments,  regardless of the strength of an external time-independent field. From a  large-distance perspective, this provide further evidence that, as far as the conservative  static sector is concerned, black holes are effectively indistinguishable  from elementary particles~\cite{Poisson:2020vap,Riva:2023rcm,Iteanu:2024dvx,Combaluzier-Szteinsznaider:2024sgb,Kehagias:2024rtz,Parra-Martinez:2025bcu}.

The situation contrasts sharply with that of material compact objects. Neutron stars possess non-zero Love numbers that encode detailed information about their internal equation of state (EoS). These coefficients are now measurable through gravitational-wave observations and provide one of the most direct probes of
nuclear matter at densities impossible to achieve in a laboratory~\cite{Flanagan:2007ix,Hinderer:2007mb, Lattimer:2012nd,Rezzolla:2018jee,Burns:2019byj,Chatziioannou:2020pqz,Lattimer:2021emm,Burgio:2021vgk,Yunes:2022ldq,Chatziioannou:2024jsr,Glendenning:1997wn,Haensel:2007yy}.
At finite frequency, the  response coefficients acquire imaginary parts associated with dissipation, further enriching the EFT description by linking the tidal response to transport properties such as viscosity and mode excitation.
The universal I-Love-Q relations~\cite{Yagi:2013awa,Yagi:2016bkt} provide EoS-insensitive handles on the tidal sector, strengthening the case for Love numbers as precision observables. 

The vanishing of black hole Love numbers is fragile. As summarized in \cref{sec:highD,sec:beyondGR}, deviations from four-dimensional, vacuum general relativity --- including higher dimensions~\cite{Kol:2011vg,Hui:2020xxx}, modified gravity theories~\cite{Cardoso:2018ptl,Barura:2024uog}, higher-curvature (bulk) extensions such as Lovelock gravity~\cite{Singha:2025xah}, 
parity-violating interactions~\cite{Cano:2025zyk}, 
quantum-corrected spacetimes~\cite{Wang:2025oek,Coviello:2025pla}, 
and non-vacuum astrophysical environments~\cite{Cannizzaro:2024fpz,Katagiri:2024fpn} --- generically induce non-zero black hole tidal responses. %
Love numbers therefore function as precision diagnostics of gravitational dynamics, capable of distinguishing matter microphysics from genuine modifications of the gravitational action. Previous observational constraints from binary black hole events 
have already placed non-trivial bounds on black hole tidal deformability~\cite{Narikawa:2021pak,Chia:2024bwc,Andres-Carcasona:2025bni}, and these will tighten considerably with next-generation detectors. 

Looking ahead, several important open problems remain. First, extending  static analyses to fully dynamical tidal responses in rotating  backgrounds is essential for precision waveform  modeling~\cite{Hinderer:2016eia,Steinhoff:2016rfi,Pratten:2021pro,Andersson:2017iav,Williams:2022vct,Chia:2024bwc,Chakraborty:2025wvs}. 
Recent computations of dynamical Love numbers for Schwarzschild and Kerr  black holes via off-shell matching~\cite{Combaluzier--Szteinsznaider:2025eoc}, scattering amplitudes, and MST  methods~\cite{Saketh:2024juq,Ivanov:2024sds,Ivanov:2026icp,Markovic:2025kvr,Caron-Huot:2025tlq} represent  significant progress. However, complete results valid for generic spin and including subleading orders in frequency are still lacking.

Second, the interplay between renormalization-group running and  post-Newtonian expansions warrants further study at higher orders. More generally, a systematic analysis of subleading tidal corrections, and their incorporation into high-precision gravitational-wave modeling, will be necessary to fully exploit the scientific potential of future interferometers. For instance, the role of nonlinear tidal effects --- both for black  holes~\cite{Poisson:2020vap,DeLuca:2023mio,Riva:2023rcm,Iteanu:2024dvx,Combaluzier-Szteinsznaider:2024sgb}  and neutron stars~\cite{Weinberg:2013pbi,Nouri:2021mvb,Yu:2022fzw,Pani:2025qxs,Pitre:2025qdf} --- is not  yet fully systematically incorporated into gravitational-wave search pipelines.
 
Third, the possible extension of the Love number symmetries to (nonlinear)  perturbations of rotating, charged~\cite{Berens:2022ebl,Rai:2024lho},  higher-dimensional~\cite{Charalambous:2023jgq,Charalambous:2025ekl,Berens:2025jfs,Parra-Martinez:2025bcu}, and beyond-GR~\cite{Sharma:2024hlz,Sharma:2025xii} black holes remains an active area of  research, with the precise connections between the various proposals yet  to be fully clarified~\cite{Hui:2021vcv,Hui:2022vbh,Charalambous:2021kcz,Charalambous:2022rre,Parra-Martinez:2025bcu,Lupsasca:2025pnt}.

Fourth, the connection between tidal Love numbers and
other probes of black hole microstructure --- including quasinormal mode
spectra, echoes, and area quantization --- deserves further
exploration \cite{Cardoso:2019rvt,Maselli:2018fay}. 

Finally, the
relationship between Love numbers and holography via the AdS/CFT
correspondence~\cite{Emparan:2017qxd,Franzin:2024cah}, and their
interpretation in terms of the membrane paradigm and near-horizon
physics~\cite{Damour:1982wm,Penna:2018gfx}, offer promising
directions for understanding the deep structure of black hole horizons.

On the observational side, next-generation detectors --- LISA, the
Einstein Telescope, and Cosmic Explorer --- will dramatically improve
sensitivity to tidal effects~\cite{AmaroSeoaneEtAl2017,Burns:2019byj}.
LISA in particular will observe extreme mass-ratio inspirals (EMRIs)
and stellar-mass binary black hole mergers at unprecedented
signal-to-noise ratios, potentially allowing separate constraints on
conservative and dissipative tidal coefficients, and on the dynamical
frequency dependence of Love numbers.
The Einstein Telescope, by improving constraints on tidal deformability and absorption effects in neutron stars by nearly an order of magnitude relative to the LVK at design sensitivity, holds the promise of tightly constraining the nuclear matter equation of state and probing the dynamical and nonlinear tidal r\'egime in high signal-to-noise neutron-star merger events~\cite{ET:2025xjr}.
Multiband observations combining
space- and ground-based detectors will further break degeneracies
between tidal and spin effects in waveform models.

Ultimately, a non-zero measurement of black hole tidal Love numbers
would signal that astrophysical black holes are not exact vacuum Kerr solutions. Whether such a deviation originates from modified gravity, new physics, %
or environmental effects remains an open question, though the EFT framework  provides a systematic language in which such discoveries can be precisely interpreted and
compared across different physical scenarios. Tidal Love numbers have thus evolved from a technical diagnostic of perturbation theory into a central observable in strong-field gravitational physics, and their continued theoretical and observational development promises to deepen our understanding of the most extreme objects in the universe.

\vspace{0.8cm}

\noindent
\textbf{Acknowledgements:} We are grateful to Thomas Apostolidis, Oscar Combaluzier-Szteinsznaider, Valerio De Luca, Rafael Porto, Jan Steinhoff, Kent Yagi, and Nicol\'as Yunes for helpful comments on the draft.
We particularly thank Lam Hui, Austin Joyce, Massimiliano Riva,  Riccardo Penco, Nikola Savi\'c, Filippo Vernizzi, and Sam Wong for years of collaboration and discussions on topics related to Love numbers.

The research of LS has been funded, in part, by the French National Research Agency (ANR) under project ANR-24-CE31-1097-01.  This work has received support under the program ``\textit{Investissement d'Avenir}'' launched by the French Government and implemented by ANR, with the reference ANR-18-IdEx-0001 as part of its program ``\textit{Emergence}.''

MJR thanks the Mitchell Family Foundation for hosting
her during the Cook’s Branch workshop where some of the research was carried out. MJR is
partially supported by the NSF grant PHY-2309270.

\newpage

\appendix

\addtocontents{toc}{\protect\newpage} %

\part*{Appendices}
\addcontentsline{toc}{part}{Appendices}

\section{Hypergeometric and related special functions}
\label{app:hypergeo}

The equations of motion of relevance to reality are typically second-order
differential equations. This is for basic physical reasons: equations of lower order correspond to constraints and are usually insufficient to obtain actual \emph{dynamics}, while the inclusion of third or higher derivatives typically signals an Ostrogradsky or ``ghost'' instability \cite{Woodard:2015zca}.

If the equation of interest has one independent variable $z$ (which may be complex), as happens if we are working in one dimension or are able to separate variables, and one dependent variable $f(z)$, it can be written generically as
\begin{equation}\label{eq:fuchs}
p_2(z)f''(z) + p_1(z)f'(z) + p_0(z)f(z) = 0,
\end{equation}
where primes denote derivatives with respect to $z$. The properties of the solutions to \cref{eq:fuchs} are determined in large part by its singularity structure. A \emph{singular point} $z=z_\star$ of \cref{eq:fuchs} is one at which one or more of the $p_i(z)$ has a pole. A singularity is \emph{regular} if $\lim_{z\to z_\star}(z-z_\star)^ip_i(z)$ exists for all $i$, and \emph{irregular} if not.\footnote{If there is a singularity at $z_\star=\infty$ then we may determine its regularity by changing to the variable $t\equiv z^{-1}$ so that $t_\star=0$. The end result is that a singular point at infinity is regular if the limit $\lim_{z\to\infty}z^ip_i(z)$ exists for all $i$.} In the case when all of the singularities of \cref{eq:fuchs} are regular, it is called a \emph{Fuchsian} differential equation.

\subsection{Hypergeometric functions}

\subsubsection{The hypergeometric equation, series, and function}

Suppose that one is working with an equation of the form \eqref{eq:fuchs} and finds that it has three singular points, each of which is regular. These can be moved to $(0,1,\infty)$ by means of a M\"obius transformation of $z$,\footnote{This is a transformation of the form
\[
z \to \frac{\alpha z + \beta}{\gamma z + \delta},
\]
with $(\alpha,\beta,\gamma,\delta)$ complex constants such that $\alpha\delta-\beta\gamma\neq0$.} after which the differential equation at hand will take the form of the \emph{hypergeometric equation},
\begin{equation}
z(1-z)f''(z) + \left[\mathfrak{c-(a+b}+1)z\right]f'(z) - \mathfrak{ab}f(z) = 0,\label{eq:hypergeo}
\end{equation}
with $\mathfrak{(a,b,c)}$ constant parameters. The singularities are at $z_\star=(0,1,\infty)$.\footnote{We emphasize that this results starting from a Fuchsian equation with three regular singularities located \emph{anywhere}.} Note that this equation is manifestly symmetric under interchange of $\mathfrak a$ and $\mathfrak b$.

The differential equation \eqref{eq:hypergeo} and its solutions are extremely well-studied.\footnote{Standard references include \rcite{DLMF,Bateman:100233,slavjanov2000special,beals_wong_2010}; see also app. B of \rcite{Hui:2020xxx} for an overview with applications to black hole Love numbers.}
The task one faces when solving a second-order differential equation like the hypergeometric equation is generally to find two linearly-independent solutions and fix their coefficients by imposing boundary conditions. When $\mathfrak c$ is not a non-positive integer (i.e., $\mathfrak c\neq0,-1,-2,...$), \cref{eq:hypergeo} is solved by the \emph{hypergeometric function} ${}_2F_1(\mathfrak{a,b;c};z)$. Within $|z|<1$ this is defined by the hypergeometric power series,
\begin{align}
{}_2F_1(\mathfrak{a,b;c};z) &= \displaystyle\sum_{n=0}^\infty \frac{(\mathfrak a)_n(\mathfrak b)_n}{(\mathfrak c)_n}\frac{z^n}{n!}\nonumber\\
&= \frac{\Gamma(\mathfrak c)}{\Gamma(\mathfrak a)\Gamma(\mathfrak b)}\displaystyle\sum_{n=0}^\infty \frac{\Gamma(\mathfrak a+n)\Gamma(\mathfrak b+n)}{\Gamma(\mathfrak c+n)n!}z^n,\label{eq:hypergeo-series}
\end{align}
where $(\mathfrak a)_n = \Gamma(\mathfrak a+n)/\Gamma(\mathfrak a)$ is the Pochhammer symbol, and $\Gamma(z)$ is the standard gamma function which analytically extends the factorial, $\Gamma(n+1)=n!$ for $n\in\mathbb{Z}$. For $|z|\geq1$, ${}_2F_1(a,b;c;z)$ can be obtained by analytic continuation,
typically by choosing the branch cut to lie along the real line $z\geq1$.

\subsubsection{Asymptotics and solution bases}
\label{app:hypergeo-sols}

The solution $f(z) = {}_2F_1(\mathfrak{a,b;c};z)$ is the one which is normalized to unity at the singular point $z_\star=0$,
\begin{equation}
{}_2F_1(\mathfrak{a,b;c};0) = 1.
\end{equation}
In fact, as one approaches any of the three singular points, \cref{eq:hypergeo} develops a scaling symmetry so a linearly-independent solution basis can be found in which both solutions are of the form $(z-z_\star)^\alpha+\mathcal{O}(z-z_\star)^{\alpha-1}$; the hypergeometric differential equation then becomes an algebraic equation for $\alpha$ with two roots. In practice it is often quite convenient to pick one of these three solution bases, particularly if one is setting boundary conditions at one or more of the singular points. The solutions with definite fall-offs around each singularity are
\begin{equation}\label{eq:scalingexp}
f_1(z) \sim
\begin{cases}
\mathrm{const.} & z\to0 \\
\mathrm{const.} & z\to1\\
z^\mathfrak a & z\to\infty \\
\end{cases}
,\qquad
f_2(z) \sim \begin{cases}
z^{1-\mathfrak c} & z\to0 \\
(z-1)^\mathfrak{c-a-b} & z\to1\\
z^\mathfrak b & z\to\infty \\
\end{cases}
.
\end{equation}
Note that the solutions with a definite scaling at one singular point do not, in general, have a definite scaling at the others, so $f_1(z)$ and $f_2(z)$ denote different functions at each $z_\star$. For generic values of the parameters $\mathfrak{(a,b,c)}$, the basis with definite fall-off at $z_\star=0$ corresponds to\footnote{The numbering for the functions $u_i(z)$ follows the standard notation of \rcite{Bateman:100233}.}
\begin{subequations}
\begin{align}
f_1(z) = u_1(z) &\equiv{}_2F_1(\mathfrak{a,b;c};z),\label{eq:u1}\\
f_2(z) = u_5(z) &\equiv z^{1-\mathfrak c}{}_2F_1(\mathfrak a-\mathfrak c+1,\mathfrak b-\mathfrak c+1;2-\mathfrak c;z),\label{eq:u5}
\end{align}
\end{subequations}
Using ${}_2F_1(\mathfrak{a,b;c};0)=1$ it is straightforward to see that these indeed have the expected behavior $f_1\sim z^0$ and $f_2 \sim z^{1-\mathfrak c}$ as $z\to0$.
The solutions with definite fall-off around $z_\star=1$ can be written for generic parameters as
\begin{subequations}
\begin{align}
f_1(z) = u_2(z) &\equiv{}_2F_1(\mathfrak a,\mathfrak b;1+\mathfrak a+\mathfrak b-\mathfrak c;1-z),\label{eq:u2}\\
f_2(z) = u_6(z) &\equiv (1-z)^\mathfrak{c-a-b}{}_2F_1(\mathfrak c-\mathfrak b,\mathfrak c-\mathfrak a;1+\mathfrak c-\mathfrak a-\mathfrak b;1-z).\label{eq:u6}
\end{align}
\end{subequations}
Again we see that these have the expected falloffs of $z^0$ and $z^\mathfrak{c-a-b}$, respectively, as $z\to1$. Finally around $z_\star=\infty$ the linearly independent solutions for generic parameters are
\begin{subequations}
\begin{align}
f_1(z) = u_3(z) &\equiv(-z)^{-a}{}_2F_1\left(a,1+a-c;1+a-b;\frac1z\right),\label{eq:u3}\\
f_2(z) = u_4(z) &\equiv (-z)^{-b}{}_2F_1\left(1+b-c,b;1+b-a;\frac1z\right),\label{eq:u4}
\end{align}
\end{subequations}
which fall off as $z^a$ and $z^b$, respectively, as $z\to\infty$.

\subsubsection{Degenerate parameters}

We are doing a lot of work with the phrase ``generic parameters.'' We have already seen that for $\mathfrak c=0,-1,-2,...$ the function $u_1(z)={}_2F_1(\mathfrak{a,b;c};z)$ is not well-defined. Two linearly-independent solutions may also coincide for certain \emph{degenerate} parameter choices, namely when at least one of $\mathfrak a$, $\mathfrak b$, $\mathfrak c-\mathfrak a$, or $\mathfrak c-\mathfrak b$ is an integer. The degenerate case is in fact typical for the hypergeometric equations which arise in computing static responses of black holes, cf. \cref{sec:love-compute}.
In these special cases one can find another of the above solutions which is linearly independent of the first solution and has the right falloff conditions. See, e.g., \rcite{DLMF,Bateman:100233,slavjanov2000special,beals_wong_2010} for a characterization of the degenerate parameter cases and the linearly independent solutions for each, and App. B.3 of \rcite{Hui:2020xxx} for a discussion specialized to black hole static responses.

In the case when $\mathfrak a$ or $\mathfrak b$ is a non-positive integer, the series \eqref{eq:hypergeo-series} truncates and the hypergeometric function becomes a finite polynomial,
\begin{equation}
{}_2F_1(-\mathfrak{a,b;c};z) = \sum_{n=0}^\mathfrak a(-1)^n\binom{\mathfrak a}{\mathfrak b}\frac{(\mathfrak b)_n}{(\mathfrak c)_n}z^n.\label{eq:hypergeo-series-poly}
\end{equation}

\subsubsection{Connection formulae}

There are many formulae connecting hypergeometric functions with one or more arguments separated by unity. Here we list some of the most useful relations for this review; for a more complete list, see, e.g., \rcite{DLMF,Bateman:100233,slavjanov2000special,beals_wong_2010}.
\begin{subequations}
    \begin{align}
        \frac{\dd}{\dd z}{}_2F_1(\mathfrak{a,b;c};z) &=\frac{\mathfrak{ab}}{\mathfrak{c}}{}_2F_1(\mathfrak{a}+1,\mathfrak b+1;\mathfrak{c}+1;z) \\
        \frac{\dd}{\dd z}{}_2F_1(\mathfrak{a,b;c};z) &= \frac{\mathfrak{a}[(\mathfrak{c-a}-1)-(\mathfrak{b-a}-1)(1-z)]}{z(\mathfrak{b-a}-1)(1-z)}{}_2F_1(\mathfrak{a,b;c};z) \nonumber\\
        &\hphantom{{}=}- \frac{\mathfrak{a(c-b)}}{z(\mathfrak{b-a}-1)(1-z)}{}_2F_1(\mathfrak{a}+1,\mathfrak{b}-1;\mathfrak{c};z) \\
        \frac{\dd}{\dd z}{}_2F_1(\mathfrak{a,b;c};z) &= \frac{\mathfrak{b}[(\mathfrak{c-b}-1)-(\mathfrak{a-b}-1)(1-z)]}{z(\mathfrak{a-b}-1)(1-z)}{}_2F_1(\mathfrak{a,b;c};z) \nonumber\\
        &\hphantom{{}=}- \frac{\mathfrak{b(c-a)}}{z(\mathfrak{a-b}-1)(1-z)}{}_2F_1(\mathfrak{a}-1,\mathfrak{b}+1;\mathfrak{c};z) \\
        \frac{\dd}{\dd z}\left[z^{\mathfrak c-1}(1+z)^\mathfrak{a+b-c}{}_2F_1(\mathfrak{a,b;c};z)\right] &= (\mathfrak c-1)z^{\mathfrak c-2}(1-z)^{\mathfrak{a+b-c}-1}{}_2F_1(\mathfrak{a}-1,\mathfrak b-1;\mathfrak{c}-1;z)
    \end{align}
\end{subequations}

\subsection{Legendre functions}
\label{app:legendre}

\subsubsection{Legendre's equation}

The hypergeometric equation \eqref{eq:hypergeo} contains as subcases many well-known differential equations. A particularly famous example, which appears in many guises in this review, is \emph{Legendre's equation},
\begin{equation}\label{eq:legendre}
(1-x^2)y''(x) -2xy'(x) +\ell(\ell+1)y(x)=0.
\end{equation}
This has regular singular points at $x=(\pm1,\infty)$, which we can map to the singular points of \cref{eq:hypergeo} by the change of variable $z=(1-x)/2$, in terms of which \cref{eq:legendre} becomes the hypergeometric equation \eqref{eq:hypergeo} with parameters
\begin{equation}
\mathfrak a = -\ell,\quad\mathfrak  b = \ell+1,\quad \mathfrak c = 1.
\end{equation}
The solutions which are regular and normalized to unity at $x=1$,
\begin{equation}
P_\ell(x) = {}_2F_1\left(-\ell,\ell+1;1;\frac{1-x}2\right),
\end{equation}
are known as \emph{Legendre functions of the first kind},
and when $\ell$ is a non-negative integer then they are called \emph{Legendre polynomials}. The first few are
\begin{equation}
P_0(x) = 1,\quad P_1(x) = x,\quad
P_2(x) = \frac12\left(3x^2-1\right),\quad P_3(x) = \frac12\left(5x^3-3x\right),\quad P_4(x) = \frac18\left(35x^4-30x^2+3\right).
\end{equation}
In this case $P_\ell(x)$ is also regular at the singular point $x=-1$.
The solutions with the opposite asymptotics at $x=1$ are Legendre functions of the second kind, labeled $Q_\ell(x)$.

\subsubsection{Orthonormality and completeness}

A particularly useful property of the Legendre polynomials is their \emph{orthonormality} over the interval $[-1,1]$,\footnote{In fact, this property can be used to define the Legendre polynomials for integer $\ell$ without recourse to a differential equation.}
that is,
\begin{equation}
\int_{-1}^1 \dd x \,P_\ell(x)P_m(x) = \frac2{2\ell+1}\delta_{\ell m},\quad \delta_{\ell m} \equiv \begin{cases}
0,&\ell\neq m\\
1,&\ell=m
\end{cases},
\end{equation}
with $\delta_{\ell m}$ the Kronecker delta. We can expand a generic function in Legendre polynomials,
\begin{equation}
f(x) = \displaystyle\sum_\ell a_\ell P_\ell(x),
\end{equation}
and then find any desired coefficient $a_\ell$ by multiplying by the corresponding $P_\ell(x)$ and integrating over the range $[-1,1]$,
\begin{align}
\int_{-1}^1\dd x\, P_{\ell} f(x)  &= \sum_{\ell'}a_{\ell'}\int_{-1}^1\dd x\, P_{\ell}P_{\ell'}(x) \nonumber \\
&= \frac2{2\ell+1}a_{\ell}.
\end{align}
In passing we note a nice compact expression for the Legendre polynomials, called \emph{Rodrigues' formula},
\begin{equation}
P_\ell(x) = \frac{1}{2^\ell \ell!}\left(\frac\dd{\dd x}\right)^\ell (x^2-1)^\ell.
\end{equation}

\subsubsection{Associated Legendre functions}

The \emph{associated Legendre equation} is
\begin{equation}\label{eq:legendre-assoc}
(1-x^2)y''(x) -2xy'(x) +\left(\ell(\ell+1)-\frac{m^2}{1-x^2}\right)y(x)=0.
\end{equation}
This reduces to \cref{eq:legendre} for $m=0$.
It is solved by the \emph{associated Legendre functions} $P_\ell^m(x)$ and $Q_\ell^m(x)$, which can be defined by the requirement that they reduce to the standard Legendre functions at $m=0$, $(P_\ell^0(x),Q_\ell^0(x))=(P_\ell(x),Q_\ell(x))$.
In physical applications $m$ is generally in the range $-\ell\leq m\leq \ell$.
It takes slightly more effort to reduce \cref{eq:legendre-assoc} to hypergeometric form. In addition to the coordinate change $z=(1-x)/2$, we also have to change variable to
\begin{equation}
f(z) \equiv \left(\frac z{1-z}\right)^{m /2} y(z),
\end{equation}
so that $f(z)$ satisfies the equation
\begin{equation}
z(1-z)f''(z) +(1-m-2z)f'(z) +\ell(\ell+1)f(z)=0.
\end{equation}
Notice that the $m$ dependence has been pushed entirely into the term $p_1(z)$ by this change of variable.
This is now of hypergeometric form \eqref{eq:hypergeo} with
\begin{equation}
\mathfrak a = -\ell,\quad\mathfrak  b = \ell+1,\quad\mathfrak  c = 1 - m.
\end{equation}
The associated Legendre function of the first kind is related to the hypergeometric function by
\begin{equation}
    P_\ell^m(x) = \left(\frac{1+x}{1-x}\right)^{m/2}{}_2F_1\left(-\ell,\ell+1;1-m;\frac{1-x}2\right).
\end{equation}

\subsection{Spherical harmonics}
\label{app:sph-harm}

The spherical harmonics are eigenfunctions of the Laplacian on the 2-sphere $S^2$,
\begin{equation}
    \nabla^2_{S^2}Y_{\ell m}(\theta,\varphi) = \frac{1}{\sin\theta}\partial_\theta(\sin\theta\, \partial_\theta Y_{\ell m}(\theta,\varphi)) + \frac1{\sin^2\theta}\partial_\varphi^2Y_{\ell m}(\theta,\varphi) = -\ell(\ell+1)Y_{\ell m}(\theta,\varphi).
\end{equation}
The solution which is regular at the poles is expressed using Legendre polynomials,
\begin{equation}
Y_\ell^m(\theta,\phi) = Ne^{im\varphi}P_\ell^m(\cos\theta)
\end{equation}
Here the normalization constant $N$ is
\begin{equation}\label{eq:N-sph-harm}
    N = \sqrt{\frac{2\ell+1}{4\pi}\frac{(\ell-m)!}{(\ell+m)!}}.
\end{equation}
This choice makes the spherical harmonics orthonormal,
\begin{equation}
    \int \dd\Omega Y_{\ell m}\bar{Y}_{\ell' m'} = \delta_{\ell\ell'}\delta_{mm'},
\end{equation}
where $\dd\Omega=\sin\theta\dd\theta\dd\varphi$ is the integration measure on the 2-sphere and the integration range is $0\leq\theta\leq\pi$ and $0\leq\varphi\leq2\pi$.

\section{Differential forms}
\label{app:diff-forms}

A differential form is an object that can be integrated on a manifold. In particular, a form of degree $p$, called a $p$-form, can be integrated over a (sub-)manifold of dimension $p$. Forms are of tremendous importance in physics, appearing naturally in the formulation of most fundamental theories. They also have a utilitarian value, as restricting the usual tensor calculus to operations that act on and give back forms can greatly simplify many computations.

In this Appendix we give a brief overview of the theory and language of forms to the extent necessary to understand its appearances in this review.

\subsection{0-forms and 1-forms}

A 0-form is just a function. Given a coordinate system $x^\mu$, the natural basis for 1-forms is the set of coordinate differentials $\{\dd x^\mu\}$. Expanding in this basis we see that the components of a 1-form are just a (dual) vector field,
\begin{equation}
    v = v_\mu \dd x^\mu.
\end{equation}
Since $v$ is a 1-form it can be integrated over a curve. In one dimension a generic 1-form can be written as $f(x)\dd x$, which is easily recognizable as the object one integrates in elementary single-variable calculus.

\subsection{Higher-degree forms and the exterior product}

To construct a basis for higher-degree forms, we need a way of multiplying the coordinate differentials $\dd x^\mu$. The \emph{exterior product}, also called \emph{wedge product} as it is denoted with the wedge symbol $\wedge$, takes a $p$-form and a $q$-form to a $(p+q)$-form. It acts on the coordinate 1-forms as an alternating product,
\begin{equation}
    \dd x^\mu\wedge\dd x^\nu = -\dd x^\nu\wedge\dd x^\mu.
\end{equation}
The coordinate differential basis for a $p$-form is then
\begin{equation}
    \{\dd x^{\mu_1}\wedge\cdots\wedge \dd x^{\mu_p}\},\qquad \mu_1<\mu_2<\cdots<\mu_p.
\end{equation}
We may expand a $p$-form $X$ in this basis as
\begin{equation}\label{eq:form-basis}
\boxed{
    X = \frac1{p!}X_{\mu_1\cdots\mu_p}\dd x^{\mu_1}\wedge\cdots\wedge\dd x^{\mu_p},}
\end{equation}
where we have included a factor of $1/p!$ to compensate for overcounting basis elements. Because $\dd x^{\mu_1}\wedge\cdots\wedge \dd x^{\mu_p}$ is totally antisymmetric, so too is the tensor $X_{\mu_1\cdots\mu_p}$,
\begin{equation}
    X_{\mu_1\cdots\mu_p} = X_{[\mu_1\cdots\mu_p]}.
\end{equation}
We see that \emph{there is a direct correspondence between $p$-forms and totally antisymmetric, rank-$(0,p)$ tensors}, given by \cref{eq:form-basis}

Let us note the consequences of this result for $p$-forms with $p\geq D$, with $D$ the dimension of the manifold. For $p=D$ there are as many index slots as there are coordinates, and since total antisymmetry implies no index can be repeated, each coordinate can only appear once. There is therefore only one independent component of a $D$-form, with the rest either vanishing or obtainable by swapping indices, picking up a minus sign each time. This is the same symmetry as that of the Levi-Civita tensor $\epsilon_{\mu_1\cdots\mu_D}$, and since each tensor has only one independent component, they must be proportional,
\begin{equation}
X_{\mu_1\cdots\mu_D} = X \epsilon_{\mu_1\cdots\mu_D}.
\end{equation}
A similar argument shows there are no non-trivial forms with $p>D$, since there will of necessity be repeated indices.

A $D$-form is of particular importance because it can be integrated over the entire manifold. Let us expand a $D$-form $f$ as
\begin{align}
    f &= \frac1{D!}f_{\mu_1\cdots\mu_D}\dd x^{\mu_1}\wedge\cdots\wedge\dd x^{\mu_D} \nonumber\\
    &= f(x^\mu) \underbrace{\frac1{D!}\epsilon_{\mu_1\cdots\mu_D}\dd x^{\mu_1}\wedge\cdots\wedge\dd x^{\mu_D}}_{\equiv \sqrt{|g|} \dd^Dx} \nonumber\\
    &= f(x^\mu)\sqrt{|g|} \dd^Dx,
\end{align}
where we have used the definition of $\dd^Dx$,
\begin{equation}
    \dd^Dx \equiv \dd x^1\wedge\dd x^2\wedge\cdots\wedge\dd x^D,
\end{equation}
and recalled that the Levi-Civita \emph{tensor} $\epsilon_{\mu_1\cdots\mu_D}$ is related to the Levi-Civita \emph{symbol} $\tilde\epsilon_{\mu_1\cdots\mu_D}$, whose values are $\{0,\pm1\}$, by $\epsilon_{\mu_1\cdots\mu_D} = \sqrt{|g|} \tilde\epsilon_{\mu_1\cdots\mu_D}$.
We see that the usual covariant integration measure $\sqrt{|g|}\dd^Dx$ is the basis element for $D$-forms. It is typical to write this as the \emph{volume form} $\epsilon$:
\begin{equation}
    \epsilon \equiv \sqrt{|g|} \dd^Dx.
\end{equation}
A familiar example of a $D$-form is the Lagrangian $L$,
\begin{equation}
    L=\mathcal{L}\epsilon,
\end{equation}
with $\mathcal L$ the scalar Lagrangian density.

It is interesting that a 0-form and a $D$-form have the same number (one) of independent components, and are therefore (in a sense we make precise below) dual to each other. By counting the number of independent components of a $p$-form,
\begin{equation}
    \frac{D!}{p!(D-p)!},
\end{equation}
we see that $p$-forms are dual to $(D-p)$-forms. We return to this in \ref{app:hodge}.

Finally let us work out the coordinate expression for the wedge product of a $p$-form $X$ and $q$-form $Y$. We expand it in two different ways and compare:
\begin{align}
    X\wedge Y &= \frac1{(p+q)!}(X \wedge Y)_{\mu_1\cdots\mu_p\nu_1\cdots\nu_q}\dd x^{\mu_1}\wedge\cdots\wedge \dd x^{\mu_p}\wedge\dd x^{\nu_1}\wedge\cdots\wedge \dd x^{\nu_q} \nonumber\\
    &= \left(\frac1{p!}X_{\mu_1\cdots\mu_p}\dd x^{\mu_1}\wedge\cdots\wedge \dd x^{\mu_p}\right)\wedge\left(\frac1{q!}Y_{\nu_1\cdots\nu_q}\dd x^{\nu_1}\wedge\cdots\wedge \dd x^{\nu_q}\right)\nonumber\\
    &= \frac1{p!q!}X_{\mu_1\cdots\mu_p}Y_{\nu_1\cdots\nu_q}\dd x^{\mu_1}\wedge\cdots\wedge \dd x^{\mu_p}\wedge\dd x^{\nu_1}\wedge\cdots\wedge \dd x^{\nu_q},
\end{align}
from which we find the coordinate form of the exterior product,
\begin{equation}
    \boxed{
(X\wedge Y)_{\mu_1\cdots\mu_p\nu_1\cdots\nu_q} = \frac{(p+q)!}{p!q!}X_{[\mu_1\cdots\mu_p}Y_{\nu_1\cdots\nu_q]}.}
\end{equation}
Notice that it is graded anticommutative,
\begin{equation}
    X\wedge Y = (-1)^{pq} Y\wedge X.
\end{equation}

\subsection{Exterior derivative}

Next we need a derivative operator acting on forms. The \emph{exterior derivative} takes $p$-forms to $(p+1)$-forms. Denoted $\dd$, the exterior derivative is a generalization of the $``\dd"$ which appears in the coordinate differentials $\dd x^\mu$ and is familiar from elementary calculus. When it acts on a 0-form $\phi$ to produce the 1-form $\dd\phi$, the components $(\dd\phi)_\mu$ can only be one thing (up to a constant factor) in order to satisfy linearity and the Leibniz rule, namely
\begin{equation}
    \dd \phi = (\dd\phi)_\mu\,\dd x^\mu = \partial_\mu\phi\,\dd x^\mu.
\end{equation}
We see that the exterior derivative of a 0-form is the usual gradient operator.

The extension to forms of arbitrary degree is
\begin{equation}
(\dd X)_{\mu_1\cdots\mu_{p+1}} = (p+1)\partial_{[\mu_1}X_{\mu_2\cdots\mu_{p+1}]}.
\end{equation}
Note in particular that acting on 1- and 2-forms in $D=3$, $\dd$ is equivalent (up to the duality between $p$- and $(D-p)$-forms) to the curl and divergence operators, respectively. We see that the exterior derivative generalizes the standard operators of vector calculus to forms of arbitrary $p$ and $D$. It satisfies the graded Leibniz rule,
\begin{equation}
    \dd(X\wedge Y) = \dd X \wedge Y + (-1)^pX\wedge\dd Y,
\end{equation}
with $X$ and $Y$ $p$- and $q$-forms, respectively.

The exterior derivative has the remarkable property that acting it twice on any objects produces a vanishing result, since this operation requires antisymmetrizing over two partial derivatives. This is usually written as the famous operator identity
\begin{equation}
    \boxed{\dd^2=0.}
\end{equation}
We say that a form $X$ is \emph{exact} if it is the exterior derivative of some potential $Y$,
\begin{equation}
    X = \dd Y,
\end{equation}
and that it is \emph{closed} if its exterior derivative vanishes,
\begin{equation}
    \dd X = 0.
\end{equation}
Clearly all exact forms are closed, $\dd X = \dd^2Y=0$. What about the converse? If a given form is closed, can it necessarily be written as the exterior derivative of a potential? The answer is provided by the \emph{Poincar\'e lemma}, which states that every closed form on an open ball of $\mathbb{R}^n$ is exact. For physics purposes this means essentially that there are closed, non-exact forms only when there is some non-trivial topology. For the spacetimes of interest in this review, every closed form is therefore exact.

An illustrative example is given by classical electromagnetism. The vector potential $A_\mu$ corresponds to a 1-form $A=A_\mu\dd x^\mu$, and the field strength $F_\mn = \partial_\mu A_\nu-\partial_\nu A_\mu$ corresponds to the form $F=\dd A$. The Bianchi identity $\partial_{[\mu}F_{\nu\alpha]}=0$ is written as $\dd F=0$, and is seen to be a trivial consequence of the identity $\dd^2=0$. This is a rather simpler calculation than its tensorial counterpart, where we must carefully keep track of the antisymmetrized indices. Poincar\'e's lemma tells us that on topologically-Minkowski spacetimes, the Bianchi identity \emph{implies} the existence of the potential $A$.

This example also illustrates why we write the components of $\dd X$ with partial rather than covariant derivatives. Consider the 2-form field strength $F$. We \emph{could} define it with covariant derivatives, but the terms involving the Christoffel symbols vanish due to antisymmetrization,
\begin{align}
    \nabla_{[\mu}A_{\nu]} &= \partial_{[\mu}A_{\nu]} - \Gamma^\rho_{[\mn]}A_\rho\nonumber\\
    &=\partial_{[\mu}A_{\nu]}.
\end{align}
This useful property generalizes straightforwardly to general $p$-forms.

Note that a form of maximal degree $p=D$ has vanishing exterior derivative, since there are no $(D+1)$-forms in $D$ dimensions.

When an exact form $\dd \alpha$ is integrated over a region $\mathcal C$ with boundary $\partial\mathcal C$, the \emph{(generalized) Stokes' theorem} tells us that we need only integrate $\alpha$ over the boundary,
\begin{equation}
    \int_{\mathcal C}\dd \alpha = \int_{\partial\mathcal C}\alpha.
\end{equation}
Special cases of this remarkable relation include the fundamental theorem of calculus in $D=1$, Green's theorem in $D=2$, and the Stokes' and divergence theorems in $D=3$.

\subsection{Hodge duality}

\label{app:hodge}

Differential forms, the exterior product, and the exterior derivative are all defined without reference to a spacetime metric. They are in this sense ``pre-geometric.'' Let us now introduce a very useful operator on forms which does see the geometry of the manifold, the \emph{Hodge star} ($\star$). This makes precise the \emph{Hodge duality} between $p$-forms and $(D-p)$-forms which we anticipated earlier based on counting degrees of freedom. Acting on a $p$-form $X$, the Hodge star produces a $(D-p)$-form with coordinate expression
\begin{equation}
(\star X)_{\mu_1\cdots\mu_{D-p}} = \frac{1}{p!}\epsilon_{\mu_1\cdots\mu_{D-p}\nu_1\cdots\nu_p}X^{\nu_1\cdots\nu_p}.
\end{equation}
As promised, the Hodge star operation depends on the curvature of the manifold, since we need a metric in order to define the Levi-Civita tensor and to perform index contractions.

Acting the Hodge star twice returns the original form, up to a possible sign:
\begin{equation}
    \star^2 = (-1)^{p(D-p)}s,
\end{equation}
where $s=\operatorname{sgn}(\det g)$ is the parity of the metric signature; in our mostly-plus convention, $s=-1$. In $D=4$ (where $s=-1$ with either sign convention), $\star^2=1$ for 1- and 3-forms and $\star^2=-1$ for 0-, 2-, and 4-forms.

The $D$-form volume element $\epsilon = \sqrt{|g|} \dd^Dx$ is dual to unity,
\begin{equation}
    \epsilon = \star 1.
\end{equation}

The Hodge star allows us to bring in the idea of contracting indices with a metric. Given a $p$-form $A$ and $q$-form $B$ ($q\geq p$), we can make a form whose components are an index contraction of $A$ and $B$ using the exterior product and Hodge star,
\begin{equation}
    \star(A\wedge\star B) = \frac{(-1)^{q(D-q)}s}{p!(q-p)!}A^{\nu_1\cdots\nu_p}B_{\mu_1\cdots\mu_{q-p}\nu_1\cdots\nu_p} \dd x^{\mu_1}\wedge\cdots\wedge x^{\mu_{q-p}}.\label{eq:form-contract}
\end{equation}

To illustrate some classic uses of Hodge duality, we return to the example of Maxwell electromagnetism in $D=4$. We have already seen that the potential $A=A_\mu\dd x^\mu$ and field strength $F=\dd A$ have natural interpretations as differential forms, and that the Bianchi identity $\dd F=0$ follows from $\dd^2=0$. What about the Maxwell equation, $\nabla^\nu F_\mn=J_\mu$? This depends on the spacetime geometry, so to write it in the language of forms we need the Hodge star,
\begin{equation}
    \dd\star F=\star J.
\end{equation}
Note that $\dd\star F$ is a 3-form; the covector $\nabla^\nu F_\mn$ appears as the components of its dual 1-form, $\star\;\dd\star F$.

We now have the ingredients to derive the famous \emph{electric--magnetic duality}, first discovered as the duality symmetry $(E,B)\to(B,-E)$\footnote{Or more generally, rotations of the vector $(E,B)$.} of the source-free ($J=0$) Maxwell equations in four dimensions. On shell (i.e., when the Maxwell equation is satisfied), the Poincar\'e lemma tells us we can write $\star F$ as the exterior derivative of a \emph{dual potential} $\tilde A$,
\begin{equation}
    \star F = \dd\tilde A.
\end{equation}
In $D=4$ both $F$ and $\star F$ are 2-forms (since $D-p=4-2=2=p$), so $A$ and its dual $\tilde A$\footnote{Not to be confused with the 3-form $\star A$.} are both 1-forms. We are free to take $\tilde A$ as the fundamental variable instead of $A$ to obtain a dual description of electromagnetism in which the roles of the Maxwell equation and Bianchi identity are interchanged: the Maxwell equation for $A$ is the Bianchi identity for $\tilde A$, and vice versa. In this picture it is $A$ whose existence follows via the Poincar\'e lemma when the equation of motion $\star\dd\tilde A=0$ is satisfied. The relation between the dual potentials $A$ and $\tilde A$ is non-local, as it requires integration to solve,
\begin{equation}
    \dd\tilde A = \star\dd A.
\end{equation}
The Maxwell equations are invariant under rotations of $(A,\tilde A)$, which are nothing other than the classical electric--magnetic duality rotations on $(E,B)$. It is also easy to see why this duality only exists in $D=4$: $\tilde A$ is a $(D-3)$-form, so it is only in four dimensions that both $A$ and $\tilde A$ are vector fields.

\subsection{General relativity in forms language}

The metric tensor $g_\mn$ does not correspond to a differential form because it is symmetric.
To describe gravity in the language of forms, in place of the metric we take as our dynamical variables a set of $D$ linearly-independent basis vector fields, collectively called a \emph{vielbein}.\footnote{In $D=4$ the terms \emph{vierbein} and \emph{tetrad} are also common, for reasons which should be apparent to readers with passing familiarity with German and/or Greek.} We denote this object by $e_a^\mu$, where the internal \emph{Lorentz indices} $(a,b,c,...)\in(0,1,2,...,D-1)$ label the individual vectors. The vielbein generalizes the notion of a coordinate basis, $e_a=\partial_{x^a}$. Consider the matrix $\eta_{ab}$ of all possible inner products of two basis vectors,
\begin{equation}\label{eq:eta-ab}
    \eta_{ab} \equiv g_\mn e_a^\mu e_b^\nu.
\end{equation}
Frequently one is interested in an orthonormal basis, in which case $\eta_{ab}=\diag(-1,1,1,\cdots)$ is indeed the Minkowski metric. Even in a non-orthonormal basis, $\eta_{ab}$ remains the metric of an internal Minkowski space, expressed in non-standard coordinates. A prominent example is the null tetrad used in Kerr perturbation theory, cf. \cref{sec:null-tetrad}; the components of $\eta_{ab}$ are given by \cref{eq:int-mink-null}.

Let us denote the matrix inverse of $e^\mu_a$ as $e^a_\mu$ and the inverse of $\eta_{ab}$ as $\eta^{ab}$, i.e.,
\begin{equation}
    e^\mu_a e^a_\nu = \delta^\mu_\nu,\qquad e^\mu_a e^b_\mu = \eta_{ac}\eta^{bc} = \delta^b_a.
\end{equation}
We can rewrite \cref{eq:eta-ab} to ``solve for'' the metric,
\begin{equation}
    \boxed{g_\mn = \eta_{ab}e^a_\mu e^b_\nu.}
\end{equation}
It is common to interpret the vielbein as the ``square root'' of the metric tensor.

The existence of a non-degenerate metric on the internal space allows us to define an inverse metric $\eta^{ab}$ with which to raise Lorentz indices. Both Lorentz and spacetime indices can be freely raised and lowered with the appropriate (inverse) metric. In particular note the object
\begin{equation}
    e^a_\mu \equiv \eta^{ab}g_\mn e^\nu_b,
\end{equation}
in terms of which the inverse Lorentz metric is
\begin{equation}
    \eta^{ab} = g^\mn e^a_\mu e^b_\nu.
\end{equation}
Even though we constructed $e^a_\mu$ by raising and lowering indices, it turns out to also be the inverse of the vielbein $e^\mu_a$,
\begin{equation}
    e^a_\mu e^\mu_b = \delta^a_b,\quad e^a_\mu e^\nu_a = \delta_\mu^\nu.
\end{equation}

\bibliographystyle{elsarticle-num-names}
\bibliography{biblio}

\end{document}